\documentclass{ws-ijmpa}

\usepackage[tableposition=top]{caption}

\usepackage{cite}
\usepackage{hyperref}

\usepackage{afterpage}
\usepackage{epsfig}
\usepackage{subfigure}
\usepackage{amssymb,amsmath}
\newcommand{\ttbar}     {$t\overline{t}$~}

\newcommand{\MET}       {\mbox{$\not\!p_T$}~}
\newcommand{\METns}     {\mbox{$\not\!p_T$}}
\newcommand{\pt}        {$p_T$~}

\newcommand{\lsim}{\mathrel{\hbox{\rlap{\lower.55ex\hbox{$\sim$}} \kern-.3em \raise.4ex \hbox{$<$}}}}
\newcommand{\gsim}{\mathrel{\hbox{\rlap{\lower.55ex\hbox{$\sim$}} \kern-.3em \raise.4ex \hbox{$>$}}}}

\newcommand{\stopone}{\tilde{t}_{1}}
\newcommand{\chargino}{\tilde{\chi}^{\pm}_{1}}
\newcommand{\charginoplus}{\tilde{\chi}^{+}_{1}}
\newcommand{\neutralino}{\tilde{\chi}^{0}_{1}}

\newcommand{\comphep}   {{\sc CompHEP}}
\newcommand{\singletop} {{\sc SingleTop}}
\newcommand{\pythia}    {{\sc pythia}}
\newcommand{\herwig}    {{\sc herwig}}
\newcommand{\alpgen}    {{\sc alpgen}}

\newcommand{\madevent}  {{\sc MadEvent}}
\newcommand{\madgraph}  {{\sc MadGraph}}
\newcommand{\mcatnlo}   {{\sc MC@NLO}}
\newcommand{\tauola}    {{\sc TAUOLA}}
\newcommand{\evtgen}    {{\sc EvtGen}}
\newcommand{\qqgen}     {{\sc QQ}}
\newcommand{\geant}     {{\sc geant}}
\newcommand{\toprex}    {{\sc TopReX}}
\newcommand{\cpsuperh}  {{\sc CPsuperH}}
\newcommand{\vecbos}    {{\sc VECBOS}}

\newcommand{\mtopwa}    {172.4  $\pm$ 1.2 GeV/c$^2$~\cite{TEWWG:2008}}

\usepackage{multirow}

\def\today{\number\day\space\ifcase\month\or January\or February\or March\or April\or May\or June\or July\or August\or September\or October\or November\or December\fi \space\number\year}

\makeatletter
\def\leaderfill{\leaders\hbox to 1em{\hss.\hss}\hfill}
\newcommand\@pnumwidth{2.55em}
\newcommand\@tocrmarg{2.55em} 
\newcommand\@dotsep{4.5}
\newcommand\tableofcontents{%
{\global
\@topnum\z@ 
\@afterindentfalse 
\if@twocolumn
\@restonecoltrue
\onecolumn 
\else 
\@restonecolfalse 
\fi 
\vspace*{10pt}
\noindent 
{\bf Contents}\par 
\vskip1em 
\nobreak} 
{\small
\@starttoc{toc}%
}\if@restonecol
\twocolumn
\fi}
\newcommand*\l@section[2]{%
\ifnum \c@tocdepth >\z@
\addpenalty\@secpenalty 
\setlength\@tempdima{1.4em}%
\begingroup 
\parindent \z@ 
\rightskip
\@pnumwidth \parfillskip -\@pnumwidth 
\leavevmode 
\advance\leftskip\@tempdima 
\hskip -\leftskip 
#1\nobreak\leaderfill\nobreak
\hb@xt@\@pnumwidth{\hss #2}\par 
\endgroup 
\fi}
\newcommand*\l@subsection{\@dottedtocline{2}{1.4em}{2.2em}}%
\newcommand*\l@subsubsection{\@dottedtocline{2}{3.6em}{3em}}%
\newcommand*\l@appendix{\@dottedtocline{2}{0em}{6.2em}}
\def\numberline#1{\hb@xt@\@tempdima{#1.\hfil}}
\makeatother

\begin{document}
\markboth{Marc-Andr\'e Pleier}
{Top Quark Measurements at the Tevatron}

\catchline{}{}{}{}{}

\title{REVIEW OF PROPERTIES OF THE TOP QUARK FROM MEASUREMENTS AT THE TEVATRON}

\author{MARC-ANDR\'E PLEIER}
\address{Physikalisches Institut, Universit{\"a}t Bonn\\
    Nussallee 12, 53115 Bonn, Germany\\
  pleier@fnal.gov}

\maketitle

\begin{history}
\end{history}

\begin{abstract}
  This review summarizes the program in the physics of the top quark
  being pursued at Fermilab's Tevatron proton-antiproton collider at a
  center of mass energy of 1.96 TeV. More than a decade after the
  discovery of the top quark at the two collider detectors CDF and D0,
  the Tevatron has been the only accelerator to produce top quarks and
  to study them directly. 

  The Tevatron's increased luminosity and center of mass energy offer
  the possibility to scrutinize the properties of this heaviest
  fundamental particle through new measurements that were not feasible
  before, such as the first evidence for electroweak production of top
  quarks and the resulting direct constraints on the involved
  couplings. Better measurements of top quark properties provide more
  stringent tests of predictions from the
  standard model of elementary particle physics. In particular, the
  improvement in measurements of
  the mass of the top quark, with the latest uncertainty of 0.7\% marking the
  most precisely measured quark mass to date, further constrains the
  prediction of the mass of the still to be discovered Higgs boson.

 \end{abstract}

\ccode{PACS numbers: 14.65.Ha, 12.38.Qk, 13.85.-t, 13.85.Rm, 13.38.Be, 12.60.-i}

\newpage

\setcounter{tocdepth}{1}
\tableofcontents

\section{Introduction}
The existence of a third and most massive generation of fundamental
fermions was unveiled in 1975 with the discovery of the $\tau$ 
lepton at SLAC-LBL~\cite{Perl:1975bf}. In 1977, the discovery of the 
bottom quark \cite{Herb:1977ek} at Fermilab extended the knowledge
of a third generation into the quark sector and immediately raised
the question of the existence of the top quark as the weak
isospin partner of the bottom quark. 

To remain self consistent, the standard model (SM) of elementary
particle physics required the existence of the top quark, and
electroweak precision measurements offered increasingly precise
predictions of properties such as its mass. The top quark's large mass
prevented its discovery for almost two decades, but by 1994 it was
indirectly constrained to be \mbox{178 $\pm$ 11 $^{+18}_{-19}$
GeV/c$^2$}~\cite{EWWG:1994}. After mounting experimental evidence
\cite{Abachi:1994je,Abe:1994st,Abe:1994xt,Abachi:1994td,Abe:1994eh,
Abachi:1995ms}, the top quark ($t$) was finally discovered in 1995 at
Fermilab by the CDF and D0 collaborations \cite{Abe:1995hr,
Abachi:1995iq} in the mass range predicted by the standard model.
The completion of the quark sector once again demonstrated the
enormous predictive power of the SM.

By now, the mass of the top quark is measured to be \mtopwa, marking
the most precisely measured quark mass and the most massive
fundamental particle known to date. The consequent lifetime of the top
quark in the SM of $\approx 5\cdot 10^{-25}$~s is
extremely short, suggesting that it decays before hadronizing. This
makes it the only quark that does not form bound states, allowing the
study of an essentially bare quark with properties such as spin
undisturbed by hadronization~\cite{Bigi:1986jk}.

The measurement of top quark pair (\ttbar) production probes our
understanding of the strong interaction and predictions from
perturbative QCD, while the decay of top quarks and the production of
single top quarks reflect the electroweak interaction. Measuring other
properties of the top quark, such as its electric charge, the helicity
of the $W$ boson in $t\to Wb$ decay, the branching fraction ${\cal
B}(t \rightarrow Wb)$, {\it etc.}, and comparing these with
predictions of the SM is a very powerful tool in searching
for new physics beyond the standard model.

The top quark can also be used to constrain the mass range of the last
yet to be observed particle of the standard model, the Higgs boson,
because their masses and the mass of the $W$ boson are linked through
radiative corrections~\cite{EWWG:2008}. The Higgs boson is a
manifestation of the Higgs mechanism~\cite{Higgs:1964ia,Higgs:1964pj,
Higgs:1966ev,Englert:1964et,Guralnik:1964eu,Kibble:1967sv},
implemented in the standard model to provide the needed breaking of
the electroweak symmetry to which the top quark may be intimately
connected because of its large mass.

Because of its fairly recent discovery, the top quark's properties
have not yet been explored with the same scrutiny as those of the lighter
quarks. However, in the ongoing data taking at Fermilab's Tevatron
proton-antiproton collider, an integrated luminosity of more than
4~fb$^{-1}$ has already been recorded by each of the collider experiments
CDF and D0, corresponding to an increase of about a factor seventy
relative to the data that was available for the discovery of the top quark.
The new data can be used to refine previous measurements to higher
precision that starts to become limited by systematic rather than
statistical uncertainties. In addition, measurements that have never
been performed become feasible, such as the first evidence for
electroweak production of single top quarks and the consequent first
direct measurement of the CKM matrix element $|V_{tb}|$, recently
published by D0 \cite{Abazov:2006gd,Abazov:2008kt} and CDF~\cite{Aaltonen:2008sy}.

This article is intended to provide an overview of the current status of
the top quark physics program pursued at the Tevatron.
Results available until the LHC startup in September 2008 have been
included, utilizing samples of data of up to 2.8~fb$^{-1}$ in
integrated luminosity. Previous reviews of the top quark
are available in Refs.\ 
\cite{Chakraborty:2003iw,Mangano:2004fq,Wagner:2005jh,Quadt:2006jk,Kehoe:2007px}.
The outline of this article is as follows: The second chapter provides
a brief introduction to the standard model, with emphasis on the
special role played by the top quark. Chapter 3 describes production
and decay modes for top quarks in the framework of the standard model.
Chapter 4 outlines the experimental setup used for the measurements
described in the following sections. Chapter 5 presents studies of the
production of top quarks, including measurements of cross section that
form the basis for other measurements of top quark characteristics.
Chapter 6 elaborates on the different results for top quark decay
properties, followed in Chapter~7 by a discussion of measurements of
fundamental attributes of the top quark, such as its charge and mass.
The final chapter (8) contains a brief summary of the achievements to
date.

\section{The Standard Model and the Top Quark}
\subsection{A brief overview of the standard model}
\label{sec:SMoverview}
The standard model of elementary particle physics describes very
successfully the interactions of the known fundamental spin
$J=\frac{1}{2}$ fermion constituents of matter through the exchange of
spin $J=1$ gauge bosons.%
\begin{table}[t]
  \caption[The known fundamental fermions]{The known fundamental
    fermions and their masses~\cite{PDG2008,TEWWG:2008}.}
  \addtolength{\tabcolsep}{-2pt}
  \begin{center}
    \begin{tabular*}{\textwidth}{@{\extracolsep{\fill}}|c|c|c|c|c|}
      \hline

      fermion & \multirow{2}{*}{} electric & \multicolumn{3}{c|}{generation}\\
       \cline{3-5} type & charge [$e$]& 1. & 2. & 3.\\ \hline\hline

      \multirow{4}{*}{quarks}& \multirow{2}{*}{$+\frac{2}{3}$} & up
      ($u$) & charm ($c$) & top ($t$)\\
      & & 1.5 - 3.3 MeV/c$^2$ & 1.27 $^{+0.07}_{-0.11}$ GeV/c$^2$ & 172.4  $\pm$ 1.2 GeV/c$^2$ \\

      \cline{2-5}& \multirow{2}{*}{$-\frac{1}{3}$}& down ($d$) & strange ($s$)& bottom ($b$)\\
      & & 3.5 - 6.0 MeV/c$^2$ & 104 $^{+26}_{-34}$ MeV/c$^2$ & 4.20 $^{+0.17}_{-0.07}$ GeV/c$^2$ \\\hline\hline

      \multirow{5}{*}{leptons} &\multirow{3}{*}{0}& $\nu_{e}$ &  $\nu_{\mu}$ &  $\nu_{\tau}$ \\
      & & $<$ 2 eV/c$^2$ & $<$ 0.19 MeV/c$^2$ & $<$ 18.2 MeV/c$^2$ \\
      & & {\tiny(95\% C.L.)} & {\tiny(90\% C.L.)} & {\tiny(95\% C.L.)}\\

      \cline{2-5}& \multirow{2}{*}{-1}&$e$ & $\mu$ & $\tau$ \\
      & & 0.511 MeV/c$^2$ & 105.658 MeV/c$^2$ & 1777 MeV/c$^2$ \\\hline
    \end{tabular*}
  \end{center}
  \label{fermions}
\end{table}

As shown in Table \ref{fermions}, both quarks and leptons occur in
pairs, differing by one unit of electric charge $e$, and are
replicated in three generations that have a strong hierarchy in mass.
The fermion masses span at least 11 orders of magnitude, with the top
quark being by far the heaviest fundamental particle, which may
therefore provide further insights into the process of mass
generation. The origin of this breaking of the flavor symmetry and the
consequent mass hierarchy is still not understood but can be
accommodated in the standard model as shown below.

The forces among the fundamental fermions are mediated by the exchange
of the gauge bosons of the corresponding quantized gauge fields, as
listed in Table \ref{bosons}. The gravitational force is not included
in the framework of the standard model, and will not be considered, as
its strength is small compared to that of the other interactions among
the fundamental fermions at energy scales considered in this article.

\begin{table}[b!]
  \caption[The known fundamental interactions and their
  properties.]{The known fundamental interactions and their
    properties~\cite{PDG2008}. Gravitation is shown separately as it is
    not included in the SM of elementary particles.}
  \addtolength{\tabcolsep}{-2pt}
  \begin{center}
    \begin{tabular*}{\textwidth}{@{\extracolsep{\fill}}|c|c|c|c|c|c|c|}
      \hline
      interaction & couples & affected & exchange &
      mass & charge & \multirow{2}{*}{spin} \\
      type & to & particles & boson & [GeV/c$^2$] & [$e$]&  \\ \hline\hline
      \multirow{2}{*}{strong} & color & quarks, & \multirow{2}{*}{gluon (g)} &
      \multirow{2}{*}{0} & \multirow{2}{*}{0} & \multirow{2}{*}{1} \\
      & charge & gluons & & & & \\ \hline
      \multirow{2}{*}{weak} & weak & quarks, $W^\pm$, & $W^+$, $W^-$ &
      80.4 & +1, -1 & 1 \\
      & charge & leptons, $Z^0$ & $Z^0$ &  91.2 & 0 & 1 \\ \hline
      electro- & electric & electrically & \multirow{2}{*}{photon ($\gamma$)} &
      \multirow{2}{*}{0} & \multirow{2}{*}{0} & \multirow{2}{*}{1} \\
      magnetic & charge & charged & & & & \\ \hline\hline
      \multirow{2}{*}{gravitation} & mass, & \multirow{2}{*}{all} & graviton &
      \multirow{2}{*}{0} & \multirow{2}{*}{0} & \multirow{2}{*}{2} \\
       & energy & & (unobserved) & & & \\ \hline
    \end{tabular*}
  \end{center}
  \label{bosons}
\end{table}

The standard model is a quantum field theory based on the local gauge
symmetries $SU(3)_{QCD}\times SU(2)_{L}\times U(1)_{Y}$. The theory of
the strong interaction, coupling three different color charges
(``red'', ``green'' and ``blue'') carried by the quarks and the eight
massless gauge bosons (gluons), is called Quantum Chromodynamics
(QCD), and is based on the gauge group $SU(3)_{QCD}$
\cite{Fritzsch:1973pi,Gross:1973ju,Weinberg:1973un,
  Weinberg:1973av,Gross:1973id,Politzer:1973fx}. This symmetry is
exact, and the gluons carry both a color and an anticolor charge. At
increasingly short distances (or large relative momenta), the
interaction gets arbitrarily weak (asymptotically free), thereby
making a perturbative treatment viable. Via the strong interaction,
quarks can form bound color-singlet states called hadrons, consisting
of either a quark and an antiquark (mesons) or three quarks
respectively antiquarks (baryons). The fact that only color-neutral
states and no free quarks are observed is referred to as the
confinement of quarks in hadrons. Due to its large mass, the top quark
decays faster than the typical hadronization time of QCD
($\Gamma_{top} \gg \Lambda_{QCD}$), and it is therefore the only quark
that does not form bound states. Its decay hence offers the unique
possibility to study the properties of essentially a bare quark.

The theory of electroweak interactions developed by Glashow
\cite{Glashow:1961tr}, Salam \cite{Salam:1964ry} and Weinberg
\cite{Weinberg:1967tq} is based on the $SU(2)_{L}\times U(1)_{Y}$
gauge group of the weak left-handed isospin~$T$ and hypercharge~$Y$.
Since the weak ($V-A$) interaction only couples to left-handed
particles, the fermion fields $\Psi$ are split up into left-handed and
right-handed fields $\Psi_{L,R} = \frac{1}{2}(1\mp\gamma_{5})\Psi$
that are arranged in weak isospin $T=\frac{1}{2}$ doublets and $T=0$
singlets:
\begin{equation*}
\begin{array}{ccccccc}
\left( \begin{array}{c} u \\ d\end{array} \right)_L&
\left( \begin{array}{c} c \\ s\end{array} \right)_L&
\left( \begin{array}{c} t \\ b\end{array} \right)_L&
\hspace{2cm}&
\begin{array}{c} u_R \\ d_R\end{array} &
\begin{array}{c} c_R \\ s_R\end{array} &
\begin{array}{c} t_R \\ b_R\end{array}\\
&&&&&&\\
\left( \begin{array}{c} \nu_e \\ e\end{array} \right)_L &
\left( \begin{array}{c} \nu_\mu \\ \mu\end{array} \right)_L &
\left( \begin{array}{c} \nu_\tau \\ \tau\end{array} \right)_L &
\hspace{2cm}&
\begin{array}{c} {\nu_e}_R \\ e_R\end{array} &
\begin{array}{c} {\nu_\mu}_R \\ \mu_R\end{array} &
\begin{array}{c} {\nu_\tau}_R \\ \tau_R\end{array} 
\end{array}
\end{equation*}
In the doublets, neutrinos and the up-type quarks ($u,c,t$) have the
weak isospin $T_3=+\frac{1}{2}$, while the charged leptons and
down-type quarks ($d,s,b$) carry the weak isospin $T_3=-\frac{1}{2}$.
The weak hypercharge Y is then defined via electric charge and weak
isospin to be $Y=2Q-2T_3$. Consequently, members within a doublet
carry the same hypercharge: $Y=-1$ for leptons and $Y=\frac{1}{3}$ for
quarks, as implied by the product of the two symmetry groups.

The $SU(2)_{L}\times U(1)_{Y}$ gauge group does not 
accommodate mass terms for the gauge bosons or fermions without
violating gauge invariance. A minimal way to incorporate these
observed masses is to implement spontaneous electroweak symmetry
breaking (EWSB) at energies around the mass scale of the $W$ and $Z$
boson, often referred to as the ``Higgs
mechanism''\cite{Higgs:1964ia,Higgs:1964pj,Higgs:1966ev,Englert:1964et,Guralnik:1964eu,Kibble:1967sv},
by introducing an SU(2) doublet of complex scalar fields $\Phi =
(\Phi^+, \Phi^0)^T$. When the neutral component obtains a non-zero
vacuum expectation value $v/\sqrt{2} \neq 0$, the $SU(2)_{L}\times
U(1)_{Y}$ symmetry is broken to $U(1)_{QED}$, giving mass to the three
electroweak gauge bosons $W^\pm, Z^0$ while keeping the photon
massless, and thereby leaving the electromagnetic symmetry $U(1)_{QED}$
unbroken. From the remaining degree of freedom of the scalar doublet,
we obtain an additional scalar particle, the Higgs boson. %

The Higgs mechanism also provides fermion masses through fermion
Yukawa couplings to the scalar field, with masses given by
$m_f=\lambda_f v/\sqrt{2}$, for a Yukawa coupling constant $\lambda_f$
for each massive fermion in the standard model. With a Yukawa coupling
close to unity, the top quark may play a special role in the process
of mass generation.

The mixing of flavor eigenstates in weak charged-current interactions
of quarks is described by the Cabibbo-Kobayashi-Maskawa (CKM) matrix
\cite{Cabibbo:1963yz,Kobayashi:1973fv}. By convention, this is a 3 x 3
unitary matrix $V_{CKM}$ that operates on the negatively-charged
flavor states $d, s$ and $b$:
\begin{equation}
  \left( \begin{array}{c} d' \\ s' \\ b' \end{array} \right)_L
  =
  \left( \begin{array}{ccc}
      V_{ud} & V_{us} & V_{ub}\\
      V_{cd} & V_{cs} & V_{cb}\\
      V_{td} & V_{ts} & V_{tb}
    \end{array} \right)
  \left( \begin{array}{c} d \\ s \\ b \end{array} \right)_L
    \equiv
    \mathbf{V_{CKM}}
    \left( \begin{array}{c} d \\ s \\ b \end{array} \right)_L .
\end{equation}
This complex matrix can have 18 independent parameters. However, to
conserve the probability, this matrix has to be unitary, which means
that there are only nine free parameters. An additional five out of
the nine can be absorbed as phases in the quark wave functions. This
results in four independent parameters in total -- three real Euler
angles and one complex phase, the latter implementing CP violation in
the standard model. Since the CKM matrix is not diagonal, charged
current weak interactions can have transitions between quark
generations (``generation mixing'') with coupling strengths of the
$W^{\pm}$ boson to the physical up and down type quarks given by the
above matrix elements.

From experimental evidence
\cite{Davis:1968cp,Cleveland:1998nv,Fukuda:1998mi,Ahmad:2001an,
  Fukuda:2002pe,Ahn:2002up,Yao:2006px}, neutrinos also have mass,
which has led, among other things, to the introduction of an analogue
leptonic mixing matrix, the Pontecorvo-Maki-Nakagawa-Sakata (PMNS)
matrix \cite{Pontecorvo:1967fh, Maki:1962mu}. It contains four
independent parameters as well if one assumes that neutrinos are not
Majorana particles.

In summary, the standard model of elementary particle physics is a
unitary, renormalizable theory \cite{Hooft:1971fh,Hooft:1971rn}, that
can be used to perturbatively calculate processes at high energies. It
incorporates 25 parameters that have to be provided through
measurement:
\begin{itemlist}
\item 12 Yukawa couplings for the fermion masses
\item 8 parameters for the CKM and PMNS mixing matrices
\item 3 coupling constants $\alpha_s, g, g'$ of $SU(3)_{QCD},
      SU(2)_{L}$ and $U(1)_{Y}$, respectively
\item 2 parameters from EWSB: $v, m_{H}$.
\end{itemlist}
At currently accessible energy scales, the standard model describes
successfully the interactions of fundamental fermions and gauge
bosons, with only the Higgs boson remaining to be observed. For a more
detailed introduction to the standard model, the reader is referred to
corresponding textbooks, Refs.\ 
\cite{Halzen:1984mc,HoKim:1998gr,Treichel:2000nj} on elementary
particle physics and topical reviews such as Ref.\  \cite{Djouadi:2005gi}.

\subsection{The need for the top quark in the standard model}
\label{sec:SMneedsTop}
\begin{figure}[t]
  \centering
  \includegraphics[width=0.5\textwidth,clip=]{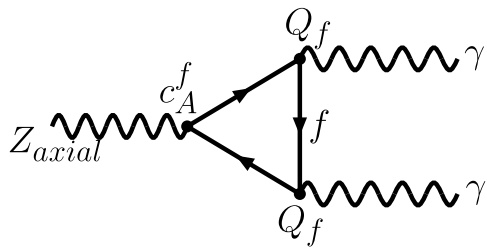}
  \caption{A ``problematic'' fermion triangle diagram that could
  introduce an anomaly.}
  \label{fig:ftriangle}
\end{figure}

The existence of the top quark was postulated well before its discovery
mainly for three reasons.
The first argument reflects the desire to have the standard model
correspond to a renormalizable theory. When expressed via a
perturbation series -- usually depicted in Feynman diagrams with first
order ``tree'' diagrams and higher order ``loop'' terms -- certain
loop diagrams cause divergences that have to cancel exactly to ensure
that the theory is renormalizable. One example is the fermion triangle
diagram, as shown in Fig.\ \ref{fig:ftriangle}.
The contribution for each such diagram is proportional to
$c_A^fQ_f^2$, with $c_A^f$ being the weak neutral current axial
coupling strength, and $Q_f$ the electric charge for the respective
fermion in the loop. Since $c_A^f=T_3$ and neutrinos do not
contribute, for the total strength of the anomaly to be cancelled, an
equal number of lepton flavors and quark-doublets $N_{families}$, and
quarks in three colors ($N_c=3$) are required~\cite{Halzen:1984mc}:
\begin{equation}
\sum_{i=1}^{N_{families}}\left(-\frac{1}{2}(-1)^2
+\frac{1}{2}N_c\left(+\frac{2}{3}\right)^2
-\frac{1}{2}N_c\left(-\frac{1}{3}\right)^2\right)
\overset{!}{=} 0.
\end{equation}
Consequently, the discovery of the $\tau$ lepton already called for an
additional quark doublet to be present to keep the standard model
renormalizable.

The second argument results from the fact that transitions that change
the flavor but not the charge of a fermion ($u\leftrightarrow c
\leftrightarrow t$ or $d\leftrightarrow s\leftrightarrow b$) are
observed to be strongly suppressed. The absence of
such flavor-changing neutral currents (FCNC) for two quark
generations could be accommodated through the GIM mechanism
\cite{Glashow:1970gm} by postulating the existence of the charm quark
-- and thereby completion of the second quark doublet -- years before its
discovery. This mechanism can be applied in a similar way for three
quark generations, requiring a sixth quark as a partner of the $b$ quark
to complete the doublet.

The third argument comes from the experimental confirmation that the $b$
quark is not a weak isospin singlet but is part of an isospin doublet
carrying the weak isospin $T_3=-\frac{1}{2}$ and electric charge
$Q_b=-\frac{1}{3}e$. 
The electric charge of the $b$ quark was measured first at the
electron-positron storage ring DORIS at DESY operating at the
$\Upsilon$ and $\Upsilon '$ resonances through a measurement of the cross
section for resonant hadron production $\sigma_h$
\cite{Berger:1978dm,Darden:1978dk,Bienlein:1978bg}. The integral over
$\sigma_h$ is related to the electronic partial width $\Gamma_{ee}$,
the hadronic partial width $\Gamma_{h}$, the total width
$\Gamma_{tot}$ and the resonance mass $M_R$ via $\int \sigma_h~dM =
6\pi^2 \Gamma_{ee} \Gamma_{h}/(M_R^2 \Gamma_{tot})$. Assuming that the
total width is dominated by the hadronic partial width
($\Gamma_{h}\approx\Gamma_{tot}$), a measurement of the integrated
cross section and the resonance mass provides the electronic
partial width $\Gamma_{ee}$ of the $\Upsilon$ and of the $\Upsilon '$. In
the framework of non-relativistic quarkonium potential models
\cite{Gottfried:1977wi,Rosner:1978cj}, this partial width can then be
related with the bound quark's charge.

The weak isospin of the $b$ quark was measured via the
forward-backward asymmetry $A_{FB}$ in the process $e^+e^-\to
b\bar{b}\to\mu^\pm+$~hadrons with the JADE detector at PETRA
\cite{Bartel:1984rg}. The asymmetry originates from electroweak
interference effects and is defined as the difference between the number of
fermions produced in the forward direction (with polar angle $\theta <
90^\circ$) and the number of fermions produced backward
($\theta > 90^\circ$), divided by their sum. $A_{FB}$
is proportional to the ratio of the weak axial to the electric charge and
vanishes for a weak isospin singlet. For a $T_3=-\frac{1}{2},
Q=-\frac{1}{3}e$ $b$-quark, the predicted asymmetry is $-25.2\%$, in
good agreement with the measurement of $-22.8 \pm$ 6.0 (stat.) $\pm$
2.5 (syst.)\%.

As a result of these measurements, the top quark's weak isospin and
electric charge within the standard model were assigned to be
$T_{3}=+\frac{1}{2}$, $Q_t=+\frac{2}{3}e$, well before its discovery.
The mass of the top quark, being a free parameter in the standard
model, could not be predicted. Nevertheless, the mass of the top quark
can indirectly be constrained by precision electroweak measurements.

\subsection{Top quark mass from precision electroweak measurements}
\label{sec:topmassewprecision}
As discussed above, the standard model comprises a set of free
parameters that are a priori unknown. However, once these
are measured, all physical observables can be expressed in terms
of those parameters. To make optimal use of the predictive power of
the theory, it is therefore crucial to measure its input parameters
with highest possible precision, and thereby probe
the self-consistency of the SM and any contributions
beyond its scope. Being a renormalizable theory,
predictions for any observable can be calculated to any order 
and checked with experiment.

Electroweak processes depend mainly on three parameters: the coupling
constants $g {\rm~and~} g'$ of $SU(2)_{L} {\rm~and~} U(1)_{Y}$,
respectively, and the Higgs vacuum expectation value $v$. Since these
input parameters have to be obtained from experiment, it is best to
substitute them with the most precisely measured quantities of the
electromagnetic fine structure constant $\alpha$ (using
electron-positron annihilations into hadrons at low center of mass
energies to measure hadronic vacuum polarization corrections
\cite{Bai:2001ct,Akhmetshin:2003zn,Aloisio:2004bu}), the Fermi
constant $G_F$ (from the muon lifetime
\cite{vanRitbergen:1999fi,Barczyk:2007hp}) and the mass of the $Z$
boson $m_Z$ (from electron-positron annihilations around the $Z$ pole
\cite{Z-Pole}).

With these input values, the theoretical framework can be used to
predict other quantities such as the $W$ boson mass. Given precision
measurements, the $W$ boson mass is sensitive to the mass of the top
quark and the mass of the Higgs boson through higher order 
radiative quantum corrections
\cite{Langacker:1989sm,Hollik:1988ii,Burgers:1989bh}.

The most precise electroweak measurements to date have been performed
at the Large Electron-Positron (LEP) Collider
\cite{Myers:1990sk,Brandt:2000xk} at CERN by the four experiments
ALEPH \cite{Decamp:1990jra,Buskulic:1994wz}, DELPHI
\cite{Aarnio:1990vx,Abreu:1995uz}, L3 \cite{Adeva:1989kxa,Adam:1996fj}
and OPAL \cite{Ahmet:1990eg,Allport:1994ec}, and at the Stanford Linear
Collider (SLC) \cite{:1980qx,Richter:1989eu} by the SLD experiment
\cite{Ash:1984rp,Abe:1997bu}. The LEP experiments have
analyzed $\approx$17 million $Z$ decays, and  a sample of $\approx$600
thousand $Z$ bosons produced with longitudinally polarized electron
beams was analyzed by SLD.

\begin{figure}[t]
  \centering
   \subfigure[]{\epsfig{figure = 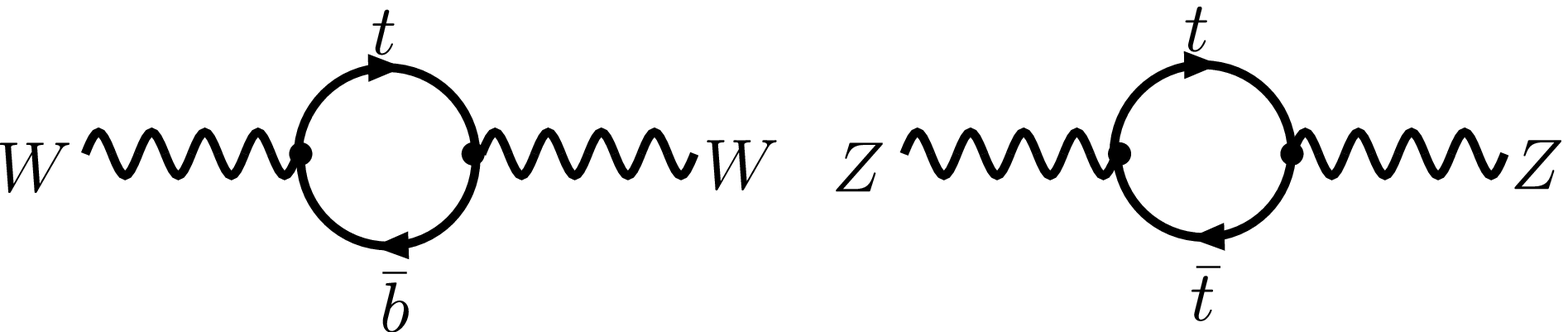, width = 0.65 \textwidth}}
   \subfigure[]{\epsfig{figure = 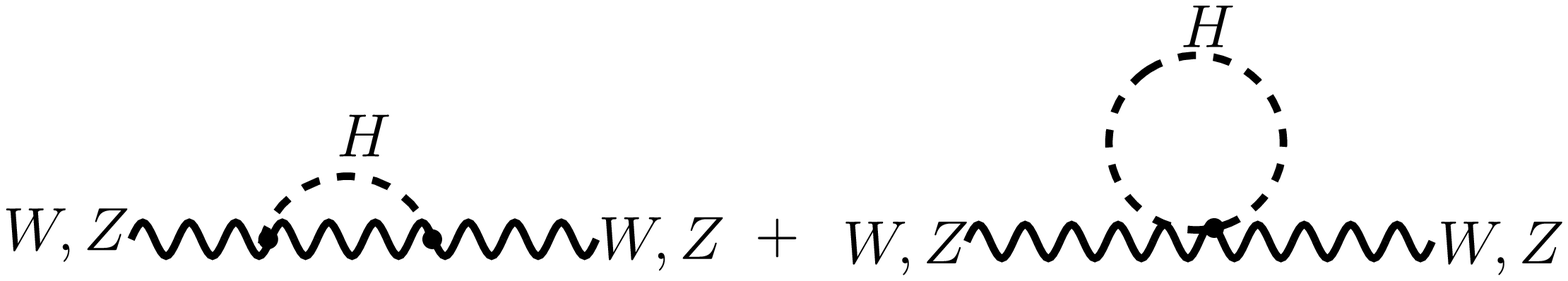, width = 0.75 \textwidth}}
  \caption{(a) Radiative contributions to the $W$ and $Z$ propagators
   from the top quark. (b) Radiative corrections to the $W$ and $Z$
   propagators from the Higgs boson.}
  \label{fig:radcorr}
\end{figure}

Defining the electroweak mixing angle $\theta_W$ via the vector boson
masses:
\begin{equation}
\frac{m_W^2}{m_Z^2} = 1 - \sin^2\theta_W ,
\end{equation}
the $W$ boson mass can be expressed as~\cite{Hollik:1988ii}:
\begin{equation}
m_W^2 = \frac{\pi\alpha}{\sqrt{2}G_F}\cdot\frac{1}{\sin^2\theta_W(1-\Delta r)} ,
\end{equation}
where the radiative correction $\Delta r$ is a directly observable quantum
effect of electroweak theory that depends on $\alpha, m_W, m_Z,
m_{H}$ and $m_t$. The contributions from single-loop insertions
containing the top quark and the Higgs boson, as depicted in Fig.\ 
\ref{fig:radcorr}, are~\cite{Grunewald:1999wn}:
\begin{eqnarray}
\Delta r^{\rm top} &=& -\frac{3\sqrt{2}G_F\cot^2\theta_W}{16\pi^2}\cdot
m_t^2  \hspace{1.5cm}({\rm for~} m_t \gg m_b)\\ 
\Delta r^{\rm Higgs} &=& \frac{3\sqrt{2}G_Fm_W^2}{16\pi^2}\cdot 
\left(\ln\frac{m_H^2}{m_W^2}-\frac{5}{6}\right)  \hspace{.5cm}({\rm for~} m_H \gg m_W) .
\end{eqnarray}
Thus, a precise measurement of $W$ and $Z$ boson masses provides
access to the mass of the top quark and the Higgs boson. The top quark
contribution to radiative corrections is large, primarily because of
the large mass difference relative to its weak isospin partner, the
$b$ quark. While the leading top quark contribution to $\Delta r$ is
quadratic, it is only logarithmic in mass for the Higgs boson.
Consequently, the constraints that can be derived on the mass of the
top quark are much stronger than for the Higgs boson mass.

In 1994, the most stringent constraints on $m_t$ were
based on preliminary LEP and SLD data, combined with measurements 
of $m_W$ in proton-antiproton experiments, and neutral to
charged-current ratios obtained from neutrino experiments, yielding
\mbox{178 $\pm$ 11 $^{+18}_{-19}$ GeV/c$^2$} \cite{EWWG:1994}. 
\begin{figure}[t]
  \centering
  \subfigure[]{\label{fig:mt1994} \epsfig{figure = 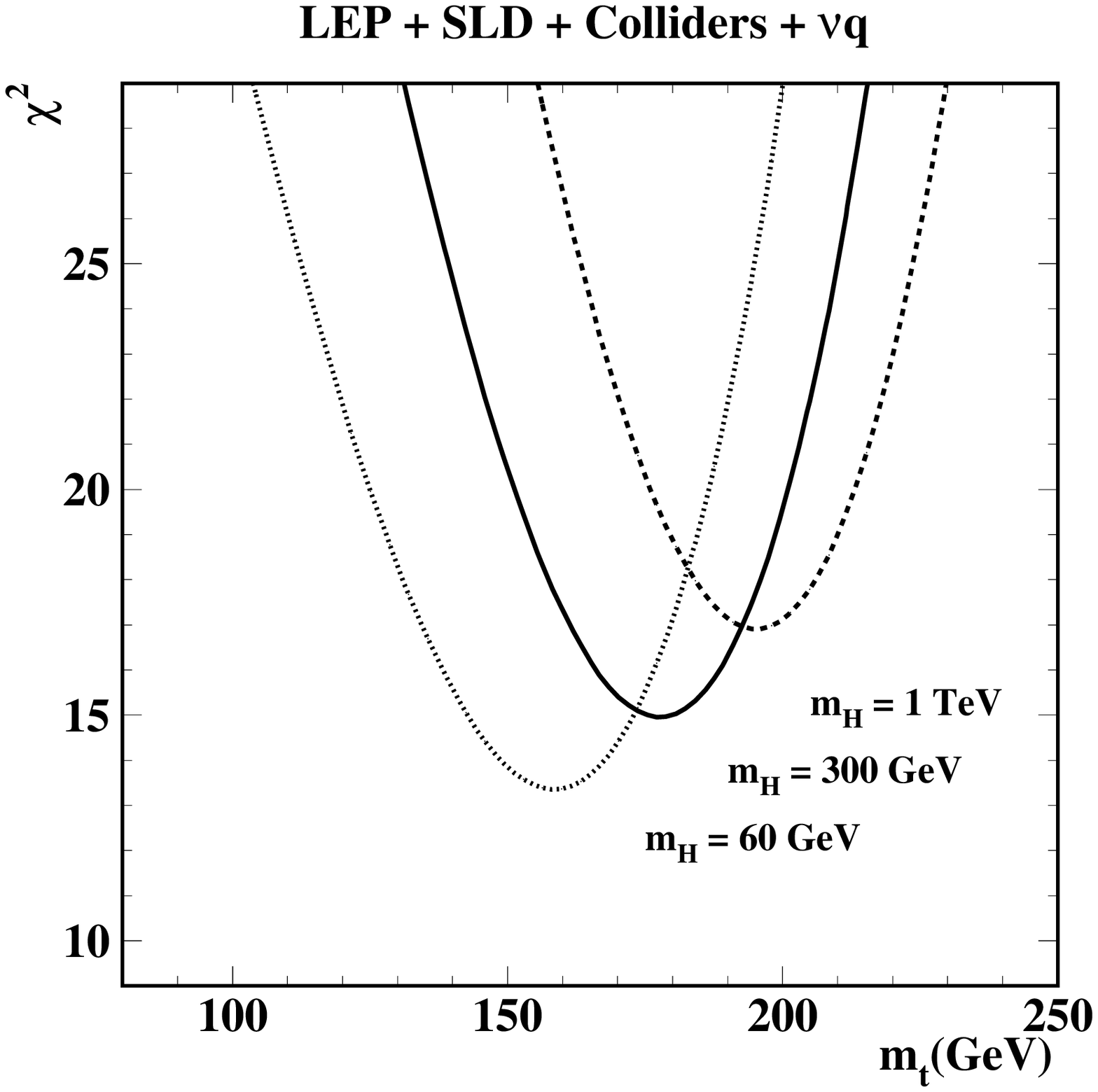, width = 0.48 \textwidth}}
  \subfigure[]{\label{fig:mthist} \epsfig{figure = 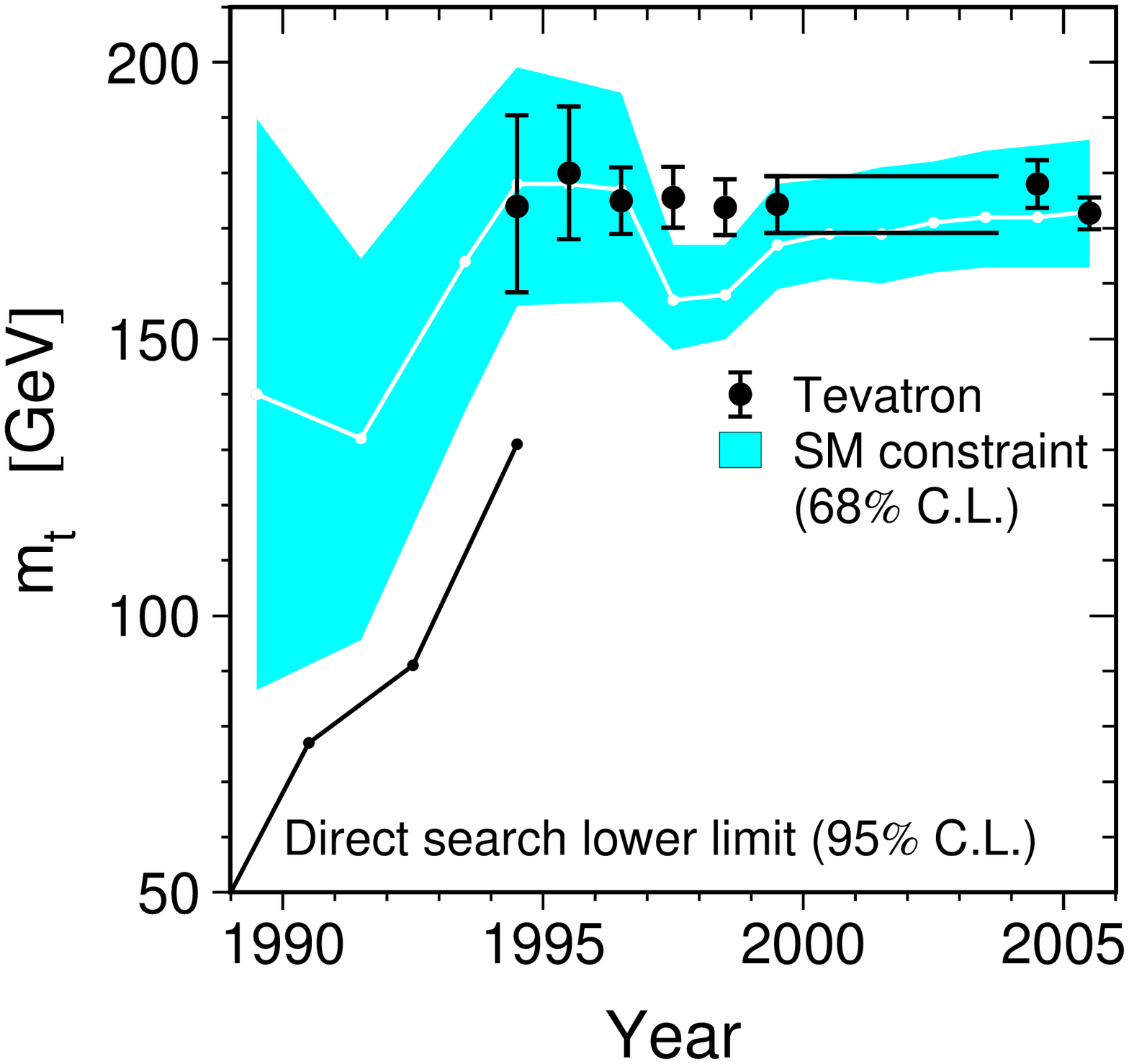, width = 0.48 \textwidth}}
  \caption{(a) $\chi ^2$ distributions of the standard model fit to
  precision electroweak data versus top quark mass for various Higgs boson
  masses \cite{EWWG:1994}. (b) Comparison of the indirect top quark
  mass measurements via radiative corrections (shaded area) with the
  direct measurements from the Tevatron (points) versus time
  \cite{Z-Pole}.}
\end{figure}
As illustrated in Fig.\ \ref{fig:mt1994}, the central value and the
first set of uncertainties are from a $\chi ^2$ fit of the SM to
precision electroweak data, assuming $m_H = 300$~GeV/c$^2$. The second
set of uncertainties stems from the impact of varying $m_H$ between 60
and 1000 GeV/c$^2$.

The good agreement between predicted and observed values of $m_t$,
shown in Fig.~\ref{fig:mthist} as a function of time~\cite{Z-Pole},
is one of the great successes of the SM. The latest prediction from
precision electroweak data yields $m_t = 179^{+12}_{-9}$ GeV/c$^2$
without imposing constraints on $m_H$~\cite{EWWG:2008}, and is in
excellent agreement with the current world average of $m_t =$ \mtopwa.

This success of the SM also gives greater confidence in the
predictions for $m_H$. Since the precision of the prediction depends
crucially on the accuracy of $m_W$ and $m_t$, it provides a strong
motivation for improving the corresponding measurements. The current
constraints on the mass of the Higgs boson will be discussed in
Section~\ref{sec:topmassaverage}.

More details on precision electroweak measurements can be found in
topical reviews such as given in
Refs.~\cite{Quast:1999sh,Grunewald:1999wn}.

\section{Production and Decay of Top Quarks}
The production of top quarks is only possible at highest center of
mass energies $\sqrt{s}$, set by the scale of $m_t$. The energies
needed for production of top quarks in the SM are currently (and will
be at least for the next decade) only accessible at hadron colliders.
The Tevatron proton-antiproton collider started operation at
$\sqrt{s}=1.8$~TeV in 1987 for a first period of data taking (``Run
0'') that lasted until 1989, with the CDF experiment recording about
4~pb$^{-1}$ of integrated luminosity. The next data taking period from
1992 until 1996 at $\sqrt{s}=1.8$~TeV (the so-called Run~I) was
utilized by both the CDF and D0 experiments and facilitated the
discovery of the top quark. For the currently ongoing data taking that
started in 2001 (Run~II), the center of mass energy has increased to
$\sqrt{s}=1.96$~TeV. The Tevatron will lose its monopoly for top quark
production only with the turning-on of the Large Hadron Collider (LHC)
that will provide proton-proton collisions at $\sqrt{s}=14$~TeV.

In the framework of the standard model, top quarks can be produced in
pairs ($t\bar{t}$) predominantly via the strong interaction and singly
via the electroweak interaction.

\subsection{Top quark pair production}
\label{sec:ttbarprod}
While hadron colliders provide the highest center of mass energies,
the collision of hadrons complicates the theoretical description and
prediction for processes such as $t\bar{t}$ production because of the
composite nature of the colliding particles. These difficulties can be
handled through the QCD factorization theorem
\cite{Collins:1985gm,Collins:1987pm} that provides a way to separate
hadron collisions into universal long-distance (small momentum
transfer) phenomena and perturbatively calculable short-distance
phenomena. The latter processes involve therefore large square of
momentum transfers $Q^2$, and consequently the production of particles
with large transverse momenta or large mass. The two components are
set apart by introducing a factorization scale $\mu^2_F$ in the
calculation.

Using this approach, the proton can be described by a collection of
partons (quarks, antiquarks, gluons) that interact at a low energy
scale $\Lambda_{QCD} < 1$~GeV, whereas the elementary collisions
between partons of the proton (or antiproton) occur on a ``hard''
energy scale characterized by large transverse momenta $\geq{\cal
O}$(100~GeV).

\begin{figure}[h]
  \centering
    \subfigure[]{\epsfig{figure = 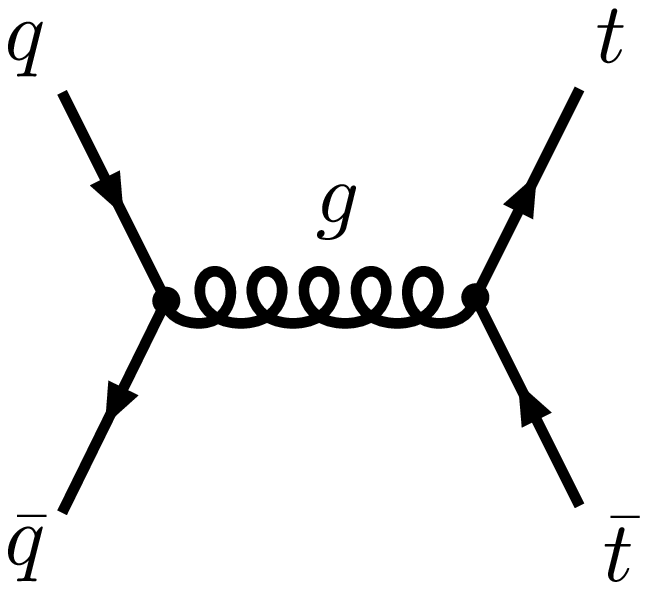, width = 0.25 \textwidth}}\hspace{1cm}
    \subfigure[]{\epsfig{figure = 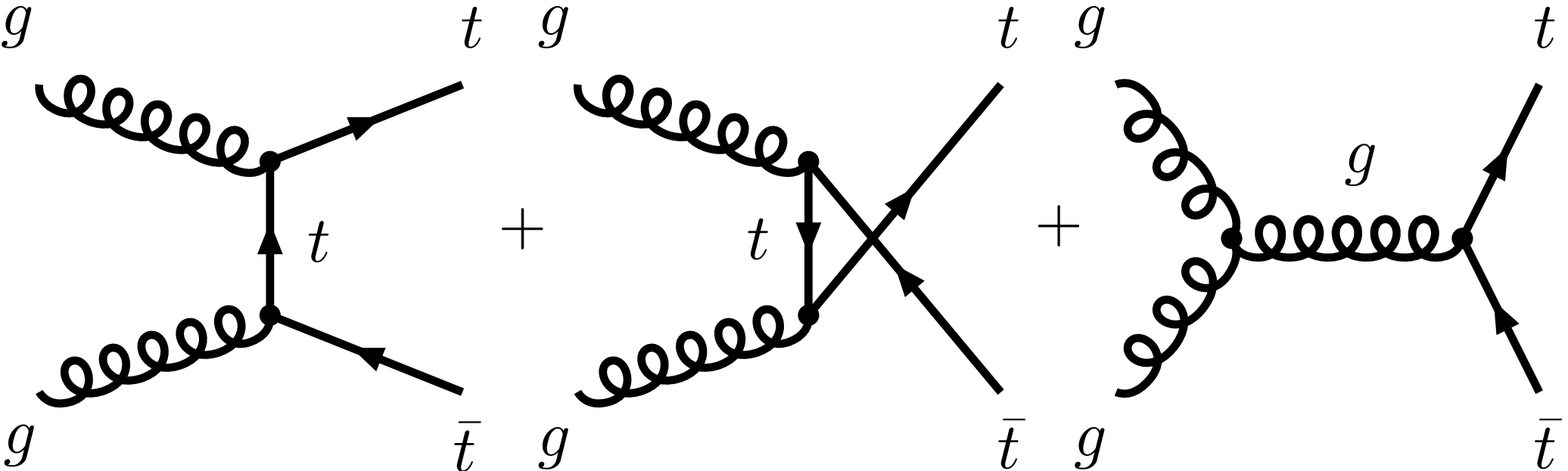, width = 0.65
        \textwidth}}
  \caption{Leading order Feynman diagrams for top quark pair
    production: (a) quark-antiquark annihilation and (b) gluon-gluon
    fusion.}
  \label{fig:ttbar-feynman}
\end{figure}

Consequently, the partons participating in any hard process ($a,b$) can
be considered quasi-free, and the partonic cross section of interest
$\hat{\sigma}_{a+b\to X}(\hat{s}, \alpha_s(\mu^2_R), \mu^2_R)$ can be
calculated using perturbative QCD, independent of the type of hadrons
containing the partons. (The hatted variables denote parton
quantities.) The regularization of divergences in higher order
calculations (such as ultraviolet divergences from loop insertions,
where the infinite range of four-momentum in the loop causes
infinities in the integration from high momentum contributions)
requires the introduction of a renormalization scale $\mu^2_R$,
along with the corresponding running coupling constant
$\alpha_s(\mu^2_R)$.
The leading order Feynman diagrams for \ttbar production are shown in
Fig.~\ref{fig:ttbar-feynman}.

The partons within the incoming proton (or antiproton) cannot be
described by perturbative QCD, as the soft energy scale corresponding
to small inherent momentum transfers implies large $\alpha_s(Q^2)$
couplings. The distribution of the longitudinal momentum of the hadron
among the partons is described by Parton Distribution Functions
(PDFs): $f_{a/A}(x,\mu^2_F)$, corresponding to the probability to find
a given parton $a$ inside hadron $A$ with momentum fraction $x$ when
probed at an energy scale $\mu^2_F$. Collinear and soft (infrared)
singularities that arise in the perturbative calculation of the
partonic cross section discussed above are absorbed in these PDFs.

The factorization theorem is used to describe the $t\bar{t}$
production cross section via an integral over the corresponding hard
scattering parton cross section, folded with the parton distribution
functions of the incident hadrons as follows:
\begin{eqnarray}
  \sigma_{A+B\to t\bar{t}}(\sqrt{s},m_t) =
  \sum_{a,b=g,q,\bar{q}} \int 
\hat{\sigma}_{a+b\to t\bar{t}}(\hat{s},\alpha_s(\mu^2_R),\mu^2_R,\mu^2_F, m_t)\notag\\
\times  f_{a/A}(x_a,\mu^2_F) f_{b/B}(x_b,\mu^2_F) dx_a dx_b .
\end{eqnarray}
The hadrons $A$ and $B$ correspond to proton and antiproton in case of the 
Tevatron and to protons in case of the LHC.

The physical cross section $\sigma_{A+B\to t\bar{t}}(\sqrt{s},m_t)$
that would emerge from the evaluation of the full perturbation series
does not depend on either of the two arbitrary scales for
factorization and renormalization $\mu^2_F, \mu^2_R$ that had to be
introduced for the calculation. However, the parton distribution
functions and the partonic cross section do depend on these scales,
and hence the result of any finite order calculation will as well.
This dependence gets weaker with the inclusion of higher order terms
in the calculation. In practical application, both scales are usually
set to the
typical momentum scale $Q^2$ of the hard scattering process, such as
the transverse momenta of the produced particles or the mass of the
produced particle, so that for $t\bar{t}$ production, typically
$\mu_F=\mu_R=\mu=m_t$. The scale dependence of the result is then
usually tested by varying the central scale by a factor of two; the
resulting variations are interpreted as systematic uncertainties that
should not be mistaken as Gaussian in nature.

The PDFs have to be determined experimentally, for example via deeply
inelastic lepton scattering on nucleons, so that they can be extracted 
from the measured cross sections using perturbative calculations
of the (hard) partonic cross sections. 
Once the parton densities $f_{a/A}(x_a,Q^2)$ have been measured as a
function of momentum fraction $x_a$ at a scale $Q^2$, their value at a
different scale can be predicted perturbatively using the DGLAP
evolution equation
\cite{Gribov:1972ri,Altarelli:1977zs,Dokshitzer:1977sg}. Since PDFs
are universal and do not depend on the process they were derived from,
they can be used to predict cross sections in other hard scattering
processes.
For consistent application, it is important that the PDFs are derived
to same perturbative order and with the same renormalization scheme as
the calculation of any prediction.
\begin{figure}[t]
  \centering
  \includegraphics[width=1.\textwidth,clip=]{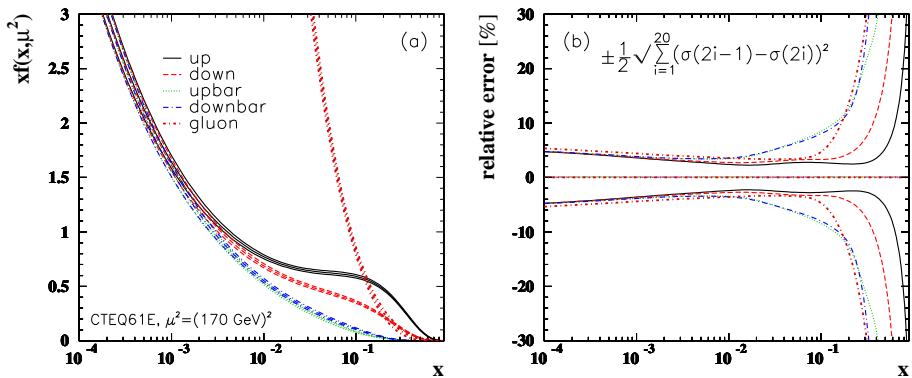}
  \caption{(a) CTEQ61~\cite{Stump:2003yu} parton distribution
    functions with their uncertainty bands, for the $m_t$ mass scale
    ($Q^2=(170{\rm~GeV})^2$) for (anti-) up quarks, (anti-) down
    quarks and gluons in the proton. (b) Relative uncertainties on the
    PDFs shown in (a).}
  \label{fig:CTEQPDF}
\end{figure}

The PDFs are extracted from global fits to the available data, as is
done, for example, by the CTEQ \cite{Pumplin:2002vw}, MRST
\cite{Martin:2002aw}, GRV \cite{Gluck:1998xa}, Alekhin
\cite{Alekhin:2002fv}, H1 \cite{Adloff:2000qk} and ZEUS
\cite{Chekanov:2002pv} groups. Different PDFs are based on different
data, different orders of perturbation theory, renormalization schemes
and fitting techniques -- see, for example, the overview given in
Ref.~\cite{Thorne:2006wq}. One commonly used set of PDFs derived at
NLO, using the $\overline{MS}$ renormalization scheme
\cite{Bardeen:1978yd}, is CTEQ61~\cite{Stump:2003yu}, which
incorporates the Tevatron Run~I data on jet production, especially
important for the gluon distribution.
CTEQ61 also includes an error analysis based on different sets of PDFs
that describe the behavior of a global $\chi^2$ function for the fit
around its minimum. The resulting error on the PDF ($\Delta f$) can be
obtained by summing over the variations $f_i^\pm$ along/against each
PDF ``eigenvector'' for every free parameter in the global fit:
$\Delta f = \pm \frac{1}{2} ( \Sigma_{i=1}^{N_{par}}
(f_i^+-f_i^-)^2)^\frac{1}{2}$.

Figure \ref{fig:CTEQPDF} shows the most important parton distributions
within protons for \ttbar production at the Tevatron or LHC, and their
corresponding uncertainties. (For antiprotons, quarks and antiquarks
have to be interchanged in Fig.\ \ref{fig:CTEQPDF}.) All PDFs vanish
at large momentum fractions $x$, and the gluon density starts to
dominate over the valence-quark densities near $x=0.13$. There is no
flavor symmetry between the up and down quark distributions, neither
on the valence nor the sea quark level (the latter is best seen at low
$Q^2$). At $x$-values below 0.1, typical relative uncertainties on the
PDFs of valence quarks and gluons are $\approx$5\%. At larger
$x$-values, these uncertainties increase drastically, especially for
gluons.

To produce a top quark pair, the squared center of mass energy at the
parton level $\hat{s}=x_ax_bs$ must at least equal $(2m_t)^2$.
Assuming \mbox{$x_a\approx x_b = x$}
yields as threshold for \ttbar production:
\begin{equation}
\langle x\rangle = \sqrt{\frac{\hat{s}}{s}}=\frac{2m_t}{\sqrt{s}} \approx
\left\{\begin{array}{cl} 0.192 & @~\mbox{Tevatron Run I,
$\sqrt{s}=1.8$~TeV}\\ 0.176 & @~\mbox{Tevatron Run II,
$\sqrt{s}=1.96$~TeV}\\ 0.025& @~\mbox{LHC, $\sqrt{s}=14$~TeV}
\end{array} \right.
\end{equation}
Since large momentum fractions are required for \ttbar production
at the Tevatron, the process is dominated by quark-antiquark
annihilation (Fig.\ \ref{fig:ttbar-feynman} (a)) of the valence
quarks. For Run~I energies, quark-antiquark annihilation contributes
roughly 90\% of the total \ttbar production rate, and for Run~II
energies this fraction is $\approx$85\% \cite{Mangano:2004fq}.

At the LHC, gluon-gluon fusion dominates (Fig.~\ref{fig:ttbar-feynman}
(b)) with a contribution of about 90\% \cite{Kidonakis:2001nj},
because a small momentum fraction suffices for \ttbar production. This
means that proton-proton collisions at the LHC have production cross
sections comparable to $p\bar{p}$ rates, thereby obviating the need
for the major technical challenge of producing an intense antiproton
beam.

The increase in the center of mass energy by $\approx$10\% between
Run~I and Run~II at the Tevatron and the correspondingly smaller
minimum momentum fraction provide an increase in the \ttbar production
rate of 30\%. At the LHC, the rate increases by roughly a factor of
100 compared to that of the Tevatron.

The highest-order complete perturbative calculations for heavy quark
pair production have been available at next-to-leading order (NLO) --
to order $\alpha_{s}^{3}$ -- since the late 1980s from Nason {\it et
al.}~\cite{Nason:1987xz} and Beenakker {\it et
al.}~\cite{Beenakker:1988bq,Beenakker:1990maa}. These calculations can
be refined by the inclusion of large logarithmic corrections
\cite{Sterman:1986aj, Catani:1989ne, Catani:1990rp} from soft-gluon
emission that are particularly important for the production of heavy
quarks close to the kinematic threshold ($\hat{s}\approx 4m^2,~x\to
1$).
The contributions of these logarithms are positive at all orders when
evaluated at the heavy quark mass scale and their inclusion therefore
increases the production cross section above the NLO level.

The impact of soft-gluon resummation on the \ttbar production cross
section has been studied by Berger and Contopanagos
\cite{Berger:1995xz,Berger:1996ad,Berger:1997gz}, Laenen, Smith and
van Neerven \cite{Laenen:1991af,Laenen:1993xr} and Catani, Mangano,
Nason and Trentadue \cite{Catani:1996dj,Catani:1996yz} at the leading
logarithmic (LL) level. Studies including even higher level
corrections as carried out by Cacciari {\it et al.}~\cite{Cacciari:2003fi},
based on work by Bonciani {\it et al.}~(BCMN) \cite{Bonciani:1998vc}, and
Kidonakis and Vogt \cite{Kidonakis:2003qe,Kidonakis:2003vs} are
summarized in Table~\ref{theory-ttbar-xsecs}.

\begin{table}[t]
  \caption[\ttbar cross section predictions]{NLO cross section
    predictions including soft-gluon resummations beyond LL for 
    \ttbar production at the Tevatron and the LHC for a top quark 
    mass of 175 GeV/c$^2$. For the different sources of the quoted
    uncertainties please refer to the text.}
  \renewcommand{\arraystretch}{1.2}
  \begin{center}
    \begin{tabular*}{\textwidth}{@{\extracolsep{\fill}}lllr}\hline
      Hadron Collider &Processes& $\sigma_{t\bar{t}}$ [pb] & Group\\
      \hline\hline
      Tevatron Run~I &90\% $q\bar{q}\to t\bar{t}$&5.19$^{+0.52}_{-0.68}$& Cacciari {\it et al.}~\cite{Cacciari:2003fi}\\
      \cline{3-4}
      ($p\bar{p},~\sqrt{s}=1.8$~TeV) &10\% $gg\to t\bar{t}$&$5.24\pm0.31$& Kidonakis {\it et al.}~\cite{Kidonakis:2003vs}\\

      \hline\hline
      Tevatron Run~II &85\% $q\bar{q}\to t\bar{t}$&6.70$^{+0.71}_{-0.88}$& Cacciari {\it et al.}~\cite{Cacciari:2003fi}\\
      \cline{3-4}
      ($p\bar{p},~\sqrt{s}=1.96$~TeV) &15\% $gg\to t\bar{t}$&$6.77\pm0.42$& Kidonakis {\it et al.}~\cite{Kidonakis:2003vs}\\
      \hline\hline
      LHC &10\% $q\bar{q}\to t\bar{t}$&833$^{+52}_{-39}$&Bonciani {\it et al.}~\cite{Bonciani:1998vc}\\
      \cline{3-4}
      ($pp,~\sqrt{s}=14$~TeV) &90\% $gg\to t\bar{t}$& 873$^{+2}_{-28}$ &Kidonakis {\it et al.}~\cite{Kidonakis:2003qe}\\
      \hline\hline
     \end{tabular*}
   \end{center}
\label{theory-ttbar-xsecs}
\end{table}

In the case of \ttbar production at the Tevatron, the inclusion of
leading and next-to-leading logarithmic (NLL) soft-gluon resummation
affects the cross sections only mildly by $\mathcal{O}(5\%)$
(indicating production occurs not too close to the threshold), while
significantly reducing the scale dependence of the predictions by
roughly a factor of two to a level of $\approx$5\%
\cite{Bonciani:1998vc}. At the LHC, \ttbar production takes place even
further away from the kinematic threshold, but since gluon fusion
dominates there, the enhancement of the total production rate from
soft-gluon resummation and the reduction of scale dependence stay at
the same level as at the Tevatron.

The results of Cacciari {\it et al.}~\cite{Cacciari:2003fi} for the Tevatron
use the NLO calculation with LL and NLL resummation at all orders of
perturbation theory as carried out by Bonciani {\it et al.}~(BCMN)
\cite{Bonciani:1998vc}, but are based on the more recent PDF sets with
error analysis CTEQ6 \cite{Pumplin:2002vw} and MRST2001E
\cite{Martin:2002aw} and also MRST2001 \cite{Martin:2001es}, which
includes varied $\alpha_{s}$ values in the PDF fit. The updated PDFs
cause an increase in the central values of about 3\% relative to Ref.
\cite{Bonciani:1998vc}. While the central values are very similar for
the MRST2001E and CTEQ6 PDFs, the uncertainties for CTEQ6 are almost
twice as large as for MRST2001E, unless the variations in
$\alpha_{s}$ in MRST2001 are also included. For the
determination of the uncertainty on the cross section, Cacciari {\it et al.}~combine 
linearly the uncertainty due to scale variation by a factor of
two with the PDF uncertainty evaluated at that scale. As central
values, the CTEQ6M results are chosen, and the maximum lower (upper) uncertainties
given stem from the CTEQ6 PDF variation 
(the $\alpha_{s}$ variation in MRST2001).
The PDF uncertainties and $\alpha_{s}$ variation
contribute about 45\% and 80\% respectively to the total quoted
uncertainty, including the scale variations, which emphasizes the
importance of considering $\alpha_{s}$ uncertainties in PDF fits. The
PDF uncertainties are in turn dominated by the uncertainty of the
gluon PDF at large $x$ values, causing, for example, the gluon fusion
contribution to the total production rate to fluctuate between 11\%
and 21\% for $\sqrt{s}=1.96$~TeV. Despite the large uncertainties on
the \ttbar production rate, the ratio of
production cross sections for the two center of mass
energies at the Tevatron is very stable and predicted with high
precision: $\sigma(1.96~\mbox{TeV})/\sigma(1.8~\mbox{TeV}) = 1.295 \pm
0.015$ for top quark masses between 170 and 180~GeV/c$^{2}$~\cite{Cacciari:2003fi}.

A prediction for the \ttbar production rate at the LHC applying the
same level of soft-gluon resummation is given by Bonciani {\it et al.}
\cite{Bonciani:1998vc} using the MRS(R$_2$) PDF
\cite{Martin:1996as}. Since no PDF uncertainties were available for
Ref. \cite{Bonciani:1998vc}, the quoted uncertainty in Table~\ref{theory-ttbar-xsecs}
comes from changing the  scale by factors of two alone. Since gluon
fusion is the dominant contribution to the total rate,
uncertainties on the gluon PDFs alone lead to an uncertainty of
$\approx 10\%$ on the total production cross section
\cite{Catani:2000xk}.

The studies performed by Kidonakis and Vogt
\cite{Kidonakis:2003qe,Kidonakis:2003vs} consider soft-gluon
corrections up to next-to-next-to-next-to leading logarithmic (NNNLL)
terms at NNLO in a truncated resummation, resulting in a reduced
sensitivity of $\approx$3\% to scale variations.
For the Tevatron, the \ttbar production cross section is evaluated
using MRST2002 NNLO \cite {Martin:2002aw} and CTEQ6M NLO
\cite{Pumplin:2002vw} parton densities. Two different parton-level
conditions are considered for the scattering process: (i) one-particle
inclusive (1PI) and (ii) pair-invariant mass (PIM) kinematics
~\cite{Kidonakis:2001nj}. While both sets of PDFs give very similar
results, the variations from the difference in kinematics are significant.
Consequently, the average of 1PI and PIM kinematics for both PDFs is
used as the central value in Table~\ref{theory-ttbar-xsecs}, while the
separate averages over the PDFs for 1PI and PIM are
quoted as uncertainties. For the predicted LHC rate, which is dominated
by gluon fusion, the 1PI kinematics is considered more
appropriate, and the value in Table~\ref{theory-ttbar-xsecs} gives the
corresponding result based on MRST2002 NNLO PDFs, using scale changes
by factors of two for estimating the uncertainties.

All results in Table~\ref{theory-ttbar-xsecs} are evaluated for a top
quark mass of 175~GeV/c$^2$, and the Run~II values serve as
the main predictions for CDF and D0. To improve comparability of 
the uncertainties on the different predictions, the calculation by Kidonakis
and Vogt has an additional uncertainty obtained from the maximum
simultaneous changes in scale and PDFs\footnote{The PDF uncertainties
in this case stem from CTEQ6 sets ``129'' and ``130'' alone.} added in
quadrature with the uncertainty due to the dependence on
kinematics~\cite{Kidonakis:2006}.

\begin{figure}[t]
  \centering
  \includegraphics[width=0.75\textwidth,clip=]{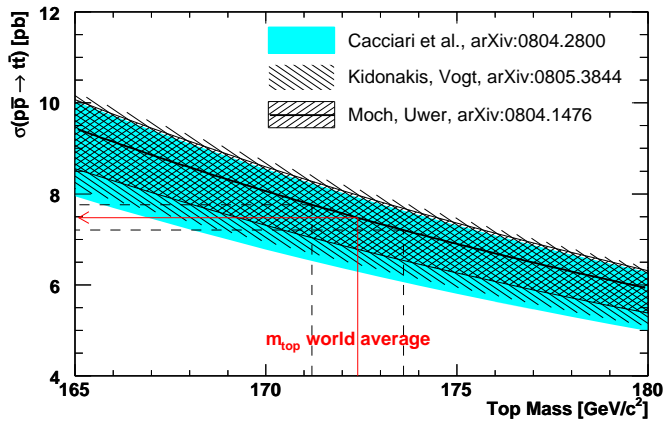}
  \caption{Dependence of the \ttbar production cross section on $m_t$
    in Run~II of the Tevatron. The current world-averaged top quark mass and
    the resulting expected \ttbar production cross section are
    indicated by the vertical and horizontal lines, respectively. The
    predictions are based on the CTEQ6.6M
    PDFs~\cite{Nadolsky:2008zw}.}
  \label{fig:xsecvsmass}
\end{figure}

In spring 2008, Cacciari {\it et al.}~\cite{Cacciari:2008zb} and Kidonakis
{\it et al.}~\cite{Kidonakis:2008mu} updated their predictions using more
recent PDFs such as CTEQ6.6M~\cite{Nadolsky:2008zw}, which had only
little impact on the results.
In addition, Moch and Uwer~\cite{Moch:2008qy} have now performed a
complete NNLL soft-gluon resummation and provide an approximation of
the NNLO cross section also based on CTEQ6.6M. To illustrate the
dependence of the predictions on the top quark mass,
Fig.\ \ref{fig:xsecvsmass} shows the central values and uncertainties
from References~\cite{Cacciari:2008zb,Kidonakis:2008mu,Moch:2008qy} for
Tevatron Run~II versus $m_t$. 
An exponential form, as suggested in Ref. \cite{Catani:1996dj}, is applied in a
fit to the central values and uncertainties for Kidonakis {\it et al.},
while third-order polynomials, as provided by the authors, are used for
the other references. The total uncertainties are obtained by linearly
combining the provided uncertainties.

For the current world-averaged top quark mass of \mtopwa, the predicted
\ttbar production cross section is $7.2 ^{+ 0.8}_{- 0.9}$~pb for
Cacciari {\it et al.}, $7.3 ^{+ 0.8}_{- 0.9}$~pb for Kidonakis {\it et al.}, and
$7.5 ^{+ 0.5}_{- 0.7}$~pb for Moch {\it et al.}. An additional uncertainty
of $\pm 0.3$~pb arises from the uncertainty on the top quark mass 
for all three predictions. It should be noted that these predictions
based on MRST 2006 NNLO PDFs~\cite{Martin:2007bv} yield about 6\%
higher central values and exhibit smaller uncertainties from PDFs.

\begin{figure}[t]
  \centering
  \includegraphics[width=0.7\textwidth,clip=]{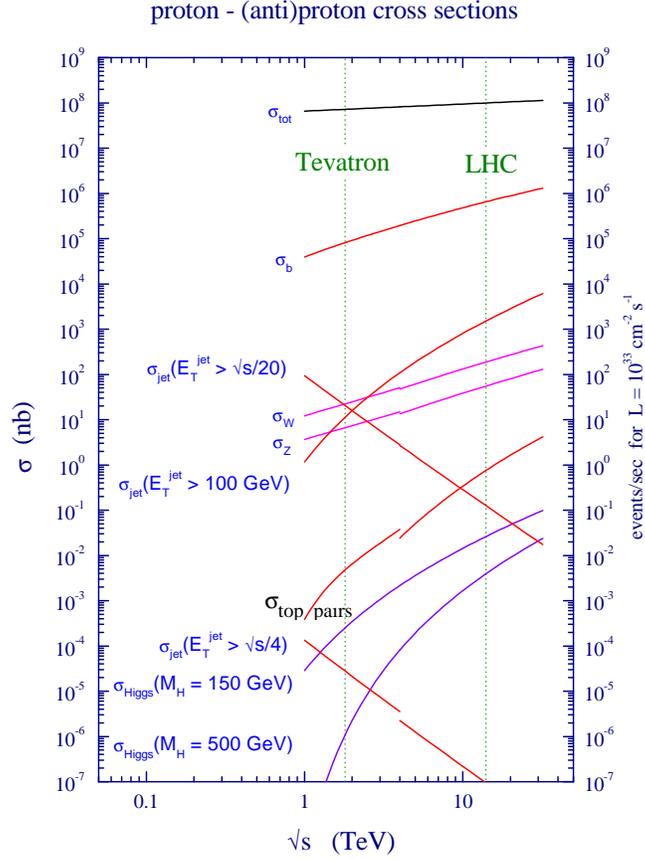}
  \caption{Cross sections for various processes at hadron colliders
    as a function of center of mass energy \cite{Catani:2000xk}. $\sigma_{\rm top pairs}$ 
    denotes the \ttbar production cross section.}
  \label{fig:catani-xsecs}
\end{figure}
A precise measurement of the \ttbar production cross section provides a
test of the predictions for physics beyond the SM.
Together with a precise mass measurement, the self-consistency of the
predictions can also be examined. Because \ttbar production is a major
source of background for single top production (to be discussed in the next
section), standard model Higgs boson production and many other phenomena
beyond the SM, its accurate understanding is crucial for such studies. 

Figure~\ref{fig:catani-xsecs} illustrates the production rates of
various processes versus center of mass energy for proton-antiproton
collisions below $\sqrt{s}= 4$~TeV and for proton-proton collisions
above $\sqrt{s}= 4$~TeV. As can be appreciated from the plot, \ttbar
production is suppressed by ten orders of magnitude relative to the total
interaction rate at the Tevatron and eight orders of magnitude at the
LHC. While the LHC is often referred to as a ``top-factory'' because of the
increased production cross section by two orders of magnitude, extraction of the
signal from the large background is a 
challenge at both hadron colliders, requiring efficient triggers and
selection methods. The \ttbar cross section measurements performed
in Run~II of the Tevatron will be described in Section~\ref{sec:ttbarxsecmeas}.

\subsection{Single top quark production}
\label{sec:singletopprod}
\begin{figure}[h]
  \centering
  \subfigure[]{\label{fig:singlet-feynman-s}\epsfig{figure = 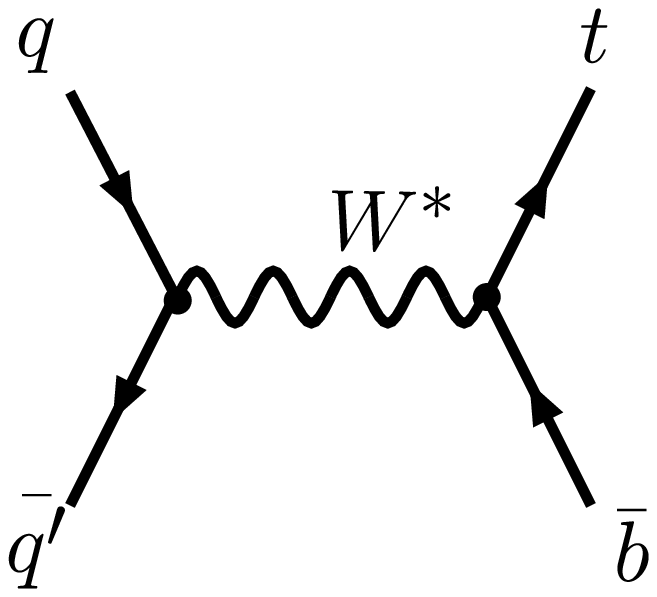, width = 0.25 \textwidth}}\\
  \subfigure[]{\label{fig:singlet-feynman-t}\epsfig{figure = 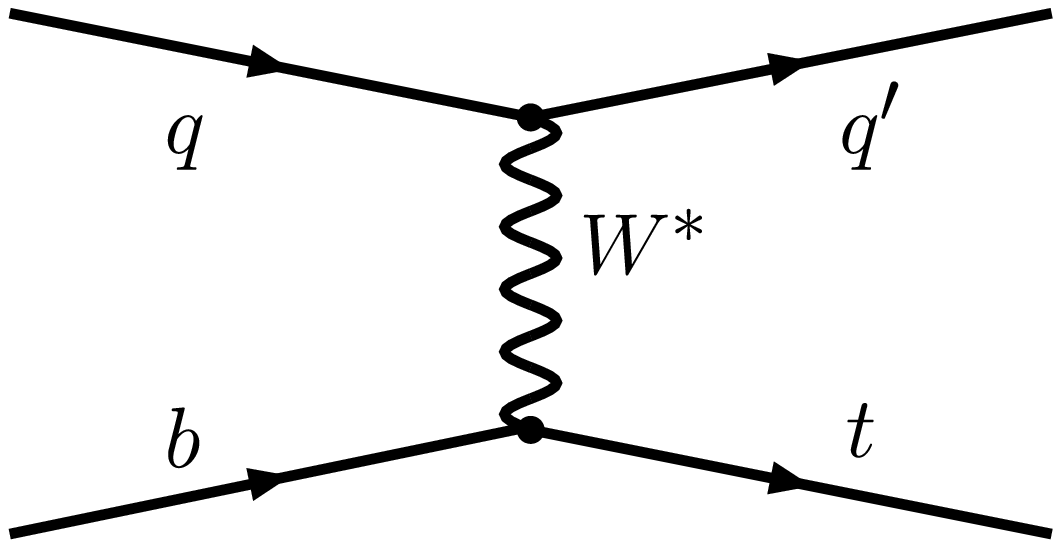, width =
      0.35 \textwidth}}\hspace{1cm}
  \subfigure[]{\label{fig:singlet-feynman-tb}\epsfig{figure = 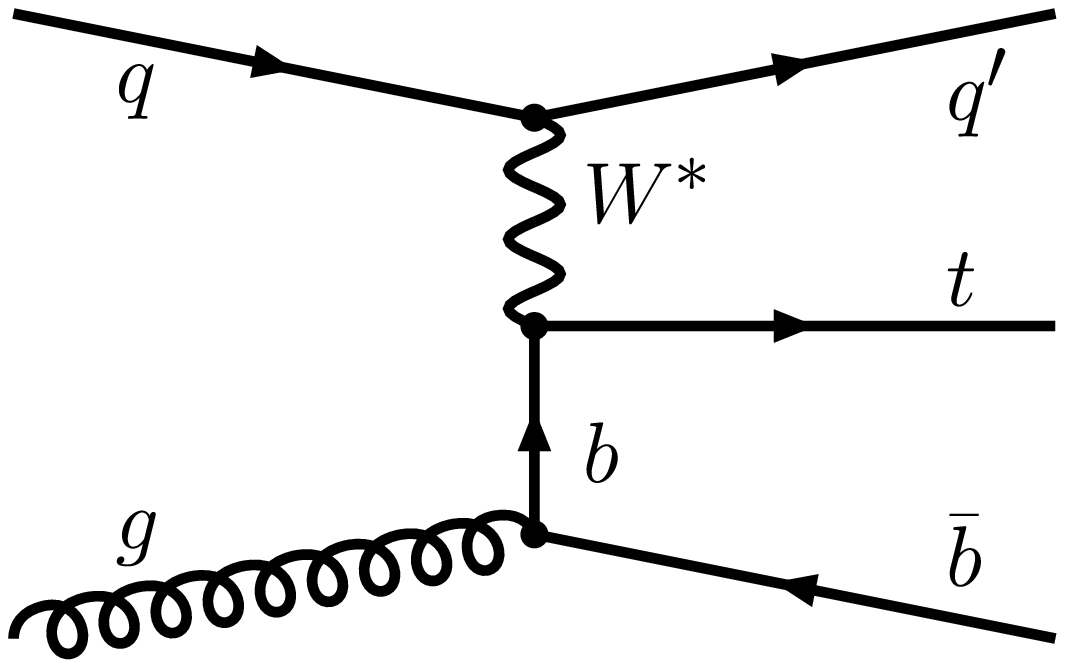, width = 0.3 \textwidth}}\\
  \hspace{1cm}\subfigure[]{\label{fig:singlet-feynman-tW}\epsfig{figure = 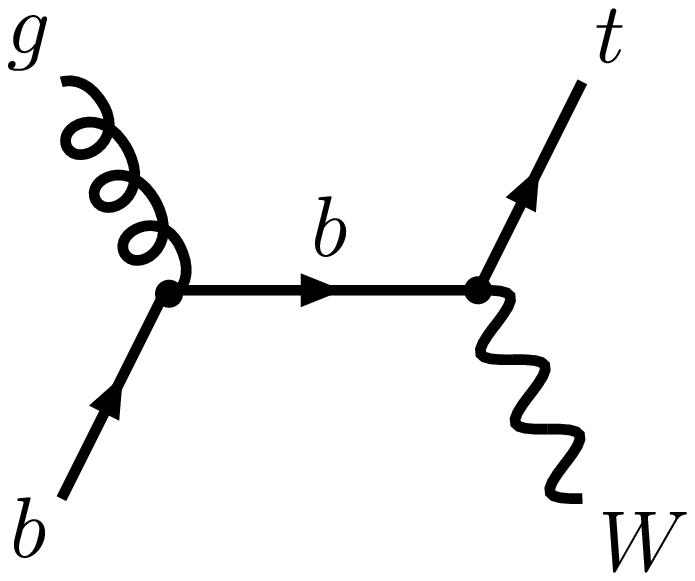, width =
      0.25 \textwidth}}\hspace{1.5cm} 
  \subfigure[]{\label{fig:singlet-feynman-tW2}\epsfig{figure = 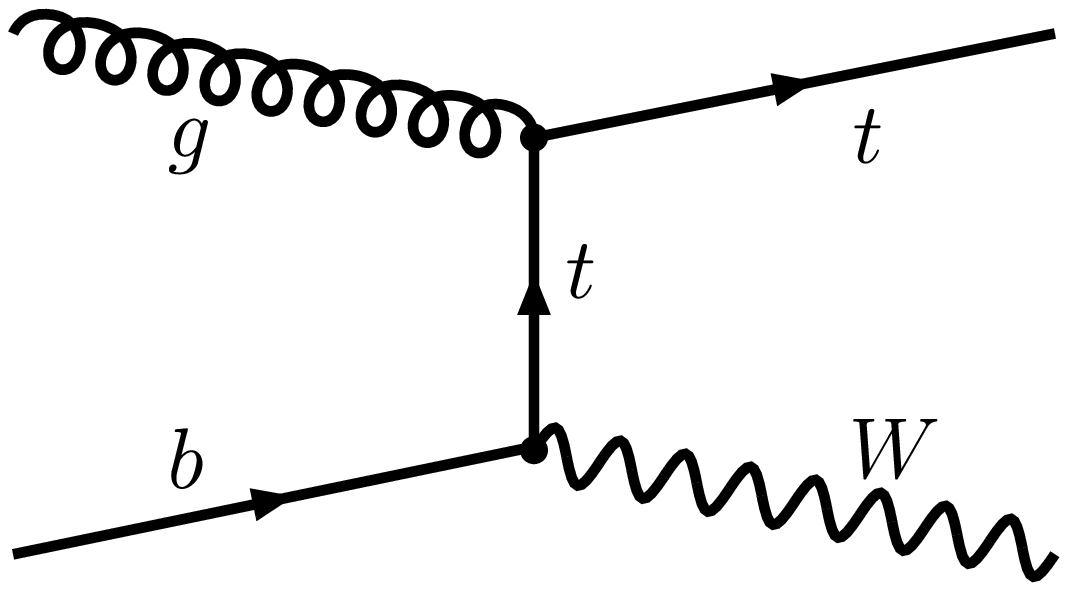, width = 0.3 \textwidth}}
  \caption{Representative Feynman diagrams for electroweak single top
    quark production: (a) $s$-channel, (b,c) $t$-channel and (d,e)
    associated production.}
  \label{fig:singlet-feynman}
\end{figure}
In addition to the strong pair production discussed in the previous
section, top quarks can also be produced singly via the electroweak
interaction through a $Wtb$ vertex (see
Fig.\ \ref{fig:singlet-feynman}). $Wts$ and $Wtd$ vertices are
strongly CKM suppressed (see Section~\ref{sec:topCKM}). There are
three different production modes, classified via the virtuality
(negative of the square of the four-momentum $q$) of the participating $W$ boson
$(Q^2_W = -q^2)$:
\begin{itemlist}

\item The Drell-Yan-like {\em $s$-channel} production proceeds via
quark-antiquark annihilation into a time-like virtual $W$ boson ($q^2
\geq (m_t+m_b)^2 > 0$), as illustrated in
Fig.\ \ref{fig:singlet-feynman-s}: $q\bar{q}'\rightarrow t\bar{b}$
\cite{Cortese:1991fw,Stelzer:1995mi}.

\item In the {\em $t$-channel} ``flavor excitation'' process, a
space-like virtual $W$ boson ($q^2 < 0$) couples to a $b$ quark from the
nucleon's sea to produce a top quark, as shown in
Fig.\ \ref{fig:singlet-feynman-t} for $qb \rightarrow q't$. A higher
order contribution of $\mathcal{O}(\alpha_s)$ comes from gluon
splitting, as depicted in Fig.\ \ref{fig:singlet-feynman-tb}, which is
also referred to as W-gluon fusion for $qg \rightarrow tq'\bar{b}$
\cite{Dawson:1984gx,Willenbrock:1986cr,Yuan:1989tc,Ellis:1992yw}.

\item In {\em associated production}, an on-shell $W$ boson ($q^2 =
m_W^2 $) is produced together with a top quark from a $b$ quark and a
gluon, as illustrated in Figs.\ \ref{fig:singlet-feynman-tW} and \ref{fig:singlet-feynman-tW2} for $gb
\rightarrow tW$
\cite{Ladinsky:1990ut,Moretti:1997ng,Heinson:1996zm,Belyaev:1998dn,Tait:1999cf,Belyaev:2000me}.
\end{itemlist}

In the above discussion, charge conjugate processes are implied for each
production mode, and $q$ represents a light-flavor
quark. All three modes differ in both their initial
and final states, and the processes are simply denoted as 
$s$-channel ($tb$), $t$-channel ($tq, ~tqb$) and associated ($tW$)
production. The corresponding signatures can be used to discriminate
between the production modes: The $s$-channel is characterized by an
additional $b$ quark accompanying the top quark, the $t$-channel by a
forward light quark, and associated production by the decay products of
the $W$ boson in addition to those of the top quark.
Due to the incoming $b$ quark and gluon, the $t$-channel and $tW$-channel
rates are especially sensitive to the corresponding PDFs, which are known with
less precision than the PDFs for the valence quarks of the proton. The 
measured cross sections will therefore provide further constraints on the $b$
quark and gluon PDFs.

The cross sections for all three modes have been evaluated at NLO,
including radiative corrections of $\mathcal{O}(\alpha_s)$:
s-channel~\cite{Smith:1996ij,Mrenna:1997wp,Harris:2002md},
t-channel~\cite{Bordes:1994ki,Stelzer:1997ns,Harris:2002md}, and
$tW$-channel~\cite{Zhu:2002uj,Zhu:2002er}. (The most recent 
references provide differential distributions.)
Subsequent calculations also include top quark decay at NLO for s-channel
\cite{Campbell:2004ch, Cao:2004ap, Cao:2004ky}, $t$-channel
\cite{Campbell:2004ch, Cao:2004ky, Cao:2005pq}, and $tW$-channel
\cite{Campbell:2005bb}, and latest NLO calculations include
higher-order soft-gluon corrections up to NNNLO at NLL accuracy
\cite{Kidonakis:2007wg,Kidonakis:2007ej,Kidonakis:2006bu}.

\begin{table}[t]
  \caption[single-top cross section predictions]{Cross sections
    for $s$-channel ($tb$), $t$-channel ($tq$) and associated
    ($tW$) single-top production at NLO (Sullivan), and NLO, including
    soft-gluon resummations (Kidonakis), expected at the Tevatron and at the LHC,
    for a top quark mass of 175 GeV/c$^2$. (For sources
    of the quoted uncertainties see the text.)}
  \renewcommand{\arraystretch}{1.2}
  \addtolength{\tabcolsep}{-4pt}
  \begin{center}
    \begin{tabular*}{\textwidth}{@{\extracolsep{\fill}}lllllr}\hline
      Hadron Collider &$t/\bar{t}$&$\sigma_{tb}$ [pb]& $\sigma_{tq}$ [pb] & $\sigma_{tW}$ [pb] & Group\\
      \hline\hline
      Tevatron Run~I & \multirow{2}{*}{$t,\bar{t}$}& \multirow{2}{*}{$0.75^{+0.10}_{-0.09}$}&\multirow{2}{*}{$1.46^{+0.20}_{-0.16}$}&\multirow{2}{*}{~~~---}&\multirow{2}{*}{Sullivan~\cite{Sullivan:2004ie}}\\
      ($p\bar{p},~\sqrt{s}=1.8$~TeV) &&&&\\
      \hline\hline
      Tevatron Run~II &\multirow{2}{*}{$t,\bar{t}$}&$0.88^{+0.12}_{-0.11}$&$1.98^{+0.28}_{-0.22}$&~~~---&Sullivan~\cite{Sullivan:2004ie}\\
      \cline{3-6}
      ($p\bar{p},~\sqrt{s}=1.96$~TeV) &&$0.98\pm 0.04$&$2.16\pm 0.12$&$0.26\pm 0.06$&Kidonakis~\cite{Kidonakis:2007wg}\\
      \hline\hline
      &$t$&$6.56^{+0.69}_{-0.63}$&$155.9^{+7.5}_{-7.7}$&~~~---&\multirow{2}{*}{Sullivan~\cite{Sullivan:2004ie}}\\
      LHC &$\bar{t}$&$4.09^{+0.43}_{-0.39}$&$90.7^{+4.3}_{-4.5}$&~~~---&\\
      \cline{3-6}
      ($pp,~\sqrt{s}=14$~TeV) &$t$&$7.2^{+0.6}_{-0.5}$&$146\pm 5$&$41\pm 4$&\multirow{2}{*}{Kidonakis~\cite{Kidonakis:2007wg}}\\
      &$\bar{t}$&$4.0\pm 0.2$&$89\pm 4$&$41\pm 4$&\\
      \hline\hline
     \end{tabular*}
   \end{center}
\label{theory-singlet-xsecs}
\end{table}

Table~\ref{theory-singlet-xsecs} summarizes the expected single-top
production cross sections at the Tevatron and the LHC for the NLO
calculations by Sullivan~\cite{Sullivan:2004ie} (based on the work of
Harris {\it et al.}~in~\cite{Harris:2002md}) and NLO results including
soft-gluon resummations by Kidonakis~\cite{Kidonakis:2007wg} (based on
his work in Refs. \cite{Kidonakis:2007ej,Kidonakis:2006bu} and matching to
the exact NLO results of Harris {\it et al.}~\cite{Harris:2002md} and
Zhu~\cite{Zhu:2002uj,Zhu:2002er}). Both results use current PDFs and
include corresponding uncertainties. 

While top and antitop production are identical at the Tevatron for all
production modes, at the LHC this is only the case for associated
production. Consequently, the results given for the Tevatron include
both top and antitop production but are given separately for the LHC.

The NLO results of Sullivan are based on CTEQ5M1
PDFs~\cite{Lai:1999wy} for their central values. The uncertainties
for PDFs are derived from CTEQ6M~\cite{Pumplin:2002vw}, and added in
quadrature with uncertainties from changes in scale by the usual factors of
two, changes in top quark mass by 4.3 GeV/c$^2$ (using an older
world-averaged $m_t = 178 \pm 4.3$~GeV/c$^2$~\cite{Azzi:2004rc}), and
uncertainties in $b$ quark mass and $\alpha_s$, the latter two being
negligible. The rate dependence on the top
quark mass is approximated as linear and is especially important
for the $s$-channel, since a change from 175~GeV/c$^2$ to the current
world-averaged
$m_t = 172.4$~GeV/c$^2$ raises the rates at the Tevatron by
7\% for the $s$-channel and 5\% for the $t$-channel. The observed 
uncertainties in scale are reduced relative to LO results, and amount to 4-6\% at the
Tevatron and 2-3\% at the LHC.

The NLO calculations of Kidonakis that include higher order soft-gluon
corrections provide single-top production cross sections based on
MRST2004 NNLO PDFs \cite{Martin:2004ir}. The quoted values are
obtained by matching the NLO cross section to the results of Harris
{\it et al.}~\cite{Harris:2002md} and Zhu~\cite{Zhu:2002uj,Zhu:2002er}, and
including the additional soft-gluon corrections up to NNNLO.
Exceptions are the $tW$ rate at the Tevatron, where no corresponding
NLO result is available, and the given value is therefore not matched, and
the $t$-channel rate at the LHC, where no soft-gluon corrections are
considered and an updated NLO result with the quoted PDFs is given
instead. The uncertainties given are derived from varying the scale by
a factor of two, and adding in quadrature PDF uncertainties derived
using the MRST2001E NLO PDFs~\cite{Martin:2002aw}. No uncertainty in $m_t$
is included. At the Tevatron, the $t$-channel uncertainty is dominated
by the uncertainty in PDFs, and corrections from soft-gluon resummations
relative to LO are small ($\approx$5\%). In contrast, the soft-gluon
corrections have a large effect ($>$60\%) for the $s$-channel at the
Tevatron and scale uncertainties dominate over those from PDFs.

At the Tevatron, $t$-channel production dominates the total rate of
single top quark production with a contribution of $\approx$65\%, followed
by $s$-channel production at $\approx$30\%. Associated ($tW$)
production at the Tevatron contributes only $\approx$5\% to the total
rate, and is usually neglected.
At the LHC, $t$-channel production again dominates at $\approx$74\%,
followed now by associated production at $\approx$23\%, while $s$-channel
production contributes only $\approx$3\% because of the missing contribution from
valence antiquarks in the collisions, which will make it difficult to
discriminate this channel from background.
Despite being an electroweak process, single top production has a
cross section of the same order of magnitude as \ttbar production
(of $\mathcal{O}(40\%)$ of the \ttbar rate at both the Tevatron and
the LHC). With only one heavy top quark to be produced, single top production is accessible at
smaller and therefore better-populated momentum fractions of the partons.
Furthermore, no color matching is required for the production.
The fact that the observed yields of single-top and $t\bar{t}$-pairs
are consistent with theory is a major triumph of the SM.

The measurement of single top production offers a check of the top
quark's weak interaction, and direct access to the CKM matrix
element $|V_{tb}|$, as the cross sections in all three production
modes are proportional to $|V_{tb}|^2$. The polarization of the top
quark at production is preserved due to its short lifetime and 
provides a test of the $V-A$ structure of the weak interaction via angular
correlations among the decay products~\cite{Carlson:1993dt,
Mahlon:1996pn,Stelzer:1998ni,Mahlon:1999gz}. All three production
modes provide different sensitivity to various aspects of physics beyond the
standard model (BSM)~\cite{Tait:2000sh}, which makes their independent
reconstruction a desirable goal. The $s$-channel is sensitive to the existence
of new charged bosons (such as $W'$ or charged Higgs) that couple to
the top-bottom weak-isospin doublet, an effect that could be detectable through
an enhancement of the observed cross section. Such effects would not
be observed in the $tW$ mode, where the $W$ boson is on-shell, or in the
t-channel, where the virtual $W$ boson is space-like and cannot go
on-shell as in the $s$-channel. The $t$-channel production rate could be
enhanced via FCNC processes involving new couplings between the
up-type quarks and a boson (Higgs, gluon, photon, $Z$). This would be
hard to observe in the $s$-channel, since there is no 
$b$ quark in the final state, which is essential for discrimination of the signal
in that production mode.
Finally, the $tW$ channel is the only mode that provides a more direct test of
the $Wtb$ vertex since the $W$ boson appears in the final state.

A thorough understanding of single top quark production will also
facilitate the study of processes exhibiting a similar signature such as 
SM $W$-Higgs production or BSM signals to which single top
production is a background process.
Despite a production rate similar to that of $t\overline{t}$, the
signature for single top quark production is much harder to separate
from background, which has delayed first measurements until
very recently. The current analyses at the Tevatron provide first
evidence for production of single top quarks, and this will be described
in Section~\ref{sec:ST}.

\subsection{Top quark decay}
\subsubsection{Top quark CKM matrix elements}
\label{sec:topCKM}
Since the mass of the top quark is larger than that of the $W$ boson,
decays $t\to Wq$, with $q$ being one of the down-type quarks $d, s, b$, 
are dominant. The contribution of each
quark flavor to the total decay width is proportional to the square
of the respective CKM matrix element $V_{tq}$. Utilizing the
unitarity of the CKM matrix and assuming three quark generations, the
corresponding matrix elements can be constrained indirectly at 90\%
confidence level to~\cite{Eidelman:2004wy}:
\begin{eqnarray}
|V_{td}|~ =& 0.0048 &- ~~0.014 \\
|V_{ts}|~ =& 0.037  &- ~~0.043  \\
|V_{tb}|~ =& 0.9990 &- ~~0.9992 . 
\end{eqnarray}
Consequently, the decay $t\to Wb$ is absolutely dominant and will be
considered exclusively throughout this article, unless noted to the contrary.
Potential deviations from the SM decay will be
discussed in Section~\ref{sec:BSMdecay}.

It should be noted that the above constraints on the CKM
matrix elements would change dramatically (especially $V_{tb}$)
if there were more than three quark generations. Assuming the
unitarity of the expanded matrix, the limits become \cite{Eidelman:2004wy}:
\begin{eqnarray}
|V_{td}|~ =& 0    &- ~~0.08      \\
|V_{ts}|~ =& 0    &- ~~0.11      \\
|V_{tb}|~ =& 0.07 &- ~~0.9993 .
\end{eqnarray}
It is therefore important to constrain these matrix elements through
direct measurements, as outlined below.

\begin{figure}[t]
  \centering
  \includegraphics[width=1.0\textwidth,clip=]{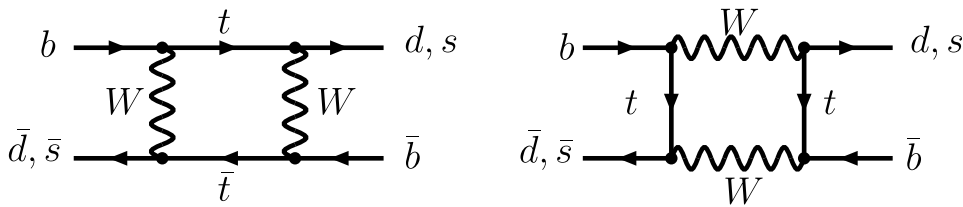}
  \caption{Feynman ``box'' diagrams for $B^0_d-\overline{B^0_d}$ and $B^0_s-\overline{B^0_s}$ mixing.}
  \label{fig:bmesonmixing}
\end{figure}

The $V_{td}$ and $V_{ts}$ matrix elements cannot be extracted from lowest-order
(tree level) top decays in the framework of the standard model, but can
be inferred from B-meson mixing, as shown in
Fig.\ \ref{fig:bmesonmixing}. While all up-type quarks can contribute
in the depicted box diagrams, the contribution from the top quark is
dominant~\cite{Buras:1984pq}. The oscillation frequency given by the
mass difference between heavy and light mass eigenstates, 
$\Delta m_d$ for $B^0_d-\overline{B^0_d}$ and $\Delta m_s$ for
$B^0_s-\overline{B^0_s}$ oscillations, is proportional to the
combination of CKM matrix elements $|V_{tb}^*V_{td}|^2$ and
$|V_{tb}^*V_{ts}|^2$, respectively.
The mass difference for the $B^0_d-\overline{B^0_d}$ system is
$\Delta m_d = 0.507\pm 0.004 {\rm~ps}^{-1}$~\cite{Barberio:2006bi}. 
Using CKM unitarity and assuming
three generations, yielding $|V_{tb}| \approx 1$, translates
into $V_{td} = (7.4\pm 0.8)\cdot 10^{-3}$~\cite{Okamoto:2005zg},
where the uncertainty arises primarily from the theoretical
uncertainty on the hadronic matrix element, which is obtained from
lattice QCD calculations. To reduce these theoretical
uncertainties, a measurement of the ratio, in which some
uncertainties cancel, is more desirable ($\Delta m_d/\Delta m_s \propto
|V_{td}|^2/|V_{ts}|^2$). With the recent first measurement of $\Delta m_s$
in $B^0_s$-oscillations by D0 and CDF at the
Tevatron~\cite{Abazov:2006dm,Abulencia:2006mq}, yielding 17~ps$^{-1} <
\Delta m_s < 21$~ps$^{-1}$ at 90\% C.L.\ and $\Delta m_s =
(17.31^{+0.33}_{-0.18} {\rm (stat.)} \pm 0.07 {\rm (syst.))\ ps}^{-1}$,
this ratio has now been measured for the first time as $|V_{td}/V_{ts}| =
0.208^{+0.001}_{-0.002} {\rm (expt.)} ^{+0.008}_{-0.006} {\rm
(theor.)}$. These results are in good agreement with SM
expectations.

The direct measurement of the $V_{tb}$ matrix element without assuming
three quark generations and unitarity of the CKM matrix is only
possible via single top quark production (described in
Section~\ref{sec:singletopprod}), because the production rate in each
channel is proportional to $|V_{tb}|^2$.
One way to assess the {\em relative} size of $|V_{tb}|$ compared
to $|V_{td}|$ and $|V_{ts}|$ is to measure the ratio $R$ 
of the top quark branching fractions, which 
can be expressed via CKM matrix elements as 
\begin{eqnarray}
R = \frac{{ \cal B}(t \rightarrow Wb)}{{ \cal B}(t \rightarrow Wq)} & = &
\frac{\mid V_{tb}\mid^2}{\mid V_{tb}\mid^2 + \mid V_{ts}\mid^2 + \mid V_{td}\mid^2}  \;.
\end{eqnarray}
Assuming three generation unitarity, the denominator in the above expression
equals one, and constraints on $|V_{tb}|$ can be inferred. The current status
of these measurements is discussed in Section~\ref{sec:Rmeasurement}.

The most precise extraction of the top quark CKM matrix elements 
proceeds via global fits to all available
measurements, imposing the SM constraints of three
generation unitarity, as done by the CKMfitter~\cite{Charles:2004jd}
or UTfit~\cite{Bona:2006ah} groups.
The CKMfitter update for summer 2008 yields~\cite{Charles:2004jd}:
\begin{eqnarray}
|V_{td}|~ &=& 0.00853 ^{+0.00034} _{-0.00027}     \\
|V_{ts}|~ &=& 0.04043 ^{+0.00038} _{-0.00116}     \\
|V_{tb}|~ &=& 0.999146 ^{+0.000047} _{-0.000016} .
\end{eqnarray}

\subsubsection{Decay width of the top quark}
\label{sec:tdecaywidth}
The decay width of the top quark in the SM, including first-order QCD
corrections, can be expressed as
follows~\cite{Jezabek:1988iv,Kuhn:1996ug}:
\begin{equation}
\Gamma_t = \lvert V_{tb} \rvert^2~{G_F~m_t^3\over 8\pi \sqrt{2}} 
\left(1-{m_W^2\over m_t^2}\right)^2
\left(1+2{m_W^2\over m_t^2}\right)
\left[ 1 - \frac{2\alpha _s}{3\pi}  \left({2\pi^2\over3}-{5\over2} \right) \right],
\label{eq:topdecaywidth}
\end{equation}
where the above formula assumes $m_b^2/m_t^2\to 0$, $m_t^2 \gg m_W^2$ and
ignores corrections of $\mathcal{O}({\alpha_s~m_W^2\over\pi~m_t^2})$
and $\mathcal{O}(\alpha_s^2)$. While the above QCD corrections lower the
width by $\approx$10\%, first-order electroweak corrections increase
the width by 1.7\%~\cite{Denner:1990ns,Eilam:1991iz}. However, the
electroweak correction is almost cancelled when the
finite width of the $W$ boson is taken into account, thereby decreasing
the width again by 1.5\%~\cite{Jezabek:1993wk,Jezabek:1994}.
Corrections to the top quark width of $\mathcal{O}(\alpha_s^2)$ have
also been evaluated~\cite{Czarnecki:1998qc, Chetyrkin:1999ju} and
reduce the width by 2\%. Including all these effects, the
decay width is predicted to a
precision of $\approx$1\%. The other SM decays, $t\to Wd$ and
$t\to Ws$, contribute negligibly to the total decay width
$\Gamma_t = \Sigma_q \Gamma_{tq}$ because of proportionality to $|V_{td}|^2$ and
$|V_{ts}|^2$.

Equation~\ref{eq:topdecaywidth} yields the top width to better than 2\% 
accuracy, and the width increases with $m_t$.
For $\alpha_s(m_Z)=0.1176$ and $G_F=1.16637\cdot 10^{-5} {({\rm
GeV/c}^2)}^{-2}$~\cite{Yao:2006px}, $\Gamma_t$ is
1.02/1.26/1.54~GeV for top quark masses of
160/170/180~GeV/c$^2$.

The resulting lifetime of the top quark $\tau_t =\hbar\, \Gamma_t^{-1}
\approx  \hbar\,(1.3{\rm~GeV})^{-1}$ is approximately \mbox{$5\cdot
10^{-25}$~s}, and significantly shorter than the
hadronization time $\tau_{\rm had} \approx \hbar\,\Lambda_{\rm QCD}^{-1} 
\approx \hbar\,(0.2~{\rm
GeV})^{-1} \approx$ \mbox{$ 3\cdot 10^{-24}$~s}. As a consequence, the
top quark decays before it can form hadrons, and in particular there
can be no \ttbar bound states (toponium), as was already pointed out
in the 1980s~\cite{Kuhn:1980gw,Kuhn:1982ua,Bigi:1986jk}.
Nevertheless, although the top quark can generally be considered as
a free quark, residual non-perturbative effects
associated with hadronization should still be present in top quark
events, and the fragmentation and hadronization processes will be
influenced by the color structure of the hard interaction. 

In electron-positron annihilation, top quark pairs are produced in 
color singlet states, so that hadronization before
decay depends mainly on the mass of the top quark and collision energy. In
hadronic \ttbar production, $t$ and $\bar{t}$ are usually produced in
color octet states and form color singlets with the proton and
antiproton remnants. The energy in the color field (or in
the string when using the picture of string fragmentation) is
proportional to the distance between top quark and the remnant. If a
characteristic length of about 1~fm is reached before the top quark
decays, light hadrons can materialize out of the string's energy. The
possibility for such string fragmentation depends
strongly on the center of mass energy in the hadron collisions. For
Tevatron energies, this can be neglected~\cite{Orr:1990nh}, while it may
be more important at LHC energies, where top quarks are produced with
sizeable Lorentz boosts. Since heavy quarks have hard fragmentation
functions and the fractional energy loss of the top quarks is therefore
expected to be small, it will be difficult to experimentally establish
these effects directly, even at the LHC. In case no string
fragmentation takes place before the top quark decays, long-distance QCD
effects will still connect the decay products of the top quark.

With top quark mass measurements aiming at uncertainties of $\leq 1$
GeV/c$^2$, it becomes more and more important to assess the
impact of such non-perturbative effects on the measurements. One
example that may play an important role in this context is the
possibility of color reconnections before hadronization, and the
corresponding modeling of the underlying event (beam-remnant
interactions)~\cite{Skands:2007zg,Wicke:2008iz}.

\subsubsection{Helicity of the $W$ boson}
\label{sec:Wheltheory}
Top quark decay in the framework of the standard model proceeds
via the left-handed charged current weak interaction, exhibiting a
vector minus axial vector ($V-A$) structure. This is reflected in the
observed helicity states of the $W$ boson, which can be exploited to
examine the couplings at the $Wtb$ vertex
\cite{Kane:1991bg,Dalitz:1991wa,Nelson:1997xd}.

\begin{figure}[t]
  \centering
  \includegraphics[width=0.75\textwidth,clip=]{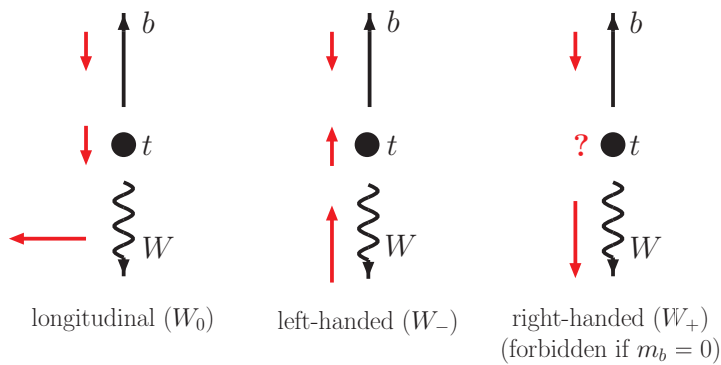}
  \caption{Angular momentum conservation in top quark decay does not
  allow right-handed $W$ bosons when $b$ quarks are assumed massless.}
  \label{fig:Whelsketch}
\end{figure}

The emitted $b$ quark can be regarded as massless compared to the top
quark, and hence expected to be predominantly of negative helicity (left-handed), meaning
that its spin points opposite to its line of flight.
The emitted $W$ boson, being a massive spin-1 particle, can assume any of
three helicities: one longitudinal ($W_0$) and two transverse states
($W_-$, left-handed and $W_+$, right-handed). To
conserve angular momentum in the $t\to Wb$ decay, the spin projection of the $W$
boson onto its momentum must vanish if the $b$ quark's spin points
along the spin of the top quark, while a left-handed $W$ boson is
needed if the $b$ quark's spin points opposite to the spin of the top
quark. In the limit of a massless $b$ quark, a right-handed $W$ boson
cannot contribute to the decay, as illustrated in
Fig.\ \ref{fig:Whelsketch}. For the decay of an antitop quark,
a left-handed $W$ boson is forbidden.

At lowest ``Born''-level, the expected fractions of decays with different $W$
boson helicities, taking the finite $b$ quark mass into account, are
given by \cite{Fischer:2000kx}:
\begin{eqnarray}\label{eq:WHel-Born}
  f_0 &=& \Gamma_0/\Gamma_t=\frac{(1-y^2)^2-x^2(1+y^2)}
  {(1-y^2)^2+x^2(1-2x^2+y^2)}\approx\frac1{1+2x^2}\\ 
  f_- &=& \Gamma_-/\Gamma_t=\frac{x^2(1-x^2+y^2+\sqrt\lambda)}
  {(1-y^2)^2+x^2(1-2x^2+y^2)}\approx\frac{2x^2}{1+2x^2}\\
  f_+ &=& \Gamma_+/\Gamma_t=\frac{x^2(1-x^2+y^2-\sqrt\lambda)}
  {(1-y^2)^2+x^2(1-2x^2+y^2)} \approx y^2\frac{2x^2}{(1-x^2)^2(1+2x^2)}
\end{eqnarray}
where the scaled masses $x=m_W/m_t$, $y=m_b/m_t$ and the
``K\"all\'en''-type function $\lambda=1+x^4+y^4-2x^2y^2-2x^2-2y^2$ were
used. Inserting $m_t=175$~GeV/c$^2$, $m_W=80.419$~GeV/c$^2$, and a
pole mass of $m_b=4.8$~GeV/c$^2$, the partial helicity rates are found
to be \cite{Fischer:2000kx}:
\begin{eqnarray}\label{eq:WHel-SMvalues}
 f_0 = 0.703,\,\, f_- = 0.297,\,\, f_+ = 0.00036.
\end{eqnarray}
$m_b\ne0$ results in a reduction of $f_0$ and $f_-$ at the
per mill level. Including one-loop QCD corrections
\cite{Fischer:2000kx}, electroweak one-loop corrections and finite
width corrections \cite{Do:2002ky} leads basically to a cancellation
of the last two corrections, as was already mentioned in
Section~\ref{sec:tdecaywidth}, and a change in partial helicity rates
$f_0$ and $f_-$ at the 1-2\% level. The right-handed
helicity fraction $f_+$ remains at the per mill level with these
corrections included. Consequently, any observation of $f_+$ at the
percent level would signal the presence of physics beyond the standard model.

Using a more general extension to the standard model
$Wtb$ interaction Lagrangian, assuming that both the $W$ boson and $b$ quark are
on-shell, leads to \cite{Kane:1991bg}:
\begin{eqnarray}
\mathcal{L}\!=\! \frac{g}{\sqrt 2}\left[ W^-_\mu \bar b
\gamma^\mu \left( f_1^L P_L \!+\! f_1^R P_R \right) t \!-\! \frac{1}{m_W}
\partial_\nu W^-_\mu \bar b \sigma^{\mu\nu} \left( f_2^L P_L \!+\! f_2^R
P_R \right) t \right]\!+\! h.c.
\label{Wtbvertex} 
\end{eqnarray}
where $P_{R(L)}$ are the right- and left-handed chiral projectors
$P_{R(L)} = \frac{1}{2} (1\pm \gamma^5)$ and $i\sigma^{\mu\nu} =
-\frac{1}{2}[\gamma^\mu,\gamma^\nu]$.
This model-independent extension has four form factors
$f_{1/2}^{R/L}$, and includes the standard model as a special case,
with $f_1^L=1$ (left-handed vector coupling) and the other form
factors (right-handed vector, and left- and right-handed
tensor couplings) vanishing. These four general couplings for the $Wtb$
vertex can be determined by measuring four
observables sensitive to this interaction: the $W$ helicity fractions
$f_0$ and $f_+$ in \ttbar events, and the single-top production cross sections in the $s$-
and $t$-channel. This model-independent determination of the parameters of the
$Wtb$ vertex can in turn be used to distinguish between
different models proposed for EWSB \cite{Chen:2005vr}.

\section{Experimental Setup}
This section describes the experimental ingredients that are needed
to study top quarks. Since this review focusses mainly on results
obtained at Run~II of the Tevatron, only the corresponding accelerator
and detector setups are discussed. The experimental setup for Run~I 
can be found, for example, in Ref.~\cite{Bhat:1998cd}.

The Tevatron collider is discussed in the first part of this chapter,
followed by a description of the two general-purpose detectors CDF and
D0 surrounding the two interaction points where protons and
antiprotons are brought to collision. Subsequently, the
reconstruction and identification of particles produced in the
collisions are briefly reviewed before the resulting experimental
signatures of top quark events are described. Finally, the Monte Carlo
(MC) simulation tools needed to model interactions in the detectors are
considered.

\subsection{The Tevatron collider}
\label{sec:Tevatron}
\begin{figure}[t]
  \centering
  \includegraphics[width=0.9\textwidth,clip=]{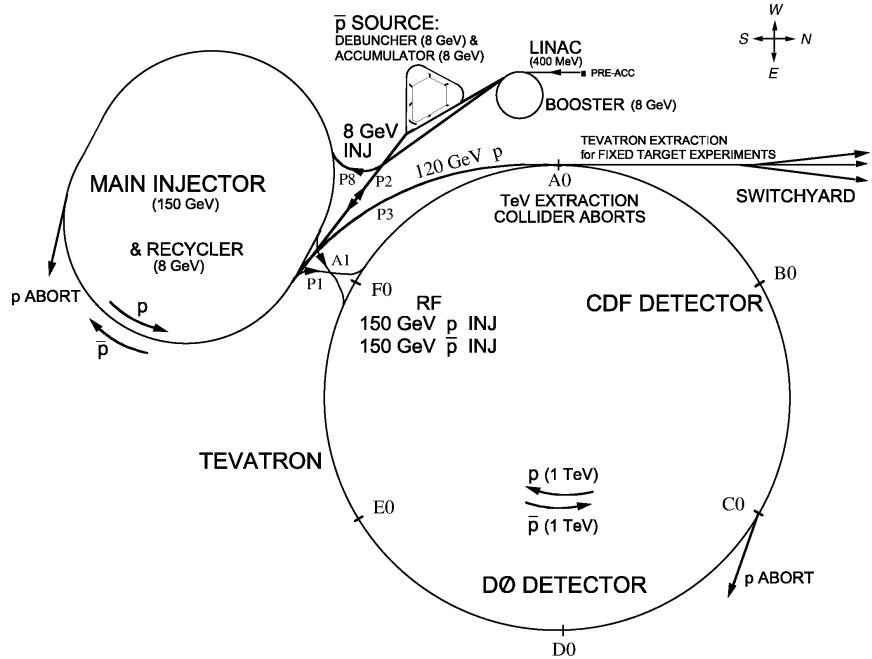}  
  \caption{The Fermilab Run~II accelerator complex consisting of the Tevatron
  $p\bar{p}$ collider and its pre-accelerators~\cite{McGinnis:2005nu,
  wwwbeamsdivision}.}
  \label{fig:Tevatron}
\end{figure}

The Tevatron collider is part of the Fermi National Accelerator
Laboratory (Fermilab, FNAL) in Batavia, Illinois, close to Chicago.
Until the Large Hadron Collider at CERN starts operation, the Tevatron
remains the particle accelerator with the highest center of mass
energy worldwide. Here, 36 bunches of protons and antiprotons with a spacing
of 396~ns are brought to collision at $\sqrt{s}= 1.96$~TeV at
the two interaction points where the multi-purpose detectors CDF and
D0 reside. As illustrated in Fig.\ \ref{fig:Tevatron}, the Tevatron
is the final stage in a chain of eight pre-accelerators and
storage rings~\cite{McGinnis:2005nu,Tan:2005dw,Moore:2007zza,
wwwbeamsdivision}.

Beam protons are generated using a magnetron surface plasma
source that produces H$^-$ ions from hydrogen gas~\cite{Moehs:2005ca}. The
H$^-$ ions are then accelerated to 750~keV in a Cockcroft-Walton
electrostatic accelerator, followed by a linear accelerator, bringing
the ions to 400~MeV. Using a carbon stripping foil, both electrons are
removed from the H$^-$ ions, and the resulting protons are then
accelerated to 8~GeV within 33~ms in the first of five synchrotrons,
called the ``Booster'', which has a ring circumference of 475~m. All these
components are often referred to as the ``Proton Source''.

Acceleration continues in the oval Main Injector synchrotron that has a
circumference of 3.3~km. Depending on their ultimate use, protons are
either brought from 8~GeV to 120~GeV within 2~s for fixed-target operation
(including the production of antiprotons) or to 150~GeV within 3~s for
injection into the Tevatron. With a ring radius of 1~km, the Tevatron
is the final and largest synchrotron at Fermilab. It accelerates
protons and antiprotons in a single beam pipe from 150~GeV to 980~GeV
in about 85~s.

For the production of antiprotons, 120~GeV protons from the Main
Injector are directed every two seconds at a Nickel target, producing
one 8~GeV antiproton for every $\mathcal{O}(50,000)$ incident protons,
in total $\mathcal{O}(10^8)$ per pulse. These antiprotons are focused
into a beamline using a Lithium lens %
and separated from the other produced particles with a pulsed
dipole magnet used as a charge-mass spectrometer. Once transferred to the
``Debuncher'' ring, the large momentum spread of the antiprotons is
reduced using radio-frequency bunch rotation~\cite{Griffin:1984yf} and
stochastic cooling~\cite{Mohl:1980jb}, before the beam is passed to
the ``Accumulator'' ring, where the antiprotons are collected (``stacked'')
and cooled further. For collider operation, approximately
30,000 such cycles are needed. The Debuncher and Accumulator are both 8~GeV
rounded, triangle-shaped, concentric synchrotrons with circumferences of
505~m and 474~m, respectively, and together with the target station
are referred to as the ``Antiproton Source''.

To operate at optimal stacking rates, antiprotons are transferred 
every few hours from the Accumulator to the ``Recycler'' (an 8~GeV storage
ring housed in the Main Injector tunnel), providing both stochastic and
electron cooling~\cite{Budker:1967sd} and thereby improved beam quality.
As the name implies, the Recycler was originally planned to
reuse antiprotons from the Tevatron, but this was abandoned
in favor of large stashes ($6\cdot 10^{12}$) of
antiprotons of high beam quality~\cite{Derwent:2007zz}. The 8~GeV
antiprotons from either Accumulator or Recycler are accelerated in the
Main Injector to 150~GeV for injection into the Tevatron, where they
together with the protons are ramped up to 980~GeV for collisions. The
bunch spacing of 396~ns corresponds to a collision rate of 2.5~MHz.
With 36 out of 53 bunches being filled, the average rate is
reduced to 1.7~MHz.

A typical store at the Tevatron contains about $9\cdot
10^{10}$ antiprotons and $26\cdot 10^{10}$ protons per bunch.
Characteristic r.m.s.\ bunch dimensions are 45~cm (50~cm) in the
longitudinal and 16~$\mu$m (28~$\mu$m) in the transverse
direction for antiprotons (protons)~\cite{PDG2008}, respectively. 
A store lasts typically 16 to 24 hours for data taking 
(governed by the remaining instantaneous
luminosity versus the one achievable with a new store) before
the beam is dumped and the Tevatron refilled (within two to three
hours). The increased antiproton stacking rates achieved recently provide
shorter overall turnaround times and store durations while raising
initial luminosities, thereby enabling the maximization of the delivered
luminosity per time period.

The Tevatron is performing well and keeps setting new world
records on peak luminosity at a hadron collider. As of July
2008, the record is $3.2\cdot
10^{32}$~cm$^{-2}$s$^{-1}$~\cite{wwwbeamsdivision}.
\begin{figure}[t]
  \centering
  \includegraphics[width=0.49\textwidth,clip=]{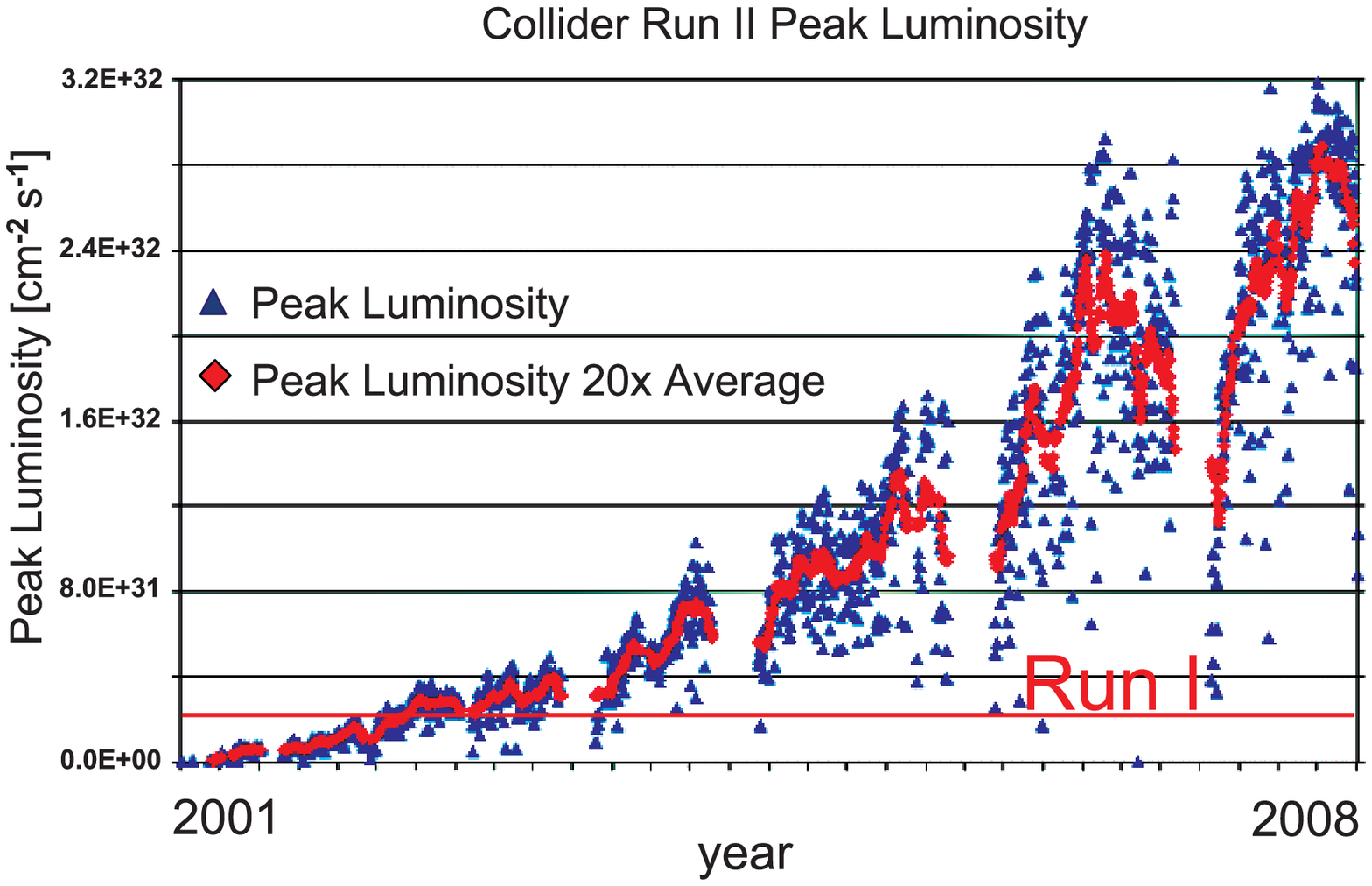}
  \includegraphics[width=0.49\textwidth,clip=]{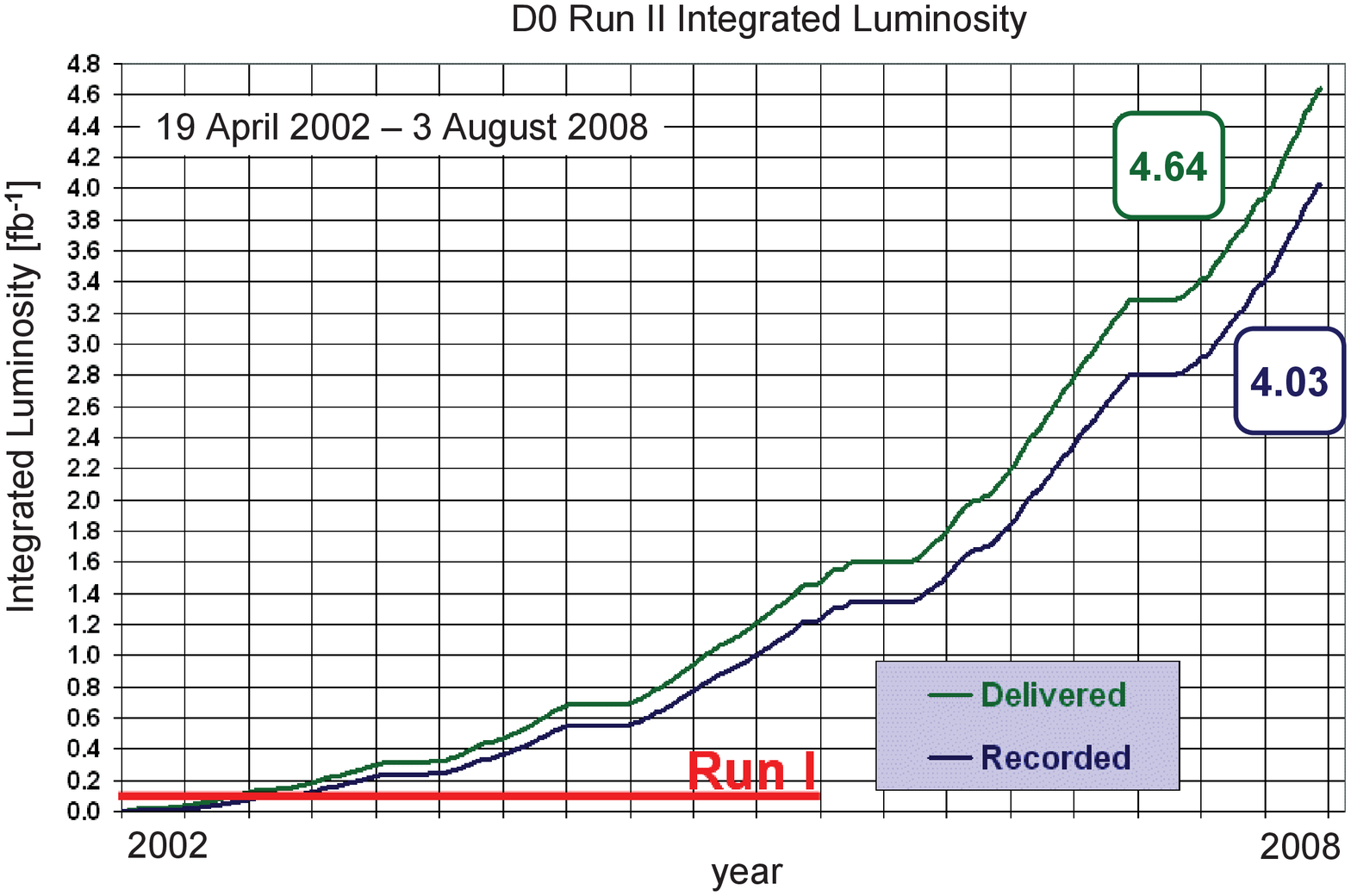}
  \caption{Left: Peak luminosities achieved at Run~II Tevatron
    versus time~\cite{wwwpeaklumi}. Right: Integrated
    luminosity delivered by the Tevatron and recorded by the D0
    experiment in Run~II versus time~\cite{wwwdellumi}.}
  \label{fig:TevLumi}
\end{figure}
For comparison, during Run~I, from 1992 to 1996
at a center of mass energy of 1.8~TeV, the record peak luminosity was
$0.2\cdot 10^{32}$~cm$^{-2}$s$^{-1}$, and both experiments recorded an
integrated luminosity of $\approx$0.1~fb$^{-1}$, respectively. As illustrated in
Fig.\ \ref{fig:TevLumi}, since the beginning of Run~II in 2001, 
each experiment has recorded more than 4~fb$^{-1}$,
and up to half of the total Run~I luminosity is now collected by
the experiments in one single week.
The analyses discussed in this review utilize datasets up to an
integrated luminosity of 2.8~fb$^{-1}$.

Until the currently scheduled end of Run~II in October 2009, the Tevatron is
expected to deliver more than 6~fb$^{-1}$ to each experiment,
with possible improvements on that value crucially dependent on the
achievable antiproton stacking rates~\cite{Moore:2007zza}. An
extension of Tevatron running into 2010 is currently being
discussed and could increase the integrated luminosity by an additional 2~fb$^{-1}$.

\subsection{The collider experiments}
 \label{sec:Detectors}
\begin{figure}[t!]
  \centering
  \includegraphics[width=0.75\textwidth]{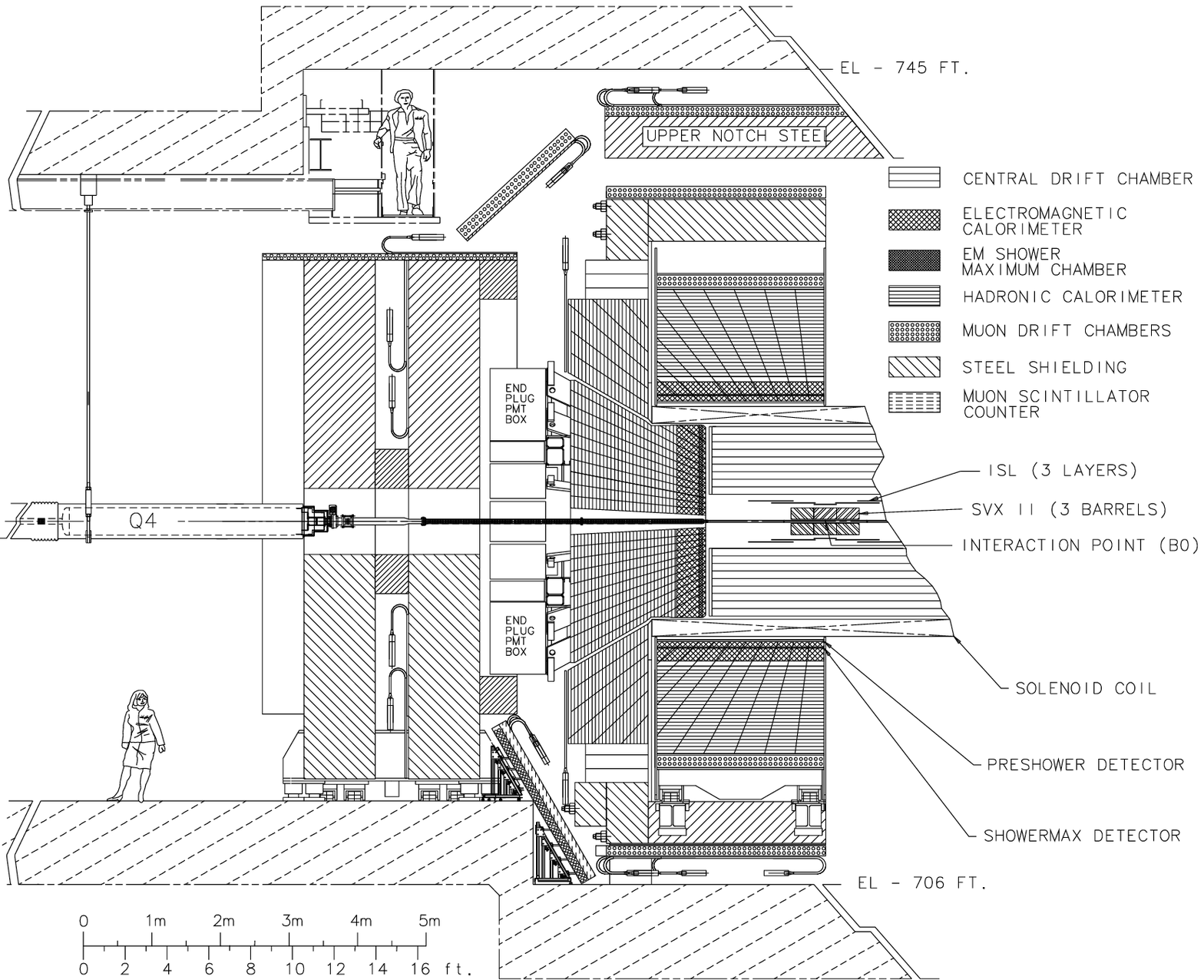}\\
  \vspace{.5cm}
  \includegraphics[width=0.85\textwidth]{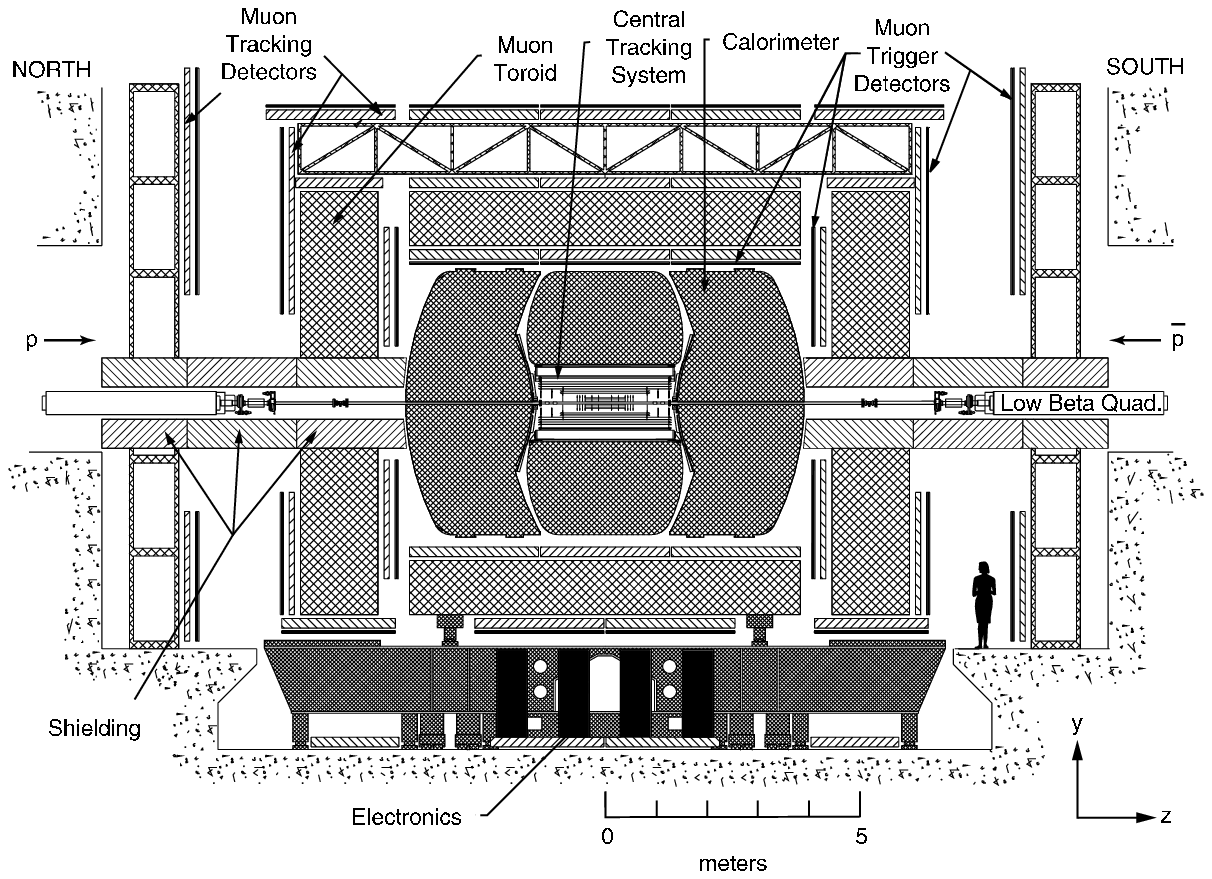}
  \caption{Cross section views of the CDF detector (top,
  \cite{Abulencia:2005ix}) and the D0 detector (bottom,
  \cite{Abazov:2005pn}).}
  \label{fig:Detectors}
\end{figure}

Both general-purpose detectors CDF and D0 follow the generic layout of
a collider detector in having their subdetectors arranged
symmetrically in layers around the interaction point and beam pipe (see
Fig.\ \ref{fig:Detectors}). The inner detectors are arranged in
concentric cylindrical layers, with charged-particle tracking systems
of low mass surrounded by solenoidal magnets defining the core. These
are enclosed by electromagnetic and hadronic calorimeters that provide
energy measurements and identification of electrons, photons and
hadrons. The outer systems are dedicated to muon detection,
relying on the penetration capabilities of muons past all inner detectors.

Both detectors use a right-handed coordinate system with the origin at the
center of the detector and the
$z$-axis pointing along the direction of the proton beam. The
transverse plane is spanned by the $y$-axis, which points vertically
upwards, and the $x$-axis that points away from the center of the
Tevatron. Positions in the transverse plane are frequently described
using the azimuthal angle $\phi$ with respect to the $x$-axis, $\phi =
\arctan \frac{y}{x}$, and radius $r = \sqrt{x^2+y^2}$. Based on the
polar angle $\theta$ relative to the $z$-axis, the pseudo-rapidity
$\eta$ is defined as $\eta = -\ln (\tan \frac{\theta}{2})$. For
massless particles (or in the ultra-relativistic case where masses can
be neglected), the pseudo-rapidity is equivalent to the rapidity $y =
\frac{1}{2} \ln [(E+p_z)/(E-p_z)]$, which is additive under
parallel Lorentz transformations, resulting in Lorentz-invariant
rapidity differences $\Delta y$. The distance of two objects in the
$\eta-\phi$ plane is usually denoted as $\Delta R =
\sqrt{\Delta\eta^2 + \Delta\phi^2}$. To
differentiate between variables calculated with respect to
$p\bar{p}$ collision point and the center of the
detector, the latter
is often denoted with a subscript ``det'' to indicate origin of the detector
coordinate system. In this review, unless indicated to the contrary, $\eta$
is used to refer to $\eta_{\rm det}$.

Enclosing the luminous region which exhibits a Gaussian width of
approximately 25~cm, both CDF and D0 have their innermost
silicon microstrip trackers. These provide vertexing and tracking
capabilities extending to pseudo-rapidities of $|\eta| \le 2$
and $|\eta| \le 3$ for CDF and D0, respectively. CDF complements its tracking system with
a cylindrical open-cell drift chamber that provides 96 track measurements
for $|\eta| \le 1$, while D0 utilizes a scintillating-fiber tracker
consisting of eight cylindrical layers with two overlapping 835~$\mu$m diameter
fiber doublets each, providing coverage for $|\eta| \lsim 1.7$.
Both tracking systems are enclosed by superconducting solenoidal
magnets that provide magnetic fields of 1.4~T (CDF) and 1.9~T
(D0) along the beamline for measuring transverse momenta ($p_T$)
of charged particles.

Supplemental particle identification systems are placed inside
and also outside of the magnet for the CDF and D0 detectors.
Within the magnet, CDF employs a Time-of-Flight detector based on
plastic scintillator panels covering $|\eta| \lsim 1$, mainly to
discriminate low-energetic ($p < 1.6$~GeV/c) charged pions from
kaons (for tagging heavy-flavor). Outside of the magnet, CDF
uses scintillator tiles for early sampling of electromagnetic showers
to improve electron and photon identification in the central detector.
D0 uses central ($|\eta| \le 1.3$) and forward ($1.5 \le |\eta| \le
2.5$) preshower detectors consisting of several layers of plastic scintillator
strips to enhance electron and photon identification.

Sampling calorimeters with an inner electromagnetic and an outer
hadronic section enclose all inner subdetectors, providing
energy measurement and identification capabilities for photons,
charged leptons and hadrons. CDF uses lead/iron-scintillator sampling
devices covering pseudo-rapidities $|\eta| \lsim 3.6$, while D0 uses
mainly depleted uranium (U$^{238}$) as absorber material and liquid argon as
active medium for nearly compensating calorimetry within $|\eta| \lsim
4.2$. Between the central and endcap calorimeter-cryostats \mbox{($1.1
\le |\eta| \le 1.4$)}, layers of scintillating tiles provide additional
sampling of showers for D0.

The outermost subdetectors serve to identify muons. These devices are based on
the fact that muons rarely interact or radiate, but rather
traverse the calorimeter as minimum ionizing
particles that rarely generate electromagnetic or hadronic showers.
CDF and D0 employ scintillators and drift tubes for muon detection
within $|\eta| \le 1.5$ and $|\eta| \le 2$, respectively. D0
has in addition 1.8~T solid-iron toroidal magnets between the detection layers
to provide stand-alone measurements of muon momentum that are
independent of the central tracking system.

The luminosity for CDF and D0 is measured, respectively, using Cherenkov and
plastic scintillation counters covering $3.6 \le |\eta| \le 4.6$
and \mbox{$2.7 \le |\eta| \le 4.4$}, respectively. To select events of
interest from the effective 1.7~MHz bunch-crossing rate, both
experiments employ three-level trigger systems of dedicated hardware
at the initial levels and commercial processor farms at the later (higher) levels.
Based on information from tracking, calorimetry and muon systems,
events are recorded at a rate of approximately 100~Hz for storage and
further processing.

More detailed descriptions of the CDF and D0 detectors can be found in
Refs. \cite{Blair:1996kx,Acosta:2004hw,Abulencia:2005ix}
and~\cite{Abazov:2005pn}, respectively.

\subsection{Object reconstruction}
\label{sec:objreco}
To analyze top quark events and to study properties of the top quark,
the fundamental objects resulting from top quark decays must first be
reconstructed. This section gives a brief overview of the objects to
be considered, and how they can be reconstructed in the detectors.
More information on such reconstruction, specific to the CDF and D0
experiments, can be found, for example, in
Refs.~\cite{Abulencia:2006kv, Abazov:2008kt}. As will become clear,
analyses of the top quark utilize all detector components and
therefore need a thorough understanding of their performance and
calibration.

\subsubsection{Primary vertices}
The point of the primary hard scatter is referred to as the primary
vertex and is determined through a fit of well-measured emerging
tracks and beamline constraints to a common origin. With increasing
luminosity, the average number of interactions per bunch crossing
increases as well, leading to the reconstruction of multiple
primary-vertex candidates, only one of which will generally be
compatible with the hard interaction of interest. The selection of the
primary vertex can be based, for example, on the presence of an
energetic lepton, the (maximal) scalar sum of $p_T$ of associated
tracks, or the (lowest) compatibility with being a ``minimum-bias''
interaction, based on track $p_T$
templates~\cite{Acosta:2004hw,Abazov:2008kt}.

The primary vertex is the origin of all objects produced in the
interaction, including those from top quark and their subsequent $W$
boson decays, both of which cannot be separated from the primary
vertex within the detector resolution due to the extremely short
lifetimes (see also Section~\ref{sec:toplifetime}). The primary vertex
is also used as the origin of coordinates for evaluating kinematic
variables of the $p\bar{p}$ collision.

\subsubsection{Charged leptons}
Leptonically decaying $W$ bosons are a source of isolated energetic
charged leptons that can be measured well with the tracking,
calorimeter and muon systems described in the previous section. Such
leptons are part of the event signatures for several top quark decay
modes (see Section~\ref{sec:topsignatures}) and key in selecting these
events at the trigger stage. However, $\tau$ leptons play a special
role due to their decay characteristics. They decay leptonically 35\%
of the time, yielding electrons or muons (and two neutrinos) that on
average have far lower momenta than $e$ or $\mu$ from direct
$W\to\ell\nu$ decay, but are otherwise hard to distinguish due to the
relatively short $\tau$ lifetime. Consequently, such decays are
usually included in the event selections for electrons and muons which
are then referred to as ``leptonic'' final states of the $W$ boson.
Decays of $\tau$ into hadrons (and a neutrino) are treated separately
and are discussed further below.

In the context of this review, the term ``leptons'' refers only to
electrons and muons, unless specified to the contrary. Their
reconstruction proceeds as follows:
\begin{itemlist}
\item {\bf Electrons} leave a track in the inner tracking system and
  form showers, mainly in the electromagnetic part of the calorimeter.
  These are reconstructed as clusters of energy deposition in the
  electromagnetic calorimeter and matched to reconstructed tracks.
  Further requirements include selections based on the fraction of
  energy deposited in the electromagnetic calorimeter, isolation from
  other energy depositions in a cone defined by the electron
  candidate, the transverse and longitudinal distribution of the
  shower, and the $E/p$ ratio of cluster energy and the reconstructed
  track momentum.
\item{\bf Muons} are identified via their characteristic penetration
  and minimum ionization along their path in the calorimeter. They are
  reconstructed by matching central tracks to track segments in the
  outer muon system and must be consistent with originating from a
  primary vertex. Cosmic-ray muons are suppressed via timing
  requirements. Isolation of muon trajectories can be imposed both on
  the charged-track candidate and within the calorimeter.
\end{itemlist}
The misidentification of isolated leptons has different origins for
electrons and muons. Assuming contributions from hadrons ``punching
through'' the calorimeter are negligible, false isolated muons arise
mainly from muons emitted in semileptonic decays of heavy quarks where
an associated jet (see below) is not reconstructed or the muon emerges
outside of the cone of the jet. While such semileptonic heavy-flavor
decays also contribute false isolated electrons, significant
contributions arise here as well, for example, from jets with large
electromagnetic fractions that can mimic electrons, photon conversions
to $e^{+}e^{-}$, or photons that overlap a random track. Such
instrumental background processes are usually estimated from data, as
realistic simulation of these effects is quite difficult.

The energy scale and resolution for leptons can be assessed by
studying the reconstructed mass of the $Z$ boson in $Z\to\ell\ell$
events.
$Z$ boson decays are also useful for studying lepton identification
efficiencies with the ``tag and probe'' method in which one lepton is
required to be well-identified (the ``tag''), thereby providing a
reasonably pure sample of $Z$ bosons, while the second lepton serves
as a probe for the efficiency being studied.

\subsubsection{Quark and gluon jets}
\label{sec:qgjets-JES}
The hadronization of quarks and gluons leads to collimated showers of
hadrons, referred to as ``jets''. The jet axis is highly correlated
with the original parton's direction. While it is not possible to
differentiate between quark and gluon jets on a per-object basis, they
can be distinguished on a statistical basis because of small
differences in shower shape (gluon jets tend to be wider and contain
more ``soft'' particles).

\begin{figure}[!t]
  \begin{center} 
    \includegraphics[width=.48\textwidth]{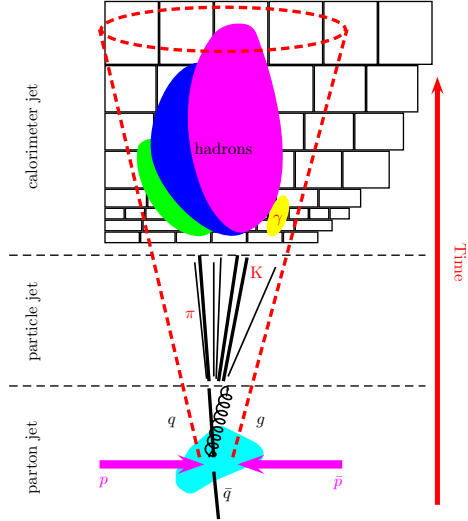}
    \caption{Illustration of the evolution of a calorimeter jet from
      an initial parton~\cite{wwwd0jes}. The dashed lines represent
      the jet cone.}
      \label{fig:JES}
  \end{center}
\end{figure}
Jets are reconstructed from their energy depositions in the cells of
the calorimeter using cone algorithms~\cite{Abe:1991ui,Blazey:2000qt}
that combine cell energies within a cone of fixed radius $\Delta R$
(see Fig.\ \ref{fig:JES}). The size of the cone radius is a compromise
between collecting a high fraction of the original parton's energy and
resolving the energy depositions of close-by partons, especially in
busy \ttbar events. D0 uses a cone size of $\Delta R = 0.5$, while CDF
uses $\Delta R = 0.4$.
The measured jet energies are converted into particle-level energies
through jet energy scale (JES)
corrections~\cite{Bhatti:2005ai,Hegeman:2008} that take into account
effects such as the presence of energy depositions not originating
from the hard scattering process (noise in the calorimeter, multiple
interactions, \dots), particles within the jet cone that deposit
energy in the calorimeter but outside of the cone or vice versa, due
to their curved trajectories in magnetic fields and showering effects,
and the calorimeter response accounting, for example, for
nonlinearities or energy loss in uninstrumented regions of the
detector.

The electromagnetic calorimeter scale determined from resonances such
as $Z\to ee$, as described above, can be transferred to the full
calorimeter by requiring a balance in $p_T$ in photon + jet events.
The intercalibration of the calorimeter is then complemented with
dijet events. A precise JES calibration is a challenging task
involving highly complex procedures to ensure understanding of all
contributions and their systematic uncertainties. D0 has achieved a
JES precision at the 1-2\% level over a wide kinematic
range~\cite{Hegeman:2008}. For this level of precision, strictly
speaking, the JES is only applicable to photon + jet samples, and
additional uncertainties need to be taken into account when
transferring the JES, for example, to top quark
samples~\cite{Harel:2008fp}. The first direct measurement of the JES
for $b$ quarks at the Tevatron based on $Z\to bb$ events has recently
been performed by CDF, reaching a precision of better than
2\%~\cite{Donini:2008nt}.

\subsubsection{$\tau$ jets}
$\tau$ leptons decay into hadrons (and a neutrino) 65\% of the time,
with $\approx$77.5\% of these decays yielding a single charged
particle (``1-prong decays'') and $\approx$22.5\% three charged
particles (``3-prong decays'')~\cite{PDG2008}. These ``hadronic''
$\tau$ decays are reconstructed as jets (that often also contain
$\pi^{0}$ mesons) and can be discriminated statistically from quark
and gluon jets via their narrow shower shape and low track
multiplicity within the jet cone \cite{Abulencia:2005et,D05484}.

\subsubsection{$b$ jets}
The identification (``tagging'') of $b$ jets is a very powerful tool
for separating the top quark signal from its background processes,
which typically exhibit little heavy-flavor content. Also, the
combinatorics for reconstructing top quark events from their
final-state objects can be reduced using this additional information.
There are in general two different approaches to identifying $B$
hadrons formed from $b$ quarks:
\begin{itemlist}
\item {\bf Lifetime Tagging:} Due to their lifetime of about 1.5~ps
  and the boost from top quark decay, B hadrons can travel several
  millimeters before they decay. The resulting charged particle tracks
  therefore originate from (point to) a vertex different than the
  primary one. This can be exploited by searching for secondary
  vertices significantly displaced relative to the primary event
  vertex ({\em secondary-vertex tagging}) or by requiring significant
  impact parameters relative to the primary vertex for tracks, without
  reconstructing a secondary vertex ({\em impact-parameter tagging}).
  A probability can also be calculated for a jet to come from the
  primary vertex based on the impact parameters of all its associated
  tracks ({\em jet-probability tagging}), or a combination of all
  information from the above tagging algorithms into a neural network
  response can be used ({\em NN tagging}). The two latter methods
  yield continuous output variables that can be used as input for
  further multivariate analysis or for selecting analysis-specific
  values as compromises between $b$-tagging efficiency and the
  fraction of light-quark jets that are misidentified as $b$ jets.
\item {\bf Soft-Lepton Tagging:} This tagging is based on semileptonic
  decays of $b$ and $c$ hadrons with branching fractions of
  $\approx$11\% and 10\%, respectively. With two $b$ quarks and two
  $W$ bosons per \ttbar decay, and the fact that about one third of
  the $W$ boson decays yield charm quarks ($c\bar{s}$), the fraction of
  events containing a soft (low $p_{T}$) lepton in a jet is about 40\%
  per lepton flavor ($e,\mu$). The isolation criteria used for
  leptonic $W$ boson decays do not work for these leptons, and their
  reconstruction within jets is quite challenging, especially for
  electrons.
\end{itemlist}
Although the mistag rate for lifetime-tagging is
usually very small for light-quark ($u,d,s$) and gluon jets, this is
not the case for charm jets. For example, a typical operating point
for D0's NN tagger yields a $b$-tag efficiency of $\approx$50\% and a
mistag rate for light jets of $\approx$0.5\%, while it
is $\approx$10\% for $c$ jets~\cite{Harel:2008fp}. More information
on $b$-tagging algorithms and their application in top quark analyses,
including the performance for $b$ jets, $c$ jets and light-quark or
gluon jets, can be found
in Refs.~\cite{Acosta:2004hw,Abulencia:2006kv,CDF9371} for CDF
and~\cite{Abazov:2006ka,Abazov:2008kt,Harel:2008fp} for D0.

\subsubsection{Neutrinos}
Neutrinos are not detected directly because of their negligible
interaction cross section. Since the energy component along the beam
axis at a hadron collider is unknown, only the transverse momentum
carried away by neutrinos (or any other ``invisible'' particles) can
be inferred from momentum conservation in the transverse plane. This
``missing'' transverse momentum (\METns) is calculated from the vector
sum of transverse energy depositions in the calorimeter, corrected for
the energy scale of reconstructed electrons and jets and for the
momenta of reconstructed muons (corrected for energy loss in the
calorimeter). The \MET resolution therefore depends strongly on the
other objects present in the event. Taking this into account, for
example, through selections on the significance of \MET rather than its
absolute value, improves performance.

\subsection{Top quark event signatures in the standard model}
\label{sec:topsignatures}
Having discussed the reconstruction of the fundamental objects from
the initially occurring particles, the experimental signatures of top
quark events can be examined. As noted in Section~\ref{sec:topCKM}, in
the framework of the standard model the top quark decays dominantly
into a $W$ boson and a $b$ quark. Consequently, the observed final
states are defined by the decay modes of the $W$ boson.

$W$ bosons decay into two fermions, either leptons (a charged
lepton-neutrino pair $\ell\bar{\nu}_\ell, \ell = e,\mu,\tau$), with
equal probability per lepton flavor at lowest order in perturbation
theory, or into quark-antiquark pairs $q\bar{q}'$ with $q=u,c$ and
$\bar{q}'=\bar{d},\bar{s},\bar{b}$. At Born level, the ``hadronic''
decay widths are enhanced over the leptonic modes by a color factor of
three (taking the three possible quark colors into account), and
scaled by the appropriate squared CKM matrix element $|V_{qq'}|^2$.
Similar to top quark decay, the off-diagonal CKM matrix elements are
greatly suppressed, and therefore only the $u\bar{d}$ and $c\bar{s}$
decay modes are considered in the following, contributing
approximately 95\% of the hadronic decay width~\cite{PDG2008}.

In summary, $W$ bosons decay leptonically with a branching fraction of
$\approx$1/9 per lepton flavor and $\approx$1/3 for each of the
hadronic decays ($u\bar{d}$ and $c\bar{s}$). The resulting
possibilities for \ttbar decays are illustrated in Fig.\
\ref{fig:ttbardecays}, where also the nomenclature for the different
decay modes is introduced. These branching fractions do not yet take
account of the leptonic decays of $\tau$ leptons and their
contributions to final states involving electrons and muons, as
discussed in Section~\ref{sec:objreco}. This is considered in the
following description of the four basic \ttbar event classes:
\begin{figure}[t]
  \centering
  \includegraphics[width=0.5\textwidth,clip=]{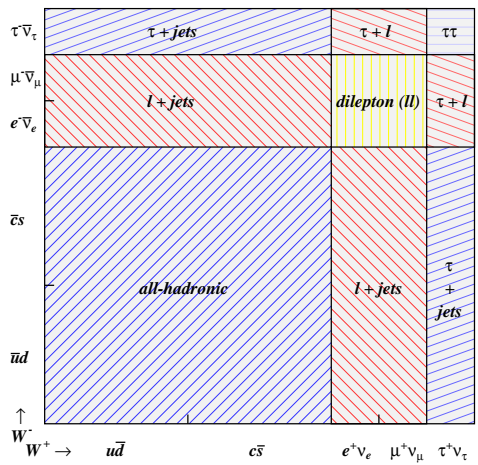}
  \caption{Illustration of the SM \ttbar decay modes and their
    branching fractions via the possible $W$ boson decays. For each
    decay mode, a $b\bar{b}$ quark pair from \ttbar decay is also
    present.}
  \label{fig:ttbardecays}
\end{figure}
\begin{romanlist}[(ii)]
\item {\bf Dilepton channels:} Both $W$ bosons decay leptonically
  ($\ell\bar{\nu}_\ell, \ell = e,\mu$), resulting in a final state
  comprised of two isolated high-\pt leptons, \MET corresponding to
  the two neutrinos, and two $b$ jets. Including leptonic $\tau$
  decays, this channel has a branching fraction of approximately
  6.5\%, shared $\approx$1:1:2 by the $ee$, $\mu\mu$ and $e\mu$ final
  states. While these channels give samples of highest \ttbar signal
  purity, they suffer from limited statistics due to the small
  branching fractions.
\item {\bf Lepton + jets channels:} One $W$ boson decays leptonically
  and the other one hadronically, yielding a final state containing
  one isolated high-\pt lepton, \METns, and four jets. Including
  leptonic $\tau$ decays, these channels exhibit a branching fraction
  of approximately 34.3\%, shared about equally by the $e$ + jets and
  $\mu$ + jets final states. These channels represent the best
  compromise between purity of sample and available statistics.
\item {\bf All-hadronic channel:} Both $W$ bosons decay to
  $q\overline{q}'$ pairs, resulting in a six-jet final state. With a
  branching fraction of $\approx$45.7\%, this channel yields the
  highest statistics of \ttbar events but also suffers from large
  background from multijet production.
\item {\bf Hadronic $\mathbf {\tau}$ channels:} Final states where at
  least one $W$ boson yields a charged $\tau$ lepton that in turn
  decays into hadrons (and a neutrino) are called hadronic $\tau$
  channels, and together comprise a branching fraction of
  $\approx$13.5\%. The nature of the decay of the second $W$ boson is
  used to differentiate between the $\tau$ + jets, $\tau$ + lepton,
  and $\tau\tau$ final states, which contribute with 9.5\%, 3.6\% and
  0.5\% branching fractions, respectively. The corresponding
  experimental signature has four/two/two jets, \METns, one/one/two
  hadronic $\tau$ final states, and no/one/no isolated high-\pt
  lepton. The identification of $\tau\to$ hadrons makes these final
  states especially challenging to reconstruct. More inclusive sample
  selections, requiring, for example, leptons and isolated tracks or
  \MET and ($b$-tagged) jets provide significant fractions of
  $\tau\to$ hadrons events, without their explicit reconstruction.
\end{romanlist}

In all of the above final states, two of the jets are $b$ jets from
\ttbar decay. The $\tau\tau$ final state of \ttbar production remains
the only channel that has not yet been explicitly analyzed. All others
are discussed in Section~\ref{sec:ttbarxsecmeas}. Properties of top
quarks have been extracted mainly from the first three of the above
channels, and especially from the lepton + jets channel.

Full kinematic reconstruction of \ttbar events is possible in the
all-hadronic final state since there are no high-\pt neutrinos
present. In the lepton + jets channel, a twofold ambiguity arises from
the determination of the neutrino $p_z$ when constraining the
invariant mass of the lepton and missing neutrino to $m_W$, while the
dilepton channel is kinematically underconstrained because two
neutrinos contribute to \METns. The unknown assignment between partons
and reconstructed objects in \ttbar events leads to various possible
combinations. The combinatorics can be reduced through identification
of $b$ jets. In particular, when both $b$ jets are identified, only
four combinations remain to be considered in the lepton + jets channel
(including the neutrino $p_z$ ambiguity), and six combinations in the
all-hadronic channel.

The experimental signature for single top quark production is
based on the top quark decay mode and the production channel: in
the $s$- ($tb$-) channel, the top quark is produced with an additional
$b$ jet, while in the $t$- ($tqb$-) channel a forward light-quark jet
accompanies top quark production, sometimes along with another
$b$ jet from the gluon splitting into $b\bar{b}$ (see
Fig.\ \ref{fig:singlet-feynman}). The $W$ boson from top quark
decay is usually required to decay leptonically ($\ell\bar{\nu}_\ell,
\ell = e,\mu$) so as to suppress multijet background. Consequently, the
final state signature of single top quark production contains an
energetic isolated electron or muon, \MET and at least two jets, with
at least one of them being a $b$ jet.

The large mass of the top quark makes it less likely that it is
produced with large kinetic energy at the Tevatron. Its decay products
therefore tend to be emitted at central rapidities, non-planar with
good angular separation, and are characterized by a large sum of
transverse energies $H_T$. Event selections usually require the
channel-characteristic objects (leptons, \METns, and ($b$ tagged)
jets) to be present with energies typically greater than 15 to 20 GeV.
Apart from selections on data quality to ensure a well-performing
detector and specific trigger selections, a well-reconstructed primary
vertex in the central detector region is also required. Variations in
observed jet multiplicities are possible as well due to, for example, jet
reconstruction thresholds, jet splitting and merging during
reconstruction, and additional gluon jets (from initial- and
final-state radiation).

More details on event selection, contributions from background, and
sample compositions in different analyses (including those concerned
with non-standard-model signatures), are given in the following
chapters.

\subsection{Monte Carlo generation}
A reliable and well-understood Monte Carlo (MC) simulation of signal
and background processes is a crucial ingredient for any top quark
analysis. MC samples are needed to understand detector response and
acceptance, selection efficiencies and kinematic distributions of
physical variables and their normalizations. This requires both good
modeling of the production process from the parton to the hadron
level, and an accurate simulation of detector response to signatures
of interest.

\begin{figure}[t]
  \centering
  \includegraphics[width=0.5\textwidth,clip=]{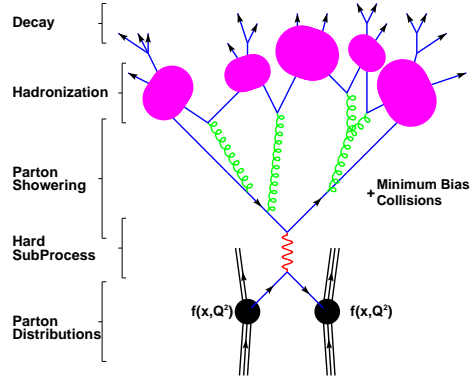}
  \caption{Illustration of a hadron collision~\cite{Dobbs:2001ck},
    indicating some of the steps involved in MC simulation.}
  \label{fig:hadroncollision}
\end{figure}

MC simulations of hadron interactions are based on the factorization
theorem discussed in Section~\ref{sec:ttbarprod}, splitting up hadron
collisions into universal long distance (small $Q^2$) phenomena and
perturbatively calculable short distance phenomena. A generic example
of steps in the simulation of a hadron collision is illustrated in
Fig.\ \ref{fig:hadroncollision}, and described below.

The non-perturbative Parton Distribution Functions (PDFs) describe the
distribution of the proton (or antiproton) momentum among its partons.
The interaction of the incoming partons in the hard process of
interest is then evaluated based on fixed-order (usually LO) matrix
elements, yielding the outgoing partons and their characteristics,
such as their momenta and colors. The ensuing parton showering adds
higher-order effects through parton splitting into pairs of softer
partons (gluon radiation, gluon-splitting, photon radiation,\dots),
until non-perturbative hadronization sets in at low $Q^2$, forming
color-neutral hadrons from the colored partons, which is based on
phenomenological models. Unstable particles and resonances are then
made to decay into the final remnants.

The colored beam-remnants of the proton and antiproton, other soft
multi-parton interactions, and color connections to the hard process
are added, and all form the ``underlying event''. Additional soft
proton-antiproton collisions from the same colliding bunch
(minimum-bias events) have to be superimposed in the MC, and depend on
the instantaneous luminosity. Finally, any overlapping interactions
from consecutive bunch crossings ``leaking'' into the current event
(pile up) must also be considered.

In principle, different programs and models can be used for every
stage in the above process, and the best choice may depend on the
process to be studied, which illustrates the complexity of these MC
simulations. Also, the parameters in some models have to be tuned to
distributions in data before they can provide adequate descriptions of
interactions~\cite{Ellis:1999ec}. General purpose, complete event
generators such as \herwig~\cite{Marchesini:1991ch} and
\pythia~\cite{Sjostrand:2000wi} include LO matrix elements for a
variety of processes, as well as models for showering and
hadronization of partons. They are widely used, either stand-alone or
in combination with other generators that simulate the hard process
and pass on the information to implement showering and hadronization.
An introduction and overview of available MC generators can be found
in Ref.~\cite{Dobbs:2004qw}, and specific simulation tools for top
quark production and decay are reviewed in
Ref.~\cite{Slabospitsky:2006gy}.
CDF and D0 have implemented different simulation chains for their top
quark analyses, and certain analyses use variants on what is described
here. CDF uses CTEQ5L~\cite{Lai:1999wy} PDFs for its generators, while
D0 employs CTEQ6L1~\cite{Pumplin:2002vw} PDFs. The \ttbar signal is
generated either with \pythia~v6.2 \cite{Sjostrand:2001yu} (CDF) or
\alpgen~v2.1 \cite{Mangano:2002ea} interfaced with \pythia~v6.3
\cite{Sjostrand:2003wg} for parton showering (D0). The latter employs
a jet-parton matching algorithm to avoid double counting of final
states that could otherwise be populated from both the hard process
and parton showering~\cite{Hoche:2006ph,Mangano:2006rw}. CDF also uses
\herwig~v6.4 \cite{Corcella:2000bw} for systematic cross checks of
modeling of signal. For single top MC, D0 uses
\singletop~\cite{Boos:2006af} based on \comphep~\cite{Boos:2004kh},
interfaced with \pythia, while CDF utilizes
\madevent~\cite{Maltoni:2002qb} and \madgraph~\cite{Stelzer:1994ta},
also interfaced with \pythia. Both experiments approximate $t$-channel
production at NLO through a combination of contributing $2\to2$ and
$2\to3$ processes. Any signal cross sections obtained at LO are scaled
up to match higher-order theoretical prescriptions (see
Sections~\ref{sec:ttbarprod} and~\ref{sec:singletopprod}). Most
analyses use a top quark mass of 175~GeV/c$^2$.
For simulation of $W$+jets and $Z$+jets background processes, both
collaborations utilize \alpgen, interfaced with \herwig~(CDF) or
\pythia~(D0) for parton showering, and both apply the above-mentioned
jet-parton matching technique. \alpgen~generates higher final-state
parton multiplicities from $2\to n$ processes, and therefore large jet
multiplicities, based on exact LO matrix elements, including the
production of heavy-flavors, which is especially important for
analyses using $b$ tagging. For the decays of $\tau$ leptons, both
collaborations use \tauola~\cite{Jadach:1990mz,Was:2004dg}. For $b$
and $c$ hadron decays \evtgen~\cite{Lange:2001uf} and
\qqgen~\cite{QQgen} are used, the latter only by CDF. Effects of
additional minimum-bias events and pile up are based on \pythia~at
CDF, and D0 uses zero-bias collider data taken by randomly sampling
filled bunch crossings, overlaid with the simulated events.

The generated events are propagated through detector simulation based
on \geant~\cite{geant-cern}, which contains a full description of
positions, geometrical shapes, and types and amounts of material
comprising the detectors. The particles in an event are tracked
through the detector volume, where they encounter energy loss and
multiple scattering that depend on particle type, the traversed
material and the particle momenta, and undergo decay corresponding to
their lifetimes. The response of the detector's readout electronics to
these interactions, including noise and inefficiencies, is then
obtained in a digitization step that yields initial simulated event
characteristics that are processed with the same reconstruction chain
as collider data.

It is not easy to obtain good agreement between any simulation and
experimental data. For example, object reconstruction, identification
and selection efficiencies tend to be higher in MC compared to data,
and must be corrected via scale factors. Such scale factors are
usually derived from comparison of efficiencies in control samples of
simulations and data in final states such as $Z\to\ell\ell$ for
leptons, and $\gamma$+jets for jets. Scale factors can be parametrized
in terms of variables sensitive to these corrections. Also, energy
scales and resolutions for reconstructed objects must generally be
adjusted. Certain effects are hard to simulate, so that at times only
data can be used to provide rates for, e.g. jets to mimic lepton
signatures.

Before searching for any signal, the background model must be verified
using data in control samples that are sufficiently depleted from
possible signal, as, for example, by requiring a reduced jet
multiplicity or no $b$ jets to be present. Sometimes the shape or
normalization of differential distributions has to be corrected, which
usually reflects not optimally tuned MC or insufficient precision in
the model.

With increasing Tevatron luminosity, data-based constraints can
improve the understanding of dominant background processes such as
vector boson + (heavy-flavor) jet production, both in terms of shapes
and normalization. This can benefit MC simulations and the precision
of measurements. For example, the production of $W$ bosons has been
investigated in terms of associated jet
production~\cite{Aaltonen:2007ip} and compared with LO and NLO
predictions, or associated heavy-flavor production was compared with
standard model expectations and found to be in
agreement~\cite{Abazov:2004bv, Abulencia:2005qa}. Nevertheless,
dedicated studies of $W$ boson + $c$
\cite{Aaltonen:2007dm,Abazov:2008qz} or $b$
jet~\cite{Abazov:2004jy,CDF9321} production have been performed as
well, with the most recent results indicating that the production
rates for these processes are currently underestimated by \alpgen.

A more detailed overview of MC simulations used in top quark analyses
at the Tevatron, both for signal and background processes, and the
remaining challenges can be found in
Refs.~\cite{Husemann:2008ad,Harel:2008px}.

\section{Measurements of Top Quark Production}
In this chapter, measurements of top quark production both via the
strong and electroweak interactions are described. Observed rates and
mechanisms of production are compared with the standard model
expectations and used to derive constraints on specific extensions
of the standard model impacting the properties under consideration.

\subsection{Top quark pair production cross section}
\label{sec:ttbarxsecmeas}
Measurements of the \ttbar production cross section are important for
several reasons. They provide a powerful test of the predictions of
perturbative QCD calculations at high transverse momenta. As shown
in Section~\ref{sec:ttbarprod}, the uncertainties on the \ttbar rate
predictions have reached the level of $\approx$10\%, a precision which has already
been matched by the measurements performed at the Tevatron.

Deviations from the standard model prediction could arise, for example, from novel
production mechanisms such as a new resonant production mode in addition
to the standard model one as discussed in Section~\ref{sec:BSMprod}.
New physics contributing to the electroweak symmetry breaking will
probably couple to particles proportional to their mass, making the
top quark and its strong coupling reflected in its production rate a
highly interesting probe for such effects. Different top quark decay
modes, such as the decay via a charged Higgs boson competing with SM
decay, as examined in Section~\ref{sec:H+topdecay}, could
cause apparent different production rates amongst the various decay
channels via modified branching fractions. Contributions of new
physics to the background samples in the various channels could have
similar effects. Analyzing different decay channels consequently helps not only
to improve statistics and studies of properties of top events, but is
also a sensitive probe for physics beyond the standard model.

To extract a cross section requires good understanding of the
reconstruction and identification of the involved objects and of the
modeling of the contributing background processes. By providing
selections for samples enriched in top quark signal and of well
characterized composition, cross section analyses form the foundation
of all further top quark property analyses.

Top quark pair production has been studied by now in all possible
decay modes -- the dilepton, lepton + jets, all-hadronic and hadronic
$\tau$ channels as defined in Section~\ref{sec:topsignatures}, with
the $\tau\tau$ decay mode being the only exception due to marginal
branching fraction and challenging separation from background processes. As
mentioned before, in the context of this review the term leptons
refers to electrons and muons alone unless indicated to the contrary.

The event selections usually require the presence of characteristic objects from
the top quark decay, namely, leptons, \METns, (heavy-flavor) jets, or
hadronically decaying $\tau$, with energies typically exceeding
15 to 20 GeV. Further kinematic characteristics can be
exploited to separate the top quark signal from the various background
processes. Due to the large mass of the top quark, its decay products
tend to be very energetic, emitted at central rapidities and
non-planar with good angular separation. In contrast to this, the jet
energy spectrum for background processes with jets from gluon
radiation is steeply falling. The observed objects are emitted less
isotropically but more back to back, and mismeasured objects giving
rise to \MET tend to exhibit characteristic angular correlations with
the reconstructed \METns.

Consequently, additional variables available for top quark signal
selection are based on the energy present in the event, such as the
sum of transverse energies ($H_T$), or the invariant mass of a combination
of reconstructed objects. Event shape variables such as sphericity and
aplanarity, derived from the eigenvalues of the normalized momentum
tensor of the objects considered~\cite{Barger:1993ww}, or centrality,
defined as the ratio of $H_T$ and the sum of the objects' energies,
provide additional discrimination. Furthermore, angular relations between
reconstructed objects (for example $\Delta\phi(\not\!\!E_T,\ell)$) and
single-object kinematic quantities (such as the jet of highest (leading) transverse momentum) are
frequently used as well.

Depending on the \ttbar decay mode considered, the use of $b$ tagging
in its different forms (see Section~\ref{sec:objreco}) is optional for
the event selection. In the dilepton and lepton + jets channels
selections based purely on topological and kinematic characteristics
suffice for a good signal to background ratio (S/B). Adding $b$
tagging improves sample purities but also implies a stronger model
dependence by relying on $b$ quarks to be present in the final state.
The
actual extraction of the signal fraction proceeds either in a counting
experiment or via template fits using the full shape information of
the sensitive variable under consideration. While the latter is
usually more sensitive, it also exhibits a stronger dependence on the
MC modeling. Using different methods with different systematic
uncertainties to measure the same quantity provides a way to assess the model
assumptions from different perspectives
and to check internal consistency, and is beneficial for combinations of
increased precision. Non-overlapping (orthogonal) sample selections
facilitate later combinations of results as independent measurements
by removing the need to evaluate the correlation amongst the
measurements from ensemble tests.

Once the sample composition is measured, the \ttbar production
cross section is calculated as follows:
\begin{equation}
\sigma_{t\bar{t}} = \frac{N_{\rm observed}-N_{\rm
background}}{\varepsilon~{\cal B}~\int{\cal L}dt},
\end{equation}
where $N_{\rm observed}$ ($N_{\rm background}$) is the 
total (background) number of events,
$\varepsilon$ is the \ttbar selection efficiency, including detector
acceptance, ${\cal B}$ is the branching fraction for the 
\ttbar decay mode in question and $\int{\cal L}dt$ is the integrated luminosity of
the used dataset.

As illustrated in Section~\ref{sec:ttbarprod}, the \ttbar cross
section depends on the top quark mass, decreasing by about 0.2~pb for
each GeV/c$^2$ increase in the $m_t$ mass range from 170~GeV/c$^2$ to
180~GeV/c$^2$. The \ttbar cross section results given in the following
sections generally refer to a top quark mass of 175~GeV/c$^2$; the few
cases where a top quark mass of 178~GeV/c$^2$ was assumed will be
pointed out explicitly. Especially for the recent measurements, a
parametrization of the obtained result versus top quark mass is
provided to allow easy projection to the current world-averaged
top quark mass. The cross section dependence on the mass can
also be turned around to provide a measurement of the top quark mass
from the cross section, which will be discussed further in
Section~\ref{sec:massfromxsec}.

In the following subsections the published and latest preliminary
Run~II results will be referenced for the different \ttbar decay modes.
Some analyses will be highlighted in more detail. The agreement
with theoretical predictions is illustrated in the summary
section, where results of combinations across channels are also given.

\subsubsection{Dilepton final state}
\label{sec:ttbarxsecmeasdil}

A typical \ttbar dilepton event selection requires two isolated
high-\pt leptons of opposite charge, \MET and at least two central
energetic jets. The dominant physics background processes exhibiting both real
leptons and \MET arise from diboson ($WW,ZZ,WZ$) production and from
$Z/\gamma^{*}$+jets processes, with $Z/\gamma^{*}\to\tau^{+} \tau^{-},
\tau\to e,\mu$. Misreconstructed \MET caused by experimental resolution in
$Z/\gamma^{*}$+jets events (with $Z/\gamma^{*}\to e^{+}e^{-}/
\mu^{+}\mu^{-}$) contributes to the instrumental background, as does
$W$+jets and multijet production (often called QCD), where one or more jets mimic the
isolated-lepton signature. The physics background processes are usually
modeled using Monte Carlo simulation, while instrumental background processes (especially
those involving false isolated-lepton signatures) typically are
estimated from data.
The signal purities of the resulting samples are usually quite good, with a
signal to background ratio (S/B) typically better than two.

The sample purity can be further enhanced by means of additional
kinematic requirements such as $H_{T}$
to be above a certain threshold, or by selecting events with at least
one identified $b$ jet. However, this reduces the already limited
statistics in this channel. To increase the signal
acceptance, the reconstruction and isolation requirements on the
second lepton can be relaxed. If the second lepton is only required to
be reconstructed as an isolated track (termed lepton + track
selection), especially 1-prong hadronic $\tau$ decays can then also contribute
to the signal.

In a recent preliminary analysis, CDF determines the \ttbar cross
section from a 2.8~fb$^{-1}$ dataset by requiring two oppositely
charged reconstructed isolated leptons with $E_T \geq 20$~GeV, \MET
$\geq 25$~GeV and at least two jets within $|\eta|<2.5$ and $E_T \geq
15$~GeV, with the leading jet fulfilling $E_T \geq 30$~GeV. The \ttbar
cross section is extracted from the resulting sample once without any
additional cuts and once after increasing the purity by requiring at
least one of the jets to be $b$ tagged. The background from
$Z/\gamma^{*}$ and diboson $WW,ZZ,WZ$ events is derived from MC, while
false isolated-lepton signatures are estimated from a dilepton dataset
where both leptons have the same charge (same sign, ``SS''), assuming
their contribution is identical in the opposite sign (``OS'') signal
selection~\cite{CDF9399}.

\begin{figure}[t]
  \begin{center} 
    \includegraphics[width=.48\textwidth]{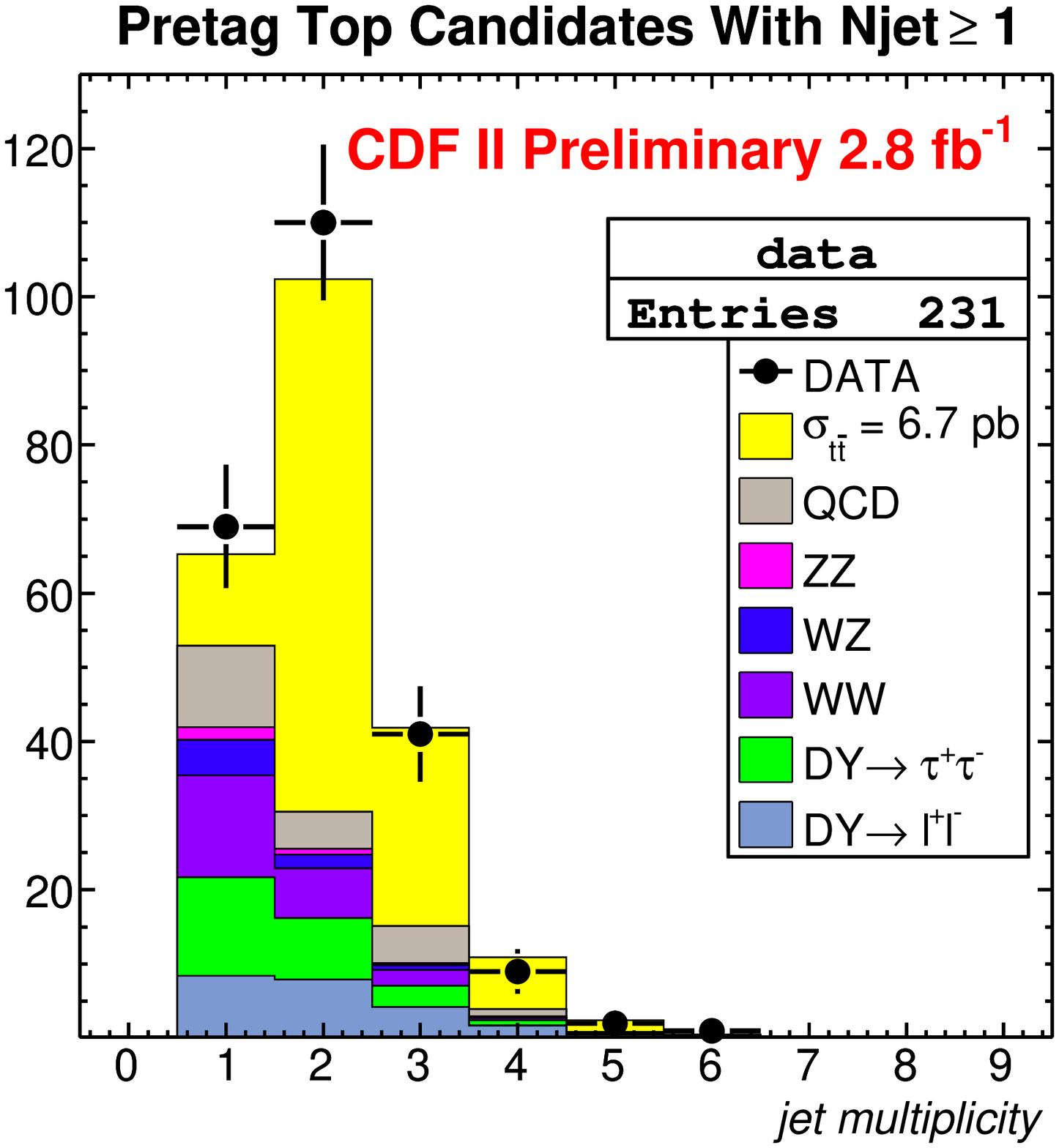}\hspace*{1mm}
    \includegraphics[width=.48\textwidth]{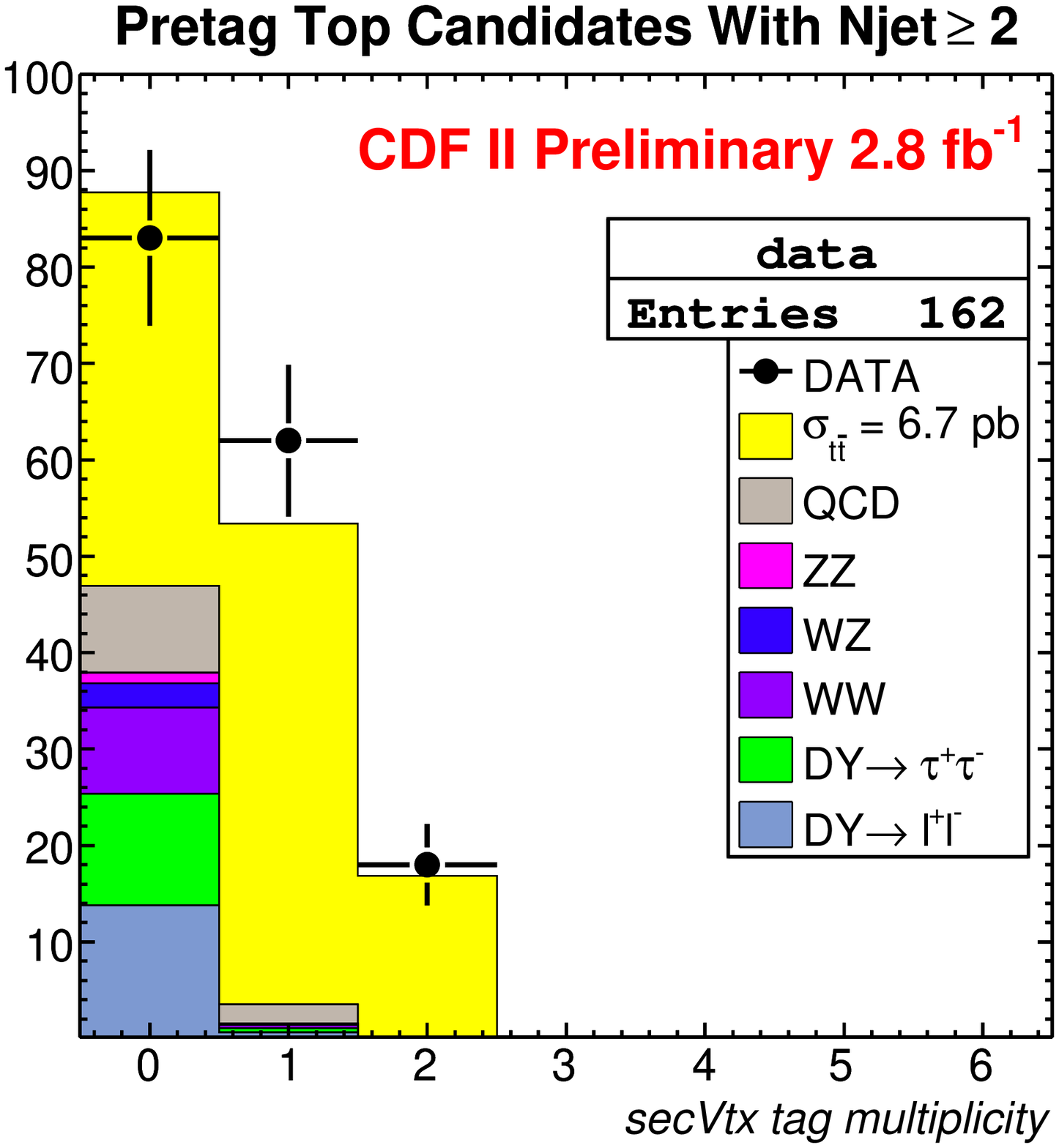} 
    \caption{Distributions in jet multiplicity (left) and number of $b$ tags (right)
      in \ttbar dilepton candidate events, respectively, for $\geq 1$
      and $\geq 2$ jets in 2.8~fb$^{-1}$ of data analyzed by
      CDF~\cite{CDF9399}.}
      \label{fig:cdfdileptonxsec}
  \end{center}
\end{figure}
\begin{table}[p]
\caption{Event yields and sample composition after \ttbar preselections in CDF's 2.8~fb$^{-1}$
dilepton dataset before and after requiring at least one $b$ tagged
jet~\cite{CDF9399}.}
\addtolength{\tabcolsep}{-2pt}
\begin{center}
\begin{tabular*}{\textwidth}{@{\extracolsep{\fill}}|l|cccc|c|} \hline
\multirow{2}{*}{Process} &\multicolumn{4}{c|}{before tagging}&$b$ tagged\\
& $e^+e^-$ & $\mu^+\mu^-$ & $e^\pm\mu^\mp$ & $\ell^+\ell^-$ &$\ell^+\ell^-$\\\hline\hline
$t\bar{t},\ \sigma=$~6.7 pb& 29.2 & 21.5 & 59.9 & 110.6                  &65.2         \\
$Z/\gamma^*\to\ell^+\ell^-$ & 9.25 & 4.79 & 0.52 & 14.6                  &0.78         \\
$Z/\gamma^*\to\tau^+\tau^-$ & 2.84 & 2.55 & 6.62 & 12.0                  &0.60         \\
$WW\to\ell^+\ell^-$ & 3.05 & 2.03 & 5.07 & 10.2                          &0.44         \\
$WZ\to\ell^+\ell^-$ & 1.52 & 0.72 & 0.67 & 2.91                          &0.09         \\
$ZZ\to\ell^+\ell^-$ & 0.80 & 0.40 & 0.26 & 1.46                          &0.10         \\\hline
Totals (MC) & 46.7 & 32.0 & 73.0 & 151.7                                 &67.2         \\\hline
$\pm\pm$ (SS) Data & 3.81 & 0.00 & 6.96 & 10.8                                      &2.00         \\\hline\hline
Sum Expected & 50.5 $\pm$ 1.7 & 32.0 $\pm$ 1.3 & 80.0 $\pm$ 2.5 & 162.5 $\pm$ 4.5 &69.2 $\pm$1.7\\
$+-$ (OS) Data & 54 & 33 & 75 & 162                                             &80           \\\hline
\end{tabular*}
\end{center}
\label{tab:CDFdileptonxsec}
\end{table}
The untagged sample yields 162 events with a total background
contribution of $51.9 \pm 4.5$, where the dominant uncertainties
arise from the estimate of false leptons and the uncertainty
on the jet multiplicity correction factors applied in the MC.
Requiring at least one $b$ tagged jet, 80 events remain, with an
expected total background of $4.0 \pm 1.7$, with the dominant
uncertainties arising again from the estimates of false leptons
and also from uncertainties on the $b$ tag modeling. The sample
composition is illustrated in Fig.\ \ref{fig:cdfdileptonxsec}, and
detailed in Table~\ref{tab:CDFdileptonxsec}. The extracted cross
sections are given in Table~\ref{tab:dilepxsecs}, together with the
other dilepton channel results obtained thus far in Run~II.
\begin{table}[p]
  \caption{\ttbar cross section measurements in dilepton final states performed thus
    far at the Run~II Tevatron with their integrated luminosities,
    data selections ($\ell\ell$ = dilepton, $\ell$+trk =
    lepton + track) and analysis methods used. The first three
    results have been published; the others are preliminary. The
    measurement marked with an asterisk refers to $m_{t}$ of
    178~GeV/c$^2$ rather than 175~GeV/c$^2$, and, unlike the other results,
    incorporates the luminosity uncertainty within the first given uncertainty.}
  \renewcommand{\arraystretch}{1.2}
  \begin{center}
    \begin{tabular}{|c|c|c|l|c|}
      \hline 
      $\int{\cal L}dt$&\multirow{2}{*}{Sel.}&\multirow{2}{*}{$b$ tag}&$\sigma_{t\bar{t}}${\footnotesize $\pm$(stat.)$\pm$(syst.)$\pm$(lumi.)}&\multirow{2}{*}{Ref.}\\ 
      ~[fb$^{-1}$] &  &  & {\hspace{1.5cm} [pb]} & \tabularnewline
      \hline
      \hline 
0.2& $\ell\ell$,$\ell$+trk& no   & $7.0^{+2.4}_{-2.1}\,^{+1.6}_{-1.1}\pm 0.4$  & \cite{Acosta:2004uw}\\
0.2& $\ell\ell$           & no   & $8.6^{+3.2}_{-2.7} \pm 1.1\pm 0.6$        & \cite{Abazov:2005yt}\\
0.4& $\ell\ell$,$\ell$+trk&no,yes& $7.4 \pm  1.4\pm  0.9\pm 0.5$             & \cite{Abazov:2007bu}\\\hline
0.4& $\ell\ell$           & no   & $8.5^{+2.6}_{-2.2}\,^{+0.7}_{-0.3}~(*)$     & \cite{Abulencia:2006mf}\\ %
1.0& $\ell\ell$           & no   & $6.8^{+1.2}_{-1.1}\,^{+0.9}_{-0.8}\pm0.4$   & \cite{D05371}\\
1.0& $\ell\ell$           & no   & $7.0^{+1.1}_{-1.0}\,^{+0.8}_{-0.6}\pm0.4$   & \cite{D05715}\\
1.0& $\ell$+trk           & yes  & $5.0^{+1.6}_{-1.4}\,^{+0.9}_{-0.8}\pm0.3$   & \cite{D05465}\\
1.0& $\ell\ell$,$\ell$+trk&no,yes& $6.2\pm 0.9^{+0.8}_{-0.7}\pm 0.4$         & \cite{D05477}\\
1.0& $\ell$+trk           & yes  & $10.1 \pm 1.8 \pm 1.1 \pm 0.6$            & \cite{CDF8912}\\
1.1& $\ell$+trk           & no   & $8.3 \pm 1.3 \pm 0.7 \pm 0.5$             & \cite{CDF8770}\\ %
2.8& $\ell\ell$           & yes  & $7.8 \pm 0.9\pm 0.7 \pm 0.4$              & \cite{CDF9399}\\ %
2.8& $\ell\ell$           & no   & $6.7 \pm 0.8 \pm 0.4 \pm 0.4$             & \cite{CDF9399}\\ %
      \hline 
    \end{tabular}
    \label{tab:dilepxsecs}
  \end{center}
\end{table}

\afterpage{\clearpage}

\subsubsection{Lepton + jets final state}
\label{sec:ttbarxsecmeaslj}
A typical \ttbar lepton + jets event selection requires exactly one
isolated high-\pt lepton, \MET and at least three central energetic
jets, allowing both lepton + jets and lepton + hadronic $\tau$
signatures to contribute. The dominant physics background in this
final state arises from $W$ boson + jets production, and the main
instrumental background comes from QCD multijet production with a jet
mimicking the isolated-lepton signature. Additional smaller background
contributions arise from $Z/\gamma^{*}$ + jets, diboson and single
top production. While for these smaller background processes shape and
normalization are determined commonly from simulation and NLO cross
sections, $W$ + jets events are usually normalized to data, and their
differential distributions derived from Monte Carlo. The QCD multijet background's shape
and normalization are typically derived from data, using, for example,
datasets fulfilling the complete event selection, except the tight
lepton isolation, for the background shape, and the rate for jets to
mimic leptons is derived from data for the normalization.

Samples selected with such a basic preselection exhibit S/B ratios below
unity, around 1/4. Signal purity can be improved significantly via
additional topological selections or by using
$b$ tagging. When no $b$ tagging is used in an analysis, then
multiple topological and kinematic event properties are usually combined in a
multivariate discriminant to yield good signal to background
separation without relying on the presence of $b$ jets in the events,
therefore being less model dependent. The sample composition can then be
determined from a template fit in that sensitive variable, providing a
higher sensitivity than a plain cut.

Requiring identified $b$ jets to be present in an event is a 
powerful tool to reject the background processes which exhibit little
heavy-flavor content. $b$ tagging algorithms based on the long
lifetime of $B$ hadrons or reconstruction of soft leptons within jets
that originate from semileptonic $B$ decays, as discussed in
Section~\ref{sec:objreco}, have been deployed for that purpose.
Using $b$ tagging, very pure \ttbar samples can be selected
that have S/B $>$ 10, if at least four jets and at least
two identified $b$ jets are required in each event.

The most precise \ttbar cross section measurement published thus far has
been performed by D0 on a 0.9~fb$^{-1}$ lepton + jets
dataset~\cite{Abazov:2008gc}. Events are selected by requiring exactly
one isolated electron or muon with $E_T > 20$~GeV, \MET $> 20$~GeV for
$e$ + jets and $> 25$~GeV for $\mu$ + jets, at least
three jets with $|\eta| < 2.5$ and $E_T > 20$~GeV, and leading-jet 
$E_T > 40$~GeV. Cuts on the azimuthal separation between lepton
and \MET are applied to suppress background from misreconstructed
objects. After these selections, the \ttbar signal contributes only about
20\% of the total sample. The \ttbar cross section is measured using
two complementary analyses.

\begin{figure}[!t]
  \begin{center} 
    \includegraphics[width=.48\textwidth]{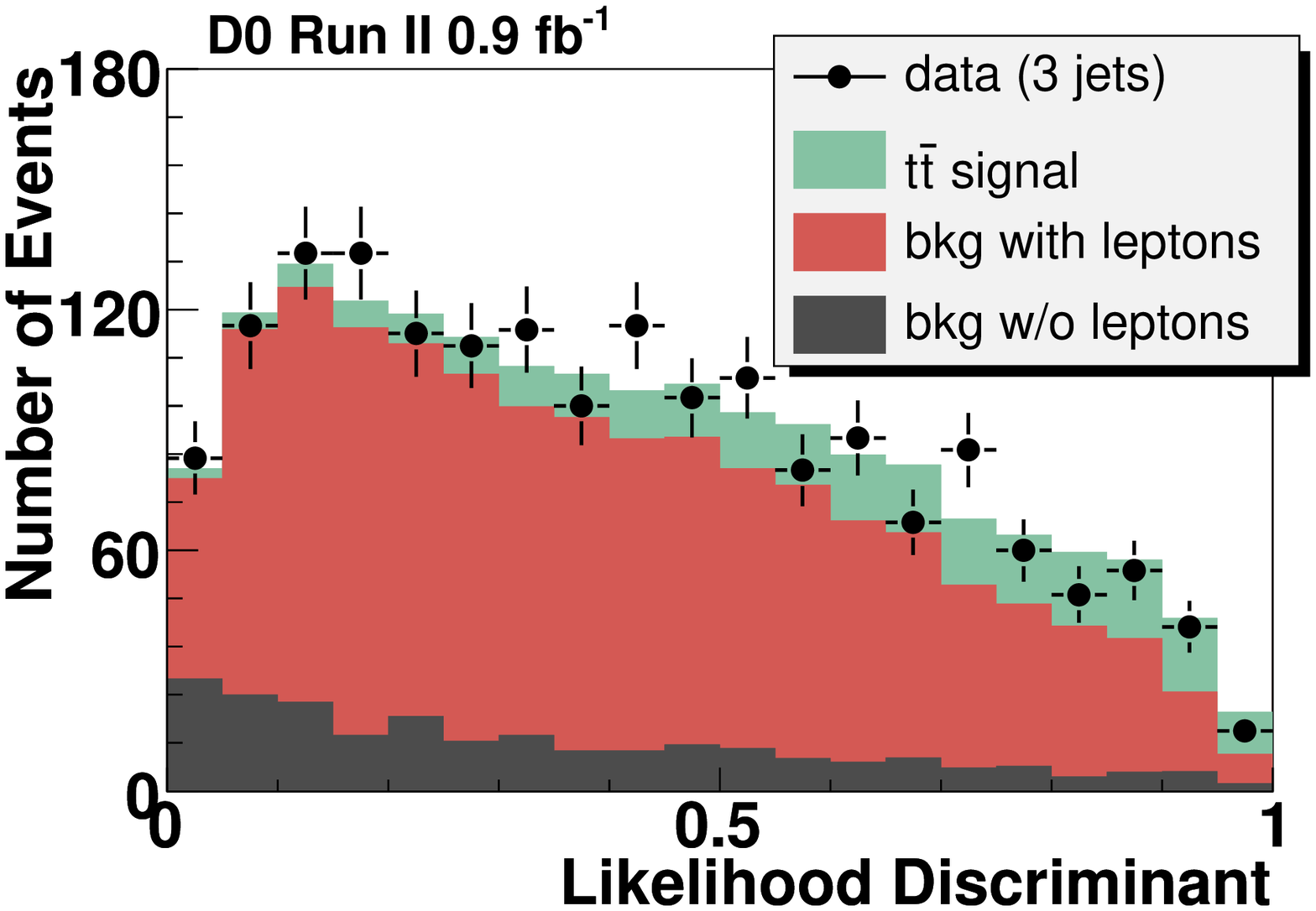}\hspace*{1mm}
    \includegraphics[width=.48\textwidth]{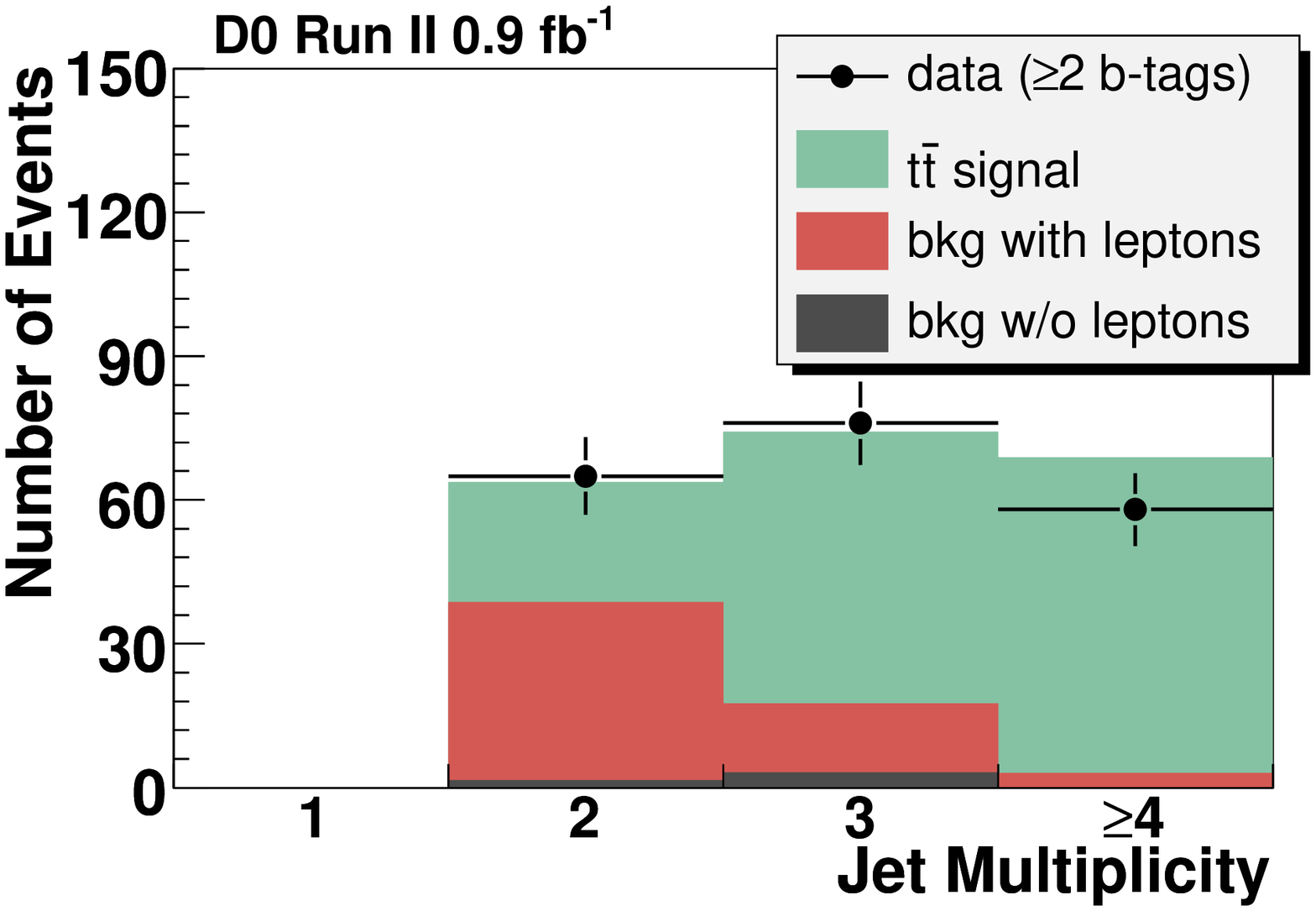} 
    \caption{Topological likelihood distribution for $\ell$ + jets
      \ttbar candidate events when three jets are required (left), and
      jet multiplicity distribution when at least two $b$ tagged jets
      are required (right) in D0's 0.9~fb$^{-1}$ lepton + jets
      dataset~\cite{Abazov:2008gc}.}
      \label{fig:D0ljetsxsec}
  \end{center}
\end{figure}

One approach is based on lifetime $b$ tagging, requiring at least one
jet in the event to be tagged and determining the \ttbar production
rate through a maximum likelihood fit to the observed event yields in the
different subchannels defined by lepton flavor, jet multiplicity and $b$
tag multiplicity. The dominant systematic uncertainties arise
from uncertainties on tagging efficiencies for $b, c, q$ and gluon
jets and the jet energy calibration. The second analysis utilizes
topological likelihood discriminants for the different subchannels based
on lepton flavor and jet multiplicity. After applying an additional
cut on jets of $H_T > 120$~GeV for three-jet events, five or six
different variables such as angular object separation, sphericity and
aplanarity (which provide good discrimination power and are well
modeled in MC) are combined into discriminants for each subchannel.
The sample composition is then determined in a maximum likelihood fit
of templates of signal and background contributions to the
observed discriminant distributions. The dominant systematic
uncertainties in this method arise from uncertainties on the selection
efficiencies and the likelihood fit uncertainty derived using
statistical fluctuations in the likelihood discriminant template
shapes. The sample compositions for both analyses are illustrated in
Fig.\ \ref{fig:D0ljetsxsec}, and detailed in
Table~\ref{tab:D0ljetsxsec}.

\begin{table}
\caption{Event yields and sample composition in D0's 0.9~fb$^{-1}$ lepton + jets
  dataset for both the topological and the $b$ tagging analysis. The
  \ttbar contribution is based on the measured cross section in the
  respective analysis~\cite{Abazov:2008gc}.}
\begin{tabular*}{\textwidth}{@{\extracolsep{\fill}} l|r@{}lr@{}lr@{}lr@{}lr@{}lr@{}l}
\hline
~& \multicolumn{2}{c}{3~jets,} & \multicolumn{2}{c}{3~jets,} & \multicolumn{2}{c}{$\geq$4~jets,} & \multicolumn{2}{c}{$\geq$4~jets,} & \multicolumn{2}{c}{3~jets,} & \multicolumn{2}{c}{$\geq$4~jets,} \tabularnewline
~& \multicolumn{2}{c}{1~tag} & \multicolumn{2}{c}{$\geq$2~tags} & \multicolumn{2}{c}{1~tag} & \multicolumn{2}{c}{$\geq$2~tags} & \multicolumn{2}{c}{topo} & \multicolumn{2}{c}{topo} \tabularnewline
\hline\hline
N$_{t\bar{t}}$   & 147&$\pm$12 & 57&$\pm$6 &130&$\pm$10  & 66&$\pm$7 & 245&$\pm$20 & 233&$\pm$19\\	
N$_{W+{\rm jets}}$   & 105&$\pm$5  & 10&$\pm$1 &  16&$\pm$2  &  2&$\pm$1 &\multirow{2}{*}{1294}&\multirow{2}{*}{$\pm$48}&\multirow{2}{*}{321}&\multirow{2}{*}{$\pm$30}\\	
N$_{\rm other}$   &  27&$\pm$2  &  5&$\pm$1 &   8&$\pm$1  &  2&$\pm$1 &&&&\\	
N$_{\rm multijet}$   &  27&$\pm$6  &  3&$\pm$2 &   6&$\pm$3  &  0&$\pm$2 &227&$\pm$28  &  70&$\pm$12\\
\hline
total  & 306&$\pm$14 & 74&$\pm$6 &159&$\pm$11 & 69&$\pm$7 &1766&$\pm$59&624&$\pm$37\\
N$_{\rm data}$ &\multicolumn{2}{c}{294}&\multicolumn{2}{c}{76}&\multicolumn{2}{c}{179}&\multicolumn{2}{c}{58}&\multicolumn{2}{c}{1760}&\multicolumn{2}{c}{626}\\	
\hline
\end{tabular*}
\label{tab:D0ljetsxsec}
\end{table}

\begin{table}[t!]
  \caption{\ttbar lepton + jets cross section measurements
    performed thus far at the Run~II Tevatron with their integrated
    luminosities, data selections and analysis methods
    used. The first eleven results have been published; the others
    are preliminary. The measurements marked with asterisks $(*)$ refer to 
    $m_{t}$ of 178~GeV/c$^2$ rather than the standard 175~GeV/c$^2$.
    Measurements marked with a double cross $(\ddagger)$ include the
    luminosity uncertainty in the systematic uncertainty,
    while measurements marked with a cross $(\dagger)$ have the first
    uncertainty represent the statistical and systematic uncertainties
    combined in quadrature and the
    second representing the luminosity.}
  \renewcommand{\arraystretch}{1.2}
  \begin{center}
    \begin{tabular}{|c|c|c|l|c|}
      \hline 
      $\int{\cal L}dt$&\multirow{2}{*}{Sel.}&\multirow{2}{*}{$b$ tag}&$\sigma_{t\bar{t}}${\footnotesize $\pm$(stat.)$\pm$(syst.)$\pm$(lumi.)}&\multirow{2}{*}{Ref.}\\ 
      ~[fb$^{-1}$] &  &  & {\hspace{1.5cm} [pb]} & \tabularnewline
      \hline
      \hline 
0.2 & $\ell$+jets & yes &$ 5.6^{+1.2}_{-1.1}\,^{+0.9}_{-0.6}~(\ddagger)     $& \cite{Acosta:2004hw}\\
0.2 & $\ell$+jets & yes &$ 6.0^{+1.5}_{-1.6}\,^{+1.2}_{-1.3}~(\ddagger)     $& \cite{Acosta:2004be}\\%
0.2 & $\ell$+jets &yes, soft-$\mu$ &$ 5.3 \pm 3.3^{+1.3}_{-1.0}~(\ddagger)$& \cite{Acosta:2005zd}\\
0.2 & $\ell$+jets & no &$ 6.6 \pm 1.1 \pm 1.5~(\ddagger)                  $& \cite{Acosta:2005am}\\
0.2 & $\ell$+jets & no &$ 6.7^{+1.4}_{-1.3}\,^{+1.6}_{-1.1}\pm 0.4          $& \cite{Abazov:2005ex}\\
0.2 & $\ell$+jets & yes &$ 8.6^{+1.6}_{-1.5}\pm 0.6~(\dagger)             $& \cite{Abazov:2005ey}\\
0.3 & $\ell$+jets & yes &$ 8.7 \pm 0.9^{+1.1}_{-0.9}~(*,\ddagger)         $& \cite{Abulencia:2006in}\\ %
0.3 & $\ell$+jets & yes &$ 8.9 \pm 1.0^{+1.1}_{-1.0}~(*,\ddagger)         $& \cite{Abulencia:2006kv}\\%
0.4 & $\ell$+jets & yes &$ 6.6 \pm  0.9\pm 0.4~(\dagger)                  $& \cite{Abazov:2006ka}\\
0.4 & $\ell$+jets & no &$ 6.4^{+1.3}_{-1.2}\pm  0.7\pm  0.4               $& \cite{Abazov:2007kg}\\
0.9 & $\ell$+jets &no, yes&$ 7.4 \pm 0.5 \pm 0.5\pm  0.5                  $& \cite{Abazov:2008gc}\\\hline
0.4 & $\ell$+jets &yes, soft-$\mu$ &$ 7.3^{+2.0}_{-1.8}\pm 0.4~(\dagger)  $& \cite{D05257}\\
0.7 & $\ell$+jets & yes &$ 8.5 \pm 0.6 \pm 1.0~(\ddagger)                 $& \cite{CDF8272}\\%
1.0 & $\ell$+jets & yes &$8.2 \pm0.5^{+0.8}_{-0.7}\pm0.5                  $& \cite{D05715}\\%
1.7 & $\ell$+jets &yes, soft-$e$&$ 7.8 \pm 2.4 \pm 1.5 \pm 0.5            $& \cite{CDF9348}\\ %
2.0 & $\ell$+jets &yes, soft-$\mu$&$ 8.7 \pm 1.1^{+0.9}_{-0.8} \pm 0.6    $& \cite{CDF9304}\\ %
2.7 & $\ell$+jets & yes &$ 7.2 \pm 0.4 \pm 0.5 \pm 0.4                    $& \cite{CDF9462}\\ %
2.8 & $\ell$+jets & no &$ 6.8 \pm 0.4 \pm 0.6 \pm 0.4                     $& \cite{CDF9474}\\ %
      \hline 
    \end{tabular}
    \label{tab:ljetsxsecs}
  \end{center}
\end{table}

Both analyses exhibit a statistical correlation of 0.31, as determined
from ensemble studies and are combined using the best linear unbiased
estimate (BLUE) approach~\cite{Lyons:1988rp,Valassi:2003mu}. The
resulting cross section is given in Table~\ref{tab:ljetsxsecs},
together with the other lepton + jets channel results obtained thus
far in Run~II.

\subsubsection{All-hadronic final state}
\label{sec:ttbarxsecmeasallh}
To select all-hadronic \ttbar decays typically requires at least
six central energetic jets per event, and no
isolated energetic leptons or significant \MET to be present. The
overwhelming background process here is QCD multijet production,
dominating over the signal by three orders of magnitude after
online selection of events using triggers on multiple jets and $H_{T}$
in the event above a certain threshold. This background is usually
modeled from the data, as the theoretical description of
final states with such high jet multiplicities has large
uncertainties and datasets even more depleted from signal can be
easily obtained, for example, by selecting a lower jet multiplicity.

After preselection, signal and background are separated further by
applying $b$ jet identification and using multivariate discriminants
based on topological and kinematic event properties.

CDF has published the most precise cross section analysis in the
all-hadronic final state to date, based on 1~fb$^{-1}$ of
data~\cite{Aaltonen:2007qf}. Events are required to have at least six
and at most eight jets with $E_{T} \geq 15$~GeV, $\Delta R \geq 0.5$
from each other and $|\eta| \leq 2$, no isolated energetic electrons
or muons as used in the leptonic \ttbar analyses, and \MET divided by
$\sqrt{H_{T}}$ of the selected jets has to be $< 3~\sqrt{\rm GeV}$.
This yields S/B of $\approx$1/370. The signal purity is then
increased using a neural network discriminant based on variables such as
$H_T$, centrality, aplanarity and minimal/maximal invariant dijet or
trijet mass values of all jet permutations. The signal is modeled using MC,
and the selected data are used directly for the background, as the
expected signal contribution in these events is very small.

At least one of the jets in each event is required to be $b$ tagged,
and the sample composition is then determined in terms of the number of
tags rather than events. The average number of tags per signal event
for a given neural network cut is determined from MC, and is used to
derive the \ttbar cross section from the observed excess in $b$ tags beyond
the background expectation obtained from data. The tagging efficiencies
of the simulation are corrected for differences relative to data. The cut
on the neural network discriminant ($N_{\rm out}$) is optimized for the
highest expected signal significance after $b$ tagging, taking both
statistical and systematical uncertainties of signal and background
into account, yielding $N_{\rm out} > 0.94$. This cut yields S/B
of $\approx$1/12 before $b$ tagging, and 1/2 after tagging.

\begin{figure}[t!]
  \begin{center} 
    \includegraphics[width=.4\textwidth]{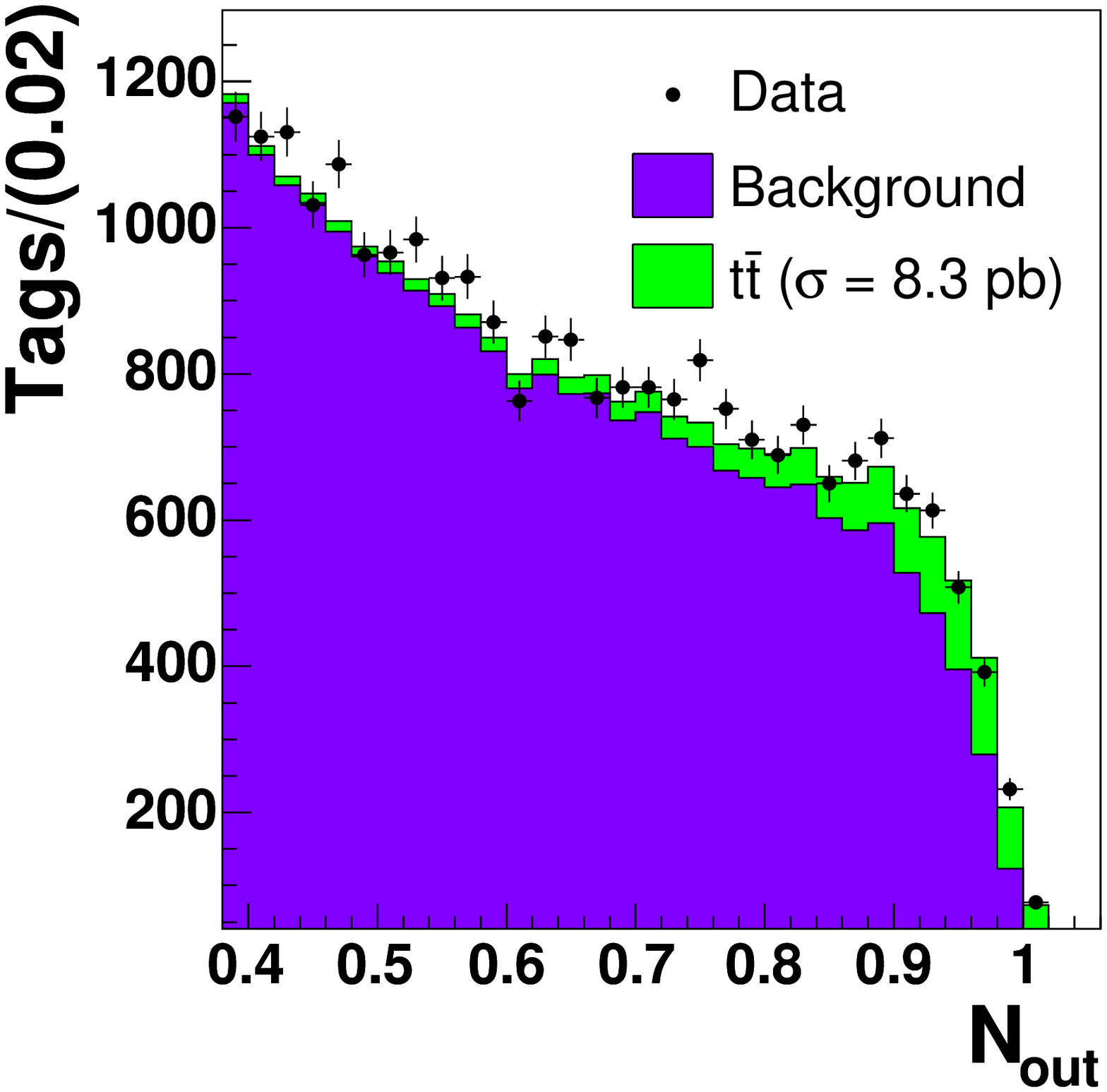}\hspace*{1mm}
    \includegraphics[width=.4\textwidth]{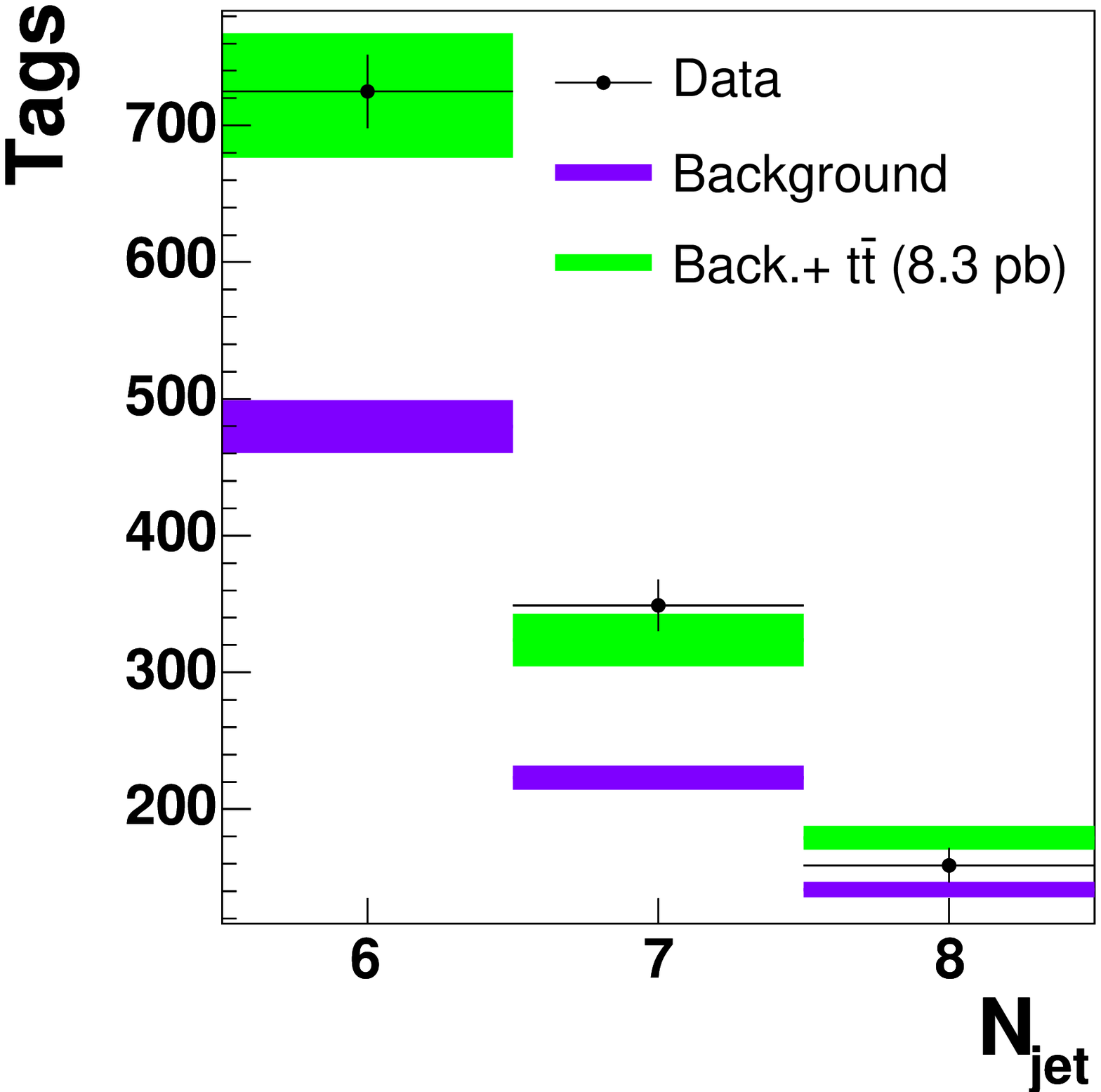} 
    \caption{Left: Number of $b$ tags in 1~fb$^{-1}$ of all-hadronic
      candidate events selected by CDF versus neural network
      discriminant ($N_{\rm out}$). Right: Number of tags versus jet
      multiplicity after requiring $N_{\rm out} > 0.94$. The \ttbar
      contributions are normalized to the measured cross section of
      8.3~pb~\cite{Aaltonen:2007qf}.}
      \label{fig:CDFalljetsxsec}
  \end{center}
\end{figure}
\begin{table}[t]
  \caption{Expected and observed yields of tags after requiring
    $N_{\rm out} > 0.94$, with uncertainties corresponding
    to statistical and systematic contributions added in quadrature.
    The corrected background (BG) contribution accounts for the
    signal contamination in the dataset used for its estimate. After
    tagging, 1020 events remain in the signal sample with 1233 tags
    and an expected background of $846 \pm 37$ tags. The average
    number of tags in \ttbar events is $0.95 \pm
    0.07$~\cite{Aaltonen:2007qf}.}
  \addtolength{\tabcolsep}{-2pt}
  \begin{center}
    \begin{tabular*}{\textwidth}{@{\extracolsep{\fill}}lcc|ccc}
      \hline
       & 4~jets & 5~jets & 6~jets & 7~jets & 8~jets \\
      \hline\hline
      BG & $16060\pm 575$ & $2750\pm 92$& $536\pm 17$& $255\pm 8$ & $146\pm 5$ \\
      BG (corrected)  & $15961\pm 677$ & $2653\pm 112$& $481\pm 20$ & $223\pm 10$ & $142\pm 7$\\
      \ttbar ($\sigma_{t \bar t}=8.3$~pb) & $120\pm 20$ & $266\pm 45$& $242\pm 41$& $101\pm 17$&$38\pm 7$\\
      \hline
      BG + $t \bar t$ & $16081\pm 677$ & $2919\pm 121$& $723\pm 46$ & $324\pm 20$ & $180\pm 10$\\
      Data & 16555 & 3139 & 725 & 349 & 159\\
      \hline
    \end{tabular*}
    \label{tab:CDFalljetsxsec}
  \end{center}
\end{table}

The dominant systematical uncertainty in the measurement arises from
the uncertainty in jet energy scale, strongly impacting both the preselection of
events and the input variables for the further neural network
selection. The sample composition of candidate tags in data is
illustrated in Fig.\ \ref{fig:CDFalljetsxsec}, and is detailed in
Table~\ref{tab:CDFalljetsxsec}. The resulting cross section is given
in Table~\ref{tab:alljetsxsecs}, together with the other all-hadronic
channel results obtained thus far in Run~II.

\begin{table}[t]
  \caption{\ttbar all-hadronic cross section measurements performed thus
    far at the Run~II Tevatron with their integrated luminosities,
    data selections and analysis methods used. All three
    results have been published. The measurement marked with an asterisk
    refers to $m_{t}$ of 178~GeV/c$^2$ rather than the standard
    175~GeV/c$^2$.}
  \renewcommand{\arraystretch}{1.2}
  \begin{center}
    \begin{tabular}{|c|c|c|l|c|}
      \hline 
      $\int{\cal L}dt$&\multirow{2}{*}{Sel.}&\multirow{2}{*}{$b$ tag}&$\sigma_{t\bar{t}}${\footnotesize $\pm$(stat.)$\pm$(syst.)$\pm$(lumi.)}&\multirow{2}{*}{Ref.}\\ 
      ~[fb$^{-1}$] &  &  & {\hspace{1.5cm} [pb]} & \tabularnewline
      \hline
      \hline 
0.3 & jets only & yes &$7.5\pm2.1^{+3.3}_{-2.2}\,^{+0.5}_{-0.4}~(*)$ & \cite{Abulencia:2006se}\\%
0.4 & jets only & yes &$4.5^{+2.0}_{-1.9}\,^{+1.4}_{-1.1}\pm  0.3$ & \cite{Abazov:2006yb}\\
1.0 & jets only & yes &$8.3\pm1.0^{+2.0}_{-1.5}\pm0.5$ & \cite{Aaltonen:2007qf}\\
      \hline 
    \end{tabular}
    \label{tab:alljetsxsecs}
  \end{center}
\end{table}

\subsubsection{Hadronic $\tau$ channels}
\label{sec:ttbarxsecmeashadtau}
By choosing a more inclusive \ttbar event selection, hadronic $\tau$
decays can be included as already mentioned in the discussion of lepton +
jets and dilepton channels for the $\tau$ + lepton case. A first
measurement without any explicit lepton identification has been
published by CDF, selecting events with at least four jets, with at
least one being $b$ tagged, and significant \MET not aligned
with any jet. Since events with isolated energetic electrons or muons
are rejected, the resulting sample is enriched in $\tau$
+ jets events~\cite{Abulencia:2006yk}.

Explicit reconstruction of hadronic $\tau$ decays is far more
demanding and usually relies on multivariate discriminants. Based on
the decay mode (1-prong or 3-prong, with or without associated
electromagnetic subclusters from neutral pions), different
discriminants can be deployed, exploiting differences between
hadronic $\tau$ decays and jets. For example, isolation in the tracking system and
calorimetry, shower shape, track multiplicity or correlations between
tracks and clusters in the calorimeter can be applied. Using such discriminants, the
$\tau$ + jets and $\tau$ + lepton \ttbar decay modes have been
studied based on their experimental signature of \METns, at least one
hadronic $\tau$ candidate, and at least four or two jets, and no
or one isolated energetic electron or muon for the two channels, respectively. $b$ jet
identification is crucial to improve sample purity in such analyses.

A first $\tau$ + jets cross section analysis has been performed by
D0 on 0.3~fb$^{-1}$ of data, deploying a preselection and $\tau$
identification as outlined above. $b$ jet identification and neural
networks based on event topology and kinematics provide further separation
of \ttbar signal and background that is mainly due to QCD multijet
production, where jets mimic $\tau$ decays~\cite{D05234}. While the measurement suffers
from large statistical uncertainties, it is a proof of principle that
will benefit greatly from the tenfold increased dataset already in
hand.

\begin{figure}
  \begin{center} 
    \includegraphics[width=.49\textwidth]{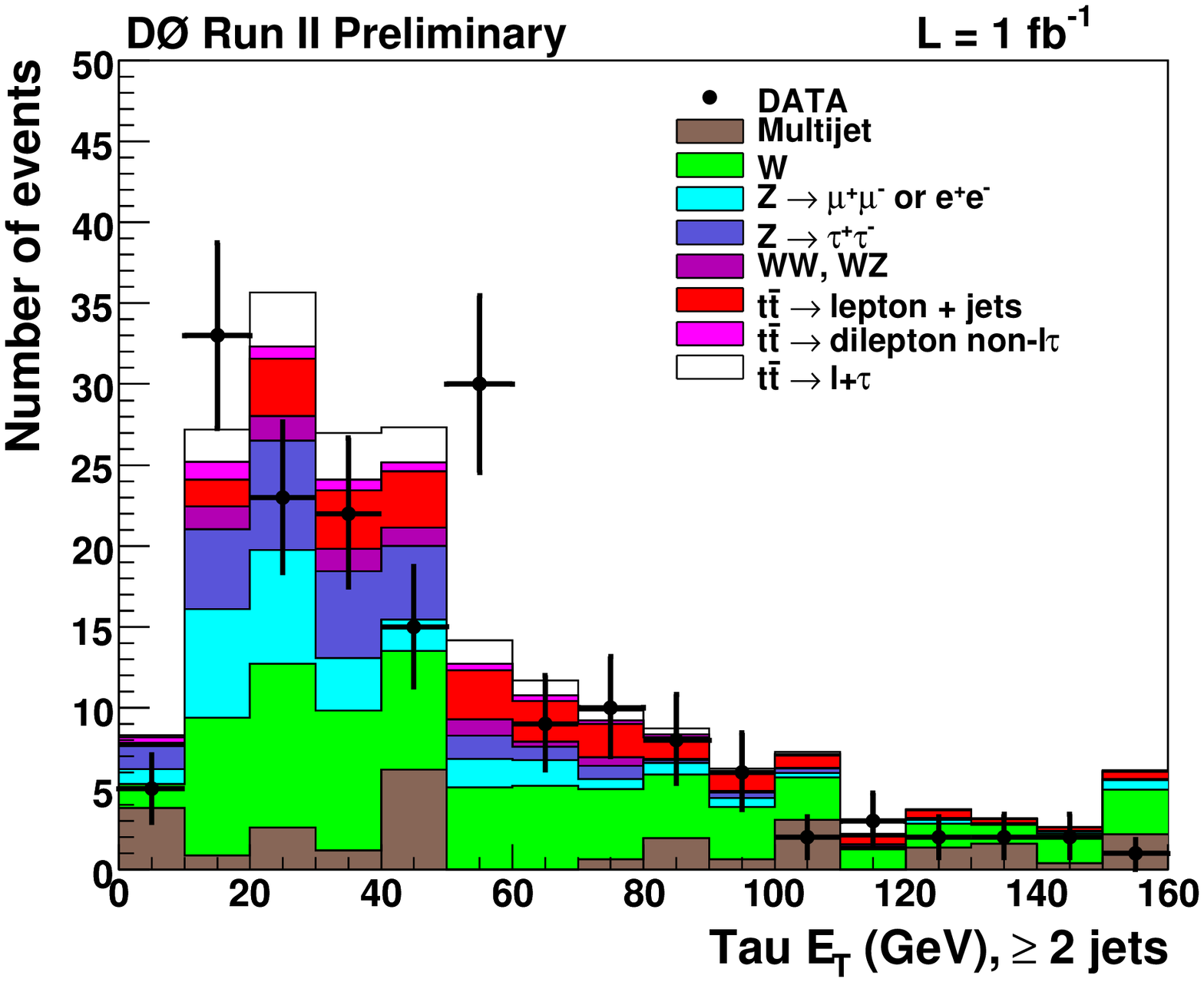}\hspace*{1mm}
    \includegraphics[width=.49\textwidth]{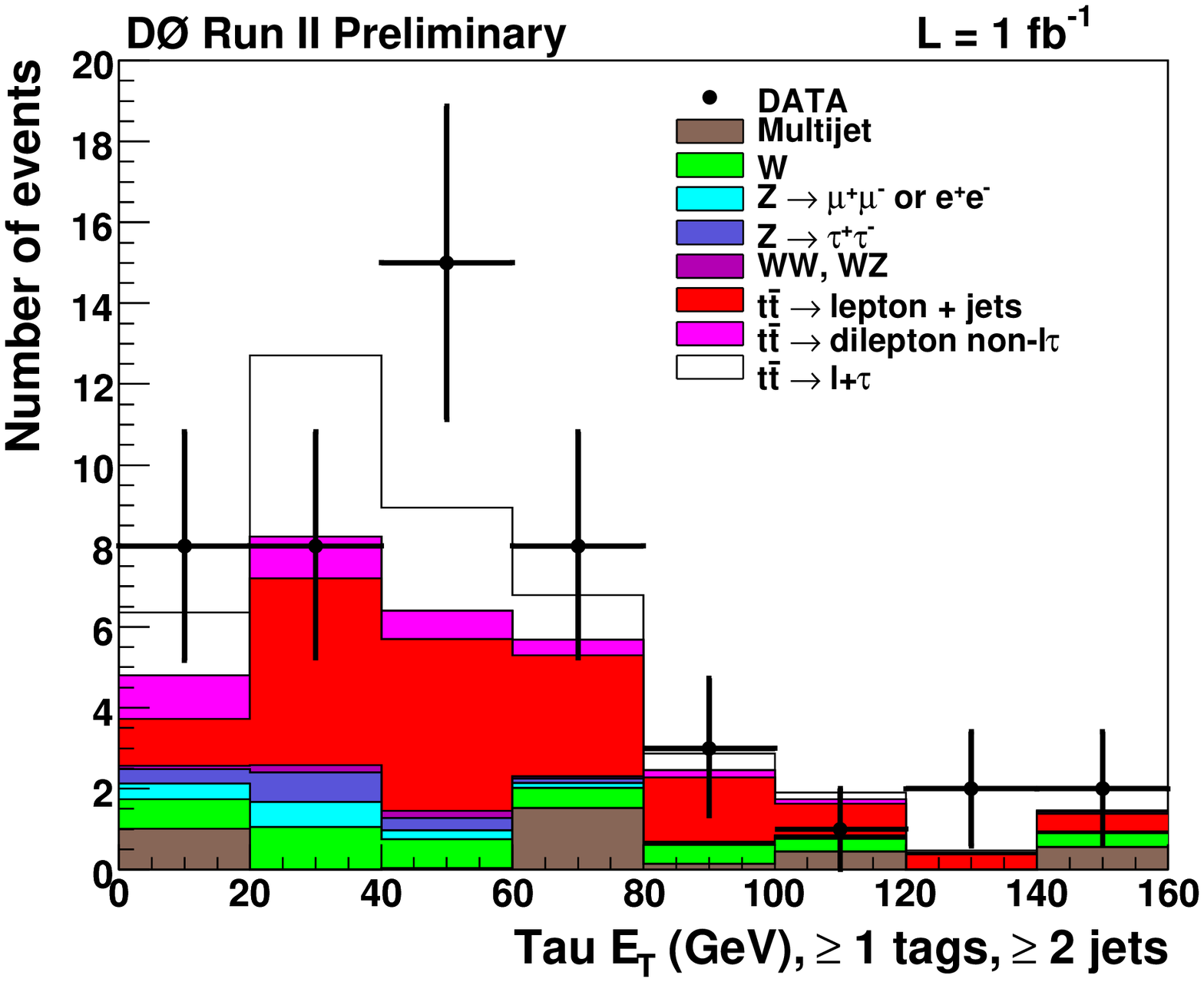} 
    \caption{Hadronic-$\tau$ $E_T$ spectrum before (left) and after
      (right) $b$ tagging is applied in the preselected 1~fb$^{-1}$
      $\ell + \tau$ sample analyzed by D0. The \ttbar signal is
      normalized to the SM expectation. The highest $E_T$ bin
      contains all overflows~\cite{D05451}.}
      \label{fig:D0ltauxsec}
  \end{center}
\end{figure}
D0 has also performed a first measurement of the \ttbar production
rate in the $\tau$ + lepton final state based on 1~fb$^{-1}$ of
data~\cite{D05451}. Events are required to have exactly one isolated
electron or muon with $E_T > 15$~GeV and $|\eta| < 1.1$ or $E_T
> 20$~GeV and $|\eta| < 2$, respectively, at least one $\tau$ candidate of opposite
charge within $|\eta| < 1$, \MET between 15 and 200~GeV, at least two
jets within $|\eta| < 2.5$ with $E_T > 20$~GeV and the leading
jet above 30~GeV, and at least one of these jets identified as $b$ jet.
Additional channel-specific background rejection is achieved by vetoing
events collected in the $\mu\tau$ analysis with invariant mass of the
isolated muon and a second non-isolated muon between 70 and
100~GeV/c$^2$. In the $e\tau$ channel, events are rejected where
electron and \MET are aligned in azimuthal angle $\phi$ by requiring
$\cos(\Delta\phi(e,\not\!\!E_T)) < 0.9$.
The resulting sample composition is illustrated in
Fig.\ \ref{fig:D0ltauxsec}, and detailed in
Table~\ref{tab:D0ltauxsec} -- there are significant contributions to signal
from the lepton + jets and dilepton channels. 

Background contributions arise from $W$ boson + jets production,
$Z/\gamma^{*}$ + jets events with $Z/\gamma^{*}\to \ell^{+}\ell^{-}
/ \tau^{+}\tau^{-}$, and diboson production, as described by MC. The
$W$ boson + jets contribution is normalized to data. Background from QCD
multijet production is estimated from data where the lepton and $\tau$
are of same charge, corrected for significant contributions from $W$ boson
+ jets and \ttbar production. The
dominant systematic uncertainties in this analysis arise from limited
background/MC statistics, the rate for jets or leptons to mimic $\tau$ decays,
modeling of $b$ tagging uncertainties, and jet energy calibration.
\begin{table}[t!]
  \caption{Event yields and sample composition before and after $b$ tagging
    in the preselected 1~fb$^{-1}$ $\ell + \tau$ sample
    analyzed by D0. The \ttbar signal is normalized to the SM
    expectation. The uncertainties are purely
    statistical~\cite{D05451}.}
  \begin{center}
    \begin{tabular*}{\textwidth}{@{\extracolsep{\fill}}l|rr|rr}
      \hline 
& \multicolumn{2}{c|}{before $b$ tagging} &\multicolumn{2}{c}{after $b$ tagging}\\
&$\mu\tau~~~~$&$e\tau~~~~$&$\mu\tau~~~~~$&$e\tau~~~~~$\\\hline\hline
$W + $jets 					&$38.0  \pm 1.7 $&$	34.1  \pm 3.5 $&$	2.31  \pm 0.22 $&$	2.13  \pm 0.27  $\\	 
$Z/\gamma^* \to ee, \mu\mu$ 		&$20.7  \pm 1.1 $&$	5.8  \pm 0.6  $&$	1.09  \pm 0.11 $&$	0.38  \pm 0.05  $\\
$Z/\gamma^* \to \tau\tau$ 		&$19.6  \pm 1.2 $&$	7.5  \pm 0.6  $&$	1.02  \pm 0.10 $&$	0.54  \pm 0.06 	$\\ 
Diboson 				&$2.8  \pm 0.1 	$&$	5.1  \pm 0.6  $&$	0.21  \pm 0.01 $&$	0.34  \pm 0.07 	$\\ 
Multijet 				&$10.6  \pm 6.3 $&$	12.7  \pm 6.6 $&$	4.52  \pm 3.01 $&$	-1.27  \pm 1.77 $\\	 
$t\bar{t} \to \ell + \tau $ 	        &$7.8  \pm 0.1 	$&$	6.67  \pm 0.1 $&$	5.64  \pm 0.04 $&$	4.70  \pm 0.05 	$\\ 
$t\bar{t} \to \ell\ell$ 		&$4.3  \pm 0.1 	$&$	0.73  \pm 0.1 $&$	3.14  \pm 0.03 $&$	0.47  \pm 0.07 	$\\ 
$t\bar{t} \to \ell +$ jets 		&$12.7  \pm 0.1 $&$	12.41  \pm 0.2$&$ 	8.40  \pm 0.11 $&$	7.88  \pm 0.12 	$\\ \hline
Total Expected 				&$116.6  \pm 6.8$&$ 	85.0  \pm 7.7 $&$	26.33 \pm 3.02 $&$	15.17  \pm 1.97 	$\\ 
Data 					&$104~~~~ 	 	$&$	69~~~~ 	      $&$	29~~~~~  	       $&$     18~~~~~                    $\\

      \hline 
    \end{tabular*}
    \label{tab:D0ltauxsec}
  \end{center}
\end{table}

\begin{table}[t!]
  \caption{\ttbar cross section measurements using hadronic $\tau$ decays
    performed thus far at the Run~II Tevatron, with their integrated
    luminosities, data selections and analysis methods
    used. The first two results have been published; the others are
    preliminary. The measurement marked with a double cross $(\ddagger)$ includes
    the luminosity uncertainty in the systematic uncertainty.}
  \renewcommand{\arraystretch}{1.2}
  \begin{center}
    \begin{tabular}{|c|c|c|l|c|}
      \hline 
      $\int{\cal L}dt$&\multirow{2}{*}{Sel.}&\multirow{2}{*}{$b$ tag}&$\sigma_{t\bar{t}}${\footnotesize $\pm$(stat.)$\pm$(syst.)$\pm$(lumi.)}&\multirow{2}{*}{Ref.}\\ 
      ~[fb$^{-1}$] &  &  & {\hspace{1.5cm} [pb]} & \tabularnewline
      \hline
      \hline 
0.2  & $\ell+\tau$ & no &$< 5.2 \cdot\sigma_{\rm SM}$ (95\% C.L.)& \cite{Abulencia:2005et}\\
0.3  & \MET + jets & yes&$6.0\pm1.2^{+0.9}_{-0.7}~(\ddagger) $& \cite{Abulencia:2006yk}\\\hline%
0.3  & $\tau$+jets & yes&$ 5.1^{+4.3}_{-3.5} \pm 0.7 \pm 0.3 $& \cite{D05234}\\
0.4  & $\ell+\tau$ & no & significance at $\approx 1$ sd& \cite{CDF8376}\\ %
1.0  & $\ell+\tau$ & yes&$ 8.3^{+2.0}_{-1.8}\,^{+1.4}_{-1.2}\pm 0.5  $& \cite{D05451}\\
1.2  & $\ell+\tau$ & yes&$6.4^{+1.8}_{-1.6}\,^{+1.4}_{-1.3}\pm 0.4$& \cite{D05607}\\
2.2  & $\ell+\tau$ & yes&$7.3^{+1.3}_{-1.2}\,^{+1.2}_{-1.1}\pm 0.5 $& \cite{D05607}\\
      \hline 
    \end{tabular}
    \label{tab:hadtauxsecs}
  \end{center}
\end{table}

The corresponding cross section is given
in Table~\ref{tab:hadtauxsecs}, together with an update from an
additional 1.2~fb$^{-1}$ of D0 data, as well as other results
involving hadronic $\tau$ final states obtained thus far in Run~II.

\subsubsection{Summary}
\label{sec:ttbarxsecmeassummary}
\begin{figure}[p]
  \centering
  \includegraphics[width=0.54\textwidth]{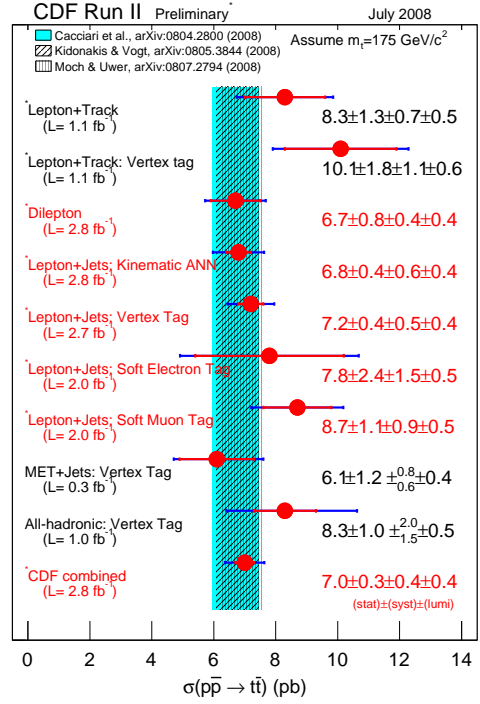}
  \includegraphics[width=0.75\textwidth]{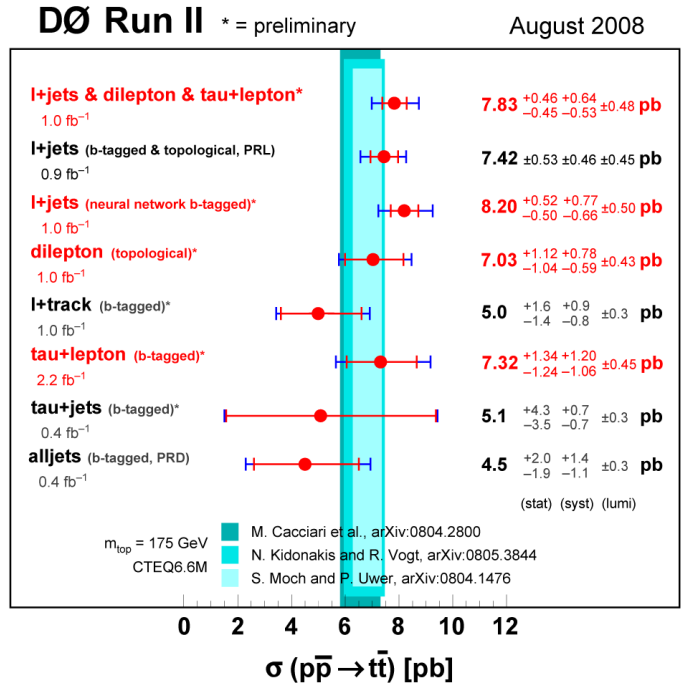}
  \caption{Current status of the \ttbar production cross section
    measurements by CDF (top,~\cite{wwwtopcdf}) and D0
    (bottom,~\cite{wwwtopd0}) compared with SM predictions
    (shown as shaded/hatched bands). The channels contributing to the
    combined results are highlighted
    \cite{CDF9448,D05715}.} %
  \label{fig:ttbarxsecsummary}
\end{figure}

An overview of the current status of \ttbar cross section measurements
performed in different decay channels is given in
Fig.\ \ref{fig:ttbarxsecsummary} for CDF and D0, showing good
agreement among channels, analysis methods and
experiments. The theoretical predictions discussed in
Section~\ref{sec:ttbarprod} are also shown for comparison as
shaded/hatched bands, and also show very good agreement with the
measurements.

CDF combines the results obtained in the lepton + jets and dilepton
channels using 1.7 $-$ 2.8~fb$^{-1}$ samples of data, achieving a
relative uncertainty on the result of 9\%~\cite{CDF9448}. The most
precise single measurement is obtained in the lepton + jets channel
using secondary-vertex $b$ tagging on 2.7~fb$^{-1}$ of data, yielding a
relative uncertainty of about 10\%~\cite{CDF9462}. D0 combines the
results from lepton + jets, dilepton and $\tau$ + lepton
channels obtained from approximately 1.0~fb$^{-1}$ of data, yielding a
relative uncertainty of about 11\%~\cite{D05715}. The most precise
single D0 measurement, with a precision of 11\%, has been published in
the lepton + jets channel, using both secondary-vertex $b$ tagging
and kinematic information in 0.9~fb$^{-1}$ of
data in a combined result~\cite{Abazov:2008gc}. For comparison, the final Run~I combined
cross section from CDF~\cite{Affolder:2001wd,Affolder:2001wderr}
and D0~\cite{Abazov:2002gy} had a precision of $\approx$25\%.
Unfortunately, no combined cross section measurement from both
experiments exists to date, unlike for top quark mass measurements
(see Section~\ref{sec:topmassaverage}). However, such
combination for the \ttbar production rate is planned in the future.

With increasing integrated luminosity, the statistical uncertainties
are becoming less important and precision is starting to be limited by
systematic uncertainties, which in turn can also be further
constrained with more data. One of the main challenges for
future measurements will be to study 
systematic uncertainties in greater detail rather than use ``conservative'' estimates. The
most precise single \ttbar production cross section measurement at the
Run~II Tevatron with the anticipated total of 8~fb$^{-1}$ of data can be expected in the lepton
+ jets channel -- probably using both kinematical and $b$ tagging
information -- at a relative precision of $\approx$8\%. This will be dominated by
uncertainties on luminosity and systematic effects. For combination of results, the
precision may ultimately be driven by the current luminosity uncertainty of
6\% for both experiments. The luminosity uncertainty could
be avoided by measuring cross section ratios, for example, relative to $Z$
boson production\footnote{After completion of this review, first such measurements became available from
CDF~\cite{CDF9474v2,CDF9616}.}. Also, with large datasets, a ratio of \ttbar
cross sections measured in different channels (such as
the lepton + jets and the dilepton channels) could be obtained with good
statistical precision while benefiting from cancellations of common
systematic uncertainties.

The precision of the measured \ttbar cross section now matches that
of the theoretical predictions, which provides stringent tests of 
perturbative QCD calculations. This furthers our
understanding of the standard model, which still provides an excellent
description of all current measurements. Based on the
observed production rate alone, severe constraints on phenomena beyond the SM
become feasible~\cite{Guchait:2007ux}. More detailed tests of the
standard model predictions for \ttbar production will be described in
the following sections. Via its mass dependence, the \ttbar production
rate can also be used to test consistency of the SM with the top quark mass
measurements performed at the Tevatron (with the benefit of easier
theoretical interpretation of the mass parameter as discussed in
Section~\ref{sec:massfromxsec}). Studying and comparing all available \ttbar
final states, including those with hadronic $\tau$ decays,
provides a probe for novel contributions that can affect the
observed final states in different ways. For example, searches for
charged Higgs bosons in top quark decays are discussed in
Section~\ref{sec:H+topdecay}.

\subsection{Top quark pair production mechanism}
\label{sec:ttggprod}
Top quark pair production at the Tevatron proceeds predominantly via
$q\bar{q}$ annihilation, as described in Section~\ref{sec:ttbarprod}.
The remaining fraction from gluon-gluon fusion ($f_{gg}$)
is $15 \pm 5 \%$, with the uncertainty mainly reflecting that
of the corresponding PDFs~\cite{Cacciari:2003fi}.

While the total \ttbar production rate has been studied extensively
(see Section~\ref{sec:ttbarxsecmeas}) and has been found to be in agreement
with the SM expectation, the production mechanism
has not yet been subject to such scrutiny. A measurement of the fraction of
\ttbar events produced via gluon-gluon fusion $f_{gg} = \sigma(gg \to
t\bar{t})/\sigma(p\bar{p} \to t\bar{t})$ provides a test of QCD,
and can contribute to a reduction in the uncertainties of
the corresponding PDFs. In addition, contributions from extensions of
the SM to \ttbar production could be
unveiled~\cite{Zhang:1999qy}, some of which may have remained undetected
because of the presence of new compensating decay mechanisms~\cite{Kane:1996ny}.

\begin{figure}[!t]
  \begin{center} 
    \includegraphics[width=.48\textwidth]{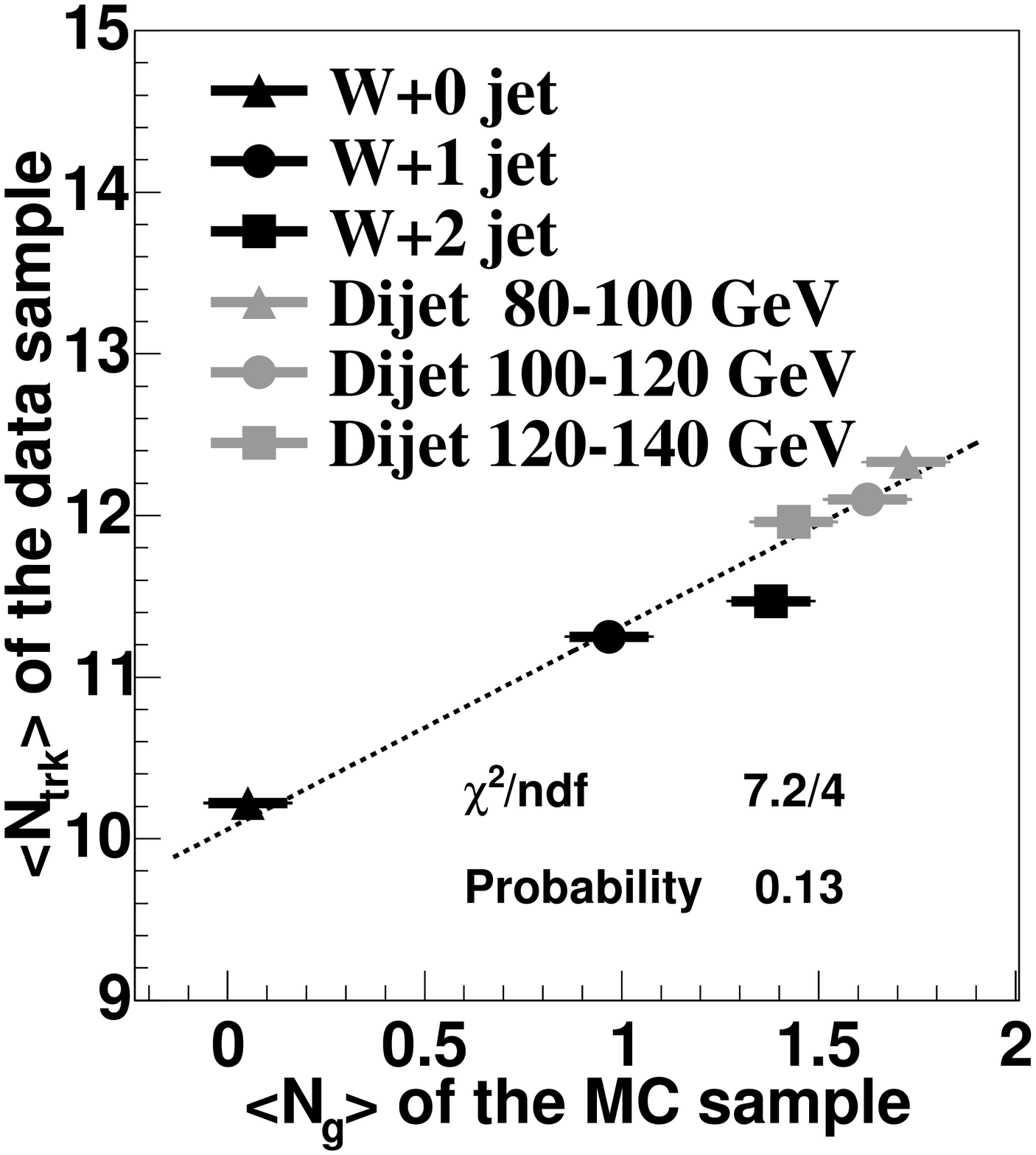}\hspace*{1mm}
    \includegraphics[width=.48\textwidth]{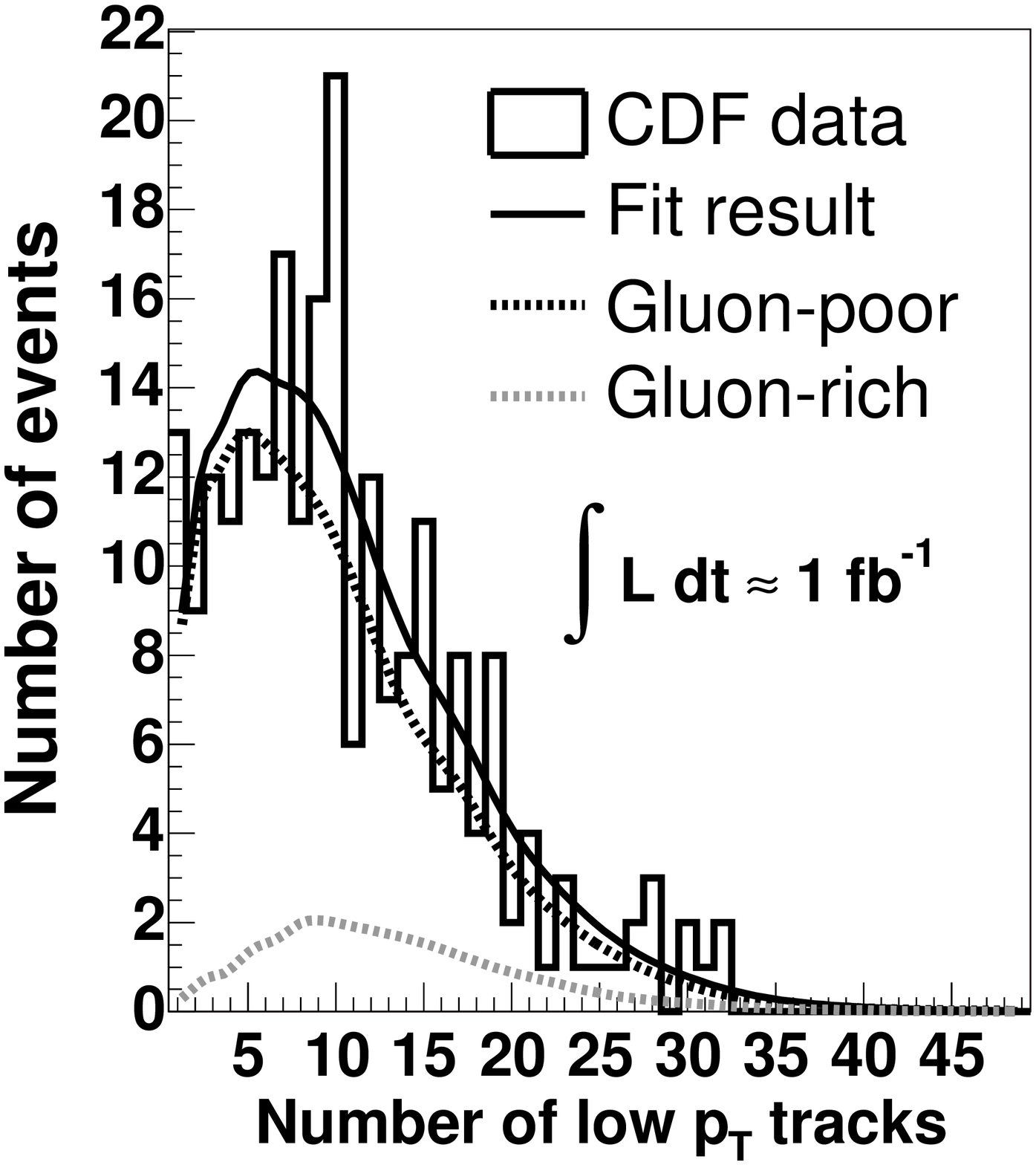} 
    \caption{Left: Correlation between the soft-track multiplicity
    observed in collider data and the average number of gluons
    in the corresponding Monte Carlo samples. Right: Fit result of the
    soft-track multiplicity distribution observed in $b$-tagged lepton
    $+$$\geq$4 jets \ttbar candidate events with a gluon-poor
    and a gluon-rich template~\cite{CDF:2007kq}.}
      \label{fig:ttggprod}
  \end{center}
\end{figure}
CDF performs a first measurement of $f_{gg}$ in a 1 fb$^{-1}$ $b$
tagged lepton + jets dataset~\cite{CDF:2007kq}. The analysis is
based on the fact that soft gluons are emitted with a higher
probability from gluons than from quarks
\cite{Gribov:1972ri,Lipatov:1974qm,Altarelli:1977zs,Dokshitzer:1977sg},
and the average number of charged particles (tracks) with low
transverse momentum should therefore be higher in $gg\to t\bar{t}$
events than in $q\bar{q}\to t\bar{t}$ events.

To avoid the large theoretical uncertainties on soft-gluon radiation
in the Monte Carlo modeling of the multiplicity of soft tracks, $W$ +
jets and dijet collider data, with well understood production mechanisms,
are used to relate the observed soft-track multiplicity to
the gluon content of a sample (see Fig.\ \ref{fig:ttggprod}). 
Templates for the soft-track multiplicity distribution in gluon-poor
and gluon-enriched events are obtained, respectively, from $W$ boson events
without jets and dijet events with a leading-jet $E_T$ of 80$-$100 GeV.
The observed soft-track multiplicity distribution in \ttbar candidate events is fitted with these
templates. From the fit result, $f_{gg}$ is extracted and found to be
$7 \pm 14\%\rm{(stat.)} \pm 7\%\rm{(syst.)}$, corresponding to a
95\% C.L.\ upper limit of 33\%.

CDF uses a complementary second method to extract $f_{gg}$ from the
same dataset, based on templates from a neural network using kinematic event
properties to separate $gg\to t\bar{t}$, $q\bar{q}\to t\bar{t}$ and
the dominant $W$ + jets background~\cite{Abulencia:2008su}, yielding a 95\%
C.L.\ upper limit for $f_{gg}$ of $61\%$. Combining both results yields
$f_{gg} = 7 ^{+15}_{-~7}\%\rm{(stat.+syst.)}$, in
good agreement with the SM expectation.

CDF has also performed a first measurement of $f_{gg}$ in 
2~fb$^{-1}$ of dilepton data, based on the variation of the azimuthal
correlation of the charged leptons for the different \ttbar
production modes~\cite{CDF9432}. This difference arises from the fact
that, close to threshold, top quark pairs are produced in a $^3S_1$
state via $q\bar{q}$ annihilation and in a $^1S_0$ state via
gluon-gluon fusion (see Section~\ref{sec:spincor}). Consequently,
the top quark spins tend to be antiparallel for \ttbar production via
gluon-gluon fusion and aligned for production via $q\bar{q}$
annihilation, which is reflected in the azimuthal correlation of the
charged leptons. The relative fraction of \ttbar production via gluon
fusion is determined in a fit of the observed $\Delta\phi$
distribution in data, with templates for $gg\to t\bar{t}$, $q\bar{q}\to
t\bar{t}$ and background arising from diboson, $Z/\gamma^{*}$ +
jets and $W$ boson + jets production, yielding $f_{gg} = 53
^{+35}_{-37}\% {\rm (stat.)}^{+7}_{-8}\% {\rm (syst.)}$,
consistent with the SM.

\subsection{Top quark charge asymmetry}
\label{sec:Afb}
The strong production of top quark pairs is symmetric under charge
conjugation at leading order, implying it does not discriminate
between top and antitop quarks. Considering that the initial proton-antiproton
state at the Tevatron is not an eigenstate of charge conjugation, this symmetry
is a coincidence. At higher orders, a charge
asymmetry arises from interference between amplitudes that are
symmetric and antisymmetric under the exchange of top and antitop
quarks~\cite{Kuhn:1998jr,Kuhn:1998kw}, leading to an excess of top over
antitop quarks in specific kinematic regions. 
One resultant observable is the 
integrated forward-backward production asymmetry for inclusive \ttbar
production at the Tevatron. This is predicted to be $5-10\%$ at
NLO~\cite{Kuhn:1998jr,Kuhn:1998kw,Bowen:2005ap,Antunano:2007da,Almeida:2008ug}, implying that top quarks
are preferentially emitted in the direction of the incoming protons.
The asymmetry depends strongly on the region of phase space being probed, and
particularly on the production of additional jets: While the
asymmetry for exclusive \ttbar production without additional jets is
predicted to be 6.4\%~\cite{Bowen:2005ap}, the inclusive \ttbar
production with one additional jet exhibits an asymmetry of about $-7\%$
at LO~\cite{Bowen:2005ap,Dittmaier:2007wz}, which is reduced drastically
to $(-1.5 \pm 1.5)\%$ at NLO~\cite{Dittmaier:2007wz}.

The size of higher order corrections for the \ttbar + jet
subprocess illustrates that higher-order evaluations of the whole
process are still necessary for the total asymmetry prediction to
converge and correctly describe the partial cancellations of the
various interference contributions. It should also be noted that in the
above theoretical predictions, the top quark decay and possible effects
on the asymmetry from reconstruction of the final-state objects
are not considered. The top quark charge asymmetry is also sensitive
to extensions of the standard model in \ttbar production involving, e.g.,
axigluons~\cite{Antunano:2007da}, technicolor~\cite{Hill:1994hp} or
additional neutral $Z'$ gauge bosons~\cite{Carena:2004xs}.
Consequently, a measurement of the asymmetry can be used to set limits
on such processes, particularly for extending the sensitivity of searches
for \ttbar production via heavy resonances (see
Section~\ref{sec:BSMprod-reso}) to include not only those of narrow
width but also wide resonances.

\begin{figure}[!t]
  \begin{center} 
    \includegraphics[width=.48\textwidth]{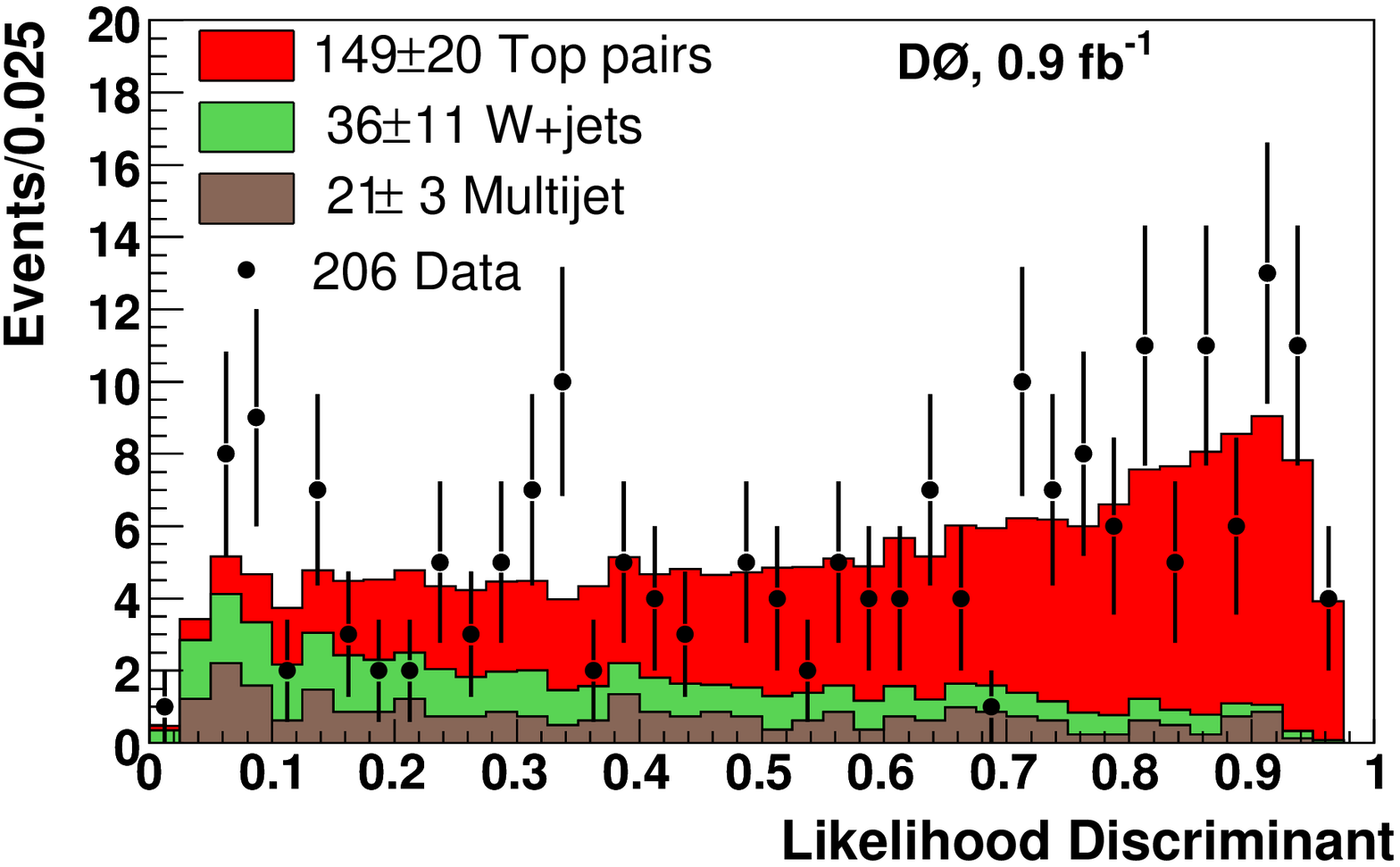}\hspace*{1mm}
    \includegraphics[width=.48\textwidth]{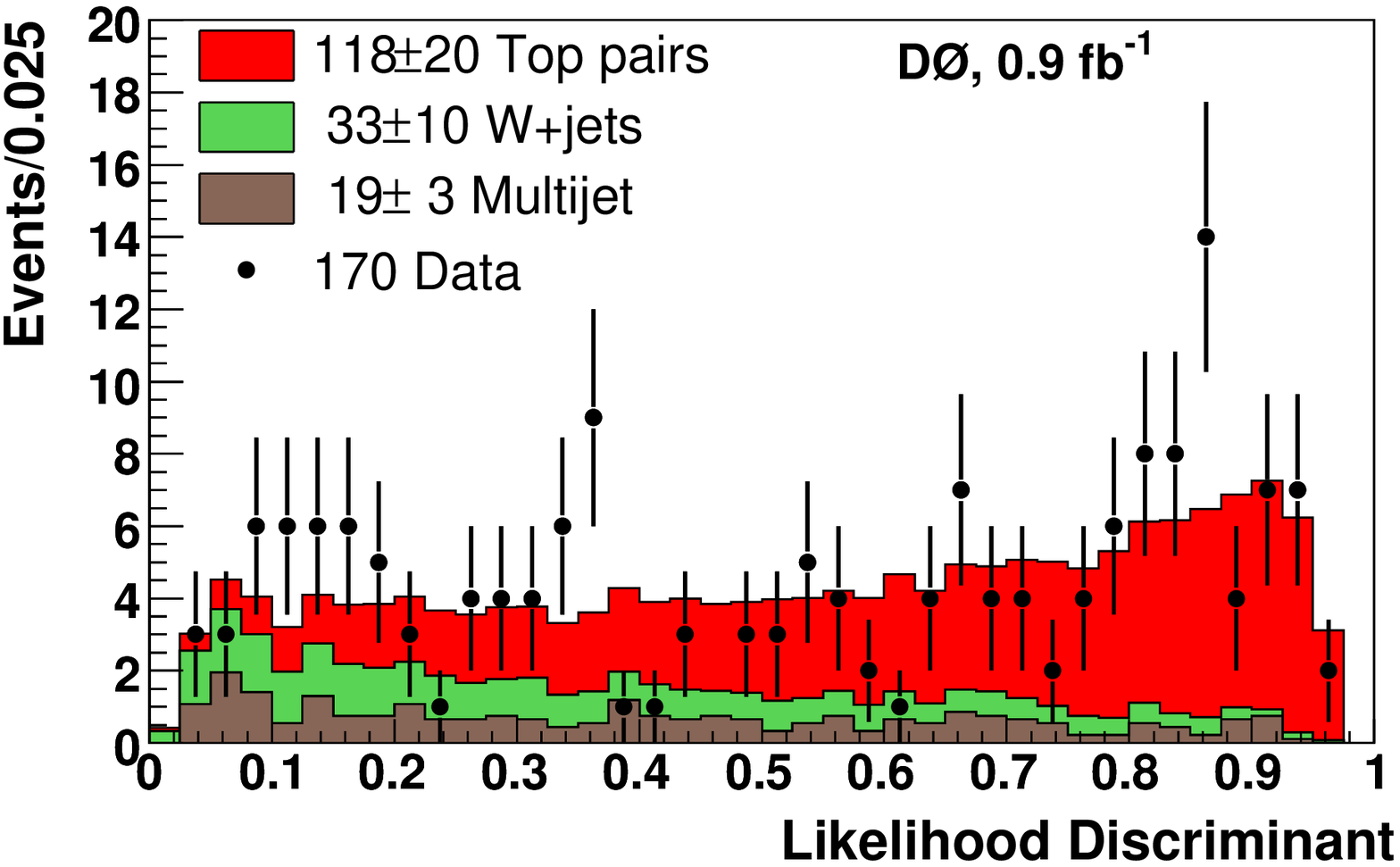} 
    \caption{Likelihood-discriminant output distribution for data with
    $\geq$4 jets, overlaid with the result of a template fit
    determining the sample composition for events with $\Delta y > 0$
    (left) and $\Delta y < 0$ (right)~\cite{Abazov:2007qb}.}
      \label{fig:Afb}
  \end{center}
\end{figure}

D0 has published the first measurement of the integrated
forward-backward charge asymmetry in \ttbar production, based on a 0.9
fb$^{-1}$ $b$ tagged lepton + jets dataset~\cite{Abazov:2007qb}.
The \ttbar system is reconstructed using a constrained kinematic fit,
where the charged lepton is used to differentiate between the top and the
antitop quark. The signed rapidity difference of the top and the antitop
quark $\Delta y = y_t - y_{\bar{t}}$ is used as an observable from which
the charge asymmetry is obtained as $A_{fb} = (N_f-N_b)/(N_f+N_b)$,
with $N_f$ ($N_b$) being the event yields with positive (negative)
$\Delta y$. The sample composition is determined in a template fit
based on a multivariate kinematic likelihood discriminant for both
signs of the reconstructed $\Delta y$ simultaneously as shown in
Fig.\ \ref{fig:Afb}.

The resulting measurement is not corrected for reconstruction and
acceptance due to the limited theoretical knowledge of the
shape of the asymmetry. Instead, a prescription is provided to model
the detector acceptance at the particle level, for ease of comparison of any
model with the measurement. For comparison with the
standard model, a slightly more detailed prescription than the one
provided in Ref.~\cite{Abazov:2007qb} is applied to the prediction from the
\mcatnlo~\cite{Frixione:2002ik,Frixione:2003ei} generator, and found to be in
agreement with the measurement for different jet
multiplicities, including a change in the sign of the asymmetry, as
shown in Table~\ref{tab:Afb}. The dominant systematic uncertainty for
the $\geq$4 jet sample arises from the relative jet energy
calibration between simulation and data, and for its
subsamples from event migration between the subsamples when splitting
the sample up into one with exactly four and one with at least five
jets. However, these systematics are negligible compared to the statistical
uncertainties. The measurement is also used to derive 95\% C.L.\ limits on the
fraction of \ttbar events that are produced via a specific $Z'$
resonance model~\cite{Hill:1994hp,Harris:1999ya} with parity-violating
couplings as a function of the resonance mass.
\begin{table}[!t]
  \caption{$A_{\text {fb}}^{\text {pred}}$: \mcatnlo\  SM
    prediction for the observed \ttbar charge asymmetry in the
    D0 detector, including uncertainties from acceptance and dilution
    (misreconstruction of sign in $\Delta y$). $A_{\text
      {fb}}^{\text {obs}}$: Uncorrected \ttbar charge asymmetry observed
    by D0~\cite{Abazov:2007qb}.}
  \begin{tabular*}{\linewidth}{l@{\extracolsep{\fill}}r@{\,$\pm$\,\extracolsep{0pt}}l@{(stat.)\,$\pm$}l@{(acc.)\,$\pm$\,}l@{(dil.)\extracolsep{\fill}}r@{}}
    \hline
    $N_{jet}$ & \multicolumn{4}{c}{$A_{\text {fb}}^{\text {pred}}$~[\%]} & \multicolumn{1}{c}{$A_{\text {fb}}^{\text {obs}}$~[\%]} \\
    \hline
    \hline	
    $\geq 4$ &   0.8  & 0.2 & 1.0 & 0.0 & $12 \pm 8\,\rm{(stat.)} \pm 1\,\rm{(syst.)}$\\
    $=4$     &   2.3  & 0.2 & 1.0 & 0.1 & $19 \pm 9\,\rm{(stat.)} \pm  2\,\rm{(syst.)}$\\
    $\geq 5$ & $-4.9$ & 0.4 & 1.0 & 0.2 & $-16 ^{+15}_{-17}\,\rm{(stat.)} \pm  3\,\rm{(syst.)}$\\
    \hline
  \end{tabular*}
  \label{tab:Afb}
\end{table}

CDF has obtained two measurements of $A_{\text {fb}}$
based on 1.9~fb$^{-1}$ $b$-tagged lepton + jets data, using
different observables after reconstruction of the \ttbar kinematics in
a constrained fit~\cite{CDF9156,CDF9169,Aaltonen:2008hc}. CDF
chooses a different approach for the measurements than D0
by providing results both
before and after background subtraction {\em and} correction for acceptance
and reconstruction effects.

The first analysis uses as observable the rapidity difference between the
hadronically and semileptonically decaying (anti-) top quark multiplied by
the lepton charge~\cite{CDF9156,Aaltonen:2008hc}. This is equivalent to $\Delta y$
in the measurement reported by D0. After background subtraction, asymmetries of $0.119
\pm 0.064, 0.132 \pm 0.075$ and $0.079\pm0.123$ are observed 
for jet multiplicities $\geq4, =4$ and $\geq5$, respectively,
which is consistent with the measurement by D0, and which
correspond to \mcatnlo\  predictions of $0.017\pm0.007, 0.038\pm0.008$ and
$-0.033\pm0.012$ (errors are statistical). The result for the
inclusive sample with at least four jets is then corrected for
reconstruction and acceptance effects, yielding $A_{\text {fb}}^{\text
{corr}} = 24 \pm 13\,\rm{(stat.)} \pm 4\,\rm{(syst.)\%}$. The
dominant systematic uncertainty comes from the shape uncertainty of
the $\Delta y$-distribution. The result is bigger than expected from
NLO predictions, but consistent within errors.

The second analysis measures the charge asymmetry using the product of
inverse lepton charge and $\cos\theta_{t_{\rm{had}}}$ as observable,
where $\theta_{t_{\rm{had}}}$ is the angle of the top quark with the
hadronic decay chain relative to the proton beam~\cite{CDF9169,Aaltonen:2008hc}.
Since this measurement is performed in the laboratory frame rather
than the parton rest frame, the asymmetry is reduced by about 30\%~\cite{Antunano:2007da}.
For $\geq4$ jets, the corrected asymmetry is 
$A_{\text {fb}}^{\text {corr}} = 17 \pm 7\,\rm{(stat.)} \pm
4\,\rm{(syst.)\%}$, with the dominant systematic uncertainty arising
from background shape and its normalization. This result is consistent
with the theoretical prediction at the level of two standard deviations (sd) for a Gaussian distribution.

It should be noted that the forward-backward asymmetry in the
laboratory frame vanishes at the LHC due to the symmetric initial
state, in contrast to the Tevatron $p\bar{p}$ collider. The
dominance of \ttbar production via the charge symmetric gluon fusion reaction at $\sqrt{s}= 14$~TeV
reduces the observable charge asymmetry at the LHC.

\subsection{Top quark pair production kinematics}
\label{sec:topkinematics}
New contributions to \ttbar production can alter the observed
event kinematics, which can be exploited in searches for such processes, as
described in Section~\ref{sec:BSMprod}. The basic kinematic properties
of leptons, jets, \MET and corresponding angular distributions are
continuously compared to the SM expectation, both in
signal-enriched datasets and signal-depleted control samples
exhibiting features similar to the signal, in all studies of top quark
properties and especially in the cross section analyses. Thus far, no
significant deviation from the SM expectation has been found
that would be indicative of new physics contributions to top quark
samples.
In Run~I, CDF and D0 observed a slight excess of the \ttbar production
rate over the SM prediction in the dilepton channel,
especially in the $e\mu$ final state~\cite{Abe:1997iz,Abachi:1997re}.
Since some of these events had rather large \MET and lepton-$p_T$,
their consistency with the SM was questioned and, for example, the
kinematic compatibility of these events with cascade decays of
heavy supersymmetric quarks was pointed out~\cite{Barnett:1996hw}.
Triggered by this, CDF performed a search for anomalous \ttbar
kinematics in Run~II, based on 0.2 fb$^{-1}$ dilepton data
yielding 13 candidate events~\cite{Acosta:2004av}. 

A priori four
kinematic event variables, including \MET and leading-lepton $p_T$, were
chosen to quantify any possible deviations of the observed distributions from
SM predictions. Using a shape comparison based on the
Kolmogorov-Smirnov statistic, no significant discrepancy was found, and
the probability of observing a sample less consistent with the SM
was determined to be 1.6\%. Including systematic uncertainties,
the $p$-value was 1.0-4.5\%, where the lowest value was obtained
by lowering the background expectation by one standard
deviation. Consequently, presence of processes beyond the SM,
with high \MET and high lepton-$p_T$, are not favored by these data.

It is also of great interest to study the kinematic
properties of the top quark itself and to compare these with the
SM expectation. In Run~I, D0 performed such an analysis
and found good agreement with the SM~\cite{Abbott:1998dc,
Abachi:1997jv}, which was then also confirmed through a dedi\-cated study of
the $p_T$ spectrum of the top quark by CDF~\cite{Affolder:2000dt}. A
corresponding analysis has not yet been published in Run~II. However,
a measurement of the differential \ttbar production cross section
$d\sigma/dM_{t\bar{t}}$ has been performed by CDF using 1.9~fb$^{-1}$
of Run~II data, as described in Section~\ref{sec:dsigma}, and also shows
good agreement with the SM~\cite{CDF9157}.

\subsection{Spin correlations in \ttbar production}
\label{sec:spincor}
Top quark pairs are expected to be produced essentially unpolarized in
hadron collisions when the incident particles are unpolarized. A small
polarization at the percent level is induced by QCD
processes~\cite{Dharmaratna:1989jr,Bernreuther:1995cx,Dharmaratna:1996xd},
and is perpendicular to the production plane, as strong
interactions conserve parity. A measurement of this effect will be
very difficult both at the Tevatron and the LHC, which in return suggests
to use a corresponding analysis to probe for non-standard
contributions in \ttbar production~\cite{Bernreuther:2003ga}. An even
smaller additional polarization in the production plane arises
from mixed strong and weak contributions to \ttbar production at order
$\alpha_s^2\alpha$~\cite{Bernreuther:2005is}.

While no observable spin
polarization in \ttbar production is predicted in the framework of the standard model, the spins of the top
and the antitop quark are expected to be correlated~\cite{Kuhn:1983ix}.
This correlation depends both on the production mode of the \ttbar pair and the
production energy. Close to threshold, the top quark pair is
produced in a $^3S_1$ state via $q\bar{q}$ annihilation and in a
$^1S_0$ state via gluon-gluon fusion
~\cite{Hara:1989yqa,Arens:1992fg}. Consequently, the top quark spins
are (anti-) parallel and the top quarks have opposite (same)
helicities for \ttbar production via $q\bar{q}$ annihilation
(gluon-gluon fusion). Above threshold, this simple picture becomes
more complicated, as effects of orbital angular momentum must be
taken into account. In the high-energy limit, where the top quark mass
can be neglected, the conservation of chirality dictates that
top and antitop quarks be produced with opposite helicities. Since
\ttbar production at the Tevatron proceeds mainly via $q\bar{q}$ annihilation,
as opposed to the LHC where the main contribution comes from
gluon-gluon fusion, the observable correlation will have opposite
signs at the two colliders~\cite{Bernreuther:2004jv}.

The short lifetime of the top quark (see
Section~\ref{sec:tdecaywidth}) assures that its spin information is passed on to
its decay products, and is reflected in their corresponding angular
distributions. This provides experimental access to spin correlations,
and a way to check whether the top quark can indeed be considered as a free
quark. The resulting indirect limits on the top quark lifetime 
(see Section~\ref{sec:toplifetime}) can provide
limits on the CKM matrix element $|V_{tb}|$, free from the assumption
of three quark families~\cite{Stelzer:1995gc} together with the
measurement of the top quark branching fractions
(Section~\ref{sec:Rmeasurement}). In addition,
spin correlations probe the dynamics of top quark
production and decay for possible contributions from physics beyond the SM.

The down-type ($T_3 = -\frac{1}{2}$) decay products of the $W$ boson
from top quark decay are most sensitive to the original top quark
spin. Their angular distribution in the top-quark rest frame is
described by $1+\cos\theta$, with $\theta$ being the angle between the
line of flight of the down-type fermion and the top spin
direction. The experimental difficulties of distinguishing between
jets from up-type and down-type quarks (charm tagging would help only
in 50\% of the cases) can be avoided by focusing on the dilepton
final state, where the charged (down-type) leptons are clearly identified.

At the Tevatron, an optimal spin-quantization basis is provided by the
``off-diagonal'' basis~\cite{Parke:1996pr,Mahlon:1997uc}, where the
spins of top and antitop quarks produced by $q\bar{q}$ annihilation
are fully aligned for all energies, and only the contribution of top
quark pairs from gluon-gluon fusion leads to a reduction of
the correlation. The off-diagonal basis is defined via the top quark's
velocity $\beta^\ast$ and scattering angle $\theta^\ast$ in
the collision rest frame of the incoming partons. The quantization
axis then forms an angle $\psi$ with the proton-antiproton beam axis,
defined by $\tan\psi = \beta^{\ast 2} \sin\theta^\ast \cos\theta^\ast /
(1-\beta^{\ast 2}\sin^2\theta^\ast)$. Consequently, in the limit of
$\beta^\ast\to0$ (top quark production at rest), the spins of top and
antitop quarks point along the beam axis in the same direction. At
very high energies, the spins are aligned with respect to the
direction of the \ttbar momenta.

Using as observables the angles $\theta_+$ and $\theta_-$ of the
down-type fermions relative to the quantization axis in the rest
frame of their respective parent (anti-) top quark, the spin
correlation is given by~\cite{Mahlon:1995zn}:
\begin{eqnarray}
{1\over \sigma}{d^2\sigma\over d(\cos \theta_+)d(\cos \theta_-)}&=&
{1+\kappa \cdot \cos \theta_+ \cdot \cos \theta_-\over 4},
\end{eqnarray}
where the correlation coefficient $\kappa$ is predicted to be +0.88 at
Run~I of the Tevatron when using the off-diagonal basis. Since the
distribution is symmetric under exchange of the two angles, an
electric charge measurement of the top decay products is not
necessary.

D0 performed a first search for evidence of spin correlations in
\ttbar production in Run~I, using a 0.1 fb$^{-1}$ dilepton data sample
containing six candidate events~\cite{Abbott:2000dt}. From the
dependence of a likelihood function on $\kappa$, at 68\% C.L.\ a lower
limit on $\kappa$ of $-0.25$ was extracted. This is in
agreement with the standard model expectation and disfavors
anti-correlation of spins ($\kappa = -1$) that would arise from \ttbar
production via a scalar particle. While this limit is rather
weak, it is a proof of principle that the analysis can be performed.
Unfortunately, there has been no result as yet from Run~II, although it
would greatly benefit from the increase in data already available.
\\

\subsection{Search for associated Higgs boson production}
\label{sec:ttH}
D0 performs a first search for associated \ttbar and standard model
Higgs boson production in the $t\bar{t}b\bar{b}$ final state in
2.1~fb$^{-1}$ of $b$ tagged lepton + jet data~\cite{D05739}. While
the observation of a significant signal in this channel is beyond the
reach of the Tevatron, this analysis can nevertheless
contribute to future combinations of the Tevatron searches for
Higgs bosons at low masses, as favored by the standard model (see
Section~\ref{sec:topmassaverage}). The investigated events exhibit
high jet and $b$ tag multiplicities that were not studied separately
before. It is therefore interesting to search for deviations from the
SM predictions that could arise, e.g., from
anomalous top-Higgs couplings~\cite{Feng:2003uv} or from a new quark singlet of
charge $\frac{2}{3} e$~\cite{AguilarSaavedra:2006gw}.
\begin{figure}[!t]
  \begin{center} 
    \includegraphics[width=.48\textwidth, height = 50 mm]{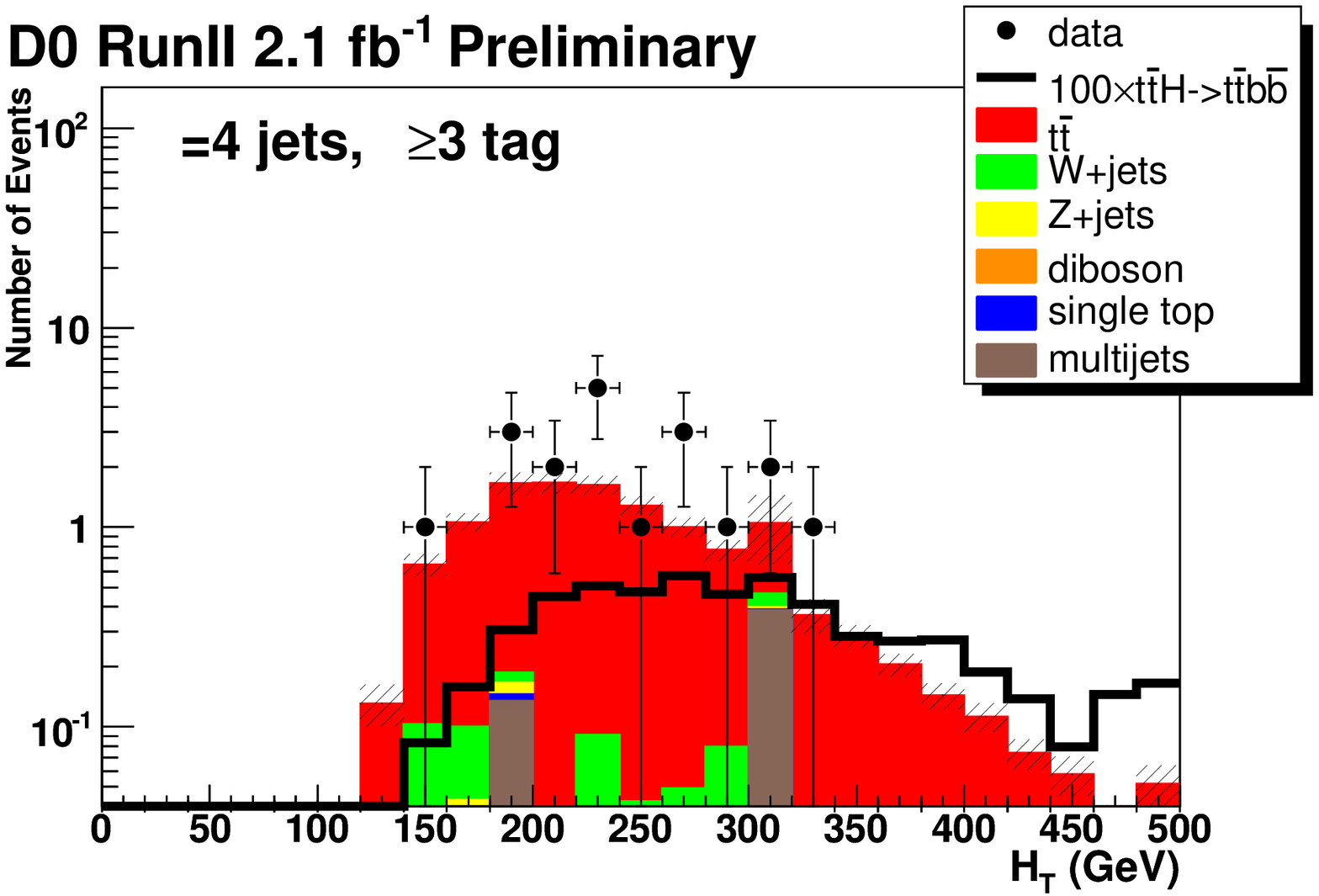}\hspace*{1mm}
    \includegraphics[width=.5\textwidth]{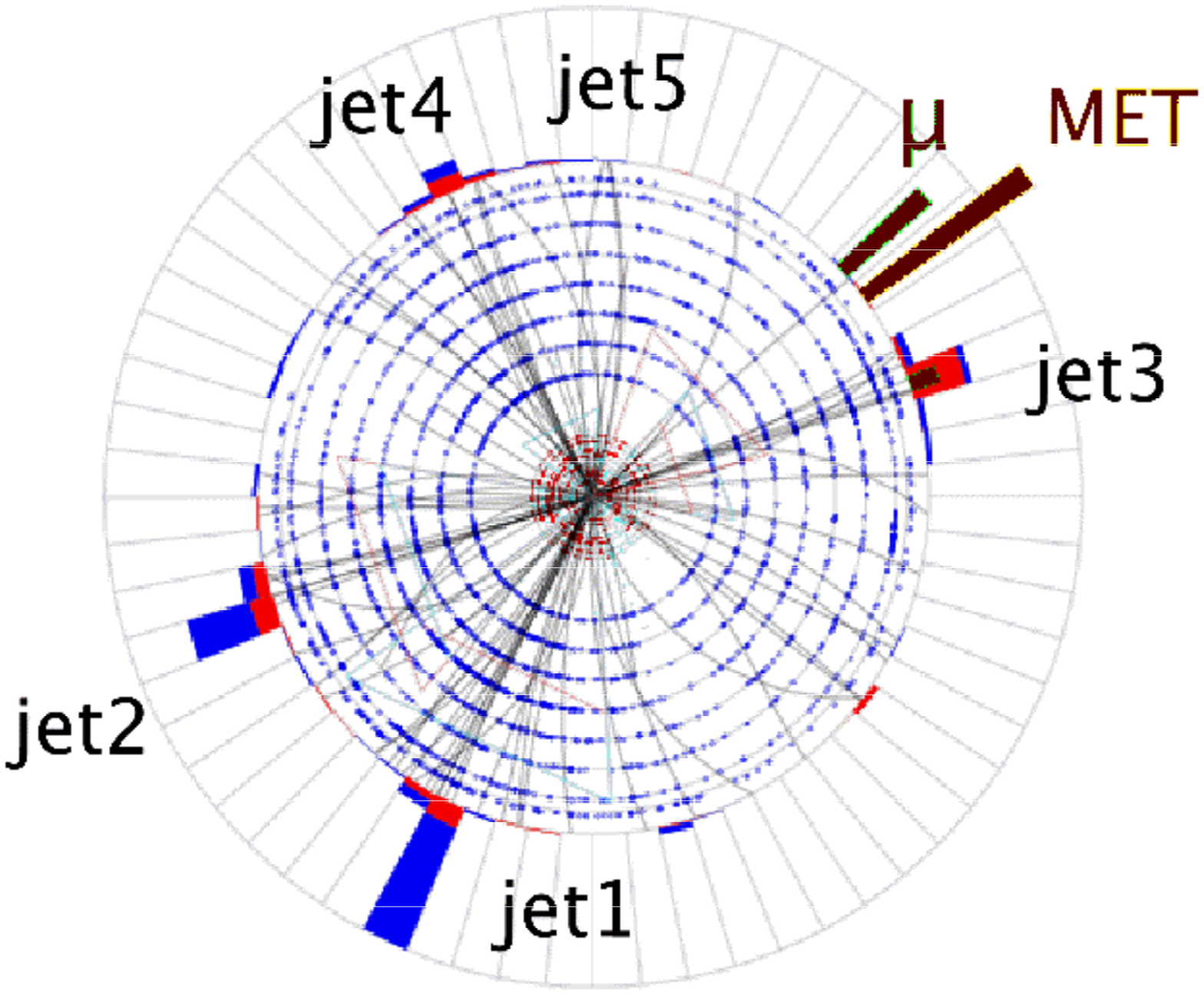} 
    \caption{Left: $H_T$ distribution observed in D0's 2.1~fb$^{-1}$
      lepton + jets data with four jets and at least three $b$
      tags, compared with expected standard model background processes and
      $t\bar{t}H$ signal scaled up by a factor of 100. Right: Event
      display ($xy$-view along the proton beam direction) for the
      triple $b$ tagged event of highest $H_T$ (444~GeV). The first
      three jets have $b$ tags~\cite{D05739}.}
      \label{fig:D0ttH}
  \end{center}
\end{figure}

The signal signature has the \ttbar lepton + jets event
characteristics, but with two additional $b$ jets from the Higgs boson
decay. The main background arises from \ttbar with additional
heavy-flavor jet production, but also $W$ boson + jets and QCD
multijet production contribute to the background. For signal
discrimination, the shape of the $H_T$ distribution of the selected
jets is used in events with four or at least five jets and one, two or
at least three $b$ tags. The observed events in all these
distributions are consistent with background expectation, which is
especially interesting for the events with at least three $b$ tags
that were studied separately for the first time.
Figure~\ref{fig:D0ttH} shows the observed $H_T$ distribution for
events with four jets and at least three $b$ tags, and an event display
for the triple $b$ tagged event of highest $H_T$.

Since no signal is observed, 95\% C.L.\ limits on $t\bar{t}H$
production multiplied by ${\cal B}(H\to b\bar{b})$ 
are derived for Higgs boson masses between 105
and 155 GeV/c$^2$. For a Higgs boson mass of 115~GeV/c$^2$, the
expected limit is a factor of 45 larger than the SM production rate, while
the observed limit is a factor of 64 above the SM
expectation. Optimization of the preselection (currently
corresponding to the standard \ttbar selection) for the $t\bar{t}H$
signal, and of the signal discrimination using additional kinematic variables, is underway.

\subsection{Search for top quark pair production beyond the standard model}
\label{sec:BSMprod}
\subsubsection{Search for a narrow-width resonance decaying into $t\bar{t}$}
\label{sec:BSMprod-reso}
The existence of yet undiscovered heavy resonances could be revealed
through their decays into top quark pairs, which would add a resonant
contribution to the standard model process. Theories beyond the
standard model predict the existence of massive $Z$-like
bosons, for example, such as Kaluza-Klein excitations of the
gluon~\cite{Lillie:2007yh} or of $\gamma$ and $Z$
bosons~\cite{Rizzo:1999en}, extended gauge
theories~\cite{Rosner:1996eb,Leike:1998wr}, massive axigluons with
axial vector couplings~\cite{Sehgal:1987wi} or
topcolor~\cite{Hill:1991at,Hill:1993hs}. 

The wealth of such models demonstrates the importance of
model-independent searches. One general way an additional
production mode can be observed -- provided the resonance $X$ decaying
to \ttbar is sufficiently heavy and narrow -- is to analyze the \ttbar
invariant mass distribution for an excess over 
expectation. In the corresponding analyses performed at the Tevatron,
no significant deviations from the SM expectation have thus far
been observed, resulting in 95\% C.L.\ upper limits on
$\sigma_X\cdot{\cal B}(X\to t\bar{t})$ as a function of resonance
mass $M_X$.

These results can be used to set lower mass limits for specific
benchmark models that provide easy comparison. For example,
topcolor~\cite{Hill:1991at,Hill:1993hs} provides a
dynamic electroweak symmetry breaking mechanism via a top quark pair
condensate~\cite{Cvetic:1997eb} $Z'$, formed by a new strong gauge
force, that couples preferentially to the third fermion generation.
Particularly, a topcolor-assisted technicolor
model~\cite{Hill:1994hp,Harris:1999ya} predicts this $Z'$ boson to
couple strongly only to the first and third generation of quarks, while
exhibiting no significant coupling to leptons. This leptophobic and
topophyllic $Z'$ boson has a significant cross section
$\sigma(p\bar{p}\to Z'\to t\bar{t})$ that is observable at the Tevatron for a
variety of masses and widths, and is used as a reference model.

CDF and D0 performed model-independent searches for narrow massive
vector bosons decaying into \ttbar already in Run~I in lepton +
jets datasets of 106~pb$^{-1}$ and 130~pb$^{-1}$, respectively. Using
the best kinematic fit to the \ttbar hypothesis in each event, the
\ttbar invariant mass distribution was reconstructed and no excess
observed above expectation. The resulting upper
limits on $\sigma_X\cdot{\cal B}(X\to t\bar{t})$ are turned into
95\% C.L.\ mass limits of $M_{Z'} > 480$~GeV/c$^2$ for
CDF~\cite{Affolder:2000eu} and $M_{Z'} > 560$~GeV/c$^2$ for
D0~\cite{Abazov:2003aw}. For these results, a width of the $Z'$
(or $X$) of 1.2\% of its mass is assumed, which is well below the
detector mass resolutions for \ttbar systems. Consequently, the
results are dominated by detector resolution and
independent of $\Gamma_{Z'}$ for values below the mass resolution of a
few percent ($\approx 0.04 M_{Z'}$ for D0 in
Run~I~\cite{Jain:2003bm}). This kind of resonance width is also used in
the Run~II measurements described below.

\begin{figure}[!t]
  \begin{center} 
    \includegraphics[width=.48\textwidth]{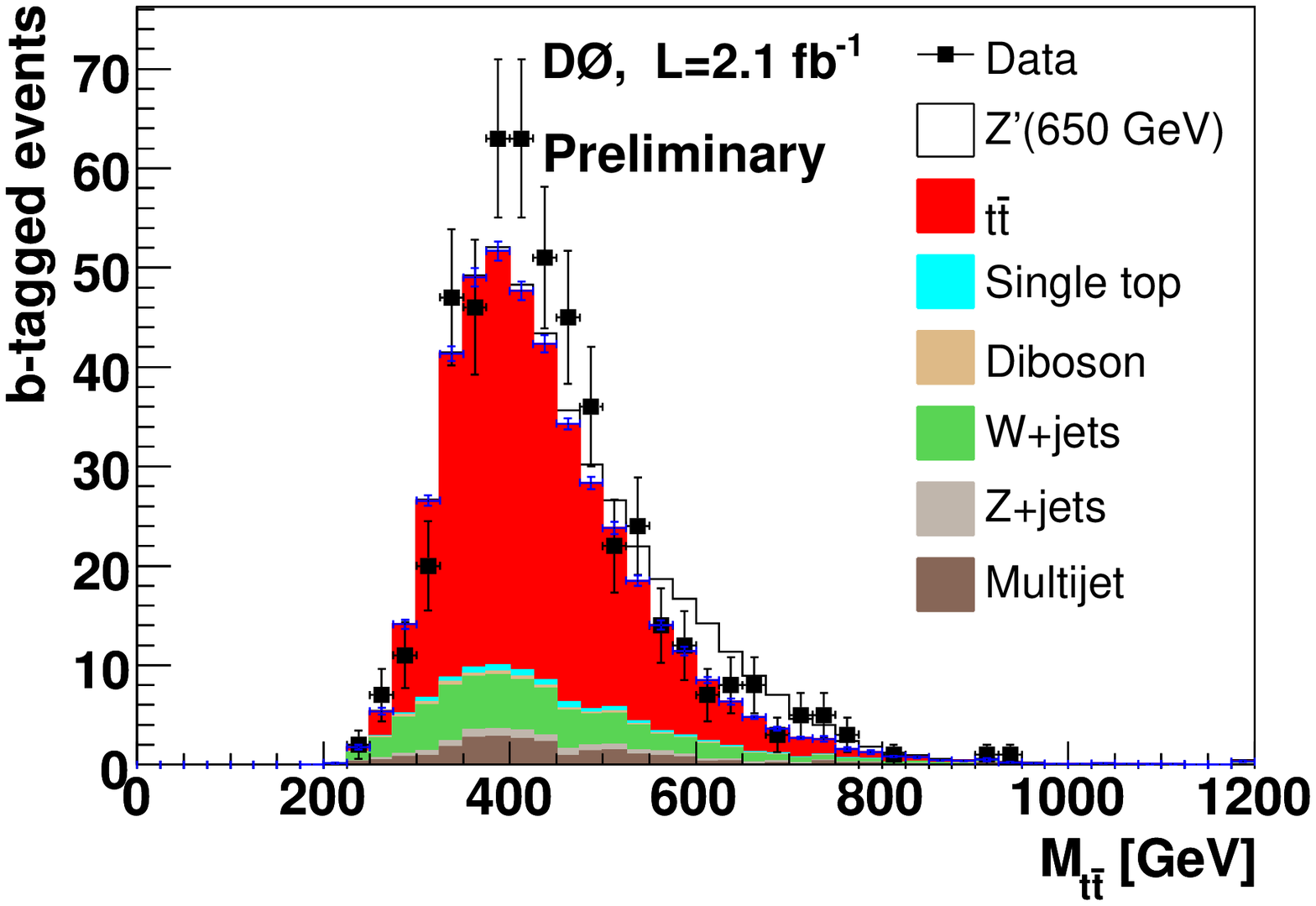}\hspace*{1mm}
    \includegraphics[width=.48\textwidth]{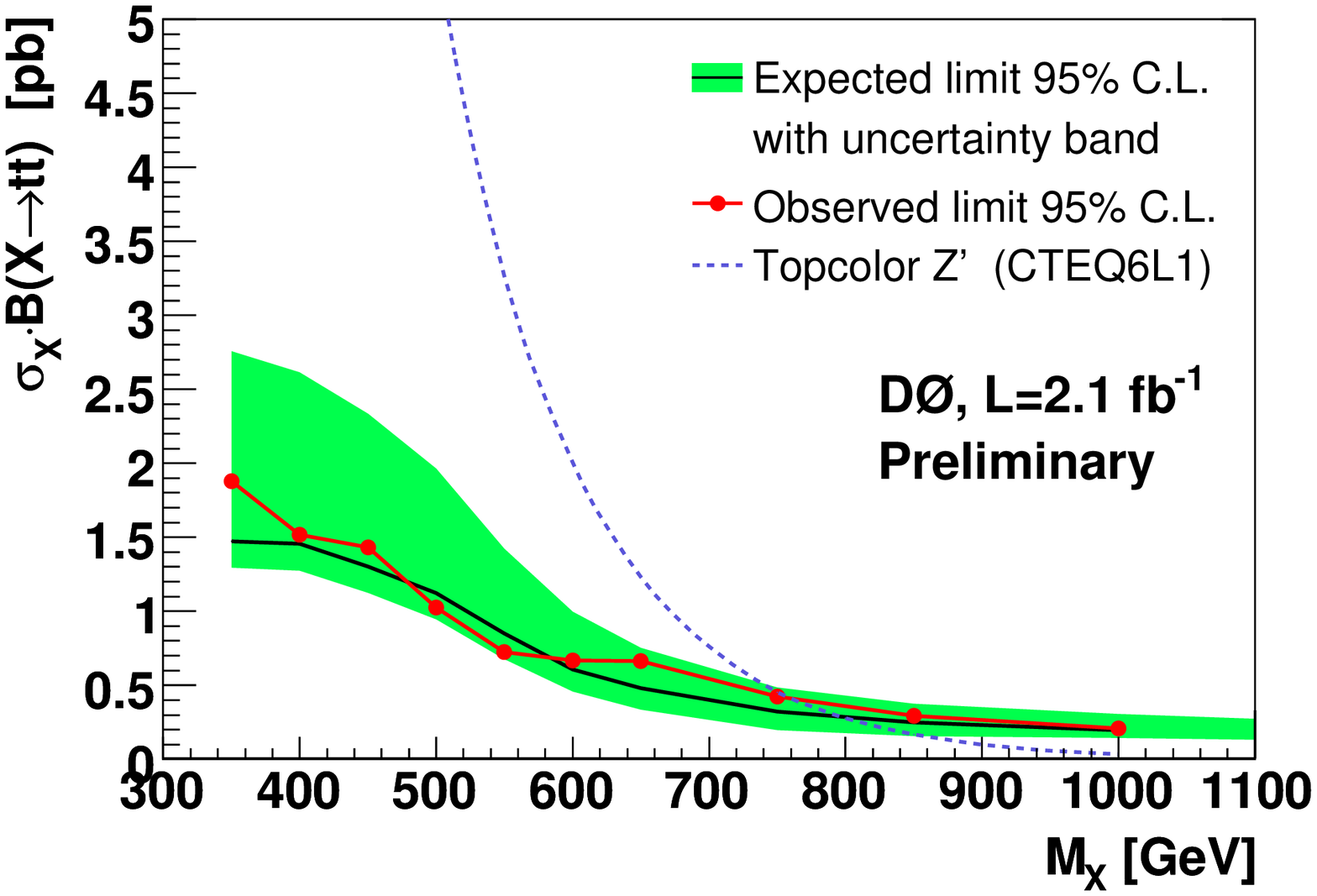} 
    \caption{Left: Expected and observed $t\bar{t}$ invariant mass
    distribution in lepton + jets events with four or more jets. Right:
    Expected and observed 95\% C.L.\ upper limits on
    $\sigma_X\cdot{\cal B}(X\to t\bar{t})$~\cite{D05600}.}
      \label{fig:ttresonance}
  \end{center}
\end{figure}
In Run~II, both CDF and D0 search for a generic heavy
resonance $X$ of narrow width ($\Gamma_X=0.012M_X$) compared to the
detector mass resolution in $b$ tagged lepton + jets datasets. The
\ttbar invariant mass spectrum is reconstructed using either the best
kinematic fit to the \ttbar production hypothesis (CDF) or directly
from the four-momenta of the up to four leading jets, the lepton and
the neutrino momentum (D0). The latter approach was shown to provide
greater sensitivity for large resonance masses than the previously used
constrained kinematic fit, and also allows for the inclusion of data with
fewer than four jets in case jets merged. As both experiments
observe no significant deviation from SM expectation, 95\% C.L.
upper limits on $\sigma_X\cdot{\cal B}(X\to t\bar{t})$ are given for
values of $M_X$ between 450 and 900 GeV/c$^2$ (CDF) and $M_X$ between 350
and 1000 GeV/c$^2$ (D0, see Fig.\ \ref{fig:ttresonance}) in
increments of 50~GeV/c$^2$.

Both experiments provide 95\% C.L.\ mass limits for the leptophobic
top\-color-assisted technicolor $Z'$ boson as a benchmark model.
With 1~fb$^{-1}$, CDF finds $M_{Z'}>$~720 GeV/c$^2$ (expected limit =
710 GeV/c$^2$)~\cite{CDFres:2007dia}, while D0 finds $M_{Z'}>$~760
GeV/c$^2$ (expected limit = 795 GeV/c$^2$)~\cite{D05600} using 2.1
fb$^{-1}$ of data, which supersedes a previous result on
0.9~fb$^{-1}$ of data~\cite{Abazov:2008ny}. CDF also obtains a result
on a subset of 0.7~fb$^{-1}$ of the data analyzed above, using an
untagged lepton + jets sample where $b$ tag information only
contributes as a way to reduce jet combinatorics in a standard model \ttbar
matrix element based reconstruction of $M_{t\bar{t}}$. This  yields a
slightly better limit than the analysis on the full 1~fb$^{-1}$ of $M_{Z'} >
725$~GeV/c$^2$~\cite{CDFres:2007dz}.

For future studies, it would be interesting to see how sensitive the
observed limits are to the assumption of $Z$ boson-like couplings used
in the analyses. The limits obtained apply to resonances of narrow
width only. Wider resonances could be detected by studying the \ttbar
differential cross section (see Section~\ref{sec:dsigma}) or the
forward-backward charge asymmetry in \ttbar production (see
Section~\ref{sec:Afb}).

\subsubsection{Search for $t\bar{t}$ production via a massive gluon}
Instead of a new color singlet particle decaying into \ttbar, as
described in the previous subsection, there could also be a new
massive color octet particle $G$ contributing to \ttbar
production~\cite{Hill:1993hs}. Such a ``massive gluon'' production
mode would interfere with the corresponding standard model production
process.
Assuming SM top decay, CDF has performed a search for a
corresponding contribution by comparing the \ttbar invariant mass
distribution in a 1.9~fb$^{-1}$ $b$ tagged lepton + jets dataset
with the standard model expectation. As the largest discrepancy 
relative to the standard model is found to be 1.7 sd for 
masses and widths of 400 GeV/c$^2$ $\leq M_{G} \leq$ 800 GeV/c$^2$ and
0.05 $\leq \Gamma_{G}/M_{G} \leq $ 0.5, respectively, 95\% C.L.\ upper and lower
limits are extracted on the corresponding coupling strengths of such
massive gluons~\cite{CDF9164}.

\subsubsection{Measurement of the $t\bar{t}$ differential cross section
  $(d\sigma/dM_{t\bar{t}})$}
\label{sec:dsigma}
\begin{figure}[!t]
  \begin{center} 
    \includegraphics[width=.48\textwidth, height=45mm]{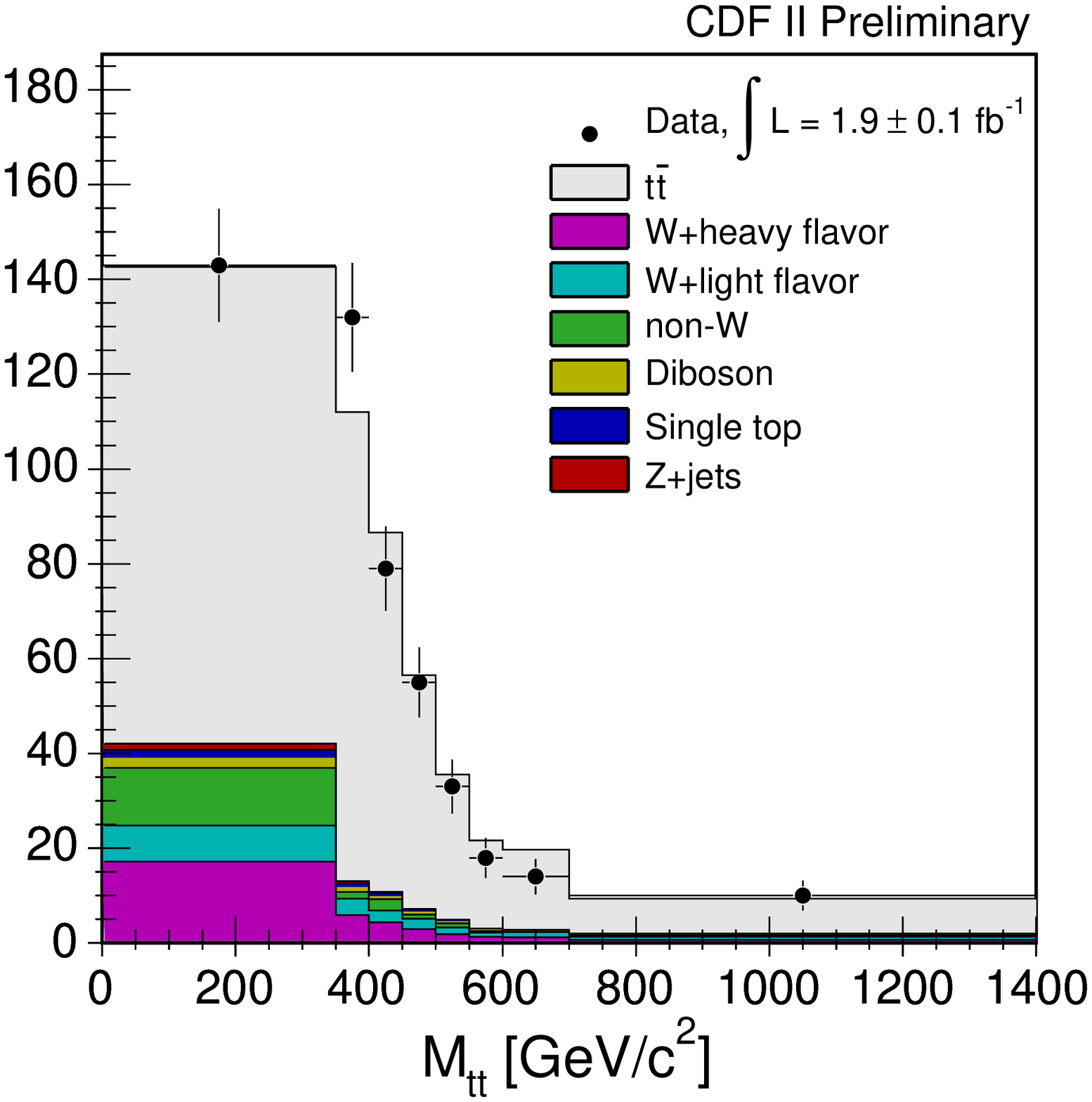}\hspace*{1mm}
    \includegraphics[width=.48\textwidth, height=45mm]{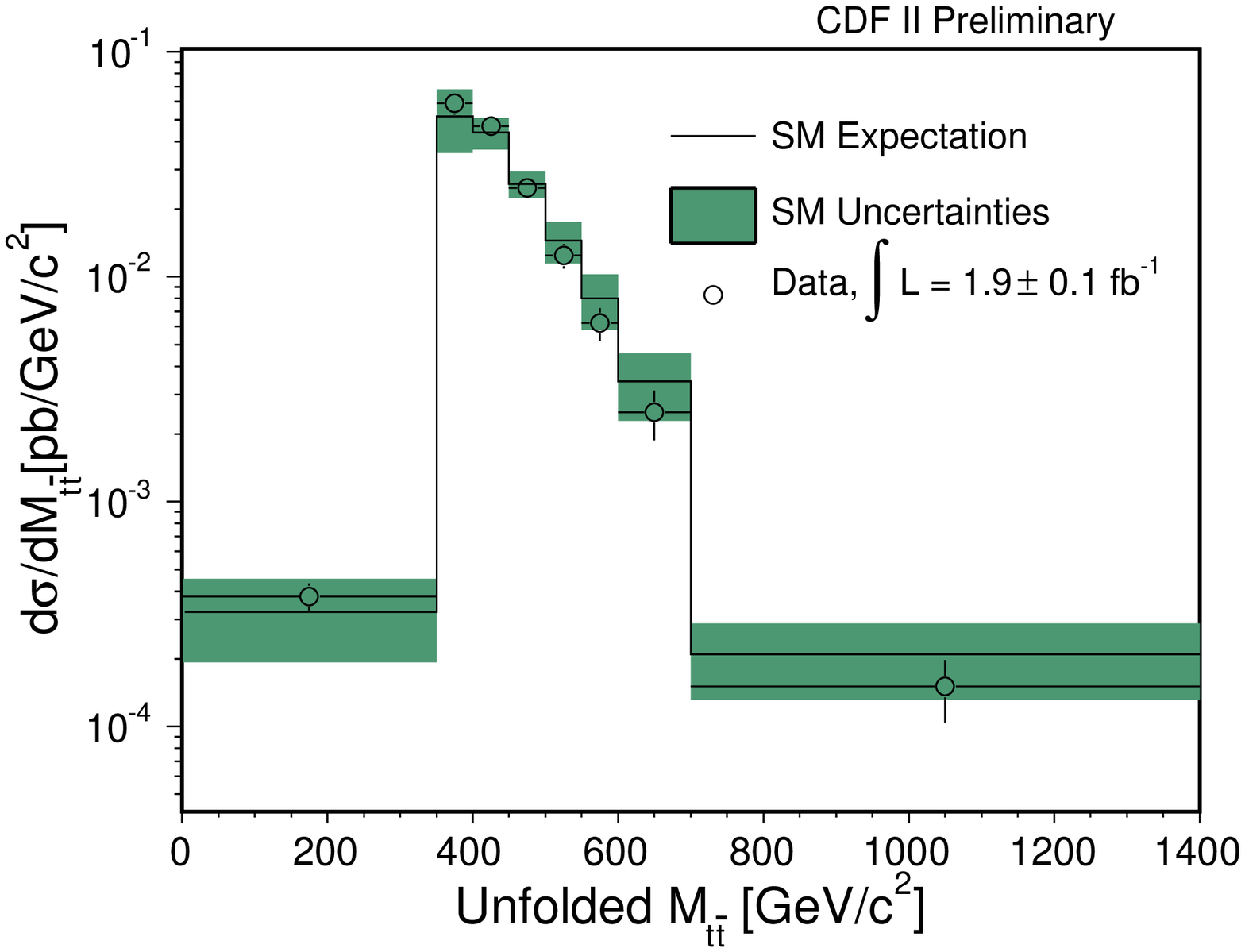} 
    \caption{Left: Expected and observed reconstructed \ttbar invariant mass
      distribution in $b$ tagged lepton + jets events with at least four jets. Right:
      Corresponding \ttbar differential cross section after
      background subtraction and unfolding,
      compared to the SM expectation~\cite{CDF9157}.}
      \label{fig:diffxsec}
  \end{center}
\end{figure}
Since new production mechanisms for top quark pairs could manifest
themselves in the \ttbar invariant mass distribution as resonances of
different widths or, more generally, as shape
distortions~\cite{Frederix:2007gi}, one approach for
detecting such contributions is to compare the shape of the observed
differential \ttbar production cross section $d\sigma/dM_{t\bar{t}}$
with the SM expectation.

CDF reconstructs the \ttbar invariant mass spectrum in a 1.9~fb$^{-1}$
$b$ tagged lepton + jets data sample (see Fig.\ \ref{fig:diffxsec}) by
combining the four-vectors of the four leading jets, lepton and
\METns. After subtracting the background processes,
the distortions in the reconstructed distribution due to detector
effects, object resolutions and geometric/kinematic
acceptance are corrected for through a regularized
unfolding technique~\cite{Hocker:1995kb}. From the unfolded distribution, the \ttbar
differential cross section $d\sigma/dM_{t\bar{t}}$ is extracted and
its shape compared with the SM expectation. The
comparison shows good agreement with the standard model, yielding an
Anderson-Darling $p$-value of 0.45~\cite{CDF9157}.

\subsubsection{Search for new heavy top-like quark pair production}
\label{sec:tprime}
The number of light neutrino species (for $m_\nu < m_Z / 2$) has been
determined to be $N_\nu = 2.9840 \pm 0.0082$ based on the
invisible $Z$ boson decay width in precision electroweak
measurements~\cite{Z-Pole}. This rules out a fourth generation of
fermions with a light neutrino $\nu_4$. However, the existence of a
fourth generation is consistent with precision electroweak data for a
fermion mass range $m_Z/2 \lesssim m_{f4} \lesssim {\cal O}
(174~{\rm GeV/c}^2)$, even without introducing new
physics~\cite{He:2001tp,Novikov:2001md}.
Fourth-generation quark masses up to 400 GeV/c$^2$ are compatible with
current measurements, and are constrained to exhibit small mass
splitting, so that decays of an up-type fourth-generation quark into
$Wq$ ($q=d,s,b$) are preferred~\cite{Kribs:2007nz}. Such an additional
generation would have a drastic impact on the phenomenology of the
Higgs boson, thereby relaxing the mass bounds obtained from the SM
up to 750 GeV/c$^2$ at 95\% C.L., and altering expected
kinematics and production rates.

The existence of a fourth chiral-fermion generation is predicted by
various extensions of the standard model, for example, in an SO(1,13)
framework unifying charges and spins~\cite{Borstnik:2006wt}, or in
models with flavor democracy~\cite{SilvaMarcos:2002bz,Arik:2005ed}.
Other models that add more exotic additional heavy quarks that can
decay via $Wq$ have been brought up as well~\cite{Frampton:1999xi}.
For example, the ``beautiful mirrors'' model~\cite{Choudhury:2001hs}
introduces mirror quark-doublets with the same quantum numbers as
their SM counterparts, but with vector couplings to the
$W$ boson. This addition helps to improve the fit of electroweak
observables by removing the observed discrepancy in the
forward-backward asymmetry of the $b$ quark.

CDF performs a search for pair production of heavy top-like quarks
($t'\bar{t}'$) that do not necessarily exhibit
SM-like up-type fourth-generation properties in terms of
charge or spin. The analysis is based on the assumptions that the $t'$
is pair-produced via the strong interaction, has a mass larger than
that of the top quark, and decays promptly into a $W$ boson and a
down-type $d,s,b$ quark with 100\% branching ratio. As a consequence,
the $t'$ decay chain is identical to that of the top quark, and
$t'\bar{t}'$ production can be sought in a lepton + jets sample
selected solely based on event kinematics to not restrict the search
to $Wb$ final states by using $b$ tagging.

\begin{figure}[!t]
  \begin{center} 
    \includegraphics[width=.495\textwidth,
      height=55mm]{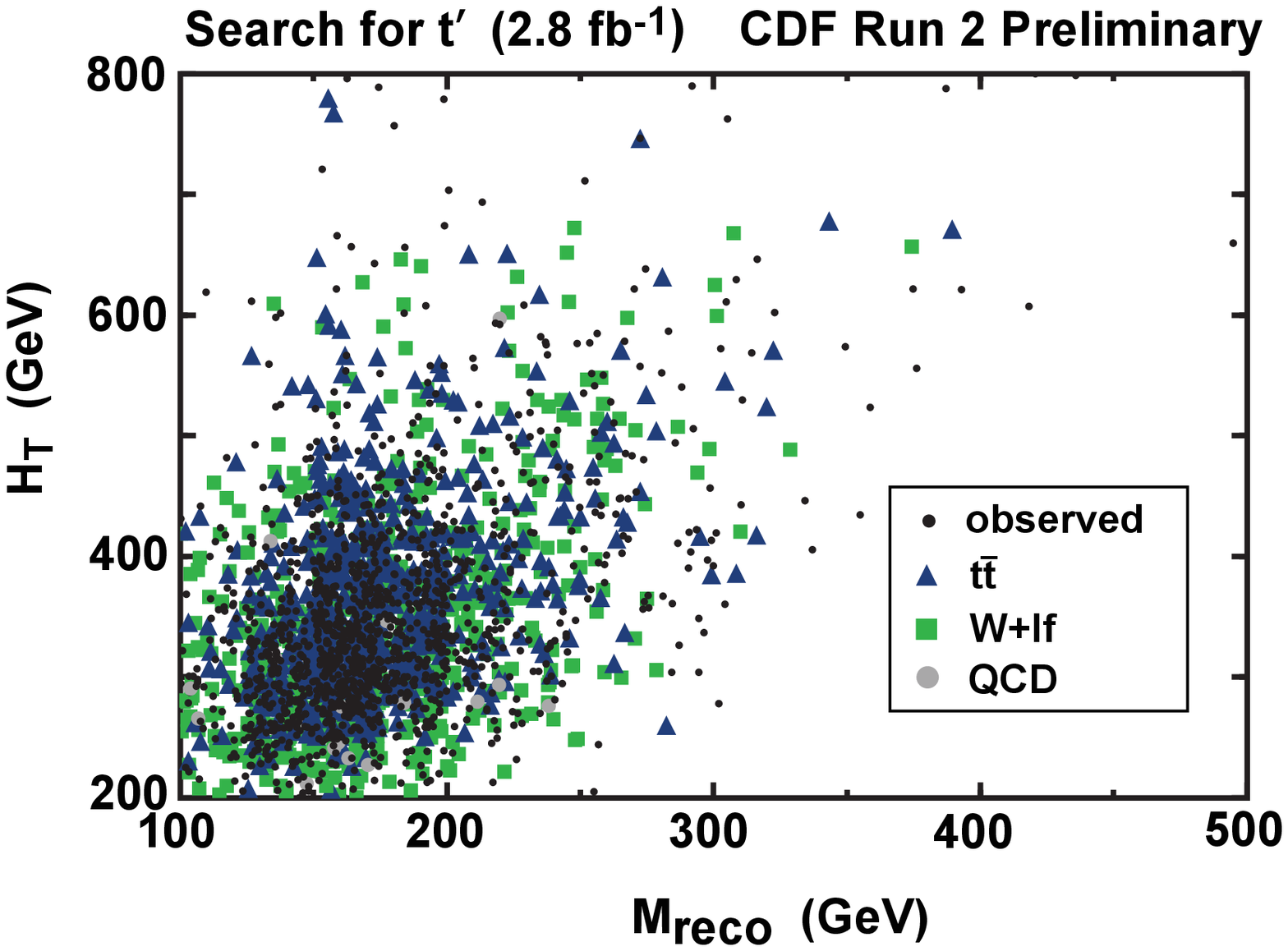}\hspace*{1mm}
    \includegraphics[width=.495\textwidth]{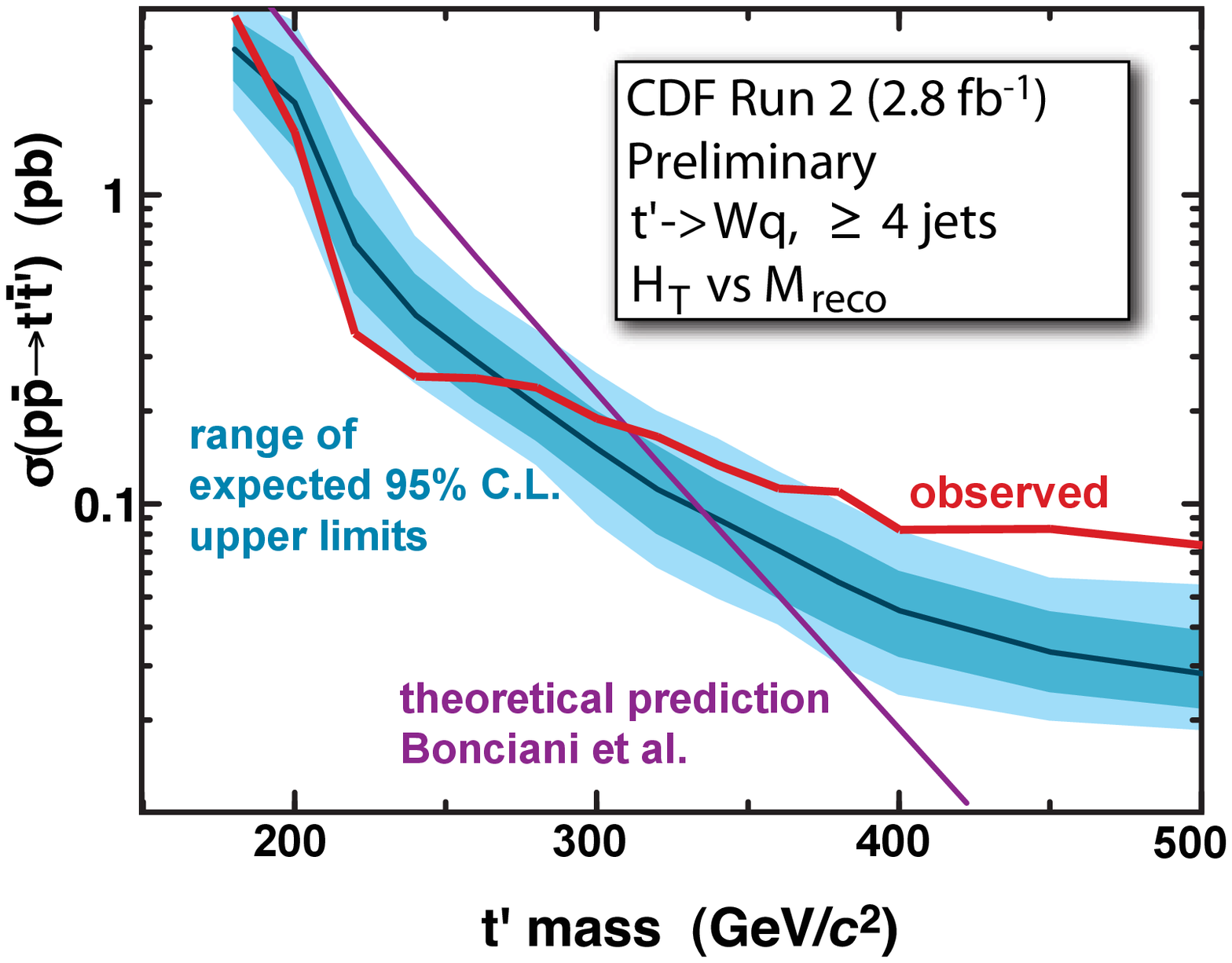}
    \caption{Left: $H_T$ versus $M_{\rm{reco}}$ distribution observed
      in data (black points) overlaid with the fitted number of
      standard model background events from \ttbar, $W$+jets
      and QCD. Right: Expected and observed 95\% C.L.
      upper limits on the $t'\bar{t}'$ production cross section,
      assuming 100\% ${\cal B}(t' \to Wq)$. The dark/light regions
      represent the 1/2 sd bands on the expected
      limit~\cite{CDF9446}.}
    \label{fig:tprimesearch}
  \end{center}
\end{figure}
The $t'$ signal can be distinguished from SM
background using, e.g., the observed distributions of total
transverse energy $H_T$ in the event based on lepton, jets and \METns,
and the reconstructed $t'$ mass ($M_{\rm{reco}}$) from the best
kinematic fit to the $t'\bar{t}'$ hypothesis in each event (see
Fig.\ \ref{fig:tprimesearch}). Superseding a previously published
result based on 0.8~fb$^{-1}$~\cite{CDFtprime:2008nf}, CDF uses a
two-dimensional binned likelihood fit in $H_T$ and $M_{\rm{reco}}$ to
separate SM background and $t'$ signal in 2.8~fb$^{-1}$
of data~\cite{CDF9446}.

Since no evidence for $t'$ production is found, 95\% C.L.\ upper limits
on the $t'\bar{t}'$ production cross section (assuming ${\cal B}(t' \to
Wq) = 100\%$) are derived for 180~GeV/c$^2$ $\leq m_{t'} \leq$
500~GeV/c$^2$. Assuming SM couplings, a 95\% C.L.\ lower
limit on the fourth-generation $t'$ mass of 311~GeV/c$^2$, based on the
calculations in Refs.~\cite{Cacciari:2003fi,Bonciani:1998vc}, is obtained, where
the largest systematic uncertainty 
arises from the jet energy scale. The deviation of the observed
limit from its expected value above $\approx$400~GeV/c$^2$
is being investigated. Using {\em a priori} defined groups of bins in $H_T$ and
$M_{\rm{reco}}$, the $p$-values to observe at least the number of events
found in data, given the SM expectation, are evaluated.
With the smallest $p$-value being 0.01, the excess in the data tails
is concluded to be not statistically significant.

\subsubsection{Search for scalar top quark production}
Many processes beyond the standard model exhibit signatures similar
to \ttbar events. Consequently, \ttbar data 
samples can, in principle, contain admixtures of such contributions.

For example, the Minimal Supersymmetric Standard Model
(MSSM)~\cite{Martin:1997ns} predicts that supersymmetric partners of
the top quarks, scalar top or ``stop'' quarks, are
predominantly produced in pairs via the strong interaction just like
SM top quarks. The stop-quark pair-production cross
section has been calculated at NLO supersymmetric-QCD and depends
mainly on the stop quark mass and very little on other supersymmetric
parameters~\cite{Beenakker:1997ut}. At a center of mass energy of
1.96~TeV, the pair production cross section for the lightest stop
quarks ($\tilde{t}_{1}\bar{\tilde{t}}_{1}$) of 175 GeV/c$^2$ mass is
0.58 pb~\cite{Beenakker:1996ed}, or roughly 10\% of the SM
\ttbar production rate (see Section~\ref{sec:ttbarprod}). The
observable final states from stop decays depend strongly on
supersymmetric parameters, especially the masses of
supersymmetric particles in the decay chain. In Run~II, the decay mode
$\stopone \to b \ell^+ \tilde{\nu}_\ell$ has been studied by
D0~\cite{Abazov:2007im} in 0.4 fb$^{-1}$ of data. The decay channel
$\stopone \to c \neutralino$, where the lightest neutralino
$\neutralino$ is the lightest supersymmetric particle (LSP), was
studied by CDF~\cite{Aaltonen:2007sw} in 0.3 fb$^{-1}$ and by
D0 in 0.4 and 1 fb$^{-1}$
of data~\cite{Abazov:2006wb,Abazov:2008rc}. 95\% C.L.\ mass-exclusion limits on the involved
supersymmetric particles were provided for both decay channels.

\begin{figure}[!t]
  \begin{center} 
    \includegraphics[width=.48\textwidth,height=45mm]{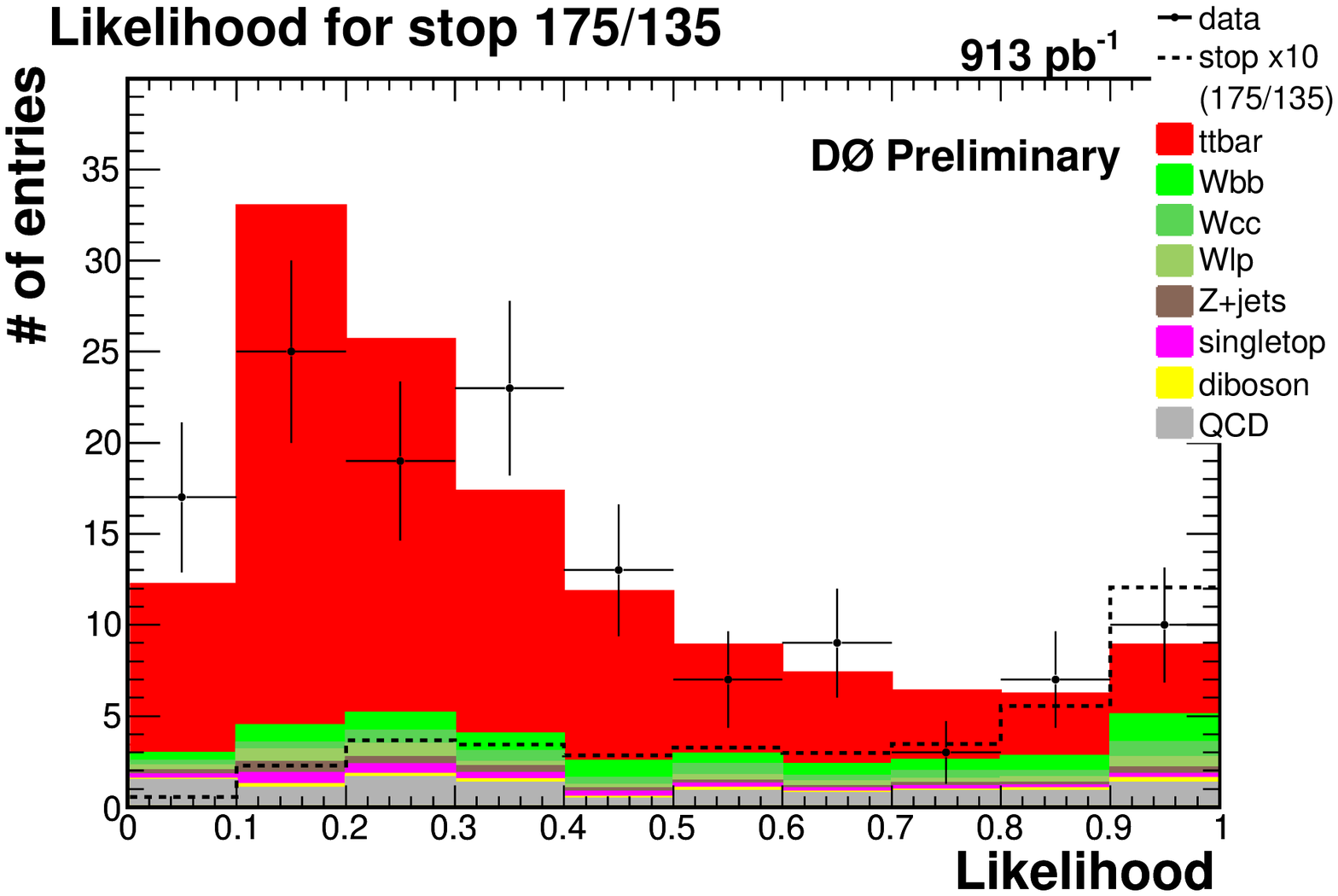}\hspace*{1mm}
    \includegraphics[width=.48\textwidth,height=45mm]{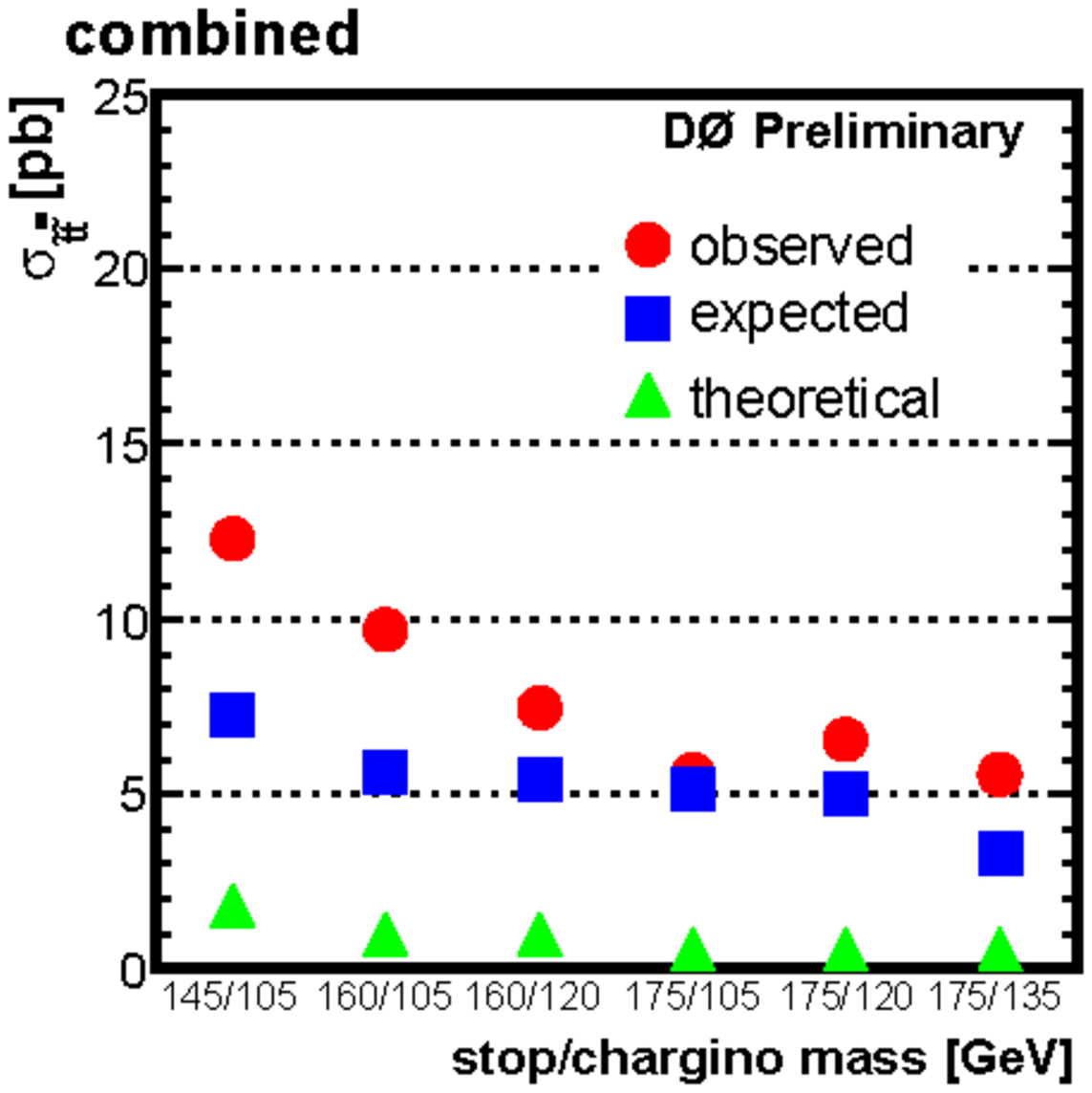}
    \caption{Left: Multivariate kinematic discriminant distribution for
    electron $+$$\geq$4 jets events in 0.9~fb$^{-1}$ of D0
    data and simulated signal and background. The dashed black line
    shows the expected signal shape for a $\stopone~(\chargino)$ mass of
    175 (135) GeV/c$^2$, enhanced by a factor of ten. Right: Expected
    and observed 95\% C.L.\ limits on the
    $\tilde{t}_{1}\bar{\tilde{t}}_{1}$ production cross section,
    together with the prediction for different stop and
    chargino mass combinations~\cite{D05438}.}
    \label{fig:D0stopsearch}
  \end{center}
\end{figure}
Another important decay mode for stop is $\stopone \to \charginoplus
b$, where the lightest chargino $\charginoplus$ decays to
$W^+\neutralino$, resulting in final states identical to those from
\ttbar decays, but with two additional neutralinos (LSPs) that
contribute to \METns. First limits for this channel were
provided by CDF in Run~I on 0.1~fb$^{-1}$ of lepton + jets
data~\cite{Affolder:1999cz}.

D0 performs the first study of this channel in Run~II in 0.9~fb$^{-1}$
$b$ tagged lepton + jets data~\cite{D05438}\footnote{An updated
version of the result has been published after completion of this
review, see Ref.~\cite{Abazov:2009ps}.}, assuming a neutralino
mass of 50~GeV/c$^2$ (slightly above the limit set by LEP~\cite{LEPSUSY})
while varying the stop and chargino masses between 145 and 175
GeV/c$^2$, and 105 and 135 GeV/c$^2$, respectively.

A possible stop admixture in the sample is searched for by employing a
multivariate discriminant based on kinematic event properties (see
Fig.\ \ref{fig:D0stopsearch}), with the main challenge being to separate the
topologically similar \ttbar background from
$\tilde{t}_{1}\bar{\tilde{t}}_{1}$ signal. Counterintuitively, the
additional neutralinos do not provide large differences in \MET that can be
exploited, but the larger chargino mass compared to that of the $W$
boson reduces the phase space for $b$ jets in
the event. 

Since no significant signal admixture in the lepton +
jets dataset is observed, 95\% C.L.\ upper limits on the
$\tilde{t}_{1}\bar{\tilde{t}}_{1}$ production rate 
are set that are a factor of $\approx7-12$ above
the prediction, as illustrated in
Fig.\ \ref{fig:D0stopsearch}. Consequently, the stop quark masses
considered cannot yet be excluded, and this analysis
should greatly benefit from the increased datasets already in hand.
The weaker observed limits relative to their expected values are driven 
by the muon + jets channel. The corresponding excess in data
was tested with pseudo-datasets to be statistically consistent
with the standard model expectation.

\begin{figure}[!t]
  \begin{center} 
    \includegraphics[width=.495\textwidth]{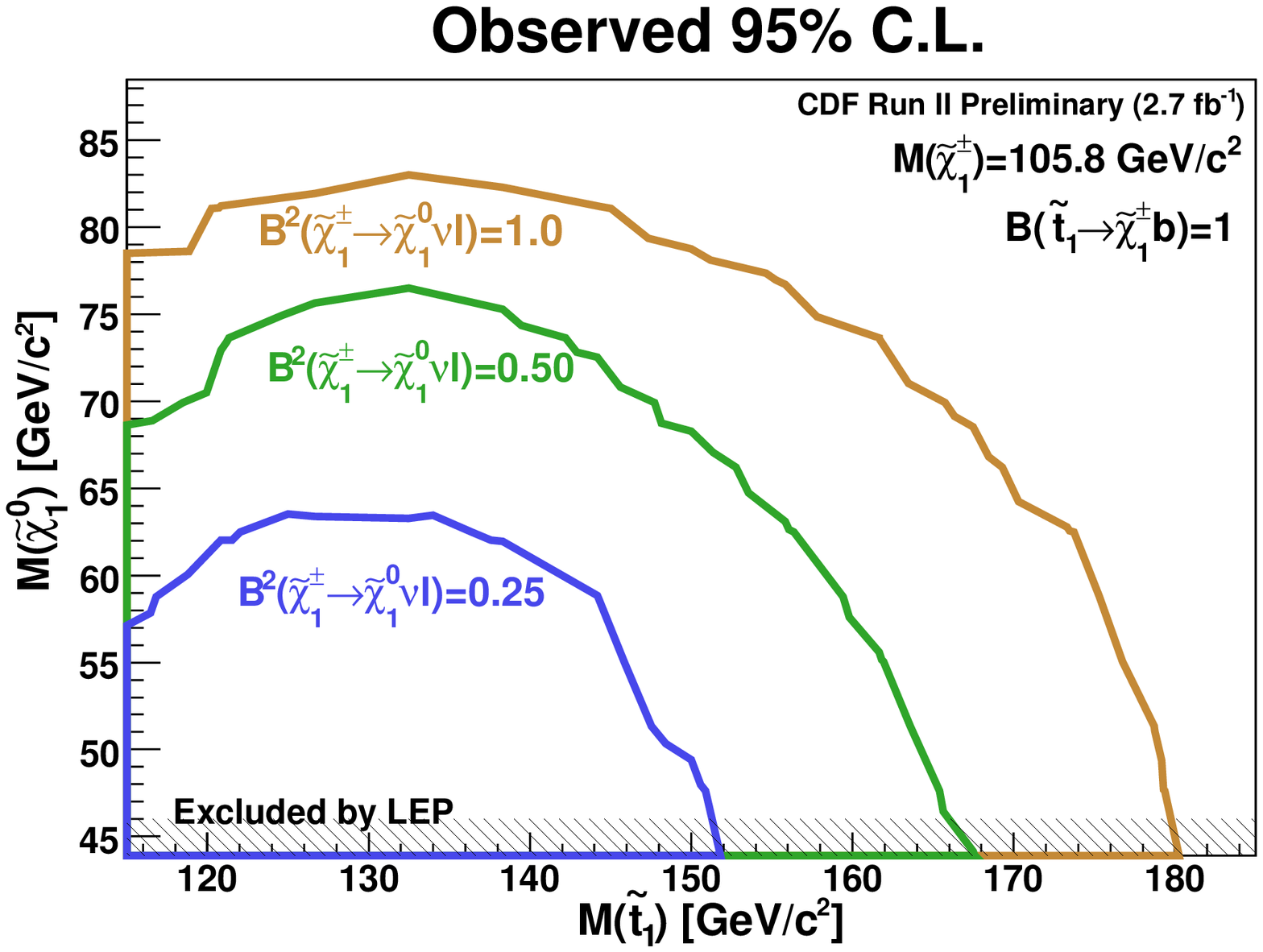}\hspace*{1mm}
    \includegraphics[width=.495\textwidth]{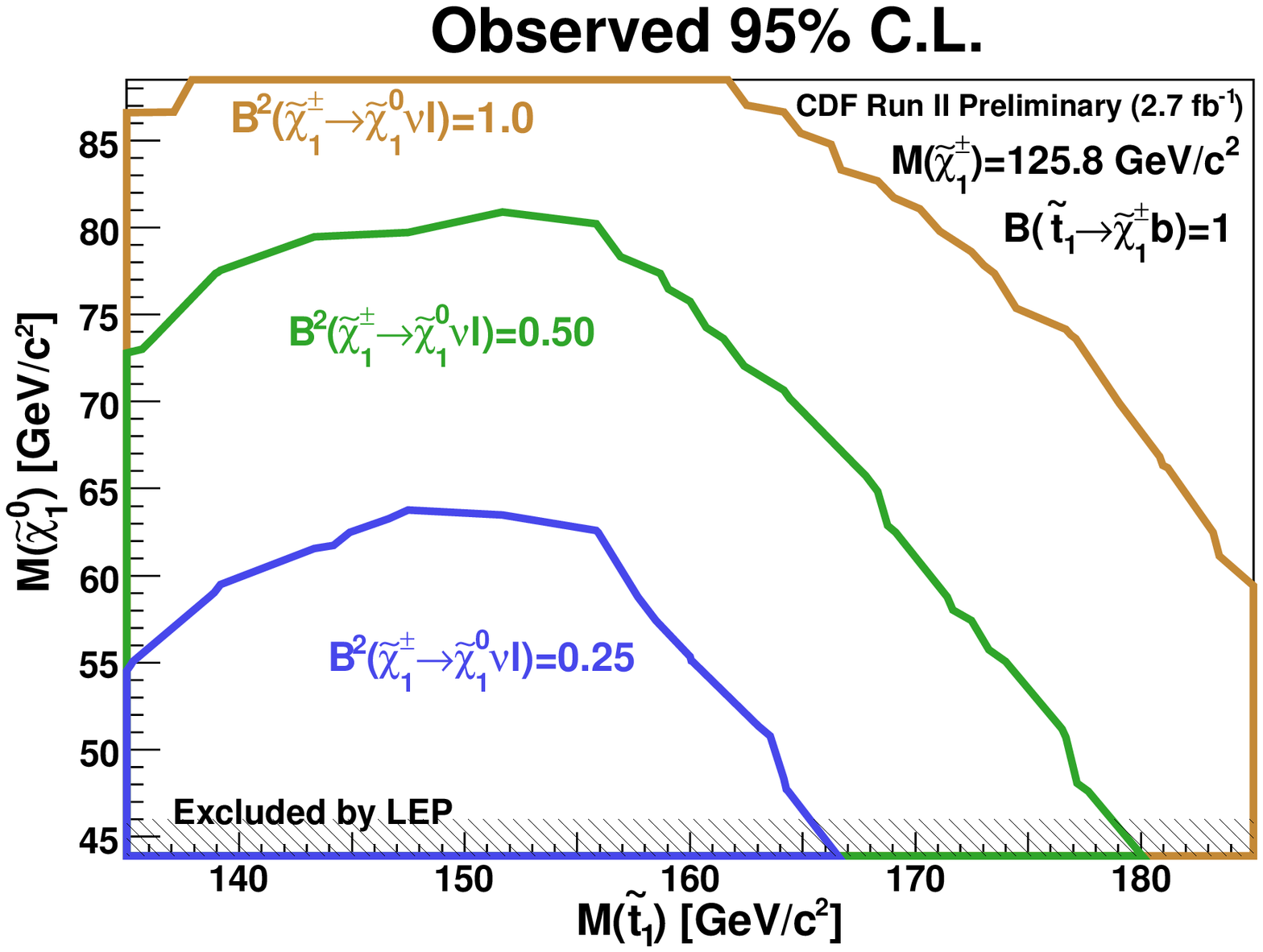} 
    \caption{Excluded areas observed at 95\% C.L.\ in the neutralino
      versus stop mass plane for various assumed dilepton branching
      fractions and two different chargino masses (left:
      105.8~GeV/c$^2$, right: 125.8~GeV/c$^2$), obtained by CDF in
      2.7~fb$^{-1}$ of dilepton data. The contributions of
      $e,\mu,\tau$ to the final state are assumed to be
      equal~\cite{CDF9439}.}
      \label{fig:CDFstopsearch}
  \end{center}
\end{figure}
CDF searches for a stop admixture in the \ttbar dilepton channel using
a 2.7~fb$^{-1}$ dilepton dataset of both $b$ tagged and untagged
events~\cite{CDF9439}. Assuming $\neutralino$ to be the LSP, heavy
sfermions, the stop mass below the top mass, and the chargino mass
smaller than the mass difference between stop and $b$ quark, the decay
$\stopone \to \charginoplus b$ obtains 100\% branching fraction. The
dilepton final state resulting from $\chargino \to \neutralino
\ell^\pm \nu$ decays is then identical to the \ttbar final state, but with
two additional neutralinos contributing to \METns. It can be reached
through a variety of chargino decay channels, resulting in variations
of the branching fraction depending on SUSY parameters.

The stop quark signal is discriminated from standard model
background using a single quantity, the reconstructed stop mass, in a
fit to the observed data distribution. The mass is reconstructed from
this underconstrained system by treating the neutralino and neutrino from
each stop decay as one massive pseudo-particle, and then applying a
standard top mass reconstruction technique in the dilepton channel,
neutrino weighting (see Section~\ref{sec:dilepmassmeasurements}).
Since the observed distributions are consistent with standard
model processes, 95\% C.L.\ limits are extracted on the
dilepton branching ratio in $\tilde{t}_{1}\bar{\tilde{t}}_{1}$
production for stop masses 115 $-$ 185~GeV/c$^2$, neutralino
masses 43.9 $-$ 88.5~GeV/c$^2$, and chargino masses
105.8 $-$ 125.8~GeV/c$^2$, as illustrated in
Fig.\ \ref{fig:CDFstopsearch}. A 100\% branching fraction for
$\stopone \to \charginoplus b$, and equal contributions from
$e,\mu$ and $\tau$ to the final state are assumed throughout.

Model-independent searches for novel admixtures in top quark samples,
via the search for anomalous event kinematics, were discussed in
Section~\ref{sec:topkinematics}.

\subsection{Single top quark production}
\label{sec:ST}
Electroweak production of top quarks without their antiparticles
can provide a direct measure of the $|V_{tb}|$ CKM matrix element, test
the $Wtb$ vertex structure, and probe for physics beyond the
standard model, such as flavor changing neutral currents or new heavy
gauge bosons $W'$ (see Section~\ref{sec:singletopprod}). A thorough
understanding of single top quark production is also important for
studies of processes with similar signatures, such as standard model
$W$-Higgs associated production, for which this process constitutes a major background. While the
single-top production rate is $\approx$40\% of the strong \ttbar
production, the signal extraction from background is very
challenging because only one top quark signature is present in the final
state. Simple kinematic selections are insufficient for such an
analysis, and sophisticated multivariate techniques have to be
deployed.

For single top quark production at the Tevatron, only $s$- and
$t$-channel production are relevant, contributing, respectively,
$0.88^{+0.12}_{-0.11}$~pb and $1.98^{+0.28}_{-0.22}$~pb to
the total rate at NLO~\cite{Sullivan:2004ie}. The experimental
signature comprises a $b$ jet and the $W$ boson decay products from
the top quark decay. In the $s$- ($tb$-) channel, one additional $b$
jet arises from the $b$ quark produced together with the top quark. In
the $t$- ($tqb$-) channel, a forward light-quark jet accompanies the production of the top 
quark, sometimes along with another $b$ jet from the 
gluon splitting into $b\bar{b}$ (see
Fig.\ \ref{fig:singlet-feynman}). In order to suppress multijet
background, the $W$ boson is usually required to decay leptonically
into an electron or muon and corresponding neutrino. Consequently, the
final state signature of single top quark production contains an
energetic isolated electron or muon, \MET and two or three jets, with
at least one of them being a $b$ jet. As usual, additional jets can arise from
initial- or final-state radiation.

\subsubsection{Production cross section and $V_{tb}$}
\label{sec:STxsec}
Searches for single top quark production were already performed in
Run~I using 0.1 fb$^{-1}$ of data by
D0~\cite{Abbott:2000pa,Abazov:2001ns} and by
CDF~\cite{Acosta:2001un, Acosta:2004er}, yielding upper limits on the
production rate that were at least a factor of six larger than the SM
expectation. In Run~II, first results were published using 0.2
fb$^{-1}$ of data by CDF~\cite{Acosta:2004bs} and
D0~\cite{Abazov:2005zz,Abazov:2006uq}, where the best observed limit
was less than a factor of three greater than the SM prediction.
Finally, D0 published first evidence for single top quark production
using 0.9 fb$^{-1}$ of data~\cite{Abazov:2006gd,Abazov:2008kt},
observing a signal of 3.6 standard deviations significance.
Preliminary results from CDF based on 2.2 fb$^{-1}$ confirmed evidence for
single top quark production~\cite{CDF9217,CDF9221,CDF9223,CDF9313}
with an observed signal significance of 3.7 standard deviations
obtained by combining three of these
analyses~\cite{CDF9251}\footnote{A slightly updated version
of the result was published by CDF~\cite{Aaltonen:2008sy}
after completion of this review.}.

\begin{table}[!t]
\caption{Expected and observed event yields of the single top
selections for $e$ and $\mu$, single and double $b$ tagged channels
combined (left for D0 based on 0.9~fb$^{-1}$~\cite{Abazov:2006gd}, and
right for CDF based on 2.2~fb$^{-1}$~\cite{CDF9223}). For the D0
result, the overall $W$ + jets background includes $Z$ + jets and diboson
events.}
\begin{minipage}[!t]{0.49\linewidth}
\centering
\vspace{-1.cm}
\begin{tabular}{l@{\extracolsep{\fill}}r@{\extracolsep{0pt}$\pm$}l@{}%
                 @{\extracolsep{\fill}}r@{\extracolsep{0pt}$\pm$}l@{}%
                 @{\extracolsep{\fill}}r@{\extracolsep{0pt}$\pm$}l@{}}
\hline
Source           & \multicolumn{2}{c}{2 jets}
                 & \multicolumn{2}{c}{3 jets}
                 & \multicolumn{2}{c}{4 jets}\\
\hline\hline	     
$tb$                 &  16  &   3  &   8  &  2  &   2  &  1  \\
$tqb$               &  20  &   4  &  12  &  3  &   4  &  1  \\
\hline                                                              
\ttbar               &  59  &   10  &  135  &  26  &  154  &  33  \\
$Wb\bar{b}$  & 261  &  55~  & 120  & 24~  &  35  &  7  \\
$Wc\bar{c}$,$Wcj~$  & 151  &  31  &  85  & 17  &  23  &  5  \\
$Wjj$              & 119  &  25  &  43  &  9  &  12  &  2  \\
Multijets         &  95  &  19  &  77  & 15  &  29  &  6  \\
\hline                                                              
BG Sum & 686  &  41  & 460  & 39  & 253  & 38  \\
Data             & \multicolumn{2}{c}{697}
                    & \multicolumn{2}{c}{455}
                    & \multicolumn{2}{c}{246}\\
\hline
\end{tabular}
\end{minipage}
\begin{minipage}[t]{0.49\linewidth}
\centering

\begin{tabular}{l@{\extracolsep{\fill}}r@{\extracolsep{0pt}$\pm$}l@{}%
                 @{\extracolsep{\fill}}r@{\extracolsep{0pt}$\pm$}l@{}}
\hline
Source           & \multicolumn{2}{c}{2 jets}
                 & \multicolumn{2}{c}{3 jets}\\
\hline\hline	     
$tb$                 &  41.2  &   5.9  &   13.5  &  1.9  \\
$tqb$               &  62.1  &   9.1  &   18.3  &  2.7  \\
\hline                                                              
\ttbar               &  146.0  &   20.9  &  338.7  &  48.2  \\
$Wb\bar{b}$  & 461.6  &  139.7  & ~~141.1  & 42.6   \\
$Wc\bar{c}$,$Wcj$  & 395.0  &  121.8  &~~  108.8  & 33.5   \\
$Wjj$              & 339.8  &  56.1  &  101.8  &  16.9    \\
Multijets         &  59.5  &  23.8  &  21.3  & 8.5   \\
Dibosons         & 63.2   &  6.3  &  21.5  & 2.2   \\
$Z$+jets         & 26.7   & 3.9   & 11.0   & 1.6   \\
\hline                                                              
BG Sum  & 1491.8  &  268.6  &~~ 754.8  & 91.3   \\
Data             & \multicolumn{2}{c}{1535}
                    & \multicolumn{2}{c}{712}\\
\hline
\end{tabular}
\end{minipage}
\label{tab:ST-evtyields}
\end{table}
The analyses yielding first evidence for electroweak top quark
production apply event selections requiring one energetic isolated
electron or muon and \METns. CDF uses events with two or three jets
and one or two $b$ tags, while D0 includes additionally events with
four jets, where the extra jet arises from initial- or final-state
radiation. The signal acceptances for the $s$- and $t$-channel are
2.8\% and 1.8\% (CDF), and 3.2\% and 2.1\% (D0).
The expected and observed event yields are shown in
Table~\ref{tab:ST-evtyields}. The dominant background contributions
come from $W$+jets production, \ttbar production in the lepton +
jets or dilepton final states, where one jet or lepton is not
reconstructed, and from multijet production. The main sources of systematic
uncertainty are background normalization, jet energy scale, and the
modeling of the $b$ tagging and triggers. As can be appreciated
from the table, the uncertainty on the background is larger than the
expected signal, which makes advanced analysis techniques necessary.

D0 applies three different multivariate analysis techniques to the
preselected data sample: boosted decision trees (BDT), Bayesian neural
networks (BNN) and matrix elements (ME), where the latter two reflect
reoptimized studies~\cite{Abazov:2008kt} of previous work~\cite{Abazov:2006gd}.
Being based on leading-order matrix elements for the description of
signal and background processes, the ME analysis does not use four-jet events. For each analysis, the combined $s$-
and $t$-channel cross sections are extracted from the peak of the
Bayesian posterior probability density derived from a binned
likelihood of the respective discriminants. The results are then
combined, yielding:
\begin{equation}
\begin{array}{lllll}
 \sigma^{\rm obs} (p\bar{p}\to tb+X,~tqb+X )
 & = & 4.9 ^{+1.4}_{-1.4}~{\rm pb} & ({\rm BDT,}&3.4 {\rm~sd}) \\
 & = & 4.4 ^{+1.6}_{-1.4}~{\rm pb} & ({\rm BNN,}&3.1 {\rm~sd}) \\
 & = & 4.8 ^{+1.6}_{-1.4}~{\rm pb} & ({\rm ME,}&3.2 {\rm~sd}) \\
 & = & 4.7 ^{+1.3}_{-1.3}~{\rm pb} & ({\rm Combined,}&3.6 {\rm~sd}),
\end{array}
\end{equation}
where the uncertainties correspond to the combination of statistical
and systematic sources. The observed production rates are in agreement
with SM expectation and with each other. The significances in the parentheses
are obtained from studies of large ensembles of pseudo-experiments. The expected
sensitivity of the combined result is 2.3 standard deviations,
indicating that the measurement benefited from a statistical upward
fluctuation. Separate measurements of the $s$- and $t$-channel cross
sections are also performed with the BDT analysis. The results are
$\sigma_s=1.0\pm0.9$~pb and $\sigma_t=4.2^{+1.8}_{-1.4}$~pb, where
the other channel (not measured) is set to its SM expectation
($\sigma_s=0.88^{+0.12}_{-0.11}$~pb and $\sigma_t=1.98^{+0.28}_{-0.22}$~pb~\cite{Sullivan:2004ie}).
The observed enhancement in the $t$-channel with respect
to the standard model prediction is not statistically significant.

CDF uses the following multivariate analysis techniques on their
preselected dataset: neural networks (NN~\cite{CDF9217}), a likelihood
function (LHF~\cite{CDF9221}), a matrix element discriminant
(ME~\cite{CDF9223}) and boosted decision trees (BDT \cite{CDF9313}).
The results are:
\begin{equation}
\begin{array}{lllll}
 \sigma^{\rm obs} (p\bar{p}\to tb+X,~tqb+X )
 & = & 2.0 ^{+0.9}_{-0.8}~{\rm pb} & ({\rm NN,}&3.2 {\rm~sd}) \\
 & = & 1.8 ^{+0.9}_{-0.8}~{\rm pb} & ({\rm LHF,}&2.0 {\rm~sd}) \\
 & = & 2.2 ^{+0.8}_{-0.7}~{\rm pb} & ({\rm ME,}&3.4 {\rm~sd}) \\
 & = & 2.2 ^{+0.7}_{-0.7}~{\rm pb} & ({\rm Combined,}&3.7 {\rm~sd})\\
 & = & 1.9 ^{+0.8}_{-0.7}~{\rm pb} & ({\rm BDT,}&2.8 {\rm~sd}),
\end{array}
\end{equation}
where the uncertainties given are both statistical and systematic. The
BDT analysis became available after the combination of
results~\cite{CDF9251}, and was therefore not included in that
compilation. The observed results agree with each other and with the
standard model. The expected sensitivity of the
combination is 5.1 standard deviations, pointing to a statistical
downward fluctuation in the data.

\begin{figure}[!t]
  \begin{center} 
    \includegraphics[width=.52\textwidth
      ]{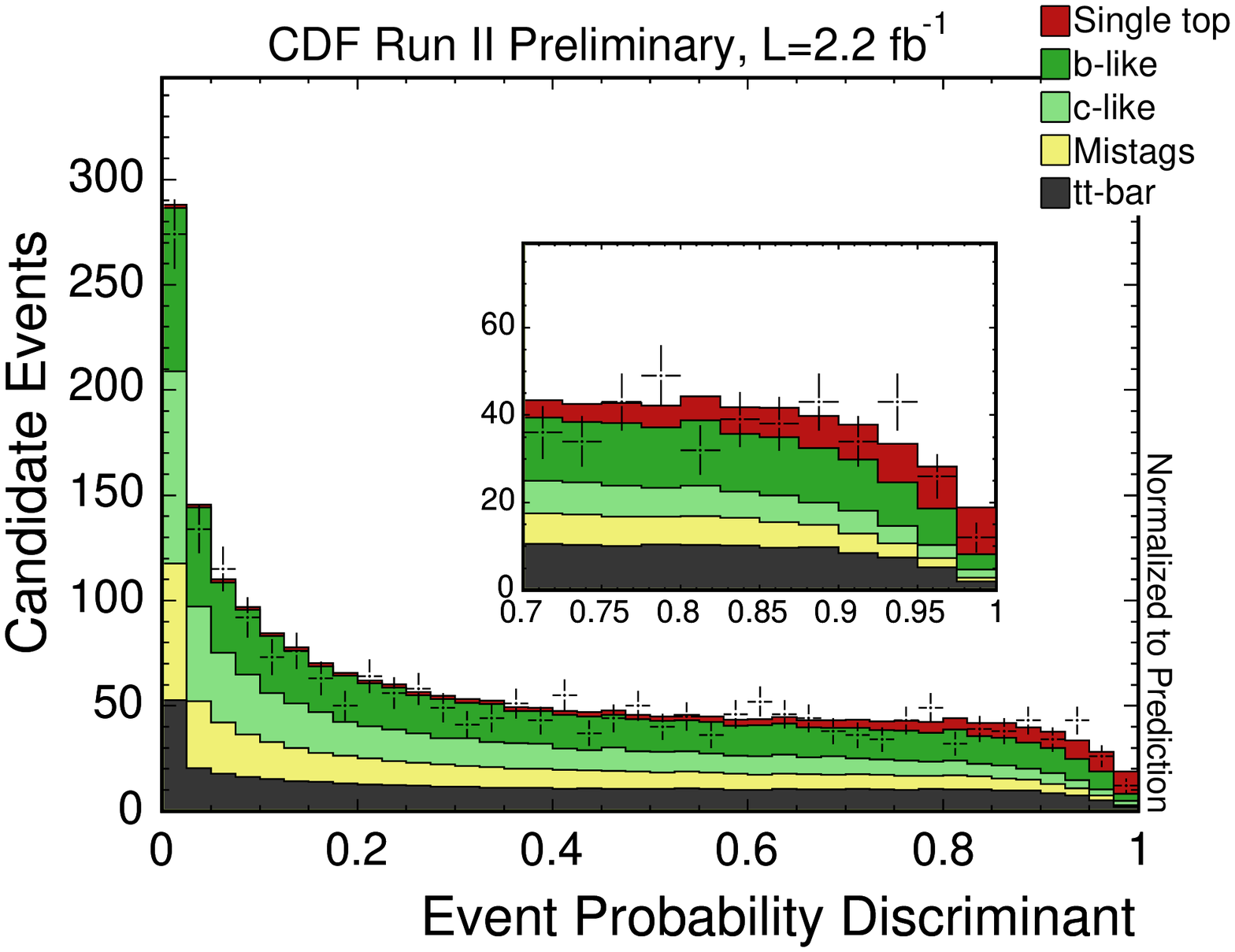}\hspace*{1mm}
    \includegraphics[width=.48\textwidth
      ]{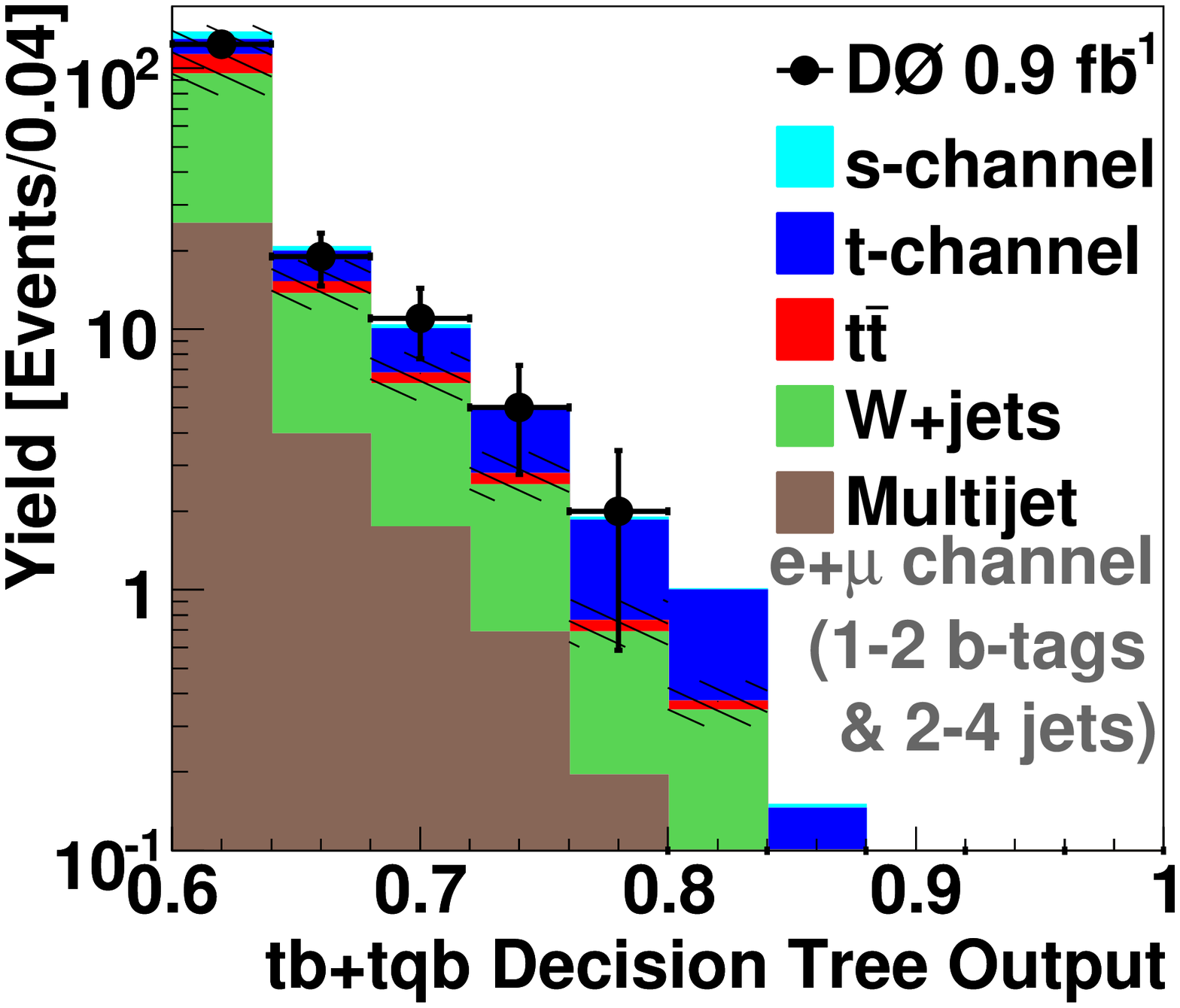}
      \caption{Multivariate discriminant outputs observed in single-top
        candidate events compared to contributions from signal and
        background processes. Left: CDF matrix element discriminant,
        with yields normalized to their SM
        predictions~\cite{CDF9223}. Right: D0 BDT
        output in the single-top signal region, with signal normalized to the
        measured cross section~\cite{Abazov:2006gd}.}
    \label{fig:STdiscriminants}
  \end{center}
\end{figure}

\begin{figure}[!h]
  \begin{center} 
    \includegraphics[width=.48\textwidth]{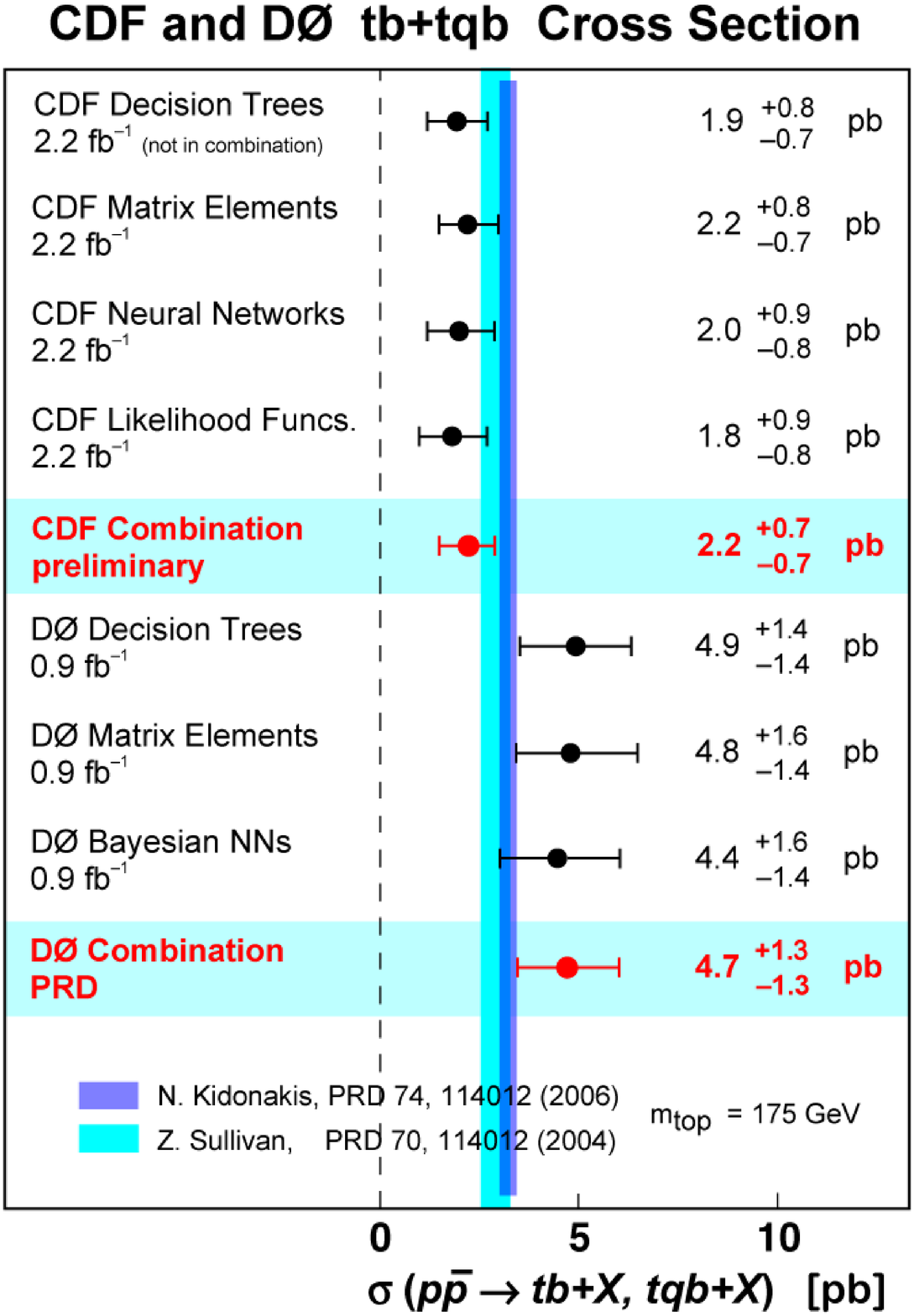}
    \caption{Cross section measurements of first evidence for
      single top production, and the combined results from CDF and
      D0 compared to the SM prediction.}
    \label{fig:STxsecs}
  \end{center}
\end{figure}

Figure~\ref{fig:STdiscriminants} shows the discriminant outputs of the
two most significant single measurements from CDF (ME) and D0 (BDT). A
graphical summary of the measurements and a comparison with the
standard model expectation is given in Fig.\ \ref{fig:STxsecs}. All
analyses assume a top quark mass of 175~GeV/c$^{2}$, ${\cal B}(t
\rightarrow Wb) = 100\%$, and the SM ratio for $s$- to
$t$-channel cross sections.
In a recent update, CDF has added 0.5~fb$^{-1}$ of data to its single
top sample, and thereby increased the observed significance for all
previous analysis techniques. The matrix element analysis yields again
the most significant single result, exceeding four standard
deviations. A combination of the measurements has not yet become
available. The results for 2.7~fb$^{-1}$, with combined statistical and
systematic uncertainties, are~\cite{CDF9479,CDF9451,CDF9464,CDF9445}:
\begin{equation}
\begin{array}{lllll}
 \sigma^{\rm obs} (p\bar{p}\to tb+X,~tqb+X )
 & = & 2.1 ^{+0.7}_{-0.6}~{\rm pb} & ({\rm NN,} &3.7 {\rm~sd})\\
 & = & 2.0 ^{+0.9}_{-0.8}~{\rm pb} & ({\rm LHF,}&2.6 {\rm~sd})\\
 & = & 2.7 ^{+0.8}_{-0.7}~{\rm pb} & ({\rm ME,} &4.2 {\rm~sd})\\
 & = & 2.4 ^{+0.8}_{-0.7}~{\rm pb} & ({\rm BDT,}&3.6 {\rm~sd}).
\end{array}
\end{equation}

Since the single top quark production rate is proportional to
$|V_{tb}|^{2}$, the observed cross sections can be turned into
measurements of $|V_{tb}|$ under the following assumptions: (i) there are
no single top quark production modes beyond the SM, (ii) single top
quark production and decay are dominated by the $Wtb$ interaction
($|V_{tb}| \gg |V_{td}|, |V_{ts}|$, as indicated by measurements of $R$
described in Section~\ref{sec:Rmeasurement}), and (iii) the $Wtb$ interaction
exhibits a $V-A$ structure and is CP conserving. The latter premise
allows for anomalous left-handed vector couplings $f_{1}^{L}$ (see
Section~\ref{sec:Wheltheory}), but not for right-handed vector or
tensor couplings. Anomalous $f_{1}^{L}$ values ($\neq 1$) do not
affect the \ttbar production rate or kinematics, nor $tb$ or
$tqb$ kinematics, but simply rescale the single-top production rate.
Consequently, $|V_{tb}f_{1}^{L}|$ extracted from single-top
production can be $>1$, and constraining the measurement
to lie between $0-1$ implies that $f_{1}^{L}=1$, as predicted by the standard model.
The measurements of $|V_{tb}f_{1}^{L}|$ and $|V_{tb}|$ are
independent from the number of fermion generations and 
unitarity of the CKM matrix.

Using the result of the BDT analysis and a positive
flat prior for $|V_{tb}|^{2}$, D0 obtains
$|V_{tb}f_{1}^{L}|=1.31^{+0.25}_{-0.21}$. Restricting the prior to
$[0,1]$ yields $|V_{tb}|=1.00^{+0.00}_{-0.12}$, with a corresponding
95\% C.L.\ lower limit of \mbox{$|V_{tb}|>0.68$}~\cite{Abazov:2008kt}.
CDF uses its combined measurement in the same way to obtain a 95\%
C.L.\ lower limit of \mbox{$|V_{tb}|>0.66$}~\cite{CDF9251}. The
matrix element analysis based on 2.7~fb$^{-1}$ yields
\mbox{$|V_{tb}|>0.71$}~\cite{CDF9464}.

\begin{figure}[!t]
  \begin{center} 
    \includegraphics[width=.48\textwidth]{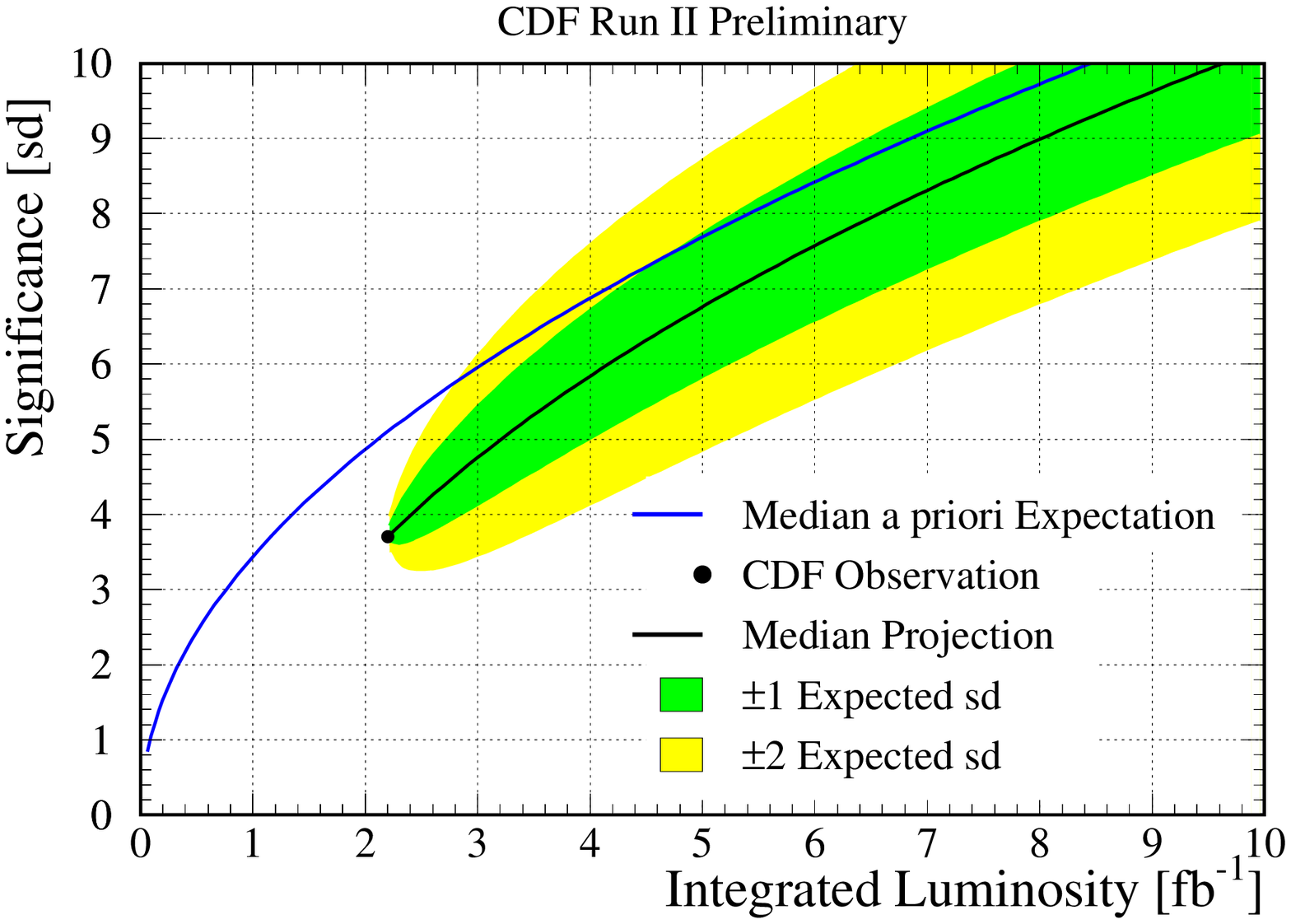}
    \includegraphics[width=.48\textwidth]{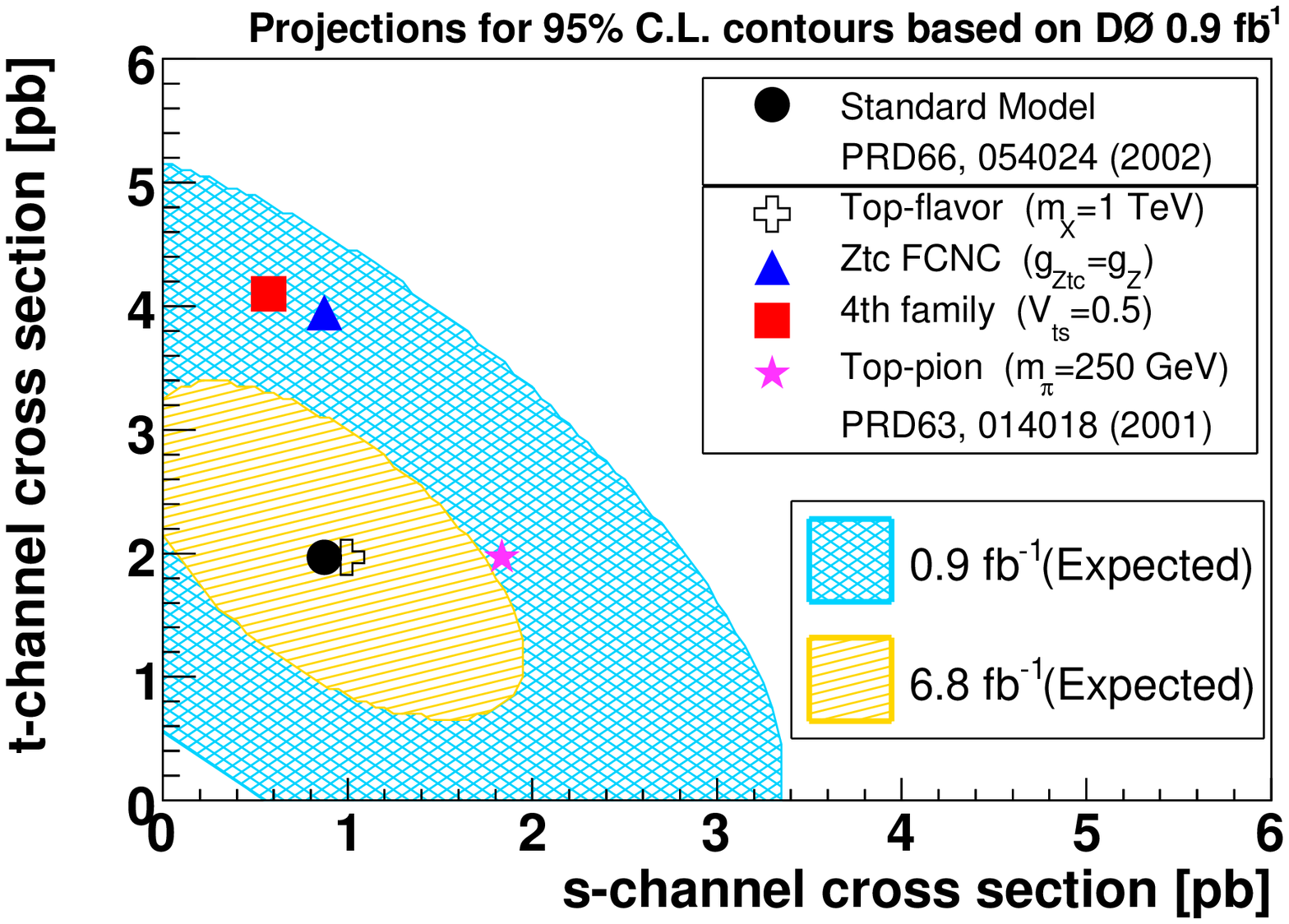}
    \caption{Left: Extrapolation of the single-top signal significance
      as a function of integrated luminosity from the 2.2~fb$^{-1}$ CDF analysis
      result. Right: Expected 95\% C.L.\ contours for a simultaneous
      measurement of the $s$- and $t$-channel single-top production
      rate by D0 for different integrated luminosities. The standard
      model expectation is shown together with various other models.}
    \label{fig:STprojection}
  \end{center}
\end{figure}

With most of the current measurements giving evidence of $>3$ sd
for single top quark production, the observation at the five
standard deviation level seems imminent at the Run~II Tevatron.
Extrapolating from the 2.2~fb$^{-1}$ result, as illustrated in
Fig.\ \ref{fig:STprojection}, CDF estimates that a $>5$ sd
significance should be reached by adding one more fb$^{-1}$ of data to
the analyses. D0 could reach that level in the 2.3
fb$^{-1}$ dataset that is currently being analyzed\footnote{After 
submission of this review, both collaborations announced 5~sd 
observations of single top quark production,
see Refs.~\cite{Abazov:2009ii,Aaltonen:2009jj}.}.

As discussed in Section~\ref{sec:singletopprod}, $s$- and $t$-channel
production exhibit different sensitivity to physics beyond the
standard model. Measuring both rates separately provides a
valuable tool to check for various exotic model contributions to single top
quark production. Figure~\ref{fig:STprojection} shows D0's
expected sensitivity to physics beyond the standard
model~\cite{Tait:2000sh} for the already analyzed data and for the
anticipated accumulation of 6.8~fb$^{-1}$ of Run~II data. With this increase in
integrated luminosity, the exclusion of certain models at 95\%
C.L.\ should be feasible. With more than 6~fb$^{-1}$, a measurement of
$|V_{tb}|$ with an absolute uncertainty below 0.07 per experiment
should be achievable as well. In addition, further refinements of the
analysis techniques should facilitate improvements in precision
beyond that expected just from the accumulation of more data.

\subsubsection{Polarization of the spin of the top quark}
\label{sec:STspin}
As opposed to top quark pair production via the strong interaction,
where the top quarks are produced essentially unpolarized (see
Section~\ref{sec:spincor}), top quarks produced singly via the
electroweak interaction are expected to be highly
left-handedly polarized~\cite{Carlson:1993dt}. The polarization of the
top quark is reflected in the kinematic distributions of its decay
products, providing a test of the $V-A$ structure of the $Wtb$
coupling~\cite{Jezabek:1994zv,Heinson:1996zm}. An observation of this
polarization would also provide limits on the top-quark decay
width and $|V_{tb}|$, since this would confirm that top
quarks decay before depolarizing through QCD interactions.

Both relevant single top quark production mechanisms at the Tevatron
($s$- and $t$-channel) exhibit up-type--down-type and $tb$ quark lines
interconnected by a $W$ boson. Since the $W$ boson couples solely to
fermions of left-handed chirality, in their rest frame single top
quarks are highly polarized along the direction of the down-type
quark~\cite{Mahlon:1996pn,Mahlon:1998uv, Mahlon:2000ze}.
(For the contributing $2\to2$ processes this polarization is
100\%. These diagrams are related to the top decay with hadronic $W$
boson decay via crossing symmetry, where the down-type decay products
of the $W$ boson exhibit optimal analyzing power.) The optimal spin
basis for studying the single-top spin polarization therefore will use the
direction of the down-type quark.

For $s$-channel production, predominantly proceeding via
$u\bar{d}\to t\bar{b}$, the antiproton beam is expected to provide the
down-type quark most of the time. Indeed, measuring the top quark spin
along the direction of the antiproton beam (``antiproton basis'')
results in 98\% of the top quark spins aligned in that direction. For
the $t$-channel, the situation is slightly more complicated, since the
down-type quark is contained in either the spectator jet or in one of
the beams. With the largest contribution to the total production rate
arising from $ug \to t\bar{b}d$, where the down-type quark produces the
(light quark) spectator jet, a reasonable choice for the spin basis is
the spectator jet direction (``spectator basis''). Since the spectator
jet is emitted in the forward direction, this basis is also still
compatible with the cases where the down-type quark is in the initial
state, resulting in 96\% of the top quarks having their spins
polarized along the light-quark jet direction.

With the top quark decaying before it hadronizes, its spin information
is passed on to its decay products. A straightforward observable is
the angular distribution of the top quark decay product $i$ in the top
quark rest-frame:
\begin{equation}
{1\over\Gamma}\thinspace { {d\,\Gamma}\over{d\,\cos\theta^t_i} } =
{1\over2} \Bigl( 1+{\cal A}_{\uparrow\downarrow}\alpha_i\cos\theta^t_i
\Bigr),
\end{equation}
where $\theta^t_i$ is the angle between the decay product and
spin-quantization axis, $\alpha_i$ is the analyzing power describing the
correlation between top quark spin and decay product and ${\cal
A}_{\uparrow\downarrow} = { (N_\uparrow - N_\downarrow) / (N_\uparrow
+ N_\downarrow) }$ is the spin asymmetry determining the magnitude of the
observable angular correlations. The analyzing power is maximal ($+1$)
for the down-type ($T_3 = -\frac{1}{2}$) decay products of the $W$
boson (charged lepton, $d$- or $s$-quark), making the
charged lepton the most sensitive and easily accessible spin analyzer.
Using the spin quantization axes described above, the expected spin asymmetry
is 0.96 for the $s$-channel and 0.93 for the
$t$-channel~\cite{Mahlon:1996pn,Mahlon:1998uv, Mahlon:2000ze}.

To perform a spin polarization measurement at the Tevatron, single top
quark production in the $t$-channel is most promising due to its
higher rate relative to the $s$-channel. The required integrated
luminosity to observe spin polarization in the $t$-channel at the
Tevatron was determined in a study, including effects of jet
resolution and acceptance~\cite{Stelzer:1998ni}. To establish the
polarization at a level of five sd, 
$\approx$5~fb$^{-1}$ of data will be needed, which should be available very
soon.

At the LHC, measurements of single-top spin polarization will benefit
from the expected high-statistics single-top datasets, and optimal spin bases
have already been explored for the two dominant production modes
($t$-channel~\cite{Mahlon:1999gz}, associated $tW$
production~\cite{Boos:2002xw}). Already with the first 2~fb$^{-1}$, a
polarization measurement with an uncertainty of 4\% should be
achievable based on the $t$-channel production
alone~\cite{Beneke:2000hk}.

\subsubsection{Search for $W'$ bosons}
\label{sec:STWprime}
Electrically charged gauge bosons that are not part of the standard
model are usually denoted as $W'$. Such bosons are predicted in a
variety of extensions of the standard model, incorporating larger gauge
groups that reduce to the standard model at
sufficiently low energies~\cite{Tait:2000sh, Boos:2006xe}.

The most stringent limit to date in a direct search on the mass of
such a $W'$ boson has been set by D0 in the leptonic final state
($W'\to\ell\nu$) using 1~fb$^{-1}$ of Run~II data~\cite{Abazov:2007bs}.
Assuming the $W'$ boson exhibits standard model $W$ boson couplings
to fermions, this search excludes the mass range below 1~TeV/c$^2$ at
95\% C.L.\ by studying the tail of the transverse mass~\cite{Smith:1983aa} spectrum
calculated from lepton transverse energy and \METns.
Indirect $W'$ mass constraints are strongly
model-dependent and range between lower limits of 549~GeV/c$^2$ and
23~TeV/c$^2$, being derived from (semi-) leptonic processes as well as
from astrophysical and cosmological constraints~\cite{Yao:2006px}.

A direct search for $W'$ bosons in the hadronic final state ($W'\to
q\bar{q}'$) provides a less model-dependent measurement since both
left-handed and right-handed $W'$ bosons can be observed in this final
state, independent of any assumption about the mass of a right-handed
neutrino $m_{\nu_R}$ in the latter case. In contrast to this, the
leptonic final state is only accessible for a right-handed $W'$ boson
if the corresponding right-handed neutrino is not too massive
($m_{\nu_R} < m_{W'}$). Searches for $W'$ bosons as resonant
structures in the dijet invariant mass spectrum have been carried out
by UA2~\cite{Alitti:1993pn} and at the Run~I Tevatron by
CDF~\cite{Abe:1997hm} and D0~\cite{Abazov:2003tj}.

Focusing on ``hadronic'' $W'$ searches using third generation quarks in
the final state, reduces the QCD multijet background compared to the (light)
dijet final state searches. Such measurements are only sensitive to
$W'$ bosons with masses above the $tb$ threshold of 
$\approx$200~GeV/c$^2$, but the low-mass range is excluded already by the
current limits on single top quark production~\cite{Abazov:2006aj}. A
$W'$ signal would be observed as peak in the invariant mass
distribution of its $tb$ decay products (as usual, $tb$ includes both $t\bar{b}$
and its charge-conjugate $\bar{t}b$).

Since the $W'\to tb$ decay mode contributes to $s$-channel single top
production (see Section~\ref{sec:singletopprod}), these searches are
based on the single-top production cross section analyses (see
Section~\ref{sec:STxsec}). For left-handed $W'$ bosons, interference
occurs with SM single top production, which is not the
case for right-handed $W'$ bosons due to the different (right-handed)
final state particles. Considering a right-handed $W'$ boson, the
decay width depends on the mass of the right-handed accompanying neutrino
in leptonic decays. If $m_{\nu_R} > m_{W'}$, only $q\bar{q}'$ final
states are accessible, resulting in a width reduced by about 25\%.
This scenario generally results in a more stringent mass limit due
to the enhanced $tb$ branching fraction. A contribution of the $W'$
boson to top quark decay is usually not considered due to its
large mass.

A first search for $W'\to tb$ was performed by CDF in Run~I, based on
0.1~fb$^{-1}$ of lepton + jets data. At 95\% C.L., lower limits on
the mass of a right-handed $W'$ boson were obtained, yielding
536~GeV/c$^2$ for $m_{\nu_R} \ll m_{W'}$ and 566~GeV/c$^2$ for
$m_{\nu_R} > m_{W'}$~\cite{Acosta:2002nu}.

D0 published a first search for $W'\to tb$ in Run~II, based on
0.2~fb$^{-1}$ of lepton + jets data and the corresponding single
top cross section result~\cite{Abazov:2005zz}. For a right-handed $W'$
boson with CKM mixing equal to that of the SM, 95\% C.L.\ lower
mass limits of 630~GeV/c$^2$ (670~GeV/c$^2$) are obtained for
$m_{\nu_R} < m_{W'}$ ($m_{\nu_R} > m_{W'}$). In addition, a first
corresponding lower mass limit for a left-handed $W'$ boson is
derived, taking the interference with SM
production into account, yielding 610~GeV/c$^2$~\cite{Abazov:2006aj}.

\begin{figure}[!t]
  \begin{center} 
    \includegraphics[width=.48\textwidth]{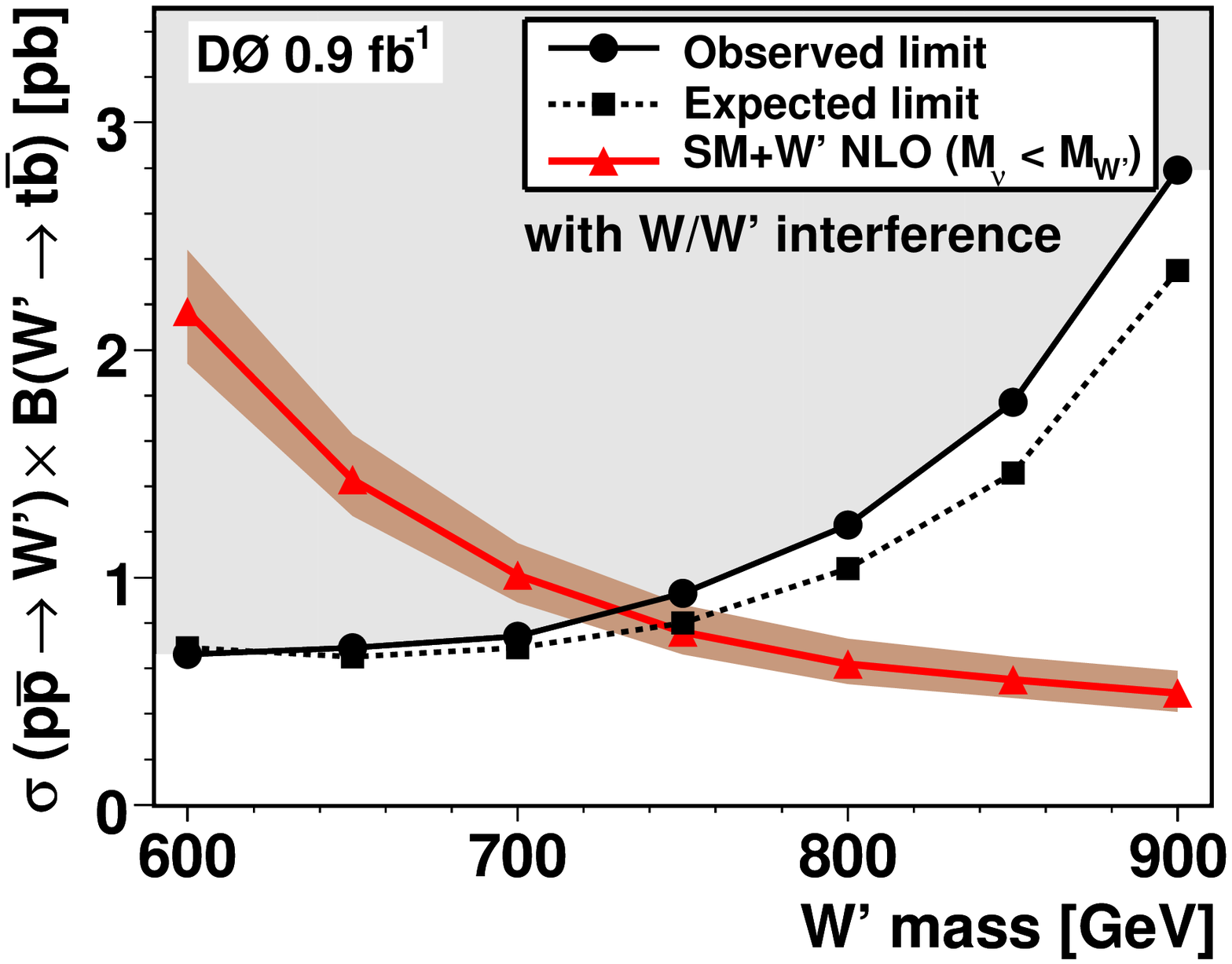}
    \includegraphics[width=.48\textwidth]{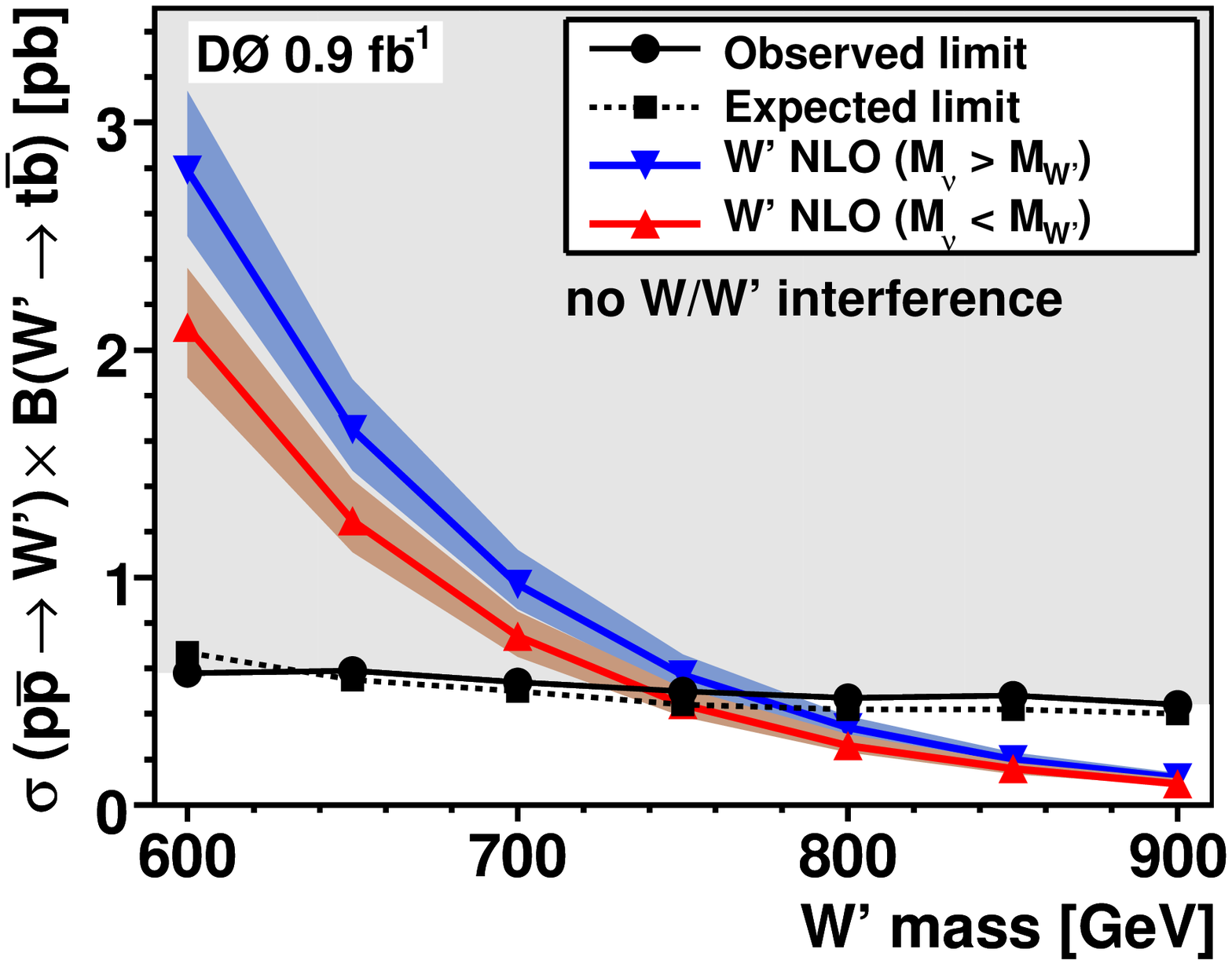}
    \caption{Theoretical prediction at NLO and 95\% C.L.\ limits for
      $\sigma(p\bar{p}\to W') \times {\cal{B}}(W'\to tb)$
      versus mass of the $W'$ boson. Left: Left-handed $W'$ boson.
      Right: Right-handed $W'$ boson~\cite{Abazov:2008vj}. 
    Results are from the D0 experiment.}
    \label{fig:STWprimeD0}
  \end{center}
\end{figure}

Based on the 0.9~fb$^{-1}$ lepton + jets dataset and the analysis
from which the first evidence for single top production was
obtained~\cite{Abazov:2006gd,Abazov:2008kt}, D0 obtains further
improved $W'$ mass limits~\cite{Abazov:2008vj}. Using the invariant
mass of charged lepton, leading two jets and neutrino as a sensitive
variable for separating signal and background, the 95\% C.L.\ lower
mass limit for a left-handed $W'$ boson interfering with SM
single-top production increases to 731~GeV/c$^2$. For a
right-handed $W'$ boson, the 95\% C.L.\ lower mass limits are
739~GeV/c$^2$ (768~GeV/c$^2$) for $m_{\nu_R} < m_{W'}$ ($m_{\nu_R} >
m_{W'}$), as illustrated in Fig.\ \ref{fig:STWprimeD0}. The latter
two cross section limits correspond to upper limits on the $W'$ gauge
coupling in units of the SM weak coupling of 0.72 (0.68)
for a $W'$ boson mass of 600 GeV/c$^2$. The dominant systematic
uncertainties included in these limits are the theoretical cross
sections (affecting the background normalization) and uncertainties on
jet energy calibration and $b$ jet simulation (affecting
background normalization and distribution in the sensitive variable).

CDF has obtained a preliminary result for their $W'\to tb$ search, based
on 1.9~fb$^{-1}$ lepton + jets Run~II data, using the invariant mass
of the reconstructed $W$ boson and the two leading jets as sensitive
variable. 95\% C.L.\ lower limits on the mass of a right-handed $W'$
boson are found to be 800~GeV/c$^2$ for $m_{\nu_R} < m_{W'}$ and
825~GeV/c$^2$ for $m_{\nu_R} > m_{W'}$. Neglecting interference
effects the former limit is considered to apply for a left-handed $W'$
boson as well. The corresponding $W'$ gauge coupling in units of the
SM weak coupling is found to be $<$0.68 and $<$0.63,
respectively, for a $W'$ boson mass of 600 GeV/c$^2$~\cite{CDF9150}.

The more general case of a $W'$ boson with an admixture of left- and
right-handed couplings to SM fermions has not been
studied thus far.

\subsubsection{Search for single top production via charged Higgs bosons}
\label{sec:STchargedH}
The standard model Higgs sector, with its single Higgs doublet of
complex scalar fields to break electroweak symmetry and generate
masses of weak gauge bosons and fermions (see
Section~\ref{sec:SMoverview}) can be easily extended to include a
second Higgs doublet, resulting in ``Two Higgs Doublet Models''
(THDM or 2HDM)~\cite{Gunion:1989we,Gunion:1992hs}. In contrast to the
single neutral scalar CP-even Higgs boson predicted by the SM, 
THDM give rise to five physical scalar Higgs bosons after
electroweak symmetry breaking. Two of these are charged bosons ($H^\pm$),
providing a unique signature for physics beyond the standard model.
Three different Higgs-fermion couplings are discerned in THDM. Type-I
models provide coupling of only one of the Higgs doublets to fermions.
In Type-II models, each of the doublets couples solely to up-type
fermions and down-type fermions, respectively, while in Type-III models
general couplings of both Higgs doublets to fermions are allowed. In
the latter case, Higgs-mediated flavor-changing neutral currents at
tree level must be sufficiently suppressed to be compatible with
experimental limits. This can be achieved through an appropriate choice of the Higgs
parameters~\cite{He:1998ie}.

If the charged Higgs boson is heavier than the top quark ($m_{H^\pm} >
m_t$), its production via quark fusion can contribute to single top
quark production through the decay into third-generation quarks:
$q\bar{q}'\to H^\pm\to tb$. Due to mass dependent couplings of the
charged Higgs boson, this decay is dominant in many models. The
signature of this process is identical to that of $s$-channel single
top-quark production, and the search for charged Higgs bosons can be
performed similar to that for $W'$ bosons, with the simplification
that interference with the SM production process can be
neglected.

\begin{figure}[!t]
  \begin{center} 
    \includegraphics[width=.48\textwidth, height = 50mm]{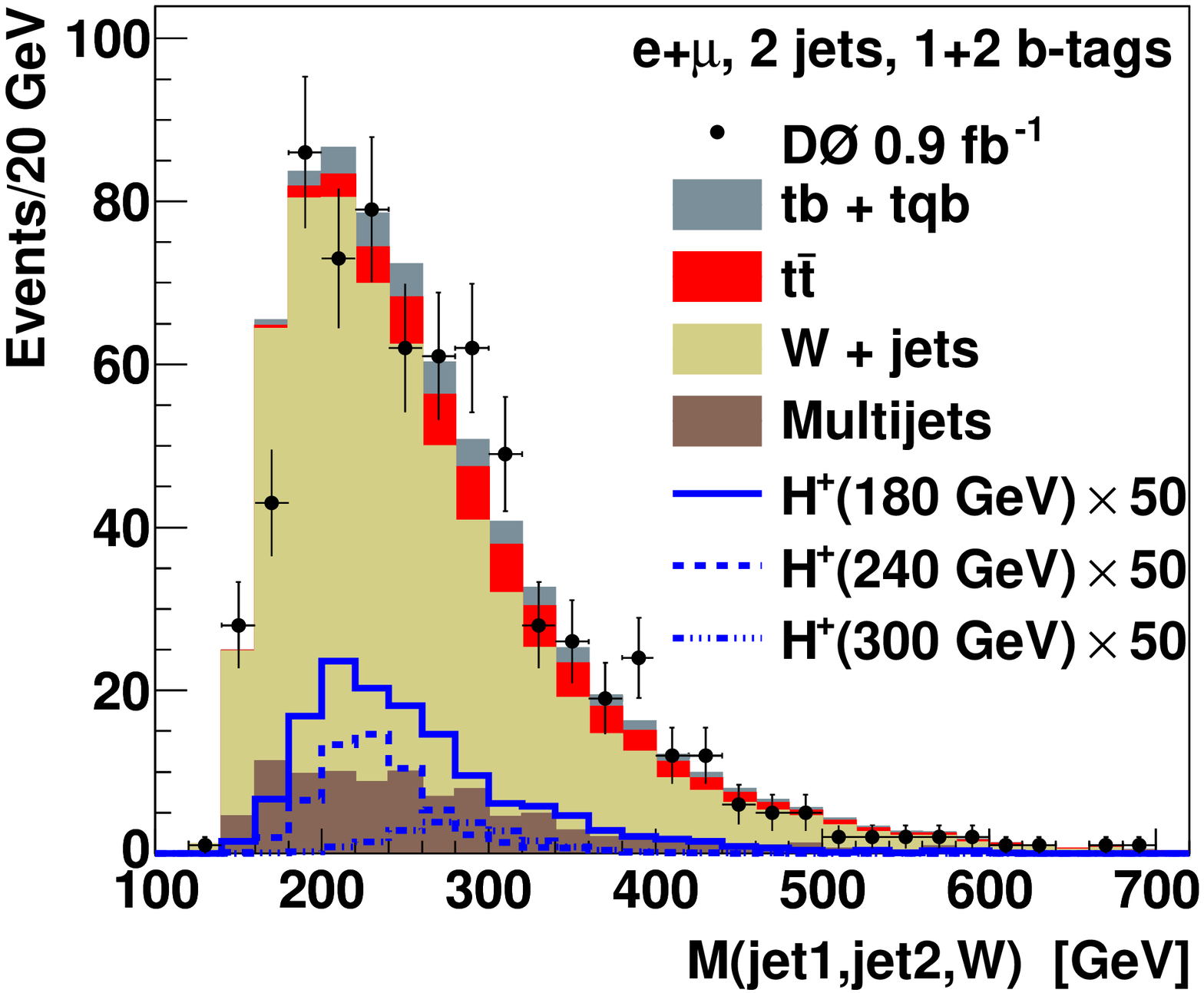}
    \includegraphics[width=.48\textwidth, height = 53mm]{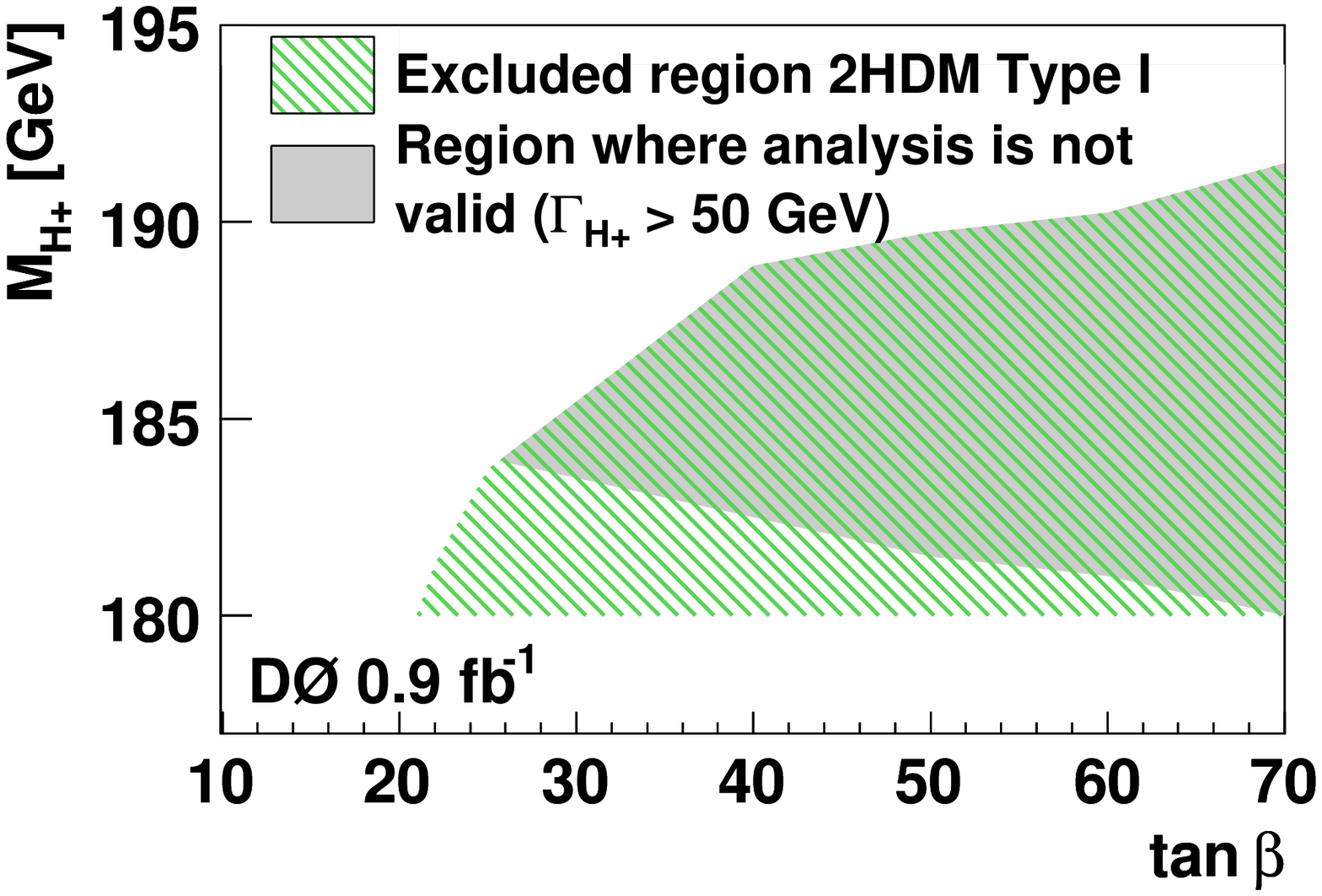}
    \caption{ Left: Distribution of the invariant mass of
      reconstructed $W$ boson and two jets in 0.9~fb$^{-1}$ lepton + jets data,
      SM background processes, and charged Higgs boson
      signal in a Type-III THDM for several $m_{H^\pm}$, with the expected rate enhanced by a
      factor of 50. Right: 95\% C.L.\ exclusion region in the
      ($m_{H^\pm},\tan\beta$) plane for a Type-I THDM. If
      $\Gamma_{H^\pm}$ exceeds 50~GeV/c$^2$, the analysis is no longer
      valid, and no limits can be derived as illustrated by the darker
      area~\cite{Abazov:2008rn}.}
    \label{fig:STchargedHD0}
  \end{center}
\end{figure}
D0 performs a first direct search for the process $q\bar{q}'\to
H^\pm\to tb \to \ell\nu b\bar{b}$~\cite{Abazov:2008rn}, based on the
analysis providing first evidence for single top
production~\cite{Abazov:2006gd,Abazov:2008kt} in a 0.9~fb$^{-1}$
lepton + jets dataset. Restricting the jet multiplicity in the
events to exactly two jets, corresponding to the $s$-channel final state, charged Higgs bosons are sought
in the mass range 180~GeV/c$^2$ $\leq m_{H^\pm} \leq$
300~GeV/c$^2$ for all three types of THDM. The sensitive variable
used to discriminate the charged Higgs boson signal from SM
background processes is the invariant mass of the reconstructed
$W$ boson and the two jets, as illustrated in
Fig.\ \ref{fig:STchargedHD0}. Since no evidence for signal is
observed in the data, 95\% C.L.\ upper limits on the charged Higgs
boson production cross section multiplied by branching fraction into third
generation quarks are provided for all three types of THDM.
The dominant systematic
uncertainties result from the jet energy scale calibration, modeling
of the $b$ jet identification and theoretical uncertainties in
modeling and normalizing the signal. For the Type-I THDM, the
limits are translated into a 95\% C.L.\ exclusion region in
($m_{H^\pm},\tan\beta$) parameter space (see
Fig.\ \ref{fig:STchargedHD0}), where $\tan\beta$ is the ratio of the
vacuum expectation values for the two Higgs doublets.

More searches for charged Higgs bosons, especially those in top quark
decays (for $m_{H^\pm} < m_t$), are described in
Section~\ref{sec:H+topdecay}.

\subsubsection{Search for single top production through neutral currents}
\label{sec:STFCNC}
Single top quark production via flavor changing neutral interactions
of light $u, c$ quarks and the $Z,\gamma, g$ gauge bosons is
possible in the standard model through higher-order radiative
corrections, but so strongly suppressed that it cannot be observed.
Consequently, searches for these production mechanisms at tree level
probe for corresponding anomalous coupling strengths
$\kappa$~\cite{Han:1998tp, Tait:2000sh} that are predicted by various
extensions of the standard model~\cite{AguilarSaavedra:2004wm}.

The processes involving photon or $Z$ boson exchange have been
extensively studied at LEP and HERA. At both accelerators top quarks
can only be produced singly at the available center of mass energies
due to the large top quark mass. 

At LEP, single top quark production proceeds via the SM
process $e^+e^-\to e^-\bar{\nu_e}t\bar{b}$, which can be ignored
in the available datasets due to its tiny production rate. All four
LEP experiments searched for single top production via $e^+e^-\to
t\bar{c}/t\bar{u}$ in both hadronic and semileptonic final states,
resulting from the different $W$ boson decay modes from the top quark
decay. While only the SM decay $t\to Wb$ is considered, a
possible reduction of its branching ratio due to possible FCNC decays is
accounted for when deriving the results. Since no evidence for single
top quark production is observed, 95\% C.L.\ upper limits on the cross
section are extracted, and corresponding model-dependent upper limits on
the anomalous coupling parameters $\kappa_\gamma$ and $\kappa_Z$ are
determined
\cite{Heister:2002xv,Abdallah:2003wf,Achard:2002vv,Abbiendi:2001wk}.
\begin{figure}[!t]
  \begin{center} 
    \includegraphics[width=.46\textwidth]{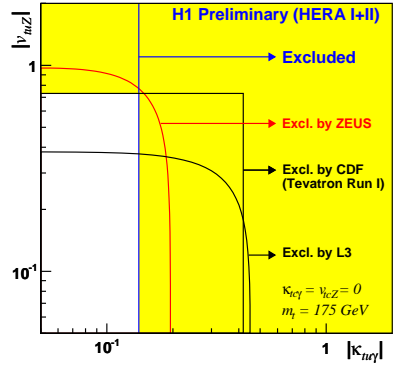}
    \caption{95\% C.L.\ upper limits on the top quark anomalous
couplings to photons and $Z$ bosons $\kappa_{tu\gamma}$ and
$v_{tuZ}$~\cite{South:2008km}. Depicted are the limits from
CDF~\cite{Abe:1997fz,Obraztsov:1997if}, L3~\cite{Achard:2002vv},
H1~\cite{South:2008km} and ZEUS~\cite{Chekanov:2003yt}.}
    \label{fig:STFCNCgamZ}
  \end{center}
\end{figure}

Single top quark production at HERA is possible via the charged
current SM process $ep\to \nu t\bar{b}X$ that has a
negligible production rate here as well. Both H1 and ZEUS have
searched for the inclusive neutral current production of top quarks in $ep\to etX$.
Because of the large $Z$ boson mass, this reaction is most sensitive
to couplings involving photons. Due to the large proton momentum
fractions needed for single top production, the $u$ quark contribution
will dominate over that from the $c$ quark (see
Fig.\ \ref{fig:CTEQPDF}), resulting therefore in highest sensitivity to
$tu\gamma$ couplings at HERA.

Using 0.1~fb$^{-1}$ of integrated luminosity, and
assuming the SM top quark decay $t\to Wb$, H1 and ZEUS
search for single top production both in the leptonic and hadronic
$W$ boson decay channel. ZEUS observes good agreement with the
SM prediction, and sets 95\% C.L.\ upper limits on single-top
production rate and on the FCNC magnetic coupling $\kappa_{tu\gamma}$
and vector coupling $v_{tuZ}$, neglecting charm
contributions~\cite{Chekanov:2003yt}. H1 observes five events in the
leptonic channel, with an expected SM background
contribution of $1.31\pm0.22$ events, while the hadronic channel
exhibits no excess over the standard model prediction. These two
channels are compatible at the 1.1~sd level, and both a combined
single-top cross section with about 2~sd significance and 95\%
C.L.\ upper limits on the cross section and on $\kappa_{tu\gamma}$ (assuming a
statistical fluctuation in the data) are provided~\cite{Aktas:2003yd}. In a recent
preliminary update of the measurement in the leptonic channel by H1,
using an integrated luminosity of 0.5~fb$^{-1}$, good agreement with
the standard model expectation is observed and improved limits on
the single-top cross section and $\kappa_{tu\gamma}$ are
obtained~\cite{South:2008km}.

Further limits on the anomalous couplings $\kappa_\gamma$ and
$\kappa_Z$ (inclusive for $u,c$ contributions) have been measured at the Tevatron by CDF via a search
for neutral-current top quark decays $t\to \gamma q$ and $t\to Zq$ as
will be discussed in Section~\ref{sec:FCNCdecay}. The most stringent
results on anomalous top quark couplings involving photons and $Z$
bosons obtained at LEP, HERA and the Tevatron are summarized in
Fig.\ \ref{fig:STFCNCgamZ}, with the exception of the latest CDF limit
on $t\to Zq$ decays~\cite{Aaltonen:2008aaa} that constrains $\kappa_Z$
better than the limit from L3.

Flavor changing neutral-current (FCNC) couplings of top quarks and gluons
have not been studied as extensively. A constraint on the anomalous
gluon coupling $\kappa_{tqg}/\Lambda$ of $< 0.52$~TeV$^{-1}$, where $\Lambda$ gives the scale
for new physics, was extracted from the observed \ttbar pair
production cross section at the Run~I Tevatron and a possible new
physics contribution that could be still accommodated within two
sd of combined experimental and theoretical uncertainties~\cite{Gouz:1998rk}.
Another limit on the anomalous gluon coupling was obtained using
the single-top production cross section limit measured by
ZEUS~\cite{Chekanov:2003yt}. Neglecting any effects that arise from the
different final states obtained in the gluon channel compared to the
original search (one additional light jet is present in the
gluon case), at 95\% C.L.\ $\kappa_{tqg}/\Lambda < 0.4$~TeV$^{-1}$ is
obtained~\cite{Ashimova:2006zc}.

\begin{figure}[!t]
  \begin{center} 
    \includegraphics[width=.48\textwidth]{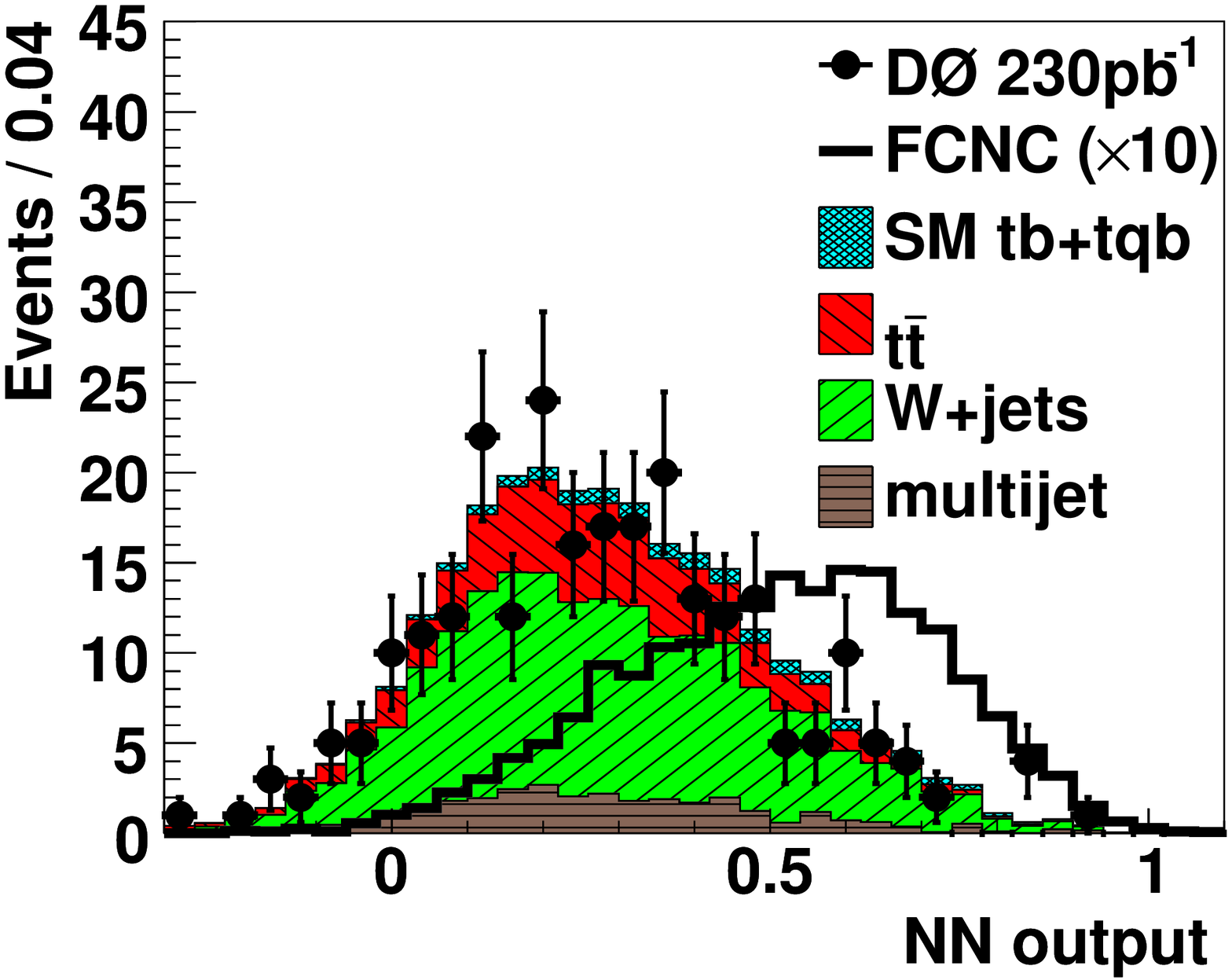}
    \includegraphics[width=.48\textwidth]{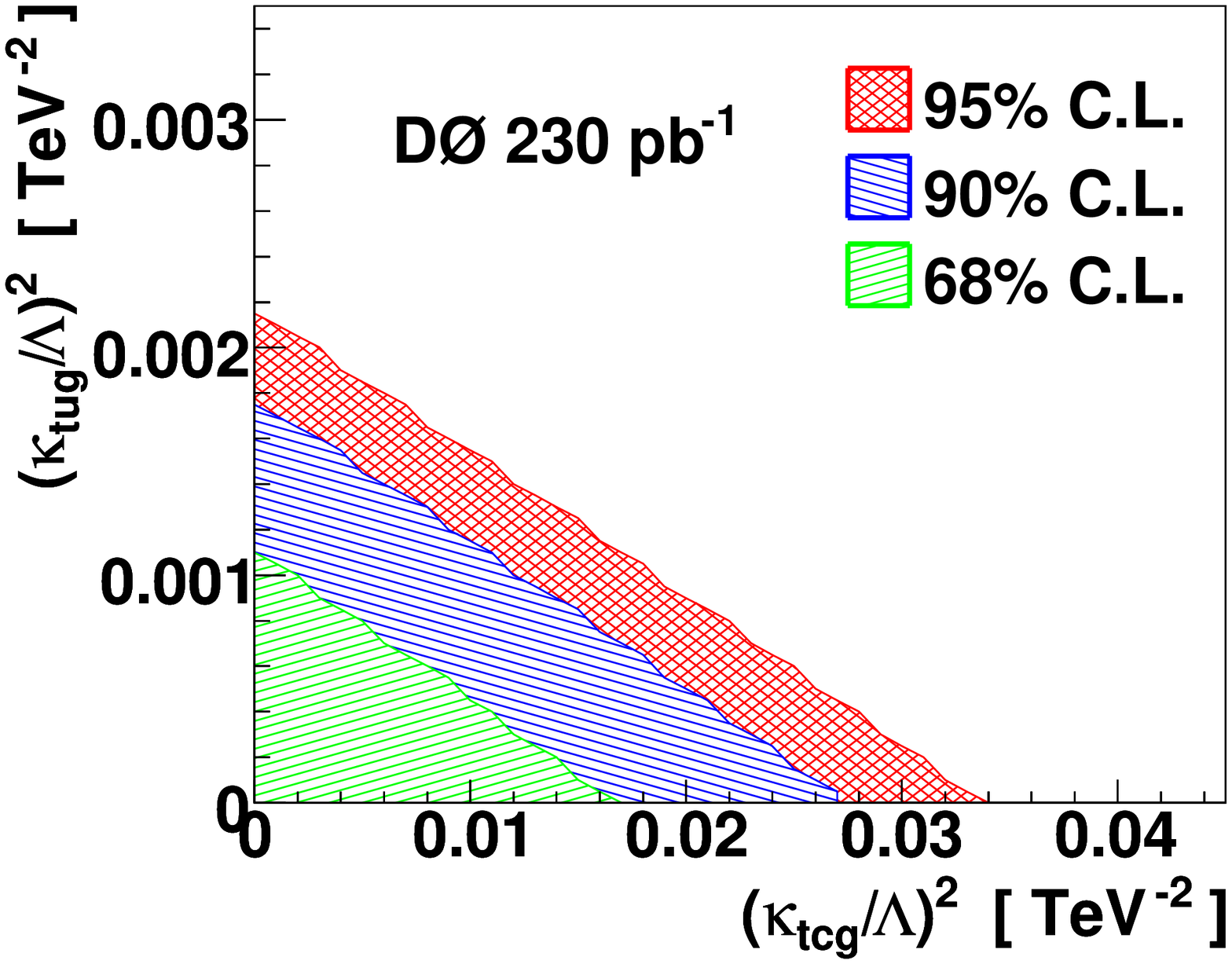}
    \caption{Left: Neural Network (NN) discriminant distribution in
0.2~fb$^{-1}$ lepton + jets data, with simulated FCNC signal increased by a factor of ten, and
SM background. The signal distribution represents the sum
of $tug$ and $tcg$ processes, evaluated for $\kappa_{tqg}/\Lambda =
0.03$~TeV$^{-1}$. Right: Exclusion contours for anomalous top-gluon
couplings for different levels of confidence~\cite{Abazov:2007ev}.}
    \label{fig:STFCNCgluon}
  \end{center}
\end{figure}
D0 has performed a first search for single top production via flavor
changing neutral-current couplings to gluons at a hadron collider,
using 0.2~fb$^{-1}$ lepton + jets data~\cite{Abazov:2007ev}. The
analysis is based on the corresponding search for SM
single-top production~\cite{Abazov:2005zz,Abazov:2006uq}, but is
restricted to events with only one $b$ tagged jet, and treats $s$- and
$t$-channel SM single-top production as background. Since
the neutral current decays $t\to gu/gc$ exhibit a negligible
branching fraction for $\kappa_{tqg}/\Lambda \lsim
0.2$~TeV$^{-1}$~\cite{Hosch:1997gz}, exclusively the standard model top quark
decay can be considered.

To separate the FCNC signal from the overwhelming SM
background, a neural network  is deployed, with ten input variables based
on global event kinematics, angular correlations and kinematics of the
individual reconstructed objects. The resulting data
distribution is shown in Fig.\ \ref{fig:STFCNCgluon}, and exhibits
good agreement with the SM prediction, which provides
limits on the FCNC couplings $\kappa_{tug}/\Lambda$ and
$\kappa_{tcg}/\Lambda$.

Systematic uncertainties affecting either the normalization or both
normalization and shape of the distributions are taken into account
when calculating the two-dimensional Bayesian posterior probability
density, resulting in the exclusion contours for the two couplings for
different confidence levels shown in Fig.\ \ref{fig:STFCNCgluon}. The
largest normalization uncertainties arise from the background cross
section uncertainties, which includes the uncertainty on the top quark mass
for \ttbar and single top samples. The largest uncertainties affecting
the shape as well arise from jet energy scale calibration and $b$ tag
modeling. 95\% C.L.\ upper limits on $\kappa_{tug}/\Lambda$ and
$\kappa_{tcg}/\Lambda$ are obtained by integrating over the other
variable, and yield 0.037~TeV$^{-1}$ and 0.15~TeV$^{-1}$ for $tug$ and $tcg$ couplings, respectively.
These limits represent a significant improvement over previous
values by up to an order of magnitude.

\begin{figure}[!t]
  \begin{center} 
    \includegraphics[width=.52\textwidth]{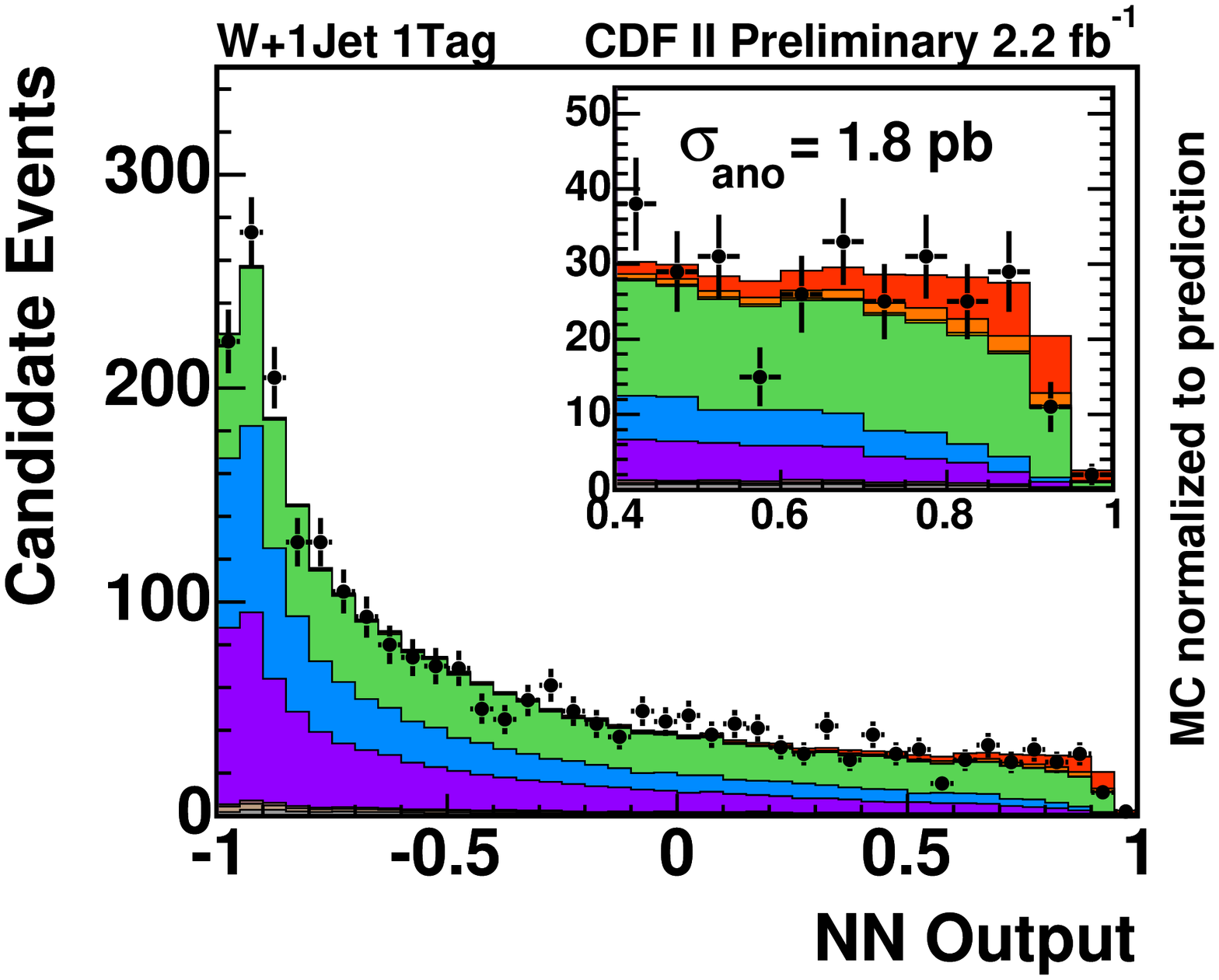}\hspace*{1mm}
    \includegraphics[width=.44\textwidth]{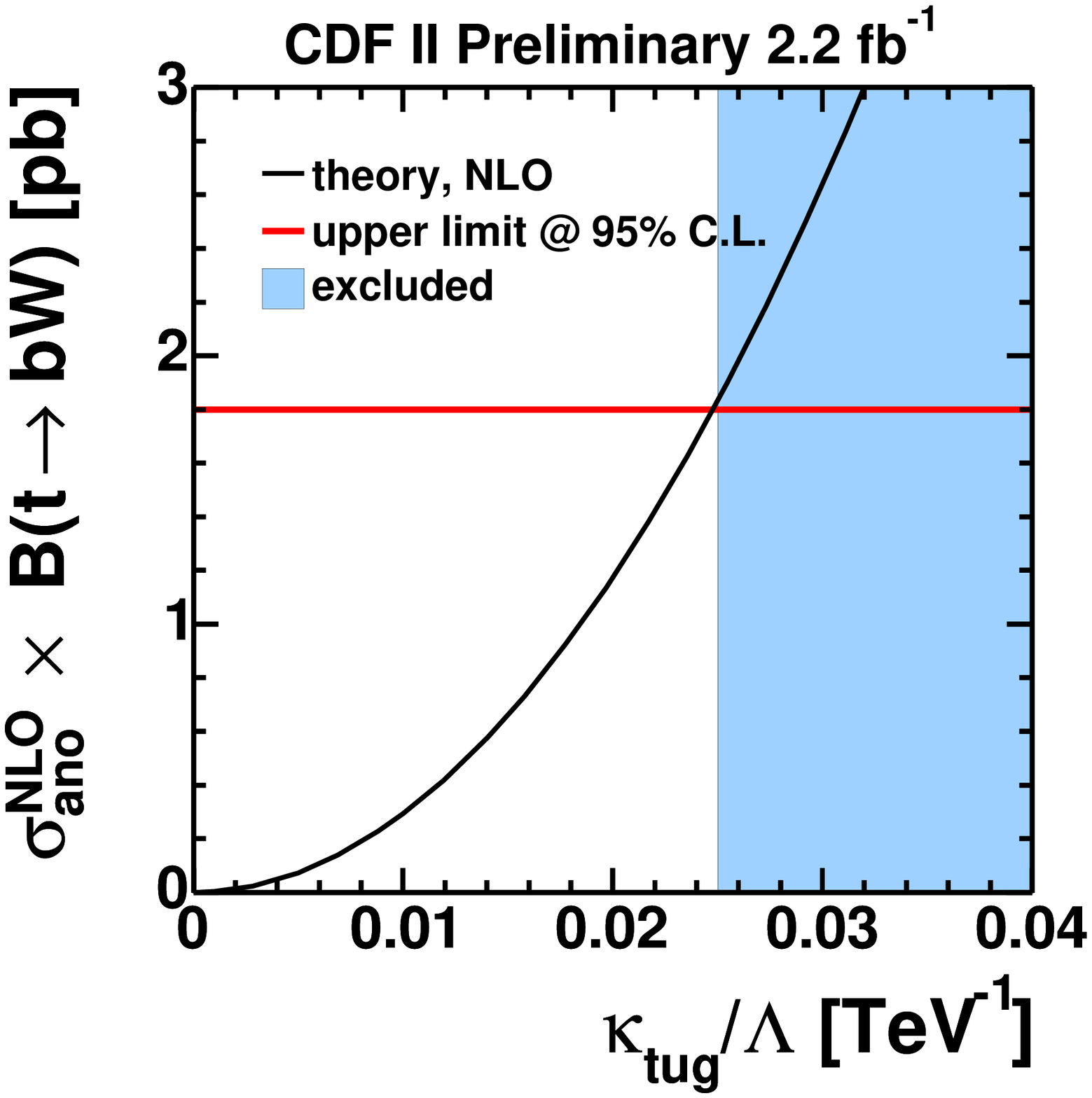} 
    \caption{Left: Neural network discriminant output in $W$ + one
      $b$ jet events observed in 2.2~fb$^{-1}$ of CDF data (black
      points) and standard model background processes. A FCNC single
      top signal with the observed 95\% C.L.\ {\em excluded} production
      rate is added in red. Right: Extraction of the
      $\kappa_{tug}/\Lambda$ limit from the intersection of the observed
      cross section limit and theoretical rate prediction, assuming
      $\kappa_{tcg} = 0$~\cite{CDF9440}.}
      \label{fig:CDFSTFCNC}
  \end{center}
\end{figure}

CDF also reports a recent preliminary search for FCNC single top 
production, based on 2.2~fb$^{-1}$ of data~\cite{CDF9440}\footnote{An updated
version of the result has been published after completion of this
review, see Ref.~\cite{Aaltonen:2008qr}.}. In distinction to
the D0 analysis, where $2\to2$ $tcg$ and $tug$ signal processes
are considered, CDF investigates the $2\to1$ processes $u(c)+g\to t$.
Since also in this analysis only SM top quark decay is
considered, events with one isolated energetic lepton, \MET and
one $b$ tagged jet are selected.

Signal and SM background processes are separated using a
Bayesian neural network based on 14 input variables containing
information from the reconstructed objects and event kinematics.
In a template fit to the observed distribution in data good agreement
is found with the SM background, as illustrated in
Fig.\ \ref{fig:CDFSTFCNC}, and a 95\% C.L.\ upper limit on FCNC single
top production via $u(c)+g\to t$ of 1.8~pb is obtained in accordance
with the expected sensitivity.

Based on LO predictions for the FCNC signal process from
\toprex~\cite{Slabospitsky:2002ag}, and using NLO
$k$-factors~\cite{Liu:2005dp,Yang:2006gs,Zhang:2008yn}, the obtained cross section
limit can be converted into limits on anomalous gluon couplings.
No two-dimensional information is used in this analysis for
contributions of $tcg$ relative to $tug$ signal processes, and
one coupling is assumed to vanish in deriving the limit for the other coupling.
The resulting 95\% C.L.\ upper limits are
$\kappa_{tug}/\Lambda <$ 0.025~TeV$^{-1}$ (see
Fig.\ \ref{fig:CDFSTFCNC}) and $\kappa_{tcg}/\Lambda <$
0.105~TeV$^{-1}$ (not shown).

\subsubsection{Anomalous $Wtb$ couplings in single top production}
The couplings between quarks and electroweak gauge bosons were
directly scrutinized at LEP~\cite{EWWG:2007}, with the exception of the
top quark. 
At the Tevatron, the couplings of the top quark and the
$W$ boson can be studied in measurements of top quark decay properties
in \ttbar production (see for example
Section~\ref{sec:Whelmeasurements}) and via single top quark
production. Physics beyond the standard model could
modify the Lorentz structure of the $Wtb$ vertex. Considering
a more general extension of the standard model $Wtb$ interaction
Lagrangian,
as discussed in Section~\ref{sec:Wheltheory}, new physics could
introduce contributions from right-handed vector ($f_1^R$)
and left- and right-handed tensor couplings ($f_2^L, f_2^R$),
in addition to the pure left-handed vector coupling ($f_1^L$) of the
standard model.

\begin{figure}[!t]
  \begin{center} 
    \includegraphics[width=.48\textwidth]{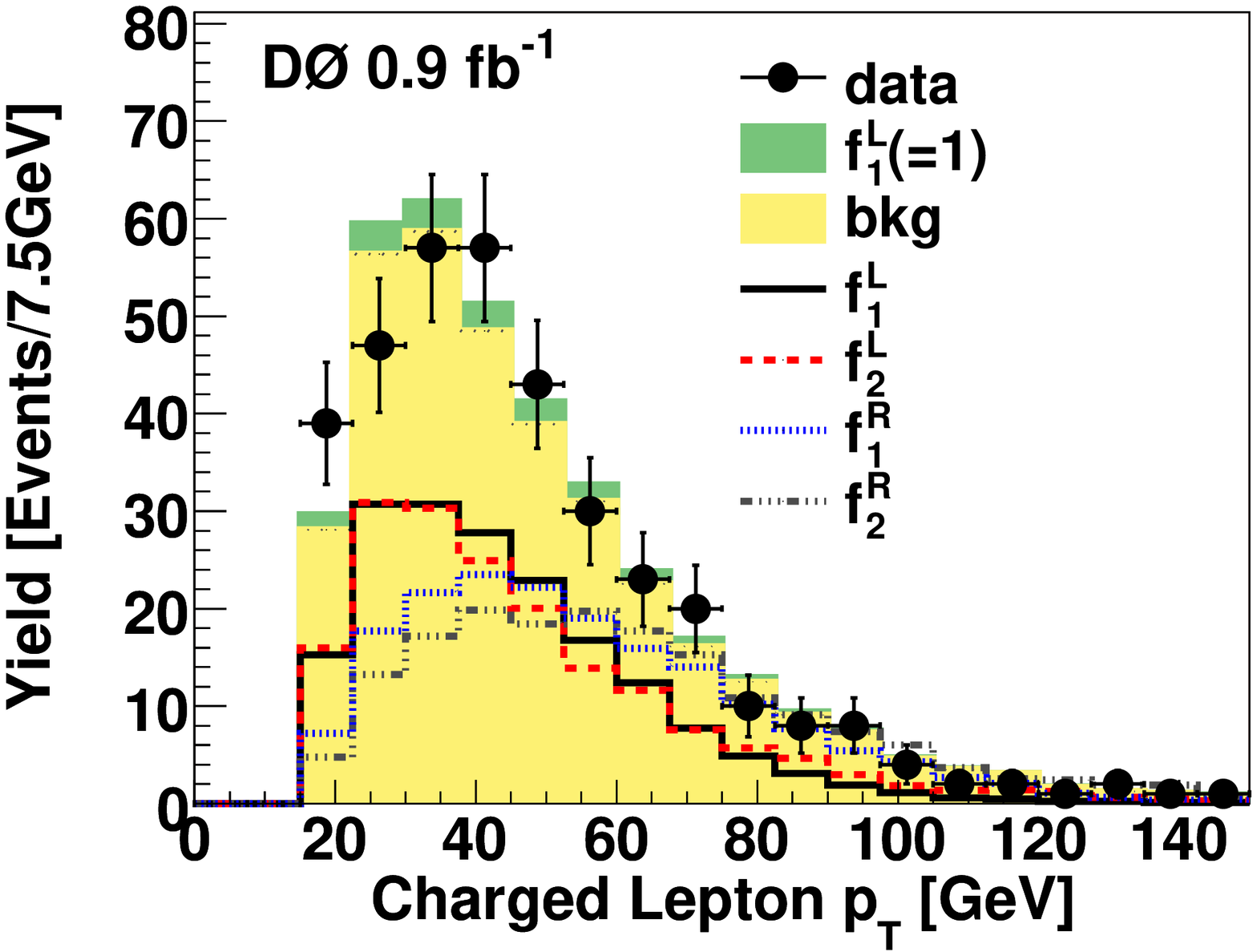}
    \includegraphics[width=.48\textwidth]{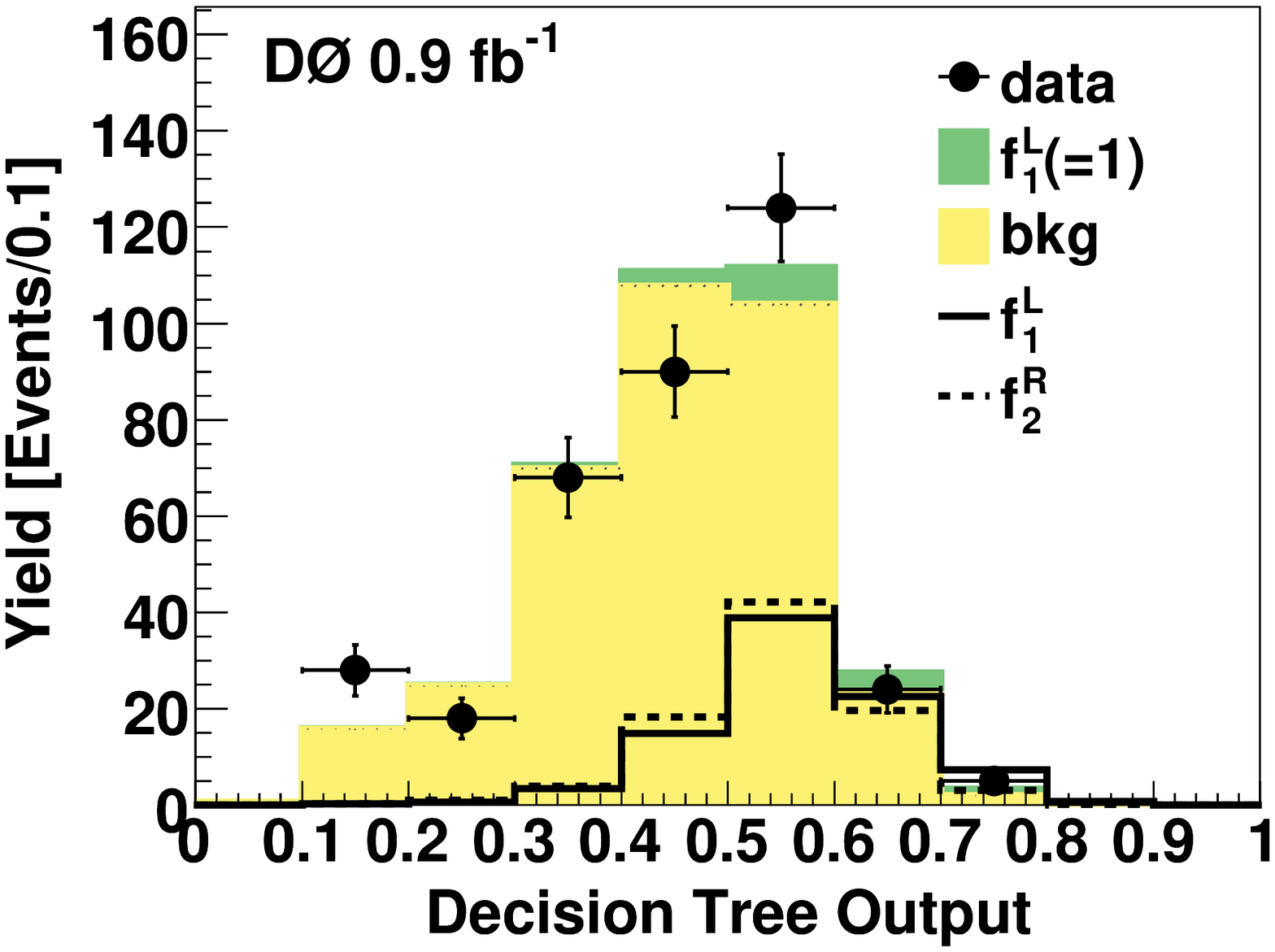}
    \caption{Left: Charged-lepton \pt distribution in
    0.9~fb$^{-1}$ lepton + jets data (two jets and one $b$ tag)
    and the corresponding SM single-top signal and
    background contributions. The effect of each of the four different $Wtb$
    couplings on the signal is also shown (calculated with the other
    couplings set to zero), with the normalization enhanced by
    a factor of ten. Right: Boosted decision tree output for the same
    data, signal and background contributions, with the ($L_1,R_2$)
    scenario couplings overlaid (normalization increased by a
    factor of five)~\cite{Abazov:2008sz}.}
    \label{fig:ST-Wtbcoupling}
  \end{center}
\end{figure}

D0 has published constraints on such
extended $Wtb$ interactions, including the first direct limits on the
left- and right-handed tensor couplings~\cite{Abazov:2008sz},
based on the analysis that provided first evidence for single top
production in 0.9~fb$^{-1}$ of lepton + jets
data~\cite{Abazov:2006gd,Abazov:2008kt}. In the analysis, single top
quark production and decay are considered to take place only via $W$ bosons,
with the dominant contribution arising from the $Wtb$
interaction, which is assumed to be $CP$ conserving. Anomalous
couplings at the $Wtb$ vertex can modify both the total single-top
production rate and the observed kinematics in the events relative
to SM expectation~\cite{Carlson:1994bg, Malkawi:1994tg,
Heinson:1996zm, Boos:1999dd}. The latter effect is illustrated in
Fig.\ \ref{fig:ST-Wtbcoupling} for the charged-lepton transverse
momentum distribution.

Since a simultaneous fit of all four couplings to data is not feasible with
the available statistics, the SM coupling and 
one additional anomalous coupling contribution at a time is considered
in varying proportions (with the remaining two other anomalous
couplings set to zero). The resulting scenarios are denoted as
($L_1,R_1$), ($L_1,R_2$) and ($L_1,L_2$). Non-negligible interference
effects in the last case are also taken into account. For signal
discrimination from SM background, boosted decision
trees are used that are based on the same variables as used
in Refs.~\cite{Abazov:2006gd,Abazov:2008kt}, but with the lepton \pt
distribution added. One example distribution for the
($L_1,R_2$) case is shown in Fig.\ \ref{fig:ST-Wtbcoupling}.

The decision tree output in data is compared with the various
single-top signal models in the twelve subchannels defined by lepton
flavor ($e,\mu$), jet multiplicity (two, three, four) and $b$ tag
multiplicity (one, two). This yields a two-dimensional Bayesian
posterior probability density, depending on $|f_1^L|^2$ and the
anomalous coupling $|f_{\rm ano}|^2$ considered in the respective scenarios.
Systematic uncertainties are taken into account, with dominant
contributions arising from background normalization, modeling of $b$
tagging and jet energy scale calibration. The latter two affect
both normalization and shape of the simulated spectra. The maxima of
the likelihoods in all three considered scenarios yield zero for
$|f_{\rm ano}|^2$, and 95\% C.L.\ upper limits on these anomalous couplings
are provided from the one-dimensional likelihood projections. These
results are summarized together with the measured single-top production rates
and $|f_1^L|^2$ values obtained from one-dimensional likelihood projections
in Table~\ref{tab:ST-Wtblimits}. The SM $Wtb$ interaction
is favored over any anomalous alternative studied. This analysis will
greatly benefit from the increased statistics already in hand and from more
expected until the end of Run~II.
\begin{table}[!t]
\caption{Total single-top production rates obtained in three
anomalous coupling scenarios, together with the corresponding
one-dimensional measurements and limits for the selected
couplings~\cite{Abazov:2008sz}.}
\renewcommand{\arraystretch}{1.2}
\centering
\begin{tabular}{lcl}
\hline
Scenario  & Cross Section ($tb+tqb$)  &  \multicolumn{1}{c}{Considered Couplings}   \\
\hline\hline
\multirow{2}{*}{$(L_1,R_1)$}  & \multirow{2}{*}{$5.2^{+2.6}_{-3.5}$~pb} & $|f^{L}_{1}|^2=1.8 ^{+1.0}_{-1.3}$\\&&$|f^{R}_{1}|^2<2.5$ (95\% C.L.)  \\
\hline
\multirow{2}{*}{$(L_1,R_2)$}  & \multirow{2}{*}{$4.5^{+2.2}_{-2.2}$~pb} & $|f^{L}_{1}|^2=1.4 ^{+0.9}_{-0.8}$ \\&&$|f^{R}_{2}|^2<0.3$ (95\% C.L.)   \\
\hline
\multirow{2}{*}{$(L_1,L_2)$}  & \multirow{2}{*}{$4.4^{+2.3}_{-2.5}$~pb} & $|f^{L}_{1}|^2=1.4 ^{+0.6}_{-0.5}$ \\&&$|f^{L}_{2}|^2<0.5$ (95\% C.L.)   \\
\hline
\end{tabular}
\label{tab:ST-Wtblimits}
\end{table}

\section{Decay Properties of the Top Quark}
\label{sec:BSMdecay}
The previous chapter demonstrated that no significant
deviations from the standard model expectations for top quark
production via the strong or electroweak interaction have thus far been
observed. In this chapter, decay properties of the top quark
will be investigated based on \ttbar data, generally assuming that top quark production proceeds
according to the standard model. %

\subsection{Measurement of the $\mathbf {W}$ boson helicity in $\mathbf {t\bar{t}}$ decays}
\label{sec:Whelmeasurements}
The helicity of the $W$ boson in top quark decays can be used to test
the $V-A$ Lorentz structure of the $Wtb$ interaction (see Section
\ref{sec:Wheltheory}). According to the expectation from the standard
model, $W$ bosons from top quark decays should be longitudinally
polarized with a fraction $f_0 \approx 70\%$ and left-handed with a
fraction $f_- \approx 30\%$. The right-handed fraction $f_+$ is
strongly suppressed, and below the per mill level~\cite{Fischer:2000kx}.
For the decay of antitop quarks, the $CP$ conjugate statement is
implied, resulting in $W^-$ bosons from $\bar{t}$ decays with either longitudinal or right-handed polarization.

A pure $V+A$ structure of the $Wtb$ interaction would result in an
observation of a right-handed fraction $f_+ = 30\%$, with negligible
left-handed contribution. Small \mbox{$V+A$} admixtures to the SM
left-handed charged-current weak interaction are
predicted, for instance, within $SU(2)_{R}\times SU(2)_{L}\times U(1)_{Y}$ extensions
of the SM~\cite{Bernreuther:2003xj,Beg:1977ti,
Beg:1977tierr, Nam:2002rq}. Such contributions would result in an
enhancement of $f_+$ while not significantly affecting $f_0$. Since
the decay amplitude for longitudinally polarized $W$ bosons is
proportional to the top quark's Yukawa coupling \cite{Kuhn:1996ug},
$f_0$ is sensitive to the mechanism of EWSB, and would be altered, for
example, in topcolor-assisted technicolor models~\cite{Chen:2005vr,
Wang:2005ra}.

The radiative decay rate $b\to s \gamma$ can be used to set indirect
limits on the $V+A$ admixture in top quark decays to below a few
percent~\cite{Fujikawa:1993zu, Cho:1993zb, Hosch:1996wu,
Jessop:2002ha}, assuming there are no contributions from gluonic
penguin diagrams in addition to the electroweak ones. This section
will discuss the direct measurements of the $W$ boson helicity
performed at the Tevatron using lepton + jets and dilepton
datasets. 

Thus far, four analysis techniques have been deployed to 
extract the $W$ boson helicity fractions, based on:
\begin{romanlist}[(ii)]
\item {\em helicity angle ($\cos\theta^{\ast}$)}: The helicity of the
$W$ boson is reflected in the angular distribution $\cos\theta^{\ast}$
of its decay products, with $\theta^{\ast}$ being the angle of the
down-type ($T_3 = -\frac{1}{2}$) decay products of the $W$ boson
(charged lepton, $d$ or $s$ quark) in the $W$ boson rest
frame relative to the top quark direction
\cite{Kane:1991bg,Dalitz:1991wa,Nelson:1997xd,Fischer:2000kx}:
\begin{eqnarray}
\frac{dN}{d\cos\theta^*} = 
f_-\cdot\frac{3}{8}(1-\cos\theta^*)^2 +
f_0\cdot\frac{3}{4}(1-\cos^2\theta^*) +
f_+\cdot\frac{3}{8}(1+\cos\theta^*)^2,
\end{eqnarray}
where $f_-$ can be replaced by $(1-f_+-f_0)$. The resulting
distributions for each helicity fraction and the superposition
expected from the standard model are shown in
Fig.\ \ref{fig:Whelcosdist}. A measurement of $\cos\theta^{\ast}$
provides the most direct measurement of the $W$ boson helicity, but
it requires the reconstruction of the momenta of the top quark and $W$ boson,
which is challenging and involves using \METns,
which has a rather poor resolution.
\begin{figure}[!t]
  \begin{center} 
    \includegraphics[height = 60mm]{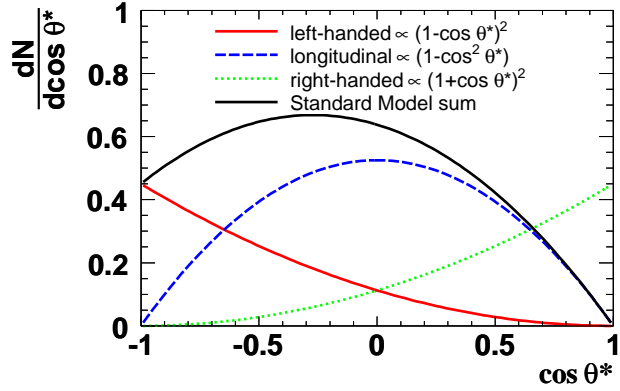}
    \caption{Helicity angle $\cos\theta^{\ast}$ distributions for
    left-handed, longitudinal and right-handed 
    $W$ bosons. The superposition expected from the
    standard model is shown as well.}
      \label{fig:Whelcosdist}
  \end{center}
\end{figure}
\item {\em charged-lepton \pt spectrum ($p_{T}^{\ell}$)}: The helicity
of the $W$ boson is correlated with the charged-lepton momentum
distribution: Since the $\nu_\ell$ from $W^+$ decays are
left-handed, while the $\ell^+$ are right-handed, in case of a
left-handed $W^+$ decay the $\ell^+$ are preferentially emitted
opposite to the momentum vector of the $W^+$. This leads to
a softer $p_{T}^{\ell}$ spectrum in the laboratory frame. Conversely,
the $\ell^+$ are preferentially emitted along the direction of the $W^+$
momentum in case of a right-handed $W^+$ decay, leading to a harder
$p_{T}^{\ell}$ spectrum. The $\ell^+$ from longitudinal $W^+$ decay
represent an intermediate case (see Fig.\ \ref{fig:Whel-lpt}).
\item {\em squared invariant mass of $b$ quark and charged lepton ($M_{\ell
b}^{2}$)}: In the limit of $m_{b}=0$, the helicity angle distribution $\cos\theta^{\ast}$ can be
approximated using the squared invariant mass of the system composed of the
$b$ quark and the charged lepton $M_{\ell b}^{2}$:
\begin{equation}
\cos\theta^{\ast} = \frac{p_\ell\cdot p_b - E_\ell
E_b}{|\mathbf{p}_\ell||\mathbf{p}_b|}
\simeq\frac{2M_{\ell b}^{2}}{m_t^2-M_W^2}-1.
\label{eq:Mlb2}
\end{equation}
This way one avoids the challenge of kinematic
reconstruction of the top quark and the application of \MET by using only
momenta measured in the laboratory frame. 
\item {\em Matrix Element method (ME)}: The Matrix Element method was
originally developed by D0~\cite{Abazov:2004cs}, yielding a very
precise mass measurement given the limited Run~I data sample (see also
Section~\ref{sec:ljmassmeasurements}). Using all the available kinematic
information in each event, a probability for the event to
correspond to a \ttbar final state as a function of the helicity of the
$W$ boson can be calculated, based on the LO matrix element.
\end{romanlist}
The following subsections will give brief examples for each method,
followed by a summary of the current status of the measurements.

\subsubsection{$p_{T}^{\ell}$ and $M_{\ell b}^{2}$}
\begin{figure}[!t]
  \centering
  \includegraphics[width = 0.48 \textwidth, height=40mm]{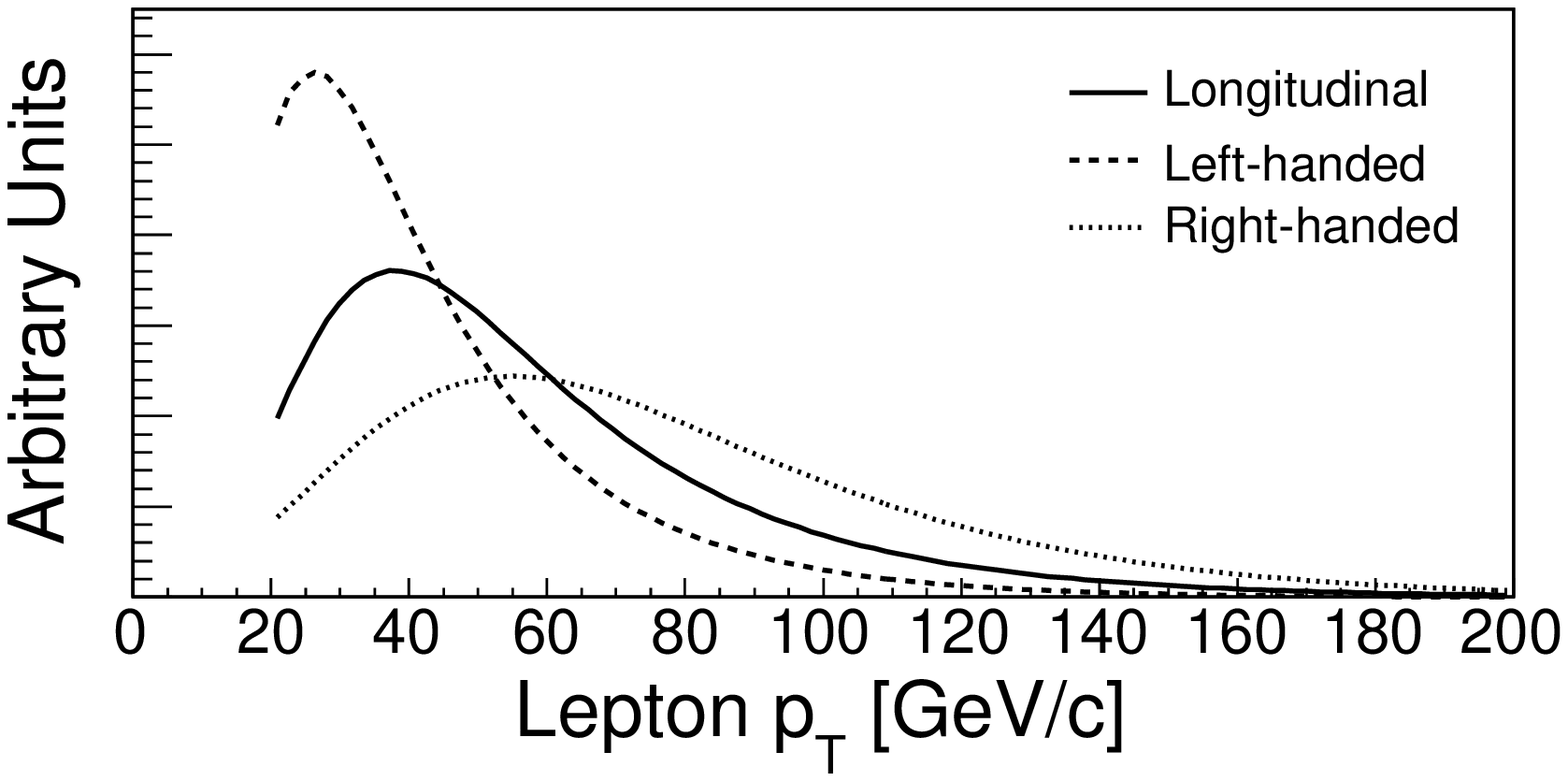}
  \includegraphics[width = 0.48 \textwidth, height=40mm]{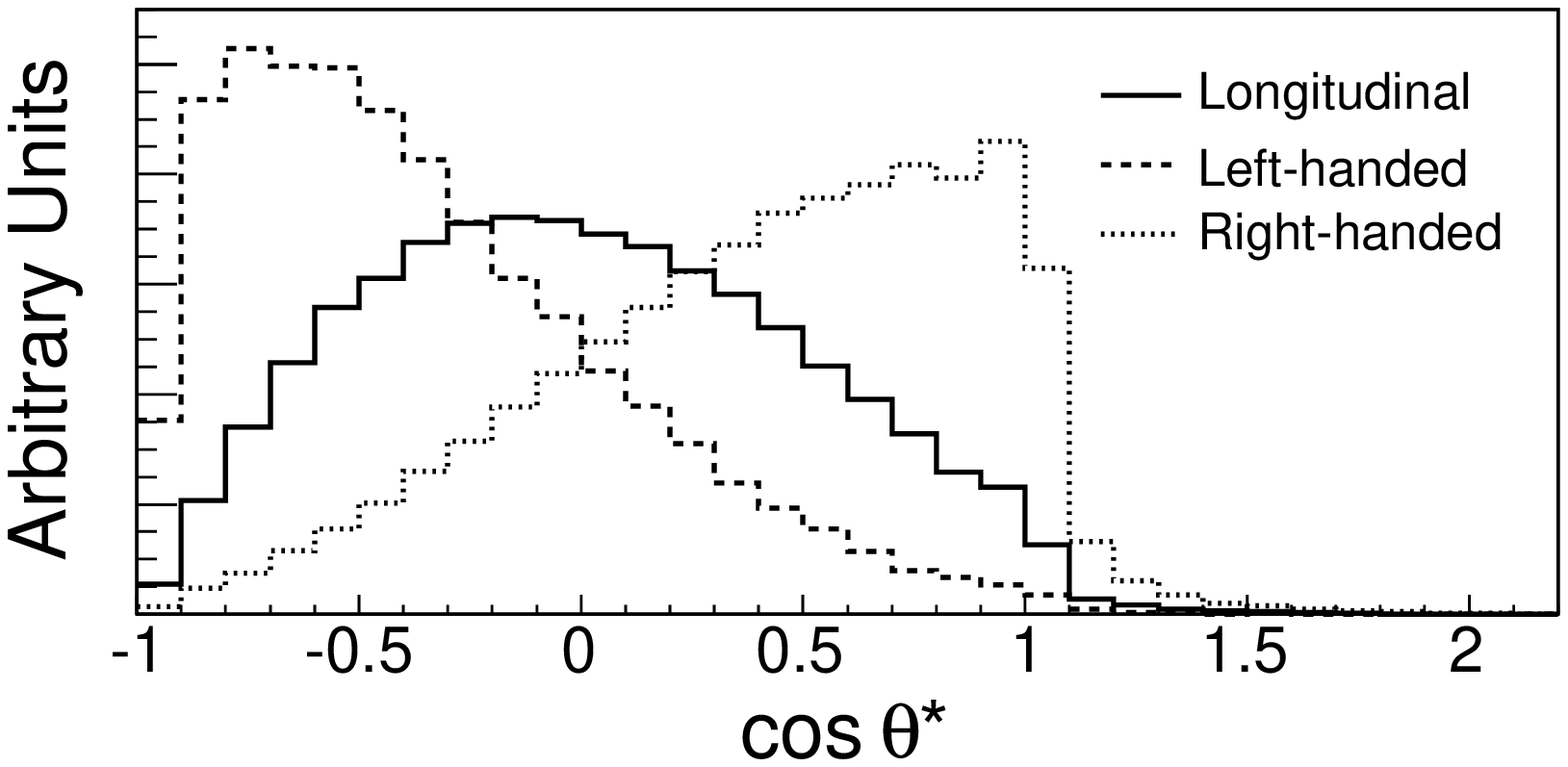}
  \caption{Distributions of reconstructed charged-lepton \pt and
    $\cos\theta^{\ast}$ (based on Eq.~\ref{eq:Mlb2})
    for top quark decays involving left-handed, right-handed and
    longitudinally polarized $W$ bosons~\cite{Abulencia:2005xf}.}
  \label{fig:Whel-lpt}
\end{figure}

CDF has measured the $W$ boson helicity in a 0.2 fb$^{-1}$ Run~II
dataset using the charged-lepton \pt ($p_{T}^{\ell}$) and the squared
invariant mass of the $b$ quark and charged lepton ($M_{\ell b}^{2}$) to
approximate $\cos\theta^{\ast}$~\cite{Abulencia:2005xf}. The
dependence of these observables on the $W$ boson helicity for a top
quark mass of 175 GeV/c$^2$ after event selection and reconstruction
is shown in Fig.\ \ref{fig:Whel-lpt}. Since the world-averaged top and
$W$ boson masses are used for calculating $\cos\theta^{\ast}$,
rather than the corresponding event-by-event reconstructed masses that
would smear out the distribution due to the larger inherent
uncertainties, values are observed outside of the physical range $-1 \leq
\cos\theta^{\ast} \leq 1$.

For the $p_{T}^{\ell}$ analysis, a $b$ tagged lepton + jets sample
is used requiring at least three jets, and yielding 57 events of which
approximately 2/3 are \ttbar signal. In addition a dilepton sample,
with a minimum value for the scalar sum of the transverse energy of jets, leptons
and \MET is analyzed, yielding 13 events with a signal fraction of
$\approx$0.79. The analysis based on $M_{\ell b}^{2}$ uses the lepton +
jets sample alone, requiring a fourth jet and a good
kinematic fit to the \ttbar hypothesis for a top mass of 175
GeV/c$^2$. This provides the lepton and appropriate
jet to form $M_{\ell b}^{2}$, and leaves 31 events for this analysis
with a signal fraction of $\approx$0.78.

For both analyses, the data distributions are fitted separately to
$p_{T}^{\ell}$ and $\cos\theta^{\ast}$ templates of signal
with the different $W$ boson helicities and background. Because of
limited statistics, the helicity fractions $f_0$ and $f_+$ cannot be
fitted simultaneously. Consequently, $f_0$ or $f_+$ are
constrained to their standard model values when fitting for $f_+$
or $f_0$, respectively. Both analyses are finally combined, taking
statistical and systematic correlations into account, and yield results
consistent with the standard model expectation, as shown in
Table~\ref{tab:Whel-lpttable}. The dominant systematic uncertainties come
from uncertainties on the top quark mass, background shape and
normalization, effects of initial- and final-state radiation (ISR/FSR)
and the PDFs.
\begin{table}
  \caption{Results of individual and combined measurements of $f_0$
    and $f_+$ using $M_{\ell b}^{2}$ and $p_{T}^{\ell}$. $N$ indicates
    the number of events or leptons used. If two uncertainties are
    given, the first is statistical and the second systematic. For the
    combined results, the statistical and systematic combined uncertainty
    is given. For the $p_{T}^{\ell}$($\ell\ell$) result, an
    observation of $\leq-0.54$ is expected 0.5\% of the time for 
    the SM $f_0$ value of 0.7~~\cite{Abulencia:2005xf}.}
\renewcommand{\arraystretch}{1.2}
  \begin{center}
    \begin{tabular}{|c|c|c|c|c|c|}
      \hline
      Analysis &$N$& $f_0$ & $f_+$  \\ \hline\hline
      $M_{\ell b}^{2}$ & 31& $0.99^{+0.29}_{-0.35}\pm0.19$  & $0.23\pm0.16\pm0.08$  \\  
      $p_{T}^{\ell}$($\ell\ell$) & 26& $-0.54^{+0.35}_{-0.25}\pm 0.16$ & $-0.47\pm0.10\pm0.09$ \\
      $p_{T}^{\ell}$($\ell j$) & 57& $0.95^{+0.35}_{-0.42}\pm0.17$  & $0.11^{+0.21}_{-0.19}\pm0.10$ \\
      $p_{T}^{\ell}$($\ell\ell,\ell j$) & 83& $0.31^{+0.37}_{-0.23}\pm0.17$  & $-0.18^{+0.14}_{-0.12}\pm0.12$ \\ \hline
      Combined & & $0.74^{+0.22}_{-0.34}$ &  $0.00^{+0.20}_{-0.19}$ \\
      95\% C.L.\ limit & & {$<0.95$},{$>0.18$} & $<0.27$ \\
      \hline
    \end{tabular}
  \end{center}
  \label{tab:Whel-lpttable}
\end{table}

CDF has also measured the fraction of right-handed $W$ bosons assuming
$f_0$ to be 0.7, using the $M_{\ell b}^{2}$ method on a 0.7
fb$^{-1}$ Run~II dataset~\cite{Abulencia:2006iy}. Using a single and
double $b$ tagged lepton + jets in addition to a dilepton dataset,
$f_+$ is extracted via maximum likelihood fits of the $M_{\ell b}^{2}$
distributions in data to $V+A$ and $V-A$ \ttbar signal Monte
Carlo and background contributions. Including uncertainties on the
\ttbar signal and background cross sections, the
lepton + jets sample yields \mbox{$f_+ =0.06 \pm 0.08$}, while the
dilepton sample gives \mbox{$f_+ = -0.19 \pm 0.11$}, corresponding to
a compatibility of the measurements at the level of 2.3~sd. A
combination of these measurements, including all systematic
uncertainties, yields \mbox{$f_+ = -0.02 \pm 0.07 \rm{(stat + syst)}$},
corresponding to $f_+ < 0.09$ at 95\% C.L. The main contributions to
the systematic uncertainty come from the jet energy scale, background
shape and normalization and limited Monte Carlo statistics.

\subsubsection{Matrix element method}

D0 has used the matrix element method that was originally employed to
measure the top quark mass~\cite{Abazov:2004cs} to extract the
longitudinal $W$ boson helicity fraction from 0.1 fb$^{-1}$ of Run~I
data~\cite{Abazov:2004ym}. The selected lepton + jets event sample
corresponds to that of the preceding mass
analysis~\cite{Abbott:1998dc} and comprises both soft muon $b$ tagged
events and untagged events, which have additional kinematic requirements,
yielding a total of 91 events.

By comparing the measured set of four-vectors in each event with the
differential cross section for \ttbar signal and the dominant $W +
\rm{jets}$ background, $f_0$ can be extracted by fixing $f_+$ to its
SM value, allowing the ratio $f_0/f_-$ to vary. The
use of both $W$ boson decays per signal event increases the
statistical sensitivity of the method. Since the calculation of signal
and background probabilities is based on leading-order matrix
elements, only events with exactly four jets are accepted, reducing
the sample to 71 events. In order to increase signal purity, a cut
on the background probability is applied, leaving 22 events to be
analyzed (as in the corresponding matrix-element mass
analysis~\cite{Abazov:2004cs}), with a signal to background ratio of
12/10.

\begin{figure}[!t]
  \centering
  \includegraphics[width = 0.68 \textwidth]{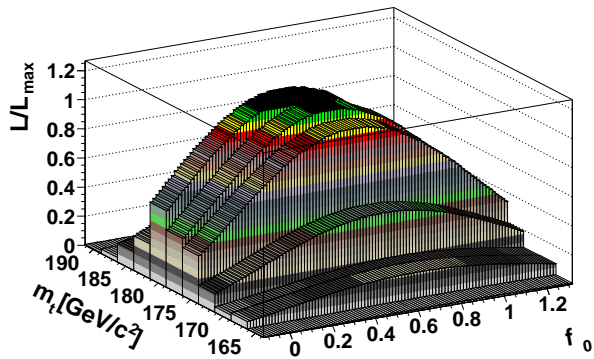}
  \caption{Two-dimensional probability density observed in 0.1 fb$^{-1}$ lepton +
  jets Run~I data as a function of $f_0$ and top quark mass
  $m_t$~\cite{Abazov:2004ym}.}
  \label{fig:Whel-ME}
\end{figure}

To take the dependence of the $f_0$ measurement on the top quark mass
into account, a two-dimensional likelihood, depending on $f_0$ and
$m_{t}$, is calculated and corrected for response deviations from unity
for different $f_0$ input values (see Fig.~\ref{fig:Whel-ME}). Since
statistics are insufficient, a simultaneous optimization for both
observables is not feasible; instead $f_0$ is evaluated by integrating
over the top quark mass, for a range between 165 and 190 GeV/$c^2$. The
maximum in the probability yields the central value of the
measurement, with the 1~sd uncertainty band corresponding to a
convolution of statistical and top quark mass uncertainties. Other
systematic uncertainties from acceptance and linearity of response or
jet energy scale are small compared to this, yielding the final result:
\begin{eqnarray}
f_0=0.56\pm0.31({\rm{stat}\oplus m_t})\pm0.07({\rm syst}).
\end{eqnarray}

CDF has obtained a preliminary result for $f_0$ using the same
method on 1.9 fb$^{-1}$ of Run~II data, assuming a fixed top quark mass
of 175 GeV/c$^2$~\cite{CDF9144}. 468 events are selected in a $b$
tagged lepton + jets sample with at least four jets (only the
leading four are used in the analysis) and a minimum value for $H_{T}$,
yielding a signal fraction of
about 0.84. Fixing $f_+$ to its SM value, the longitudinal
$W$ boson helicity fraction is found to be
$f_0=0.64\pm0.08({\rm{stat.}})\pm0.07({\rm syst.})$, with the
dominant systematic uncertainty coming from the Monte Carlo generator
used (\pythia\ versus \herwig) for the calibration of the measurement.
$f_0$ is found to change by $\mp 0.035$ for a $\pm 2.5$ GeV/c$^2$
variation in the top quark mass. Thus far, the analysis has not yet been
extended to measure $f_0$ and $f_+$ or $f_0$ and $m_t$ simultaneously.

\subsubsection{Helicity angle $\cos\theta^\ast$}
D0 has published a first model-independent measurement of the $W$
boson helicity fractions by comparing the
$\cos\theta^\ast$ distribution in data to templates of background and
right-handed, left-handed or longitudinal $W$ bosons in \ttbar signal,
using $f_+$ and $f_0$ as freely floating parameters and \mbox{$f_- = 1
- f_+ -f_0$} \cite{Abazov:2007ve}. In a 1~fb$^{-1}$ dataset, lepton
$+\geq 4$ jets and dilepton events are selected. The signal
purity is increased in each subsample by a cut on an individually
optimized multivariate likelihood discriminant based on event
kinematics and the output of a neural network $b$-tagging
discriminant. The cut values are chosen in each subsample to yield
the best expected precision for the helicity measurement.

The statistical sensitivity of the analysis is further improved by
about 20\% by including the $W\to q\bar{q}'$ decays in the lepton
+ jets sample in the measurement. This is accomplished through
picking one of the $W$ boson daughter jets
at random for the calculation of $\cos\theta^{\ast}$, which
introduces a sign ambiguity. Consequently, only
$|\cos\theta^{\ast}|$ is considered which does not permit to
discriminate left- from right-handed $W$ bosons, but still adds
information on $f_0$.

\begin{figure}[!t]
  \begin{center} 
    \includegraphics[width=.48\textwidth]{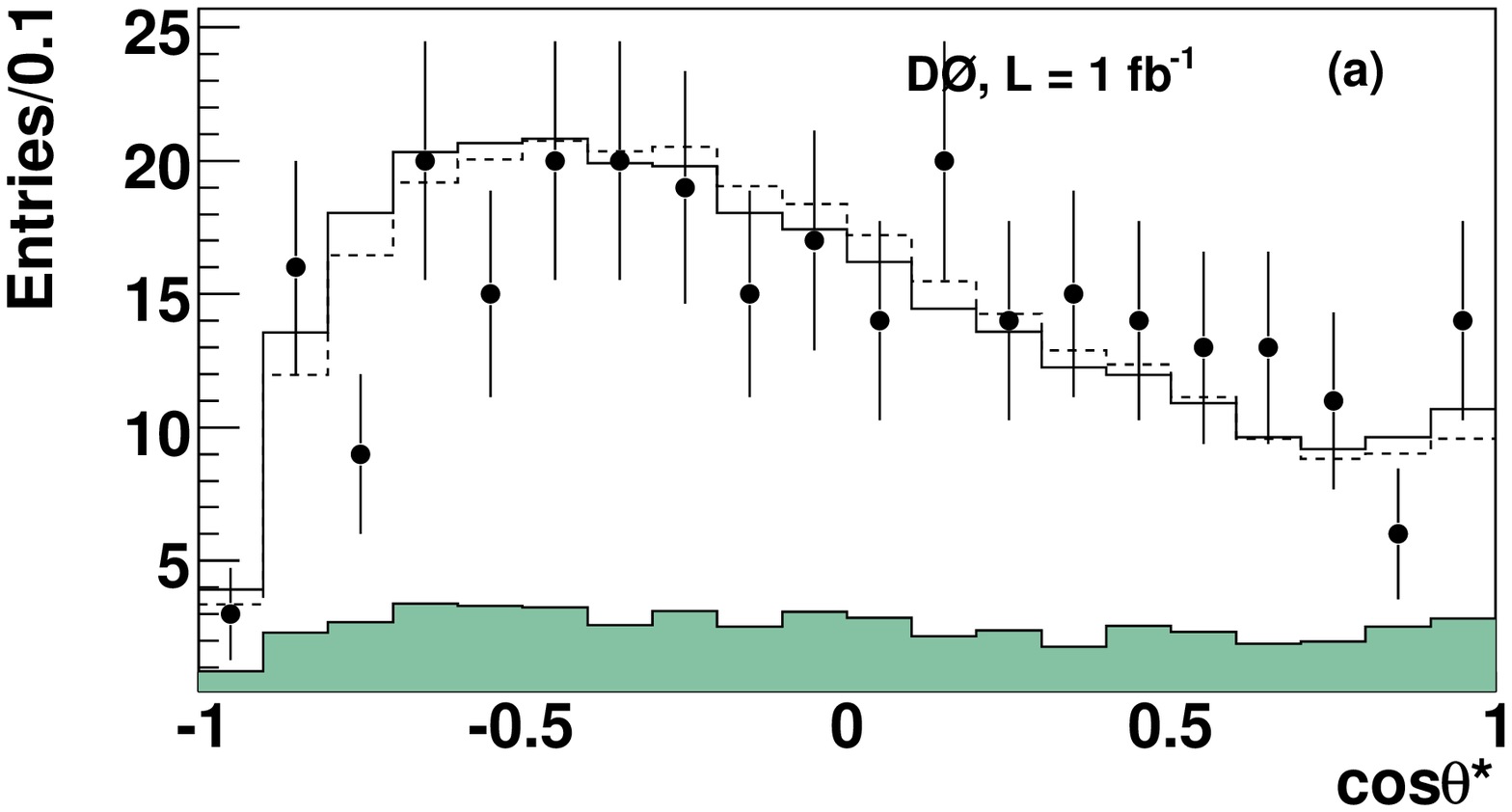}
    \includegraphics[width=.48\textwidth]{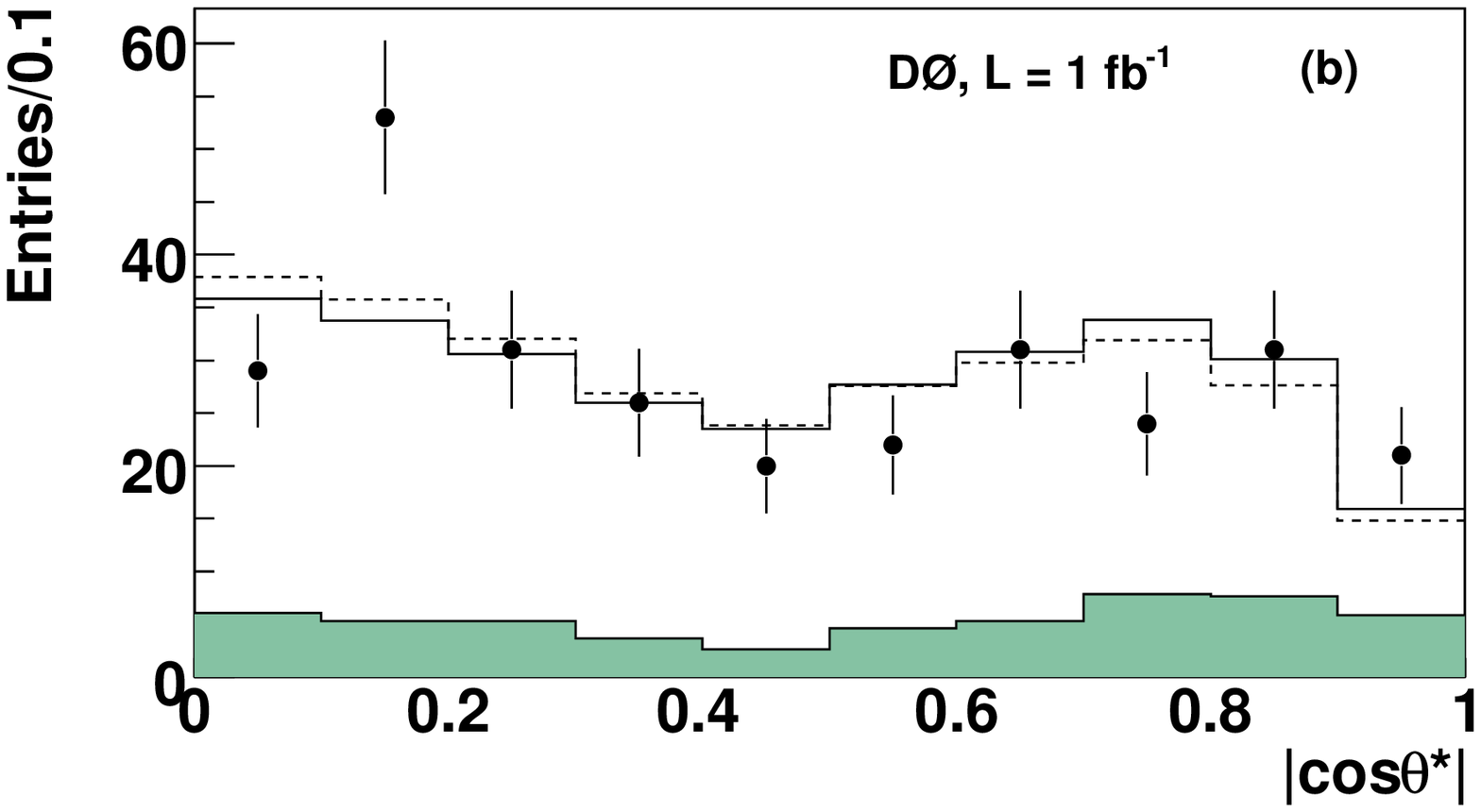}
    \includegraphics[width=.48\textwidth]{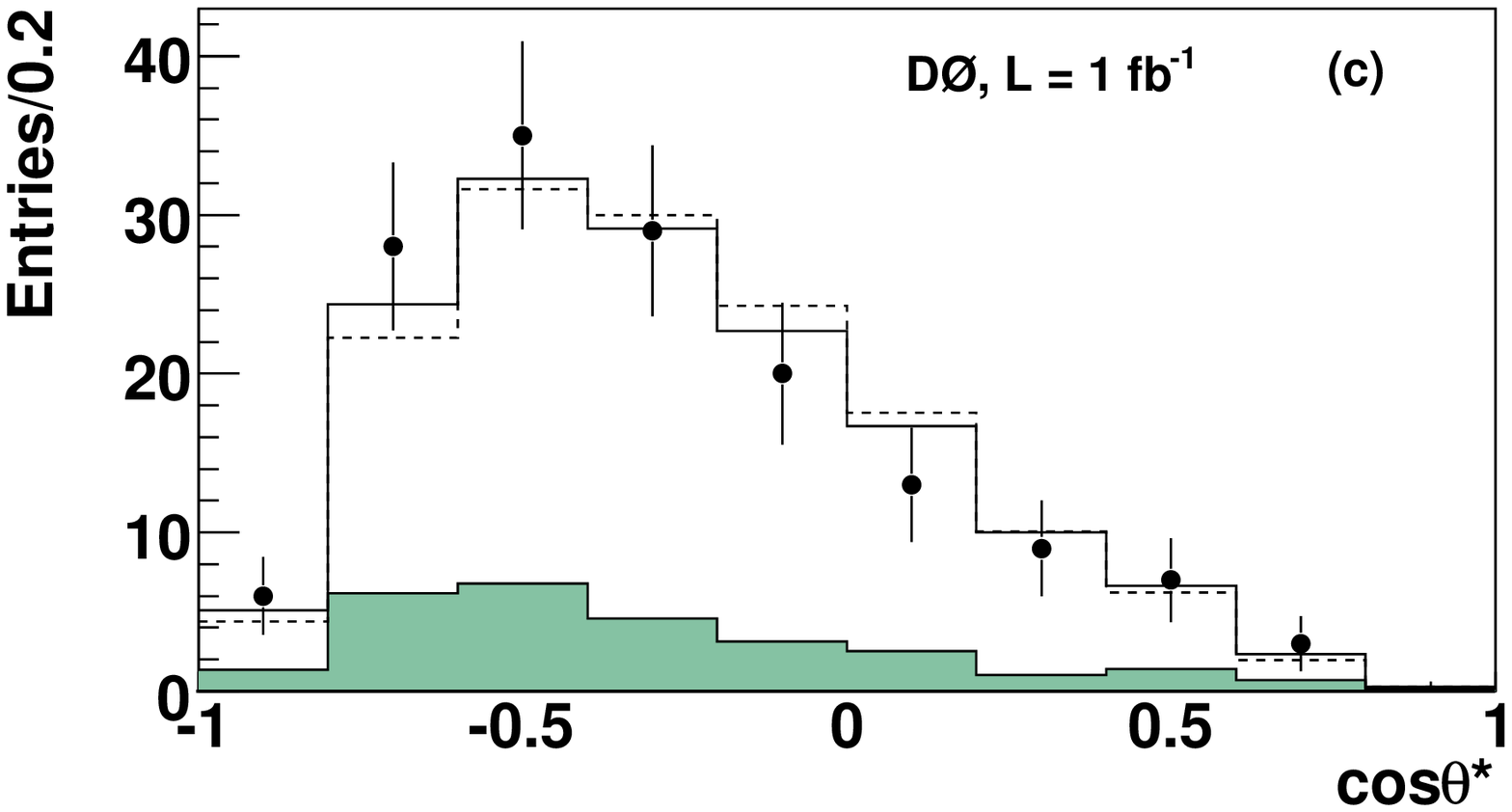}
    \caption{Helicity angle distributions in lepton + jets (a,b)
      and dilepton events (c). Points with error bars represent 1~fb$^{-1}$ of D0 data.
      The solid open histograms show the result of the
      model independent fit described in the text, while the dashed open histograms show the
      standard model expectation. The filled histograms represent
      the background contribution~\cite{Abazov:2007ve}.}
    \label{fig:D0-Whel-1fb-costs-dists}
  \end{center}
\end{figure}

The four-momenta of the top quarks and $W$ bosons in the lepton +
jets sample are reconstructed based on the best kinematic fit to a top
quark event hypothesis for $m_t = 172.5$ GeV/c$^2$, using the leading
four jets to obtain $\cos\theta^{\ast}$ and
$|\cos\theta^{\ast}|$. For the kinematically less constrained dilepton
events, a top quark mass of $m_t = 172.5$ GeV/c$^2$ is assumed and the
kinematics solved up to a four-fold ambiguity in addition to the
two-fold ambiguity from the lepton-jet pairing (only the leading two
jets are used). Jet and lepton energies in each event are smeared
within their resolutions to explore the phase space
consistent with the observed values. The average of the obtained
$\cos\theta^{\ast}$ values is then used for each charged lepton,
providing two measurements per event. The resulting distributions are
shown in Fig.\ \ref{fig:D0-Whel-1fb-costs-dists}. 
Note that due to reconstruction effects the shape of the standard
model expectation differs from the theoretical prediction in
Fig.\ \ref{fig:Whelcosdist}.

A template fit of these distributions yields
$f_0=0.425\pm0.166\rm{(stat.)}\pm0.102\rm{(syst.)}$ and
$f_+=0.119\pm0.090\rm{(stat.)}\pm0.053\rm{ (syst.)}$. The result is
compatible with the standard model expectation at 30\% C.L. 
It should be noted that the
individual measurements in the lepton + jets and dilepton channels
differ by about 2.1~sd~\cite{priv:2008a}.
The major systematic uncertainties on the measurement are summarized
in Table~\ref{tab:D0Whelsyst}, with the largest uncertainty arising from
\ttbar signal modeling, evaluated through varying the Monte Carlo
generators used (\pythia\ versus \alpgen), from changing underlying event
models to estimate the effects of gluon radiation, and restricting the
samples to contain only one primary vertex to study the sensitivity of
the measurement to variations in instantaneous luminosity.
\begin{table}[!t]
  \caption{Major systematic uncertainties on the simultaneous 
    measurement of $f_0$ and $f_{+}$ by D0 in 1~fb$^{-1}$ of data~\cite{Abazov:2007ve}.}
  \begin{center}
  \begin{tabular}{|l|c|c|}
    \hline
    Source & Uncertainty ($f_0$) & Uncertainty ($f_+$)  \\ \hline\hline
    Top mass               & 0.009 & 0.018 \\
    Jet reconstruction eff.                & 0.021 & 0.010 \\
    Jet energy calibration       & 0.012  &  0.019\\
    $b$ fragmentation      & 0.016 &  0.010 \\ 
    \ttbar ~model          & 0.068 & 0.032 \\
    Background model       & 0.049   & 0.016  \\
    Template statistics   & 0.049 & 0.025 \\ \hline
    Total                  & 0.102  &   0.053 \\ \hline
  \end{tabular}
  \label{tab:D0Whelsyst}
 \end{center}
\end{table}
\begin{figure}[h!]
  \begin{center} 
    \includegraphics[ height=56mm]{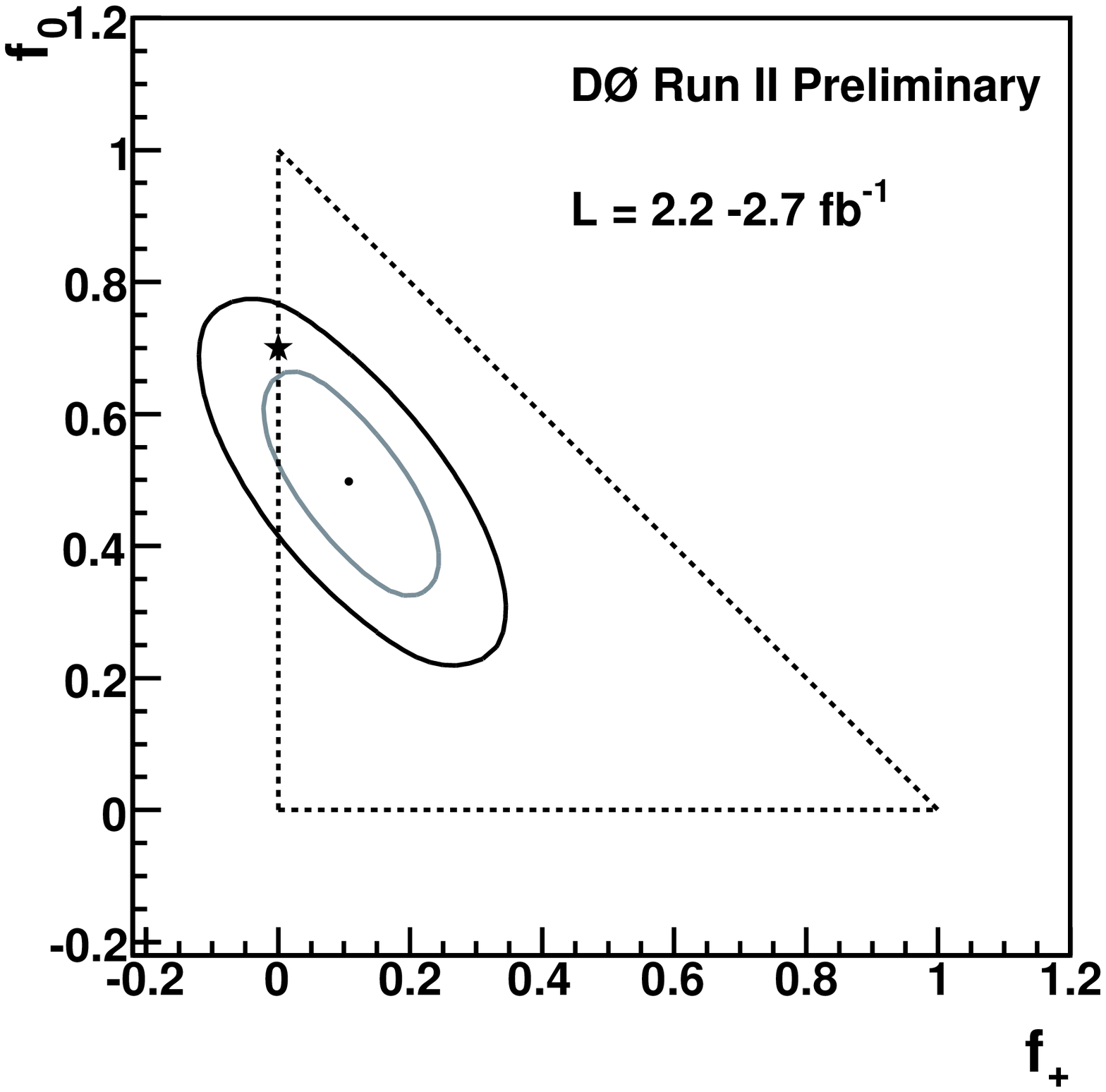}
    \includegraphics[ height=51mm]{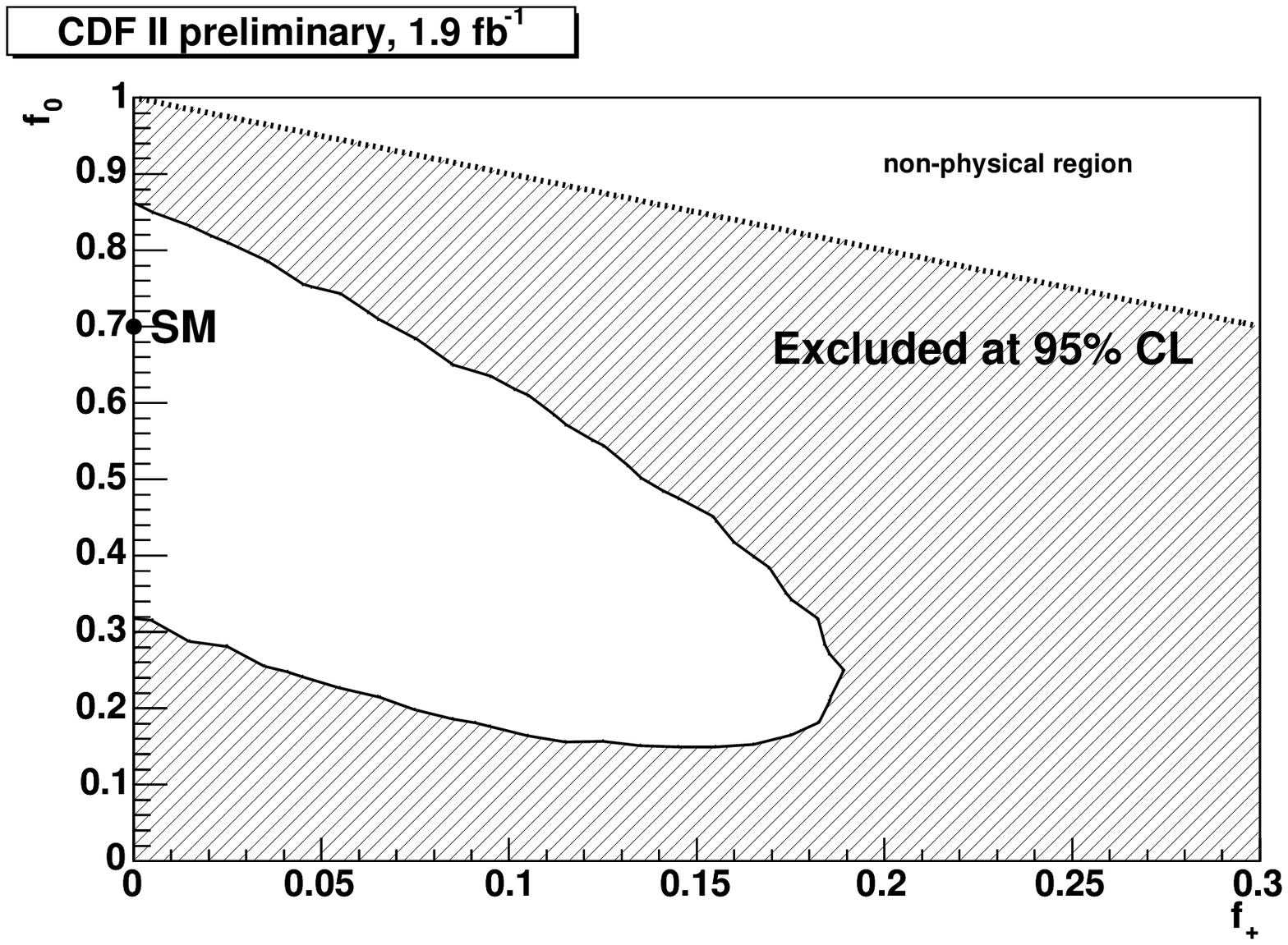}
    \caption{Left: Result of the model-independent $W$ boson helicity
      fit by D0 \cite{D05722}. The ellipses show the 68\% and
      95\% C.L.\ contours around the measured data point. The star
      shows the SM expectation, and the triangle denotes the physically
      allowed region where $f_0$ and $f_+$ sum to $\leq 1$. Right:
      Two-dimensional 95\% C.L.\ exclusion area in the $(f_+, f_0)$
      plane measured by CDF~\cite{CDF9215}.}
    \label{fig:Whel-costs}
  \end{center}
\end{figure}
Constraining $f_{0}$ or
$f_{+}$ to their SM values, when fitting for $f_{+}$ or for $f_{0}$, respectively,
yields $f_{0} = 0.619 \pm 0.090\:{\rm(stat.)} \pm
0.052\:{\rm(syst.)}$ and $f_{+} = -0.002 \pm 0.047\:{\rm(stat.)} \pm 0.047
\:{\rm(syst.)}$, in agreement with expectations from the SM.

In a recent preliminary update, D0 has added 1.2~fb$^{-1}$ lepton +
jets and 1.7~fb$^{-1}$ dilepton ($e\mu$ only) data~\cite{D05722}
to the above analysis. The model-independent fit for the combined data
yields $f_0=0.490\pm0.106\rm{(stat.)}\pm0.085\rm{(syst.)}$ and
$f_+=0.110\pm0.059\rm{(stat.)}\pm0.052\rm{ (syst.)}$, consistent at
23\% C.L.\ with the standard model (see
Fig.\ \ref{fig:Whel-costs}). The results from the lepton + jets
and dilepton channels remain marginally consistent with a $p$-value of
1.6\%.

CDF has obtained two preliminary results for a model-independent extraction of $W$
boson helicity in 1.9 fb$^{-1}$ of data. These are based on using only the charged
lepton in $b$ tagged lepton $+\geq 4$ jets events to obtain
$\cos\theta^{\ast}$~\cite{CDF9114, CDF9215}. A combination of both
results also has become available~\cite{CDF9431}\footnote{This result has been published after completion of this
review, see Ref.~\cite{Aaltonen:2008ei}.}. The measurements are
compatible with the standard model expectation, with each other, and with the D0
measurements presented above and are summarized together with other
results in Table~\ref{tab:Whelsummary}. The two-dimensional 95\%
C.L.\ exclusion area in the $(f_+, f_0)$ plane measured by
CDF~\cite{CDF9215} is shown in Fig.\ \ref{fig:Whel-costs}.

\begin{table}[t!]
    \caption{$W$ boson helicity measurements in \ttbar events performed thus far at the
Tevatron, with their integrated luminosities, data selections
($\ell j$ = lepton + jets, $\ell\ell$ = dilepton) and
analysis methods used. Model independent results are indicated by a yes in
the ``Ind. fit'' column. The three analyses using 0.1 fb$^{-1}$ are from
Run~I; the analyses using more than 1~fb$^{-1}$ are preliminary.
Ref. \cite{CDF9431} provides a combination of the results from Refs.~\cite{CDF9114,CDF9215}.}
\renewcommand{\arraystretch}{1.2}
\addtolength{\tabcolsep}{-4pt}
  \begin{center}
    \begin{tabular}{|c|c|c|c|c|c|c|c|}
      \hline 
       $\int{\cal L}dt$  &\multirow{2}{*}{Sel.}  & \multirow{2}{*}{$f_{0}$} & \multirow{2}{*}{$f_{+}$} & Ind. & $f_{+}<$ & \multirow{2}{*}{Method} & \multirow{2}{*}{Ref.}\\ 
      ~[fb$^{-1}$] &  &  &  & fit & {\tiny (95\% C.L.)} &  & \tabularnewline
      \hline
      \hline 
      0.1 & $\ell j,\ell\ell$ & $0.91\pm0.37\pm0.13$ & $0.11\pm0.15$ & no & 0.28 & $p_{T}^{\ell}$  & \cite{Affolder:1999mp}\tabularnewline
      \hline 
      0.1 & $\ell j,\ell\ell$ & --- & $-0.02\pm0.11$ & no & 0.18 & $M_{\ell b}^{2}$, $p_{T}^{\ell}$ & \cite{Acosta:2004mb}\tabularnewline
      \hline 
      0.1 & $\ell j$ & $0.56\pm0.31$ & --- & no & --- & ME & \cite{Abazov:2004ym}\tabularnewline
      \hline 
      0.2 & $\ell j,\ell\ell$ & $0.74_{-0.34}^{+0.22}$ & $0.00_{-0.19}^{+0.20}$ & no & 0.27 & $M_{\ell b}^{2}$, $p_{T}^{\ell}$ & \cite{Abulencia:2005xf}\tabularnewline
      \hline 
      0.2 & $\ell j$ & --- & $0.00\pm0.13\pm0.07$ & no & 0.25 & $\cos\theta^{\ast}$ & \cite{Abazov:2005fk}\tabularnewline
      \hline 
      0.3 & $\ell j$ & $0.85_{-0.22}^{+0.15}\pm0.06$ & $0.05_{-0.05}^{+0.11}\pm0.03$ & no & 0.26 & $\cos\theta^{\ast}$ & \cite{Abulencia:2006ei}\tabularnewline
      \hline 
      0.4 & $\ell j,\ell\ell$ & --- & $0.06\pm0.08\pm0.06$ & no & 0.23 & $\cos\theta^{\ast}$ & \cite{Abazov:2006hb}\tabularnewline %
      \hline 
      0.7 & $\ell j,\ell\ell$ & --- & $-0.02\pm0.07$ & no & 0.09 & $M_{\ell b}^{2}$ & \cite{Abulencia:2006iy}\tabularnewline
      \hline 
      \multirow{2}{*}{1.0}  &\multirow{2}{*}{$\ell j,\ell\ell$}  & $0.62\pm0.09\pm0.05$ &  $0.00\pm0.05\pm0.05$ & no &\multirow{2}{*}{---} & \multirow{2}{*}{$\cos \theta^\ast$} & \multirow{2}{*}{\cite{Abazov:2007ve}}\\%
      &&$0.43\pm0.17\pm0.10$&$0.12\pm0.09\pm0.05$&yes&&&\tabularnewline
      \hline 
        1.9  & $\ell j$ & $0.64\pm0.08\pm0.07$  & ---  & no & --- & ME& \cite{CDF9144}\tabularnewline%
      \hline
      \multirow{2}{*}{1.9}  &\multirow{2}{*}{$\ell j$}  &$0.59\pm0.11\pm0.04$  &  $-0.04\pm0.04\pm0.03$ & no & 0.07 & \multirow{2}{*}{$\cos \theta^\ast$} & \multirow{2}{*}{\cite{CDF9215}}\\
      &&$0.65\pm0.19\pm0.03$&$-0.03\pm0.07\pm0.03$&yes&---&&\tabularnewline
      \hline 
      \multirow{2}{*}{1.9}  &\multirow{2}{*}{$\ell j$}  &$0.66\pm0.10\pm0.06$  &  $0.01\pm0.05\pm0.03$ & no & 0.12 & \multirow{2}{*}{$\cos \theta^\ast$} & \multirow{2}{*}{\cite{CDF9114}}\\
      &&$0.38\pm0.21\pm0.07$&$0.15\pm0.10\pm0.05$&yes&---&&\tabularnewline
      \hline 
      \multirow{2}{*}{1.9}  &\multirow{2}{*}{$\ell j$}  &$0.62\pm0.11$  &  $-0.04\pm0.05$ & no &\multirow{2}{*}{---}  & \multirow{2}{*}{$\cos \theta^\ast$} & \multirow{2}{*}{\cite{CDF9431}}\\
      &&$0.66\pm0.16$&$-0.03\pm0.07$&yes&&&\tabularnewline
      \hline 
      2.7 & $\ell j,\ell\ell$ & $0.49\pm0.11\pm0.09$ & $0.11\pm0.06\pm0.05$ & yes & --- & $\cos\theta^{\ast}$ & \cite{D05722}\tabularnewline
      \hline 
    \end{tabular}
    \label{tab:Whelsummary}
  \end{center}
\end{table}

\subsubsection{Summary}
All available measurements of the $W$ boson helicity performed thus far in \ttbar events at the
Tevatron are compatible with the standard model expectation, and are summarized
in Table~\ref{tab:Whelsummary}. 

The sensitivity of the measurements in Run~I and initial studies in Run~II only
allowed model-dependent measurements of any single
helicity fraction to be performed at a time, while fixing the other fraction to its
standard model value. However, with the large amount of data available by now,
a simultaneous extraction of  $f_{+}$ and $f_{0}$ 
for $W$ bosons is possible without constraint, except for unitarity ($f_{+}+f_{0}+f_{-}=1$).
Such measurements will also clearly
benefit from increased luminosity. A combination with
the measurement of the single-top production cross section in the $s$-
and $t$-channel will help to fully specify the $tWb$
coupling~\cite{Chen:2005vr} (see Section~\ref{sec:Wheltheory})\footnote{A first such measurement was published by
  D0~\cite{Abazov:2009ky} after completion of this review.}.

The model-dependent measurements, where one of the helicity fractions
is fixed to its standard model value, have reached a considerable
precision with statistical approaching systematic
uncertainties.

It is interesting to note that discrepancies of $>2$~sd
between results from the dilepton and lepton + jets samples have
been observed both at CDF and D0 using different analysis techniques.
This deserves further scrutiny in future analyses.

\subsection{Measurement of \boldmath{${\cal B}(t \rightarrow Wb) / {\cal B}(t \rightarrow Wq)$}}
\label{sec:Rmeasurement}
As described in Section~\ref{sec:topCKM}, in the standard model
framework the top quark decays almost exclusively into a $W$ boson
and a $b$ quark due to the dominant corresponding CKM matrix element
$V_{tb}$. The ratio $R$ of the top quark branching fractions can be
expressed via the CKM matrix elements as:
\begin{eqnarray}
R = \frac{{ \cal B}(t \rightarrow Wb)}{\displaystyle\sum_{q=d,s,b} {\cal{B}}(t
\rightarrow Wq)} & = &
\frac{\mid V_{tb}\mid^2}{\mid V_{tb}\mid^2 + \mid V_{ts}\mid^2 + \mid V_{td}\mid^2}  \;.
\label{eq:Rdefinition}
\end{eqnarray}
Measuring $R$ provides therefore the relative size of $|V_{tb}|$
compared to $|V_{td}|$ and $|V_{ts}|$, with the current measurements
indicating $|V_{tb}| \gg |V_{td}|, |V_{ts}|$. While a direct
measurement of the $V_{tb}$ matrix element is only possible through single
top quark production, as described in Section~\ref{sec:singletopprod},
model-dependent constraints on $V_{tb}$ can also be inferred from a
measurement of $R$: Assuming the validity of the standard model,
specifically the existence of three fermion generations, unitarity of
the CKM matrix and insignificance of non-$W$ boson decays of the top
quark (see Sections~\ref{sec:FCNCdecay}-\ref{sec:H+topdecay}), the denominator in 
Eq.~\ref{eq:Rdefinition} equals one. $R$ then simplifies to $|V_{tb}|^{2}$, and is therefore
strongly constrained to $0.9980 < R < 0.9984$ at 90\%
C.L.\ by global CKM fits~\cite{Eidelman:2004wy}.

Deviations of $R$ from unity could, for example, be caused by the
existence of a fourth heavy quark generation, non standard model top
quark decays, or non standard model background processes.
Consequently, precise measurements of R probe for physics
beyond the standard model, and provide a required ingredient for the
model-independent direct determination of the $|V_{tq}|$ CKM matrix
elements from electroweak single top production~\cite{Alwall:2006bx}.

\begin{figure}[!t]
  \begin{center} 
    \includegraphics[width=.48\textwidth]{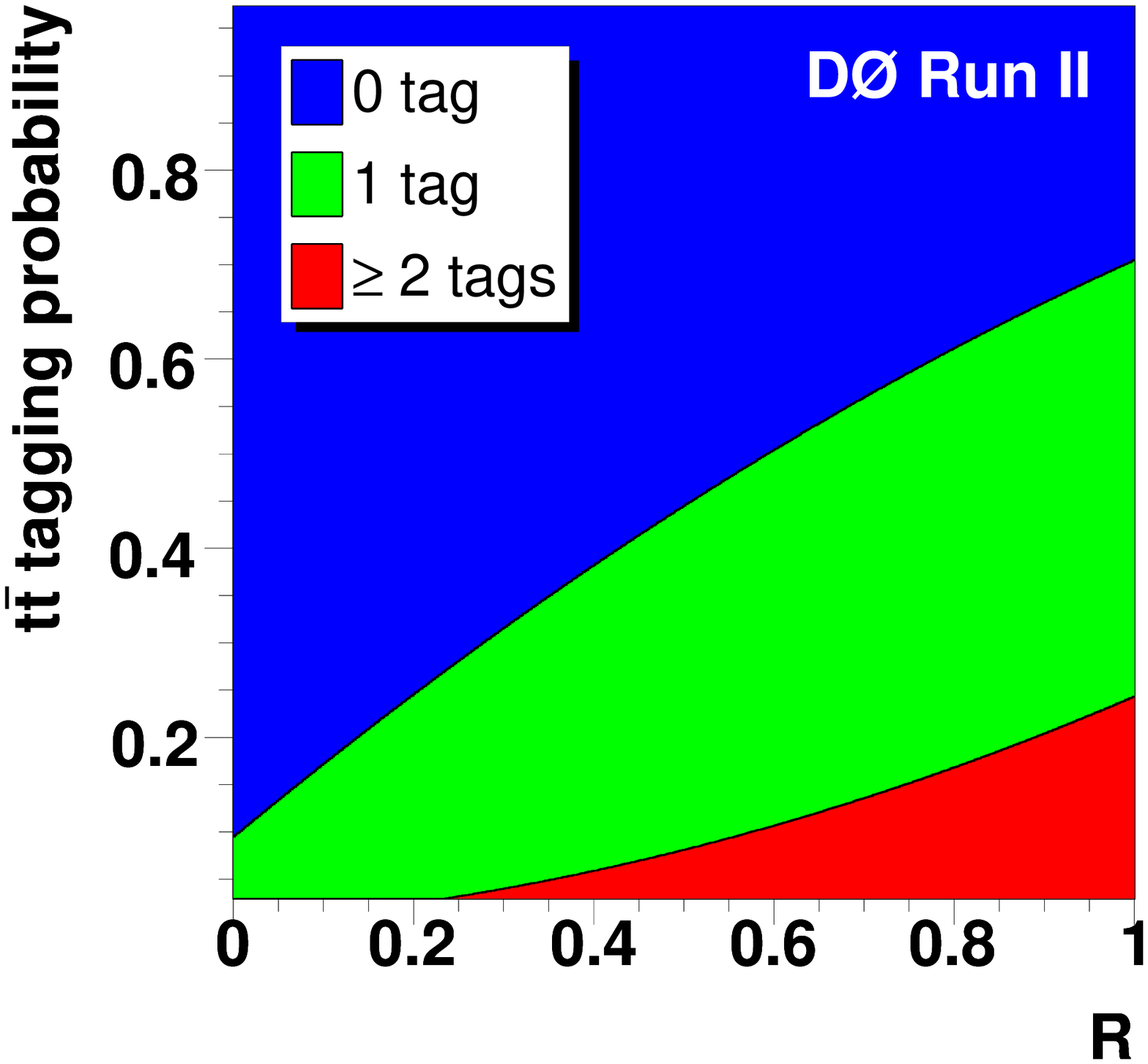}\hspace*{1mm}
    \includegraphics[width=.48\textwidth]{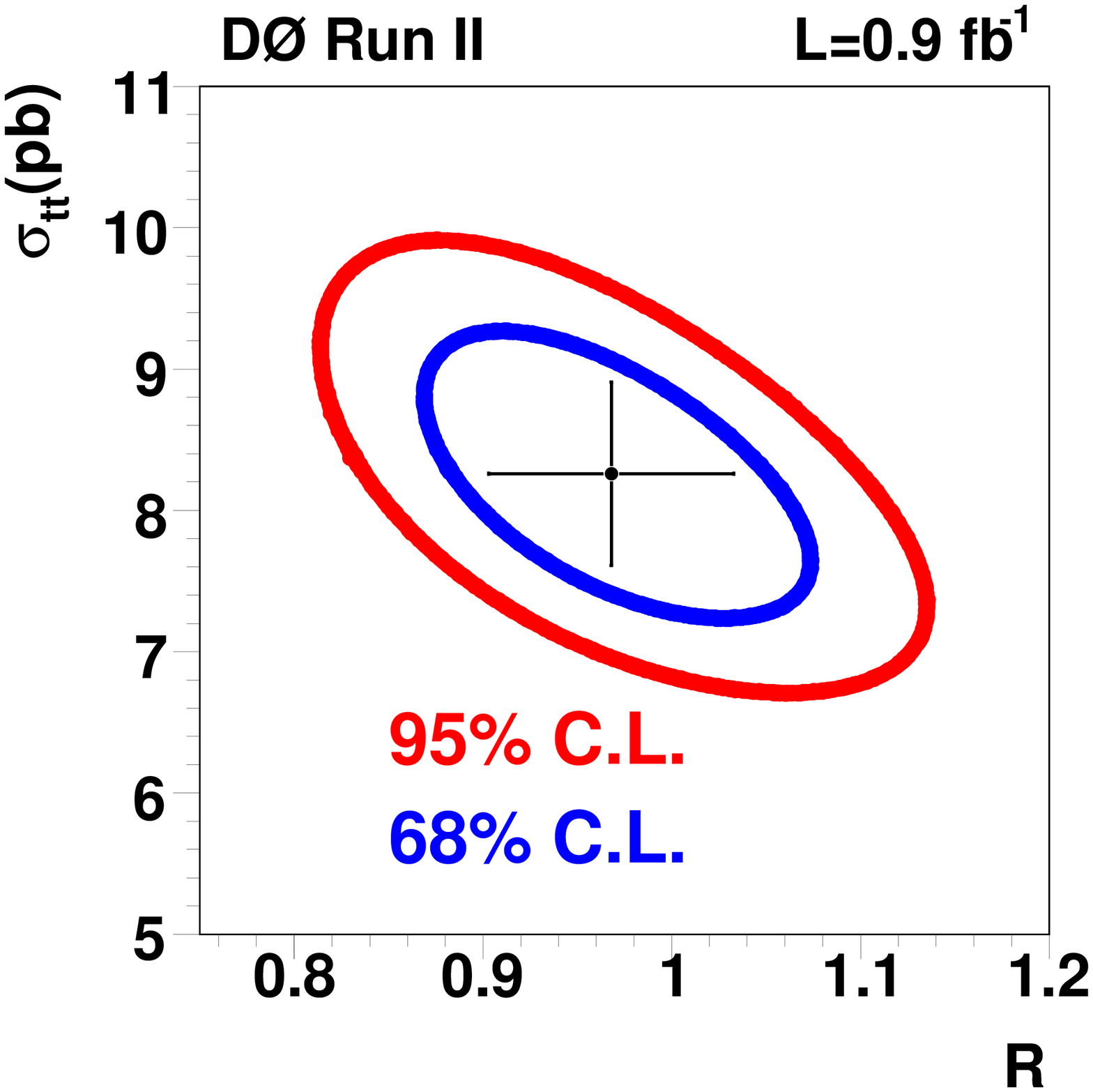} 
    \caption{Left: Fractions of events with 0, 1 and $\geq 2$ $b$ tags
      for \ttbar events with $\geq$4 jets as a function of $R$.
      Right: 68\% and 95\% C.L.\ statistical uncertainty contours in
      the R vs. $\sigma_{t\bar{t}}$ plane around the measured point 
      in 0.9~fb$^{-1}$ of D0 data~\cite{Abazov:2008yn}.}
      \label{fig:D0Rmeas}
  \end{center}
\end{figure}
The most precise measurement of $R$ thus far has been performed by D0
in the lepton + jets channel using data corresponding to an integrated
luminosity of 0.9~fb$^{-1}$~\cite{Abazov:2008yn}, superseding the
previously published measurement based on
0.2~fb$^{-1}$~\cite{Abazov:2006bh}. The \ttbar signal sample
composition depends on $R$ in terms of the number of $b$ jets present
in the sample, as illustrated in Fig.\ \ref{fig:D0Rmeas}. By comparing
the event yields with zero, one and two or more $b$ tagged jets, and
using a topological discriminant to separate \ttbar signal from
background in events without $b$ tags, $R$ can be extracted simultaneously
with the \ttbar production cross section $\sigma_{t\bar{t}}$.
This approach yields a measurement of
$\sigma_{t\bar{t}}$ without assuming ${\cal B}(t\to Wb)= 100\%$, and
exploits the different sensitivity of the two measured quantities to systematic
uncertainties, thereby improving overall precision.

A maximum likelihood fit to the sample composition observed in data
gives
\begin{eqnarray}
R &=& 0.97^{+0.09}_{-0.08}~\rm{(stat.+syst.)~and}\\ \sigma_{t\bar{t}}
&=& 8.18^{+0.90}_{-0.84}~\rm{(stat.+syst.)}~\pm 0.50~\rm{(lumi)~pb}
\end{eqnarray}
(see Fig.\ \ref{fig:D0Rmeas}) with a correlation of $-58\%$ for a top
quark mass of $175$~GeV/c$^{2}$, in agreement with the standard model
prediction. From the measurement 95\% C.L.\ limits are extracted, yielding
$R > 0.79$ and $|V_{tb}| > 0.89$, the latter being model-dependent as
mentioned above. $R$ exhibits no significant dependence
on the top quark mass within $\pm10$~GeV/c$^{2}$, while
$\sigma_{t\bar{t}}$ varies by $\mp 0.09$~pb per $\pm1$~GeV/c$^{2}$ in
the same mass range. The total uncertainty on $R$ in this measurement
is 9\%, dominated by the statistical uncertainty of
$^{+0.067}_{-0.065}$ and the largest systematic uncertainty from the
$b$ tagging efficiency estimation of $^{+0.059}_{-0.047}$. The cross
section measurement yields a result similar but not identical to the
measurement on the same dataset~\cite{Abazov:2008gc}, presented in
Section~\ref{sec:ttbarxsecmeaslj}. This is due to the assumption of $R=1$, and
slightly different event selection in the latter analysis.
\begin{figure}[!t]
  \begin{center} 
    \includegraphics[ height=65mm]{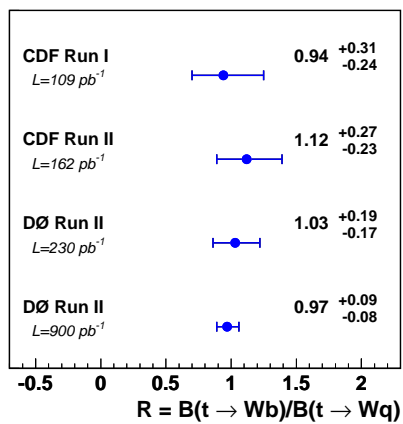}
    \caption{Summary of the branching ratio $R$ measurements and their
      total uncertainties obtained at the Tevatron by
      CDF~\cite{Affolder:2000xb,Acosta:2005hr} and
      D0~\cite{Abazov:2008yn,Abazov:2006bh}.}
    \label{fig:Rsummary}
  \end{center}
\end{figure}

CDF performed the first measurement of $R$ in Run~I using both
dilepton and lepton + jets events on 0.1~fb$^{-1}$ of
data~\cite{Affolder:2000xb}, and repeated the analysis in Run~II on
0.2~fb$^{-1}$ of data~\cite{Acosta:2005hr}, also finding good agreement
with the standard model expectation.

All measurements of $R$ performed thus far at the Tevatron are
summarized in Fig.\ \ref{fig:Rsummary}.

\subsection{Search for neutral-current top decays}
\label{sec:FCNCdecay}
Flavor changing neutral interactions of the top quark with a light
quark $q = u, c$ through gauge ($Z,\gamma, g$) or Higgs ($H^0$) bosons are
forbidden at lowest order and are suppressed by the GIM mechanism
\cite{Glashow:1970gm} at higher orders in the standard model framework.
Consequently, the corresponding FCNC top quark decays are expected to
occur, at most, with branching ratios at ${\cal O}(10^{-12})$
\cite{AguilarSaavedra:2004wm}, well out of reach of sensitivity of the
Tevatron or the LHC. Any observation of such FCNC decays would therefore signal
physics beyond the standard model.

Many extensions of the standard model predict the occurrence of FCNC
interactions, affecting both electroweak single top production
(see Section~\ref{sec:STFCNC}) and top quark decay. The branching
fractions of FCNC top decays can increase by many orders of magnitude
in such models, as for example in Supersymmetry
\cite{deDivitiis:1997sh, Lopez:1997xv,Eilam:2001dh},
additional broken symmetries~\cite{Fritzsch:1999rd},
dynamical EWSB~\cite{Peccei:1989kr,Arbuzov:1998th} including
topcolor-assisted technicolor~\cite{Yue:2001cy} or extended Higgs
models such as Two Higgs Doublet Models~\cite{Eilam:1990zc,Eilam:1999zc,
Hou:1991un,Atwood:1995ud}. 
Overviews of such models and their impact on top couplings
are given in Refs.~\cite{Han:1996ce,Han:1998yr,delAguila:1998tp,AguilarSaavedra:2004wm}.

\begin{figure}[!t]
  \begin{center} 
\subfigure[]{\epsfig{figure = 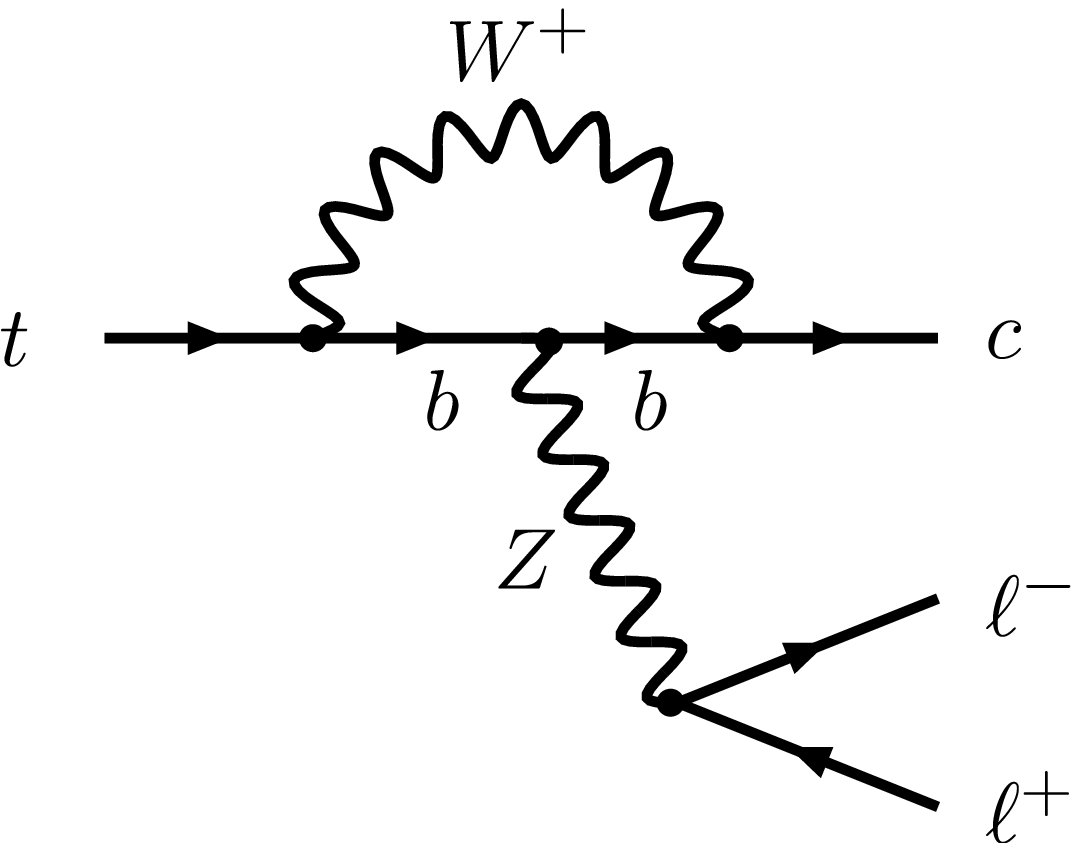, height=40mm}}\hspace{1cm}
\subfigure[]{\epsfig{figure = 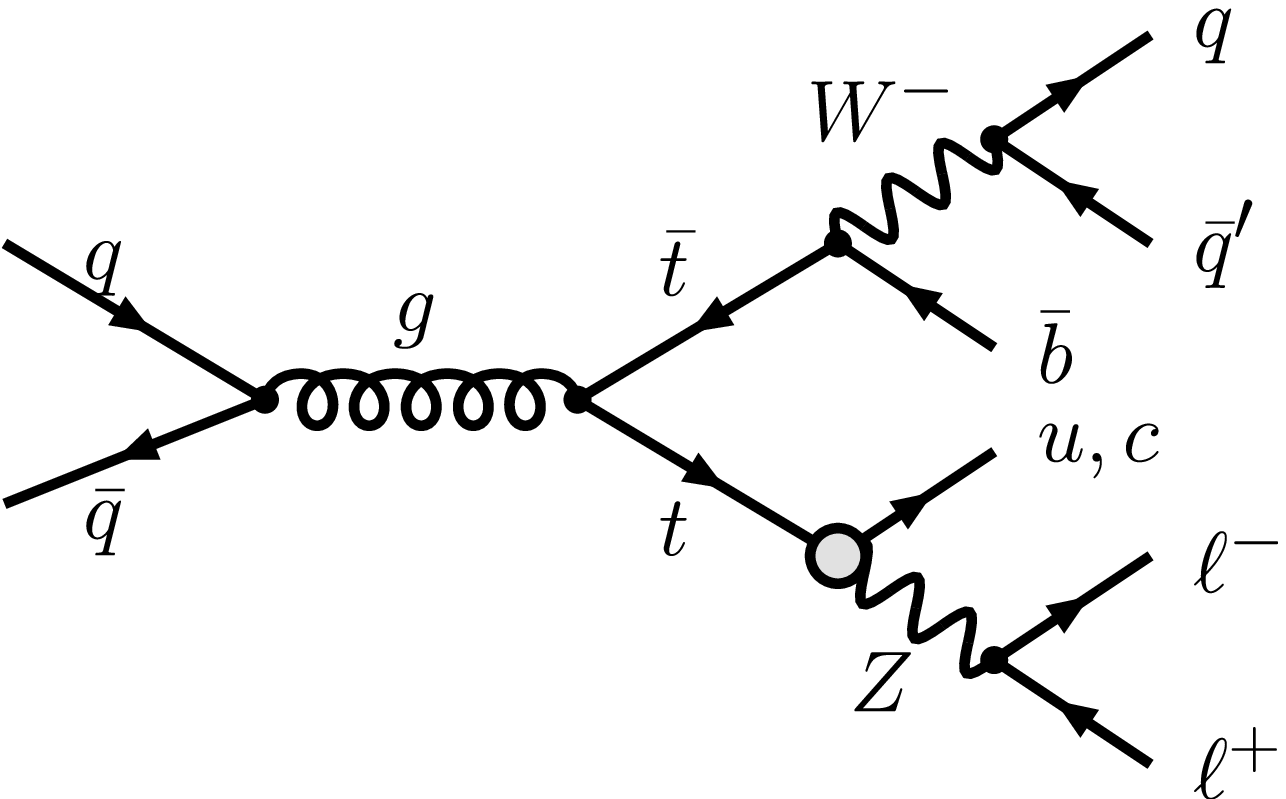, width = 0.45 \textwidth}}
    \caption{(a) Feynman ``penguin'' diagram for the FCNC decay $t\to
      Zc$ with $Z\to \ell^+\ell^-$. Including the corresponding diagrams
      with a $d$ and $s$ quark in the loop, the process is nearly
      cancelled in the standard model. (b) Event signature for top
      quark pairs containing one FCNC $t\to Zq$ decay and one
      $W\to q\bar{q}'$ decay from $t\to Wb$, resulting in a final
      state that contains a $Z$ boson and four jets.}
      \label{fig:ttbarFCNCdecay}
  \end{center}
\end{figure}
A search for the top quark FCNC decay $t\to Zq$ at the Tevatron is
considered especially interesting due the large top quark mass and
very distinct experimental signature (see
Fig.\ \ref{fig:ttbarFCNCdecay}). It was already suggested in
1989~\cite{Fritzsch:1989qd}, well before the discovery of the top
quark. The expected sensitivity for such a branching ratio measurement
is ${\cal O}(10^{-2})$ at the Run~II Tevatron and ${\cal O}(10^{-4})$
at the LHC \cite{Han:1995pk}, while the largest expected branching
fractions from SM extensions reach up to ${\cal
O}(10^{-2})$~\cite{Arbuzov:1998th} and ${\cal
O}(10^{-4})$~\cite{AguilarSaavedra:2004wm}, respectively. The best published
limit before Run~II on ${\cal B}(t\to Zq)$ was obtained at LEP by the L3
Collaboration via a search for single top quark production, where no
significant deviation from the SM background expectation
was observed, yielding ${\cal B}(t\to Zq) < 13.7 \%$ at 95\%
C.L.~\cite{Achard:2002vv}.

In Run~I, the CDF Collaboration performed a search for the FCNC decays
$t\to Zq$ and $t\to \gamma q$ on a dataset with an integrated
luminosity of 0.1 fb$^{-1}$~\cite{Abe:1997fz}. For the $t\to \gamma q$
search, a photon is reconstructed as an energy cluster in the
electromagnetic calorimeter, either without an associated track or with a single soft
track (presumably a random overlap) carrying less than 10\% of the
photon energy pointing to the cluster. Two event signatures are
considered, where the $W$ boson from the standard-model-like second top
decay branches either leptonically into $e\nu_e\ \rm{or}\ \mu\nu_\mu$, or
hadronically into $\bar{q}q'$ quarks. Consequently, these samples are
selected by requiring either a charged lepton ($e\ \rm{or}\ \mu$),
\METns, at least two jets and a photon, or by requiring at least four
jets and a photon. In both samples, a photon-jet combination must
yield a mass between 140 and 210 GeV/c$^2$ and the SM-like
top decay must contain one $b$ tag. 40\% of the $t\to \gamma q$
acceptance comes from the photon + multijet selection, while the
lepton + photon mode contributes 60\%. After all selections,
one event remains in the leptonic channel and none in the photon +
multijet channel, with an expected background of about 0.5 events
mainly from $W\gamma$ production with additional jets in each channel.
This translates into a 95\% C.L.\ upper limit on the branching fraction
of ${\cal B}(t\to c \gamma) + {\cal B}(t\to u \gamma) < 3.2\%$.

In the $t\to Zq$ search, $W\to q\bar{q}'$ decays from the
SM-like second top decay are considered together with a
leptonically decaying $Z$ boson into $e^+e^-$ or $\mu^+\mu^-$. Using
the leptonic $W$ boson decay as well does not substantially increase
the acceptance and consequently does not improve the limit. The
resulting event signature therefore contains four jets and two leptons with an
invariant mass consistent with that of a $Z$ boson, as illustrated in Fig.\
\ref{fig:ttbarFCNCdecay}. Since the branching ratio of $Z \to
\ell^+\ell^-$ is small, this search is less sensitive than the
$t\to \gamma q$ one. One $Z \to \mu^+\mu^-$ event passes the
selection, with an expected background of $\approx$0.6 events from $Z + $
multijet and \ttbar production. This corresponds to a 95\% C.L.\ upper
limit on the branching fraction of ${\cal B}(t\to c Z) + {\cal B}(t\to
u Z) < 33\%$. These measurements can be translated into limits on the
FCNC couplings $\kappa_\gamma$ and $\kappa_Z$ at 95\%
C.L.~\cite{Obraztsov:1997if}, which are $\kappa_\gamma^2 < 0.176\ \rm{and}\
\kappa_Z^2 < 0.533$ (see Section~\ref{sec:STFCNC} and Fig.\
\ref{fig:STFCNCgamZ}).
\begin{figure}[!t]
  \begin{center} 
    \includegraphics[height = 50mm]{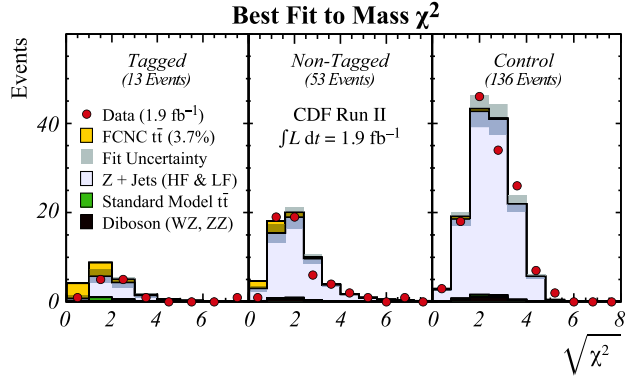}
    \caption{Expected and observed mass $\chi^{2}$ distributions in
      $Z+\geq4$ jets events in signal samples with and without $b$
      tags and in a background-enriched sample used to ascertain uncertainties
      in the background shape and normalization in 1.9~fb$^{-1}$ of CDF
      data~\cite{Aaltonen:2008aaa}. The expected FCNC $t\to Zq$
      signal with the observed 95\% C.L.\ upper limit on the branching
      fraction is shown as well.}
      \label{fig:CDFFCNC}
  \end{center}
\end{figure}
In Run~II, the CDF Collaboration has performed a search for the FCNC
decay $t\to Zq$ on a dataset with an integrated luminosity of 1.9
fb$^{-1}$~\cite{Aaltonen:2008aaa}. Events consistent with a
leptonically decaying $Z$ boson to $e^+e^-$ or $\mu^+\mu^-$ are
selected together with at least four jets, one of which can be $b$
tagged (see Fig.\ \ref{fig:ttbarFCNCdecay}). The event selection was
optimized in a preceding version of this analysis with 1.1 fb$^{-1}$ of
data~\cite{CDF8888}, which was a ``blind'' cut-based counting experiment. By
requiring only one well-identified lepton for $Z$ reconstruction,
while the second lepton can be formed from an isolated track, the
acceptance is doubled compared to using only fully identified leptons.
The sensitivity of the search is further increased by dividing the
data into two subsamples, one $b$ tagged and one not tagged. The
best discriminant found to separate signal from background
is a mass $\chi^{2}$ variable that combines the kinematic constraints
present in FCNC decays: Two jets in the event have to form a $W$ boson,
and together with a third jet a top quark, while the $Z$ boson has to
form a top quark with the fourth jet. Because the event signature does
not contain neutrinos, the events can be fully reconstructed. The
signal fraction in the selected dataset is determined via a template
fit in signal samples with or without $b$ tags, and a
background-enriched control sample is used to constrain uncertainties on the
background shape and normalization (see Fig.\ \ref{fig:CDFFCNC}).

Since the observed distributions are consistent with the standard
model background processes, a 95\% C.L.\ upper limit is extracted on the branching
fraction ${\cal B}(t\to c Z) + {\cal B}(t\to u Z)$ of $<
3.7\%$. The expected limit in absence of signal is 5\%. This is the
best limit on ${\cal B}(t\to Zq)$ to date, starting to constrain
predictions from a dynamic EWSB model~\cite{Arbuzov:1998th}.

\subsection{Search for invisible top decays}
\label{sec:invisibletopdecay}
Apart from the direct search for $t\to (Z/\gamma)~q$ decays, as described
in the previous section, one can also perform an indirect search for
``invisible'' top quark decays by comparing the predicted \ttbar
production cross section with the observed yield in data. In order to
be sensitive to novel top decay modes with this method, these decays
must exhibit a significantly different acceptance from the standard
model top quark decay.

Based on a 1.9~fb$^{-1}$ doubly $b$ tagged lepton + jets dataset,
CDF searches for deviations of the observed \ttbar production rate
from the theoretical prediction~\cite{Cacciari:2008zb} due to the
decays $t\to Zc$, $t\to g c$, $t\to \gamma c$ and $t\to$
``invisible'' states~\cite{CDF9496}. These decays exhibit a relative
acceptance ${\mathcal R}_\textrm{WX/WW}$, where one novel and one
SM top quark decay occur, normalized to the standard model
\ttbar decay acceptance, from 32\% down to no acceptance (for decays to invisible states).

With an observed \ttbar production cross section of 8.8~pb, and a
prediction of 6.7~pb for a top quark mass of 175 GeV/c$^2$, the
obtained limits on the novel top quark decay modes are all lower than
expected, but statistically consistent with expectation. The
results are summarized in Table~\ref{tab:CDFinvisibletop} for
top quark masses of 170, 172.5 and 175 GeV/c$^2$.

\begin{table}[!t]
  \caption{Relative signal acceptances and observed 95\% C.L.\ upper
    limits on the branching fractions for several non-SM
    top quark decay modes as a function of assumed top quark mass in a
    1.9~fb$^{-1}$ doubly $b$ tagged lepton + jets
    dataset~\cite{CDF9496}.}
  \begin{center}
    \addtolength{\tabcolsep}{-2pt}
    \begin{tabular*}{\textwidth}{@{\extracolsep{\fill}}lrrrr} 
      \hline
      \multirow{2}{*}{Decay} &\multirow{2}{*}{${\mathcal R}_\textrm{WX/WW}$~[\%]}&Limit~[\%] &Limit~[\%] &Limit~[\%] \\ 
      & & {\footnotesize (175 GeV/c$^2$)}&{\footnotesize (172.5 GeV/c$^2$)}&{\footnotesize (170 GeV/c$^2$)}\\
      \hline\hline
      ${\mathcal B}(t\to Zc)$                 &  32 & 13 & 15 & 18 \\
      ${\mathcal B}(t\to g c)$                &  27 & 12 & 14 & 17 \\
      ${\mathcal B}(t\to \gamma c)$          &  18 & 11 & 12 & 15 \\
      ${\mathcal B}(t\to \textrm{invisible})$      &   0 &  9 & 10 & 12 \\ 
      \hline
    \end{tabular*}
    \label{tab:CDFinvisibletop}
  \end{center}
\end{table}

\subsection{Search for top decays to charged Higgs bosons}
\label{sec:H+topdecay}
As indicated previously, the standard model incorporates one Higgs doublet of
complex scalar fields to break electroweak symmetry and to generate
masses of weak gauge bosons and fermions (see
Section~\ref{sec:SMoverview}). As a consequence, one obtains a single
neutral scalar CP-even particle that still remains to be discovered,
the Higgs boson $H$. An extension of the
standard model Higgs sector introduces a second Higgs
doublet, referred to in Section~\ref{sec:STchargedH} as Two Higgs Doublet Models
(THDM or 2HDM)~\cite{Gunion:1989we,Gunion:1992hs}. These models provide
five physical scalar Higgs bosons after electroweak symmetry breaking, namely
two neutral CP-even Higgs bosons $H^0$ and $h^0$, one neutral
pseudoscalar CP-odd Higgs particle $A^0$, and two charged Higgs bosons
$H^\pm$. The observation of charged Higgs bosons would therefore offer clear
evidence for physics beyond the standard model.

Three choices of Higgs-fermion couplings are
differentiated in THDM. In Type-I models only one of the two Higgs
doublets couples to fermions, while in Type-II models one doublet
couples to up-type fermions and the other doublet to down-type
fermions. Type-III models have general Higgs-fermion Yukawa
couplings of both Higgs doublets, leading to Higgs-mediated FCNC at
tree level, which requires tuning of the Higgs parameters to ensure
sufficient suppression of FCNC to be compatible with current experimental limits.
One example of a Type-II THDM is the Minimal Supersymmetric Standard
Model (MSSM)~\cite{Martin:1997ns} which is frequently used as reference in the
analyses described below. The relevant model parameters
in searches for charged Higgs bosons are the ratio of the 
vacuum expectation values of the two Higgs doublets ($\tan\beta$)
and the mass of the charged Higgs boson $m_{H^\pm}$.

The inclusive single charged Higgs boson production rate
$\sigma(p\bar{p}\to tH^-X)$ reaches a maximum of
$\mathcal{O}(\rm{1~pb})$ at the Tevatron where the charged Higgs boson
can be produced via the decay of a top quark. The corresponding
inclusive cross section for pair production of charged
Higgs bosons $\sigma(p\bar{p}\to H^+H^-X)$ is below
$\mathcal{O}(\rm{0.1~pb})$~\cite{Borzumati:1999th,Carena:2002es}.

\begin{figure}[!t]
  \begin{center} 
    \includegraphics[width=0.48\textwidth,height = 50mm]{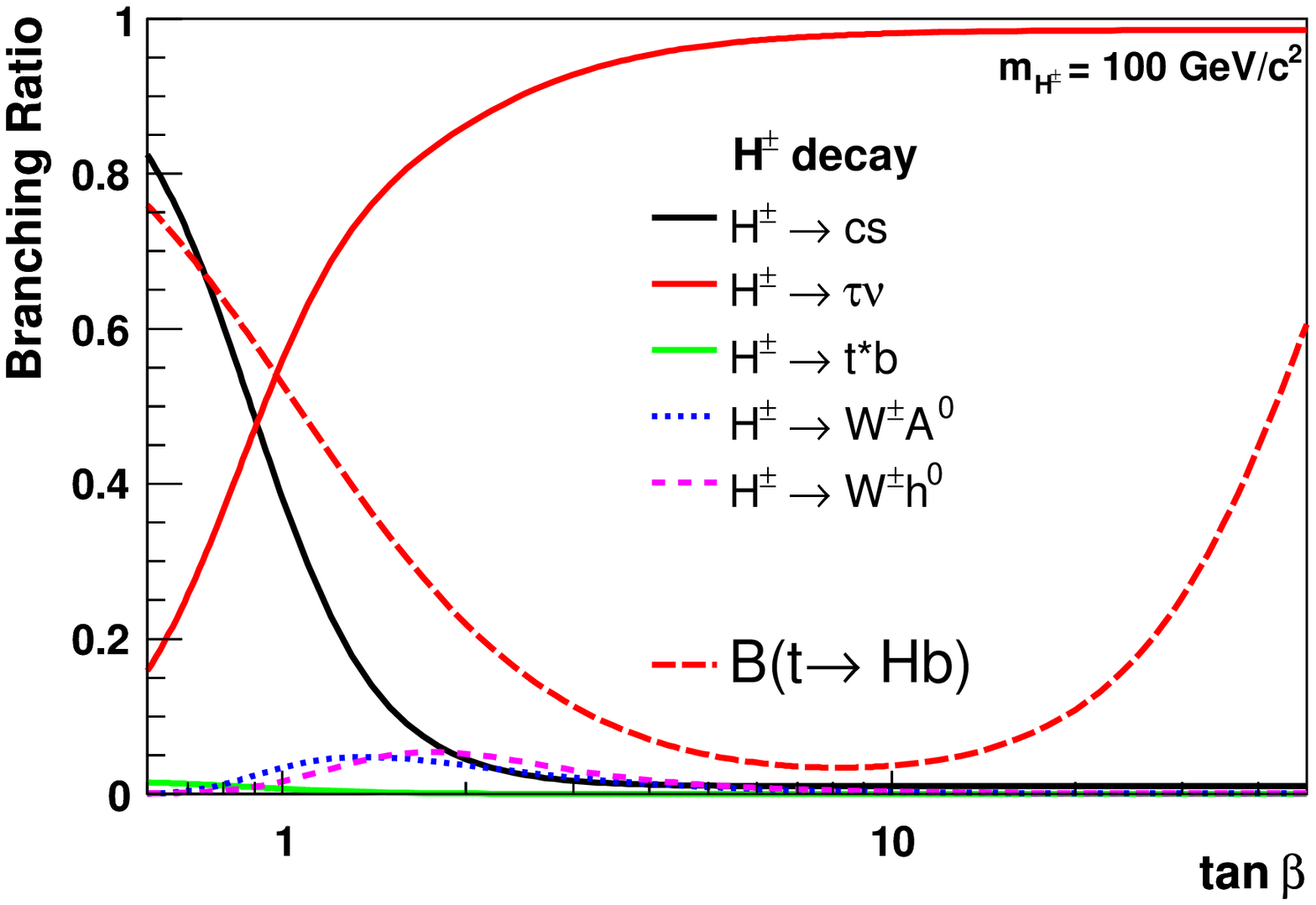}
    \includegraphics[width=0.48\textwidth,height = 50mm]{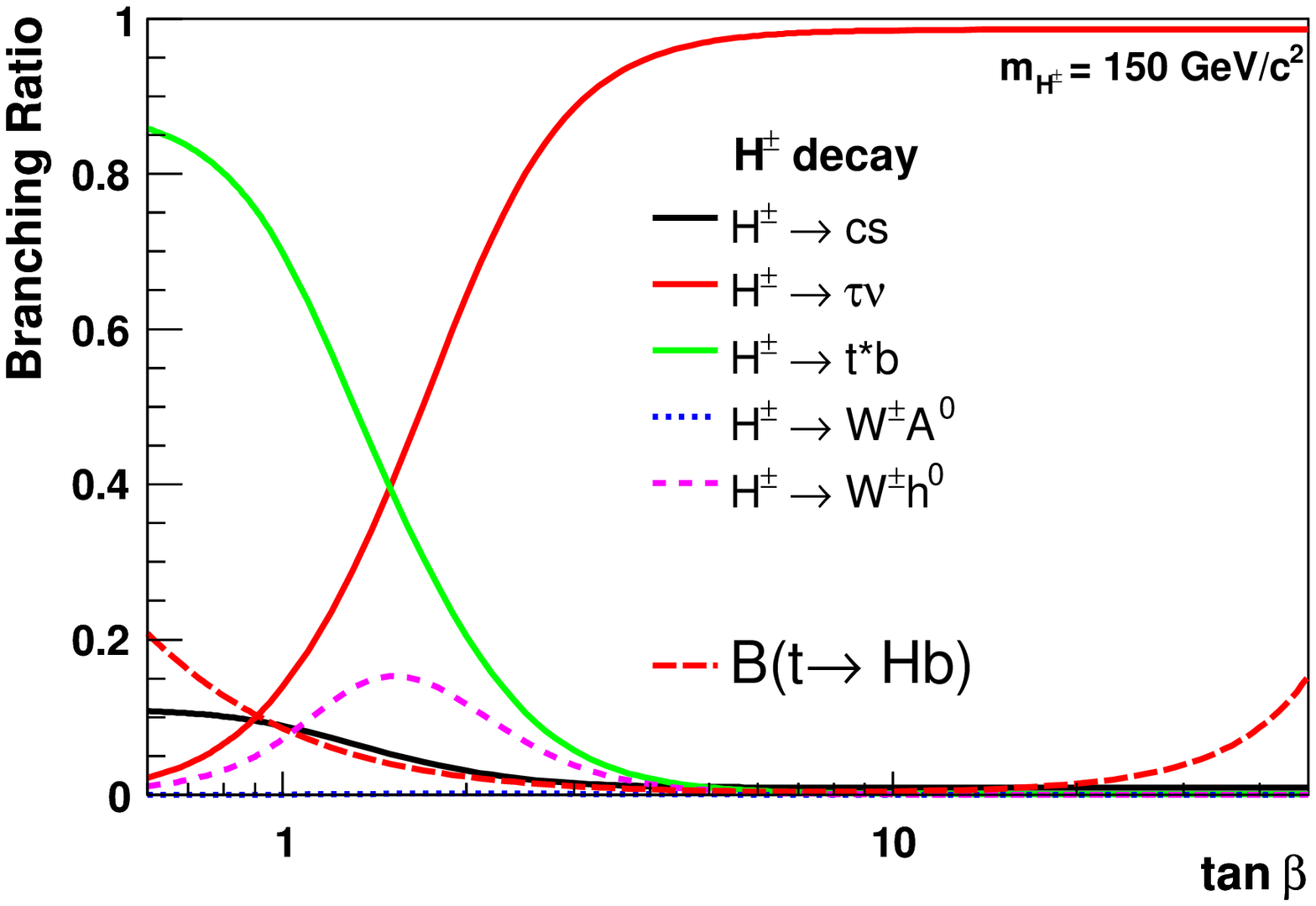}
    \caption{Branching ratios for top quark and charged Higgs boson
      decays versus $\tan\beta$ in the MSSM framework as simulated
      with \cpsuperh~\cite{Lee:2003nta} for $m_{H^\pm} = 100$~GeV/c$^2$
      (left) and $m_{H^\pm} = 150$~GeV/c$^2$ (right).}
      \label{fig:MSSMBrH+}
  \end{center}
\end{figure}

The decay mode $t\to Hb$ is kinematically accessible if the mass of
the charged Higgs boson is less than the difference between top and $b$
quark masses $m_{H^\pm} < (m_t - m_b)$, and will then compete with the
standard model decay $t\to Wb$. The distinct top quark decay signature
provides an additional handle for background suppression compared to
direct production of charged Higgs bosons. The branching fraction of
$t\to Hb$ depends on $\tan\beta$ and $m_{H^\pm}$. As illustrated in
Fig.\ \ref{fig:MSSMBrH+} for the MSSM, the branching ratio of $t\to
Hb$ increases significantly both for small $\tan\beta\lsim 1$ and large
$\tan\beta\gsim 40$, for a given $m_{H^\pm}$. The standard model decay
is assumed to account for the difference of ${\cal B}(t\to H b)$ from unity. For a given
$\tan\beta$, the branching ratio of $t\to Hb$ decreases with increasing
$m_{H^\pm}$. The decay of $H^\pm$ is dominated by $H^\pm\to\tau\nu$
for large $\tan\beta$, independent of $m_{H^\pm}$, which would result in
an excess of \ttbar events in the $\tau$ decay channel relative to
standard model expectation. At small $\tan\beta$, the decay $H^\pm\to cs$
is enhanced for small $m_{H^\pm}$, while $H^\pm\to t^\ast b$ dominates for
$m_{H^\pm}$ close to the top quark mass.
Consequently, searches for charged Higgs bosons focus on these three
fermionic decay modes.

In the early 1990s (before the discovery of the top quark) first
searches for $t\to Hb$ in the $H^\pm\to\tau\nu$ decay mode
assuming specific branching fractions were performed, and limits derived in the
$m_{t}$ versus $m_{H^\pm}$ parameter space by the UA1 and UA2
experiments~\cite{Albajar:1990zs,Alitti:1992hv} at the CERN
$Sp\bar{p}S$ collider, and by CDF at the Run~I Tevatron 
\cite{Abe:1993mr,Abe:1994ng}. All four LEP experiments searched
for pair-production of charged Higgs bosons in $e^+e^-\to H^+H^-$,
assuming that only the decays $H^\pm\to\tau\nu$ and $H^\pm\to cs$ can occur
\cite{Heister:2002ev,Abdallah:2003wd,Achard:2003gt,Abbiendi:1998rd}, as
favored by Type-II THDM. The dominant background in the resulting
three decay modes is pair production of $W$ bosons, yielding similar
final states. 95\% C.L.\ lower mass limits, independent of the $H^\pm$
decay mode, yielded $m_{H^\pm} >$~78.6~GeV/c$^2$ in a preliminary
combination of all four experiments~\cite{LEP:2001xy}. This was superseded
by a more stringent limit obtained by ALEPH of
79.3~GeV/c$^2$~\cite{Heister:2002ev}. 

Indirect limits on the mass of the charged Higgs boson can be obtained
from measurements of the $b\to s\gamma$ FCNC process at $B$ factories,
since the involved loop diagrams are sensitive to contributions from
new particles such as $H^\pm$. For a Type-II THDM scenario, a 95\% C.L.
lower limit of $m_{H^\pm} >$~295~GeV/c$^2$ can be
derived~\cite{Misiak:2006zs} if the used theoretical description is
assumed to be complete. Direct searches are less model dependent, and
therefore serve as important tools to scan for new physics beyond the regions
of parameter space excluded by the corresponding direct analyses
described above. The direct searches for $t\to Hb$ performed at the
Tevatron are based on Type-II THDM scenarios.

After the discovery of the top quark, the first searches for $H^\pm$
in top decays $t\bar{t}\to H^\pm W^\mp b\bar{b}, H^\pm H^\mp b\bar{b}$
focused on the decay $H^\pm\to\tau\nu$, which corresponds to large
$\tan\beta$. CDF published an analysis superseding and extending a
previous result~\cite{Abe:1995pj}, requiring inclusive final states
with \METns, a hadronically decaying tau lepton ($\tau_h$), and (i) two
jets and at least one additional either lepton or jet, or (ii) a second
energetic $\tau_h$~\cite{Abe:1997rk}. Another search investigated the
dilepton channels $e\tau_h, \mu\tau_h$, with accompanying \MET and at least two
jets~\cite{Affolder:1999au}. D0 performed a first $H^\pm$ analysis in
Run~I based on a ``disappearance'' search in the lepton + jets channel,
sensitive to $H^\pm$ fermionic decay modes, by looking for a
discrepancy in the event yields relative to the standard model
predictions~\cite{Abbott:1999eca}. This analysis was complemented
through a direct search for $H^\pm\to\tau\nu$ with a hadronically decaying
$\tau$ reconstructed as narrow jet in a dataset with events containing
\MET and at least four, but no more than eight,
jets~\cite{Abazov:2001md}. All analyses observe good agreement with
SM expectation, and provide limits in the $\tan\beta ,
m_{H^\pm}$ plane.

It should be noted that the above limits are based on tree level MSSM
calculations of branching fractions depending on
$\tan\beta$. By now, it has become clear that higher-order radiative
corrections, which strongly depend on model parameters, modify
these predictions significantly~\cite{Coarasa:1999da,Carena:1999py}.
Also, non-fermionic (bosonic) $H^\pm$ decay modes can have non-negligible
contributions at small $\tan\beta$, as illustrated in
Fig.\ \ref{fig:MSSMBrH+}, affecting the limits derived in that area
without taking this into account. Independent of these issues, one can
still provide upper limits on ${\cal B}(t\to H b)$ %
based on the observed production rate for any specified $H^\pm$
branching ratio. For example, for a purely tauonically decaying charged
Higgs boson, 95\% C.L.\ upper limits on ${\cal B}(t\to H b)$ are found by CDF
to lie between 0.5 and 0.6 for 60~GeV/c$^2$ $\leq m_{H^\pm} \leq$
160~GeV/c$^2$~\cite{Affolder:1999au}. The combined D0 result
corresponds to ${\cal B}(t\to H b) < 0.36$ at 95\% C.L.\ for $m_{H^\pm}
<$ 160~GeV/c$^2$ and $0.3<\tan\beta<150$, which is the full range where the
leading-order MSSM calculation is valid~\cite{Abazov:2001md}.

\begin{figure}[!t]
  \begin{center} 
    \includegraphics[width=0.495\textwidth,height = 50mm]{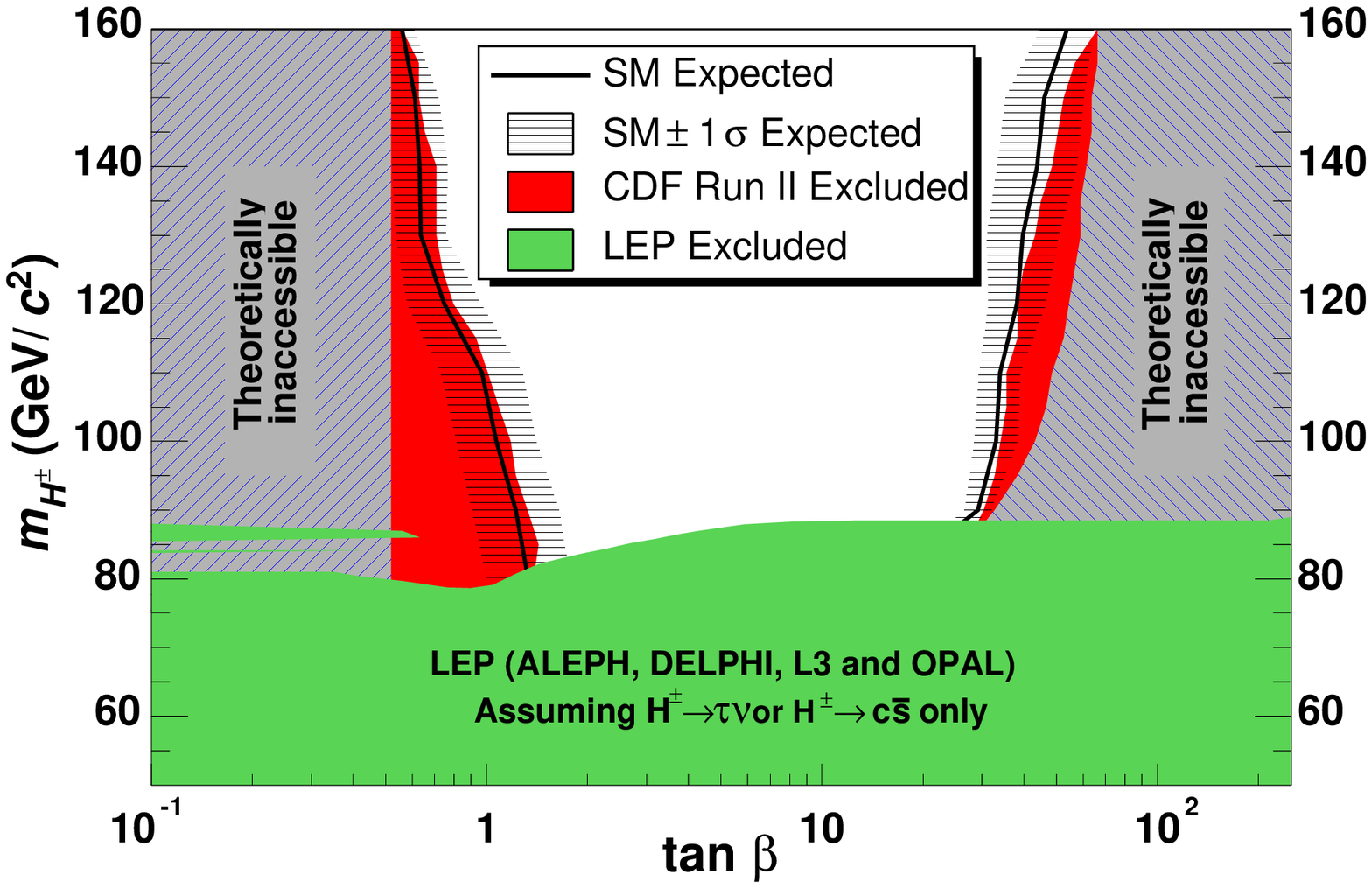}
    \includegraphics[width=0.495\textwidth,height = 50mm]{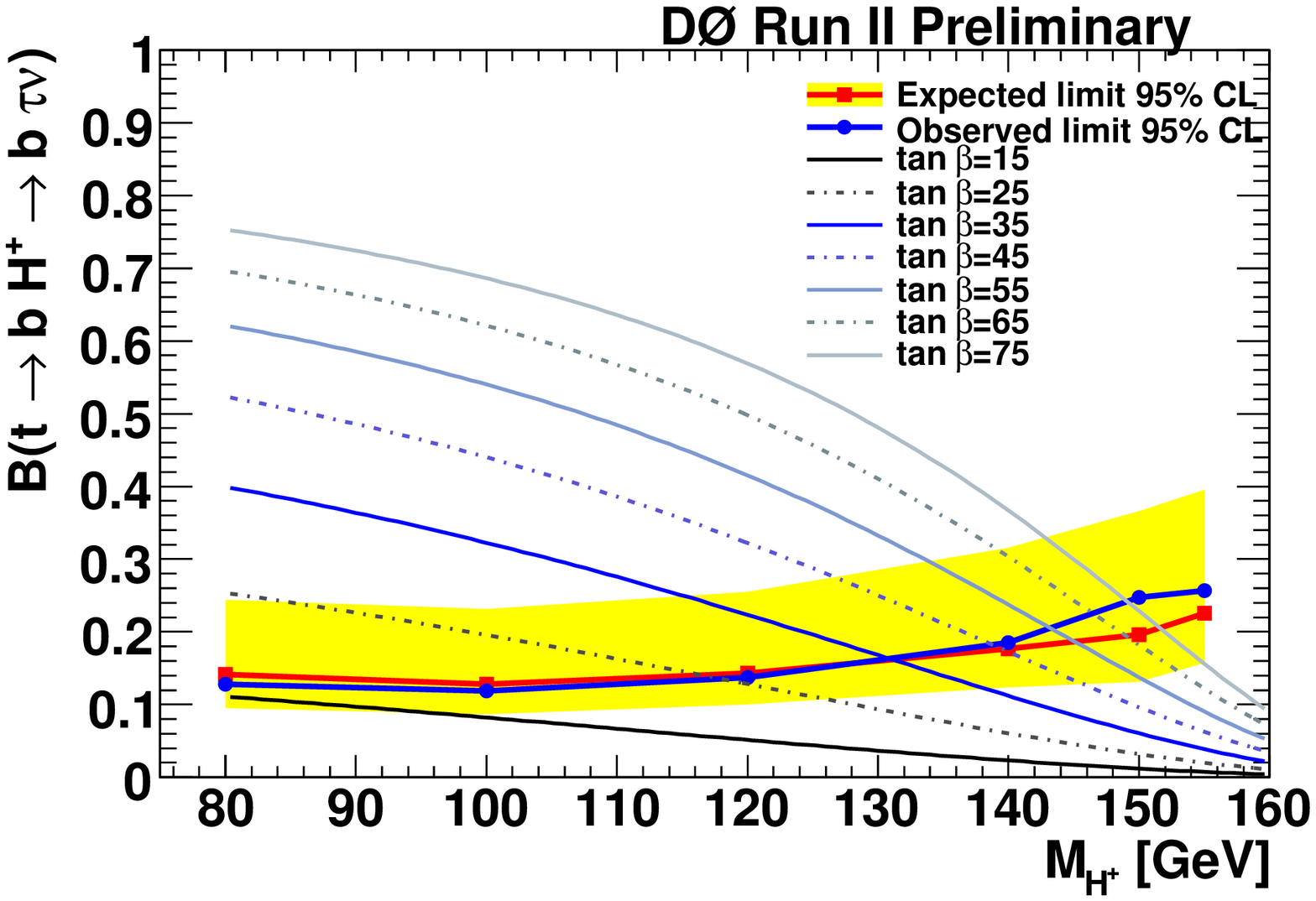}
    \caption{Left: Expected and observed 95\% C.L.\ exclusion limits
    for charged Higgs bosons in the ($m_{H^\pm}, \tan\beta$) plane
    obtained by CDF for the MSSM benchmark scenario discussed
    in Ref.~\cite{Abulencia:2005jd}. Right: Expected and observed 95\% C.L.
    limits on ${\cal B}(t\to H b)$ versus $m_{H^\pm}$ found by D0
    for a tauonic charged Higgs model using a simultaneous fit of
    ${\cal B}(t\to H b)$ and the \ttbar production cross
    section~\cite{D05715}. MSSM tree-level predictions for several
    $\tan\beta$ values are shown as well.}
      \label{fig:H+MSSMntauonic}
  \end{center}
\end{figure}

CDF published a first search for $t\to Hb$ in Run~II using
0.2~fb$^{-1}$ of integrated luminosity~\cite{Abulencia:2005jd}. The
search is based on the corresponding \ttbar cross section
analyses~\cite{Acosta:2004hw,Acosta:2004uw,Abulencia:2005et}
in the topology \MET + jets + $\ell$ + X, where $\ell$
corresponds to an electron or muon and X to either $\ell$ (dilepton
channel), $\tau_h$ (lepton + $\tau$ channel) or one or more $b$ tagged
jets (lepton + jets channels). Dropping the assumption of ${\cal
B}(t\to H b) = 0$, and avoiding overlaps of the channels,
the observed yields can be compared with the expected deficits or
excesses in the channels relative to the standard model prediction,
depending on the top quark and $H^\pm$ branching fractions. Apart from
the standard model top quark decays, $t\to Hb$ is considered with
$H^\pm$ decaying to $\tau\nu$, $cs$, $t^\ast b$ or $Wh^0$, with $h^0
\to b\bar{b}$. The \ttbar production rate is assumed to be not
affected by the extension of the Higgs sector. Since no $H^\pm$ signal
is observed, 95\% C.L.\ upper limits on ${\cal B}(t\to H b)$ are
obtained, for example, for a tauonic Higgs model (${\cal B}(H^\pm\to
\tau\nu) = 1)$ to be 0.4 for 80~GeV/c$^2$ $\leq m_{H^\pm} \leq$
160~GeV/c$^2$. 95\% C.L.\ limits are also obtained in the ($m_{H^\pm},
\tan\beta$) parameter space in the framework of the MSSM for certain
benchmark settings of parameters~\cite{Abulencia:2005jd}, taking radiative corrections into
account. While the excluded area for large $\tan\beta$ strongly
depends on the different benchmarks investigated, this is not the case
for small $\tan\beta$. An example result is shown in
Fig.\ \ref{fig:H+MSSMntauonic}.

D0 performs a similar analysis based on the
same \ttbar final states in 1~fb$^{-1}$ of integrated
luminosity~\cite{D05715}. Two models for the decay mode of
the charged Higgs boson are studied: (i) a ``tauonic'' Higgs model with ${\cal
B}(H^\pm\to \tau\nu) = 1$ that would give an enhancement of the
lepton + $\tau$ channels and a deficit in the lepton + jets and
dilepton channels, and (ii) a ``leptophobic'' model with ${\cal B}(H^\pm\to cs) =
1$ that would yield an enhancement of the all-hadronic channel and
a deficit in all channels considered in this analysis. For both model
assumptions, good agreement with the standard model prediction is
observed, and 95\% C.L.\ upper limits on ${\cal B}(t\to H b)$ are
provided for 80~GeV/c$^2$ $\leq m_{H^\pm} \leq$ 155~GeV/c$^2$,
yielding 0.16 - 0.20 for the tauonic and 0.2 for the leptophobic
model. The dominant systematic uncertainties arise here from
uncertainties on the \ttbar cross section and the luminosity.

For the tauonic model, an improvement of the obtained limits by about
30\% in the low $m_{H^\pm}$ range ($\lesssim$ 100~GeV/c$^2$) is possible when the \ttbar cross
section is allowed to float in the fit rather than be fixed to the
SM value. The resulting limits are displayed in
Fig.\ \ref{fig:H+MSSMntauonic}, and range from 0.12 to 0.26 in the
indicated $ m_{H^\pm}$ range. Assuming the standard model scenario of
${\cal B}(t\to H b) = 0$, a combination of the \ttbar cross sections
from the analyzed final states is also obtained, as was discussed in
Section~\ref{sec:ttbarxsecmeassummary}.

As illustrated in Fig.\ \ref{fig:H+MSSMntauonic}, the results
obtained in Ref.~\cite{Abulencia:2005jd} leave room for improvement,
particularly close to $\tan\beta\approx 1$ and $m_{H^\pm}$ above
the $W$ boson mass. For this range of parameters, the MSSM predicts
a significant branching fraction for the decay $H^\pm\to cs$.
CDF has searched for $t\to H b$ in this decay channel in
2.2~fb$^{-1}$ double $b$ tagged lepton + jets data~\cite{CDF9322}.
Both standard model and exotic decay exhibit the same final state, but
can be distinguished via the dijet invariant mass, where the two untagged
leading jets are assigned to $W^\pm/H^\pm$ decay products.
A binned likelihood fit using $W^\pm/H^\pm$ dijet mass templates yields
no significant excess over the SM prediction, and 95\% C.L.
upper limits on ${\cal B}(t\to H b)$ are provided for 90~GeV/c$^2$
$\leq m_{H^\pm} \leq$ 150~GeV/c$^2$, assuming a leptophobic Higgs
model. The limits range from 8\% for $m_{H^\pm} = 130$~GeV/c$^2$
to 32\% for $m_{H^\pm} = 90$~GeV/c$^2$, complementing the analysis
by D0 described above for the mass range above 100~GeV/c$^2$.

A first direct search for charged Higgs boson production in the mass
range beyond $m_{t}$ via the process $q\bar{q}'\to H^\pm\to
tb$ has been performed by D0, as was discussed previously in
Section~\ref{sec:STchargedH}.

\section{Top Quark Properties}
In the past two chapters, it was demonstrated that both top quark production and
decay thus far have been found to be consistent with the standard model
expectations. No new particles or anomalous couplings have been observed
yet. To actually confirm the top quark's standard model identity, its
fundamental quantum numbers need to be measured and their
self-consistency in the standard model framework needs to be confirmed as well.

In this chapter, measurements of the top quark's electric charge, lifetime and mass
performed thus far at the Tevatron are described.
Again, top quark pair events are used for this, because these events provide
higher statistics and favorable sample purities compared to single
top events. First direct measurements of the $V_{tb}$ matrix element
in electroweak single top quark production have already been discussed
in Section~\ref{sec:STxsec}.
\subsection{Top quark electric charge}
\label{sec:topcharge}
The electric charge of quarks can be determined, for example, in
electron-positron collisions via the ratio of the hadronic cross
section to the muon cross section $R = \sigma(e^+e^-\to \rm{hadrons})
/ \sigma(e^+e^-\to\mu^+\mu^-)$, which is proportional to the sum of the
squared electric charges of the quark flavors accessible at the chosen
center of mass energy. Due to the large top quark mass,
such a direct measurement could not yet be performed at past and
present electron-positron colliders. Also, a direct measurement of
photon radiation in \ttbar events at hadron colliders, which would give
access to the top quark's charge and its electromagnetic coupling, is
unrealistic due to limited statistics at the
Tevatron~\cite{Baur:2001si}. Consequently, the top quark is the only
quark whose fundamental quantum numbers of weak isospin and electric
charge could thus far be determined only indirectly in the framework of
the standard model from measurements of its weak isospin partner, the
$b$ quark, to be $T_{3}=+\frac{1}{2}$, $Q_t=+\frac{2}{3}e$ (see
Section~\ref{sec:SMneedsTop}).

Information on the electric charge of the top quark can also be
inferred from the electric charges of its decay products. However,
there is an inherent ambiguity in $p\bar{p}\to
t\bar{t}\to W^+W^-b\bar{b}$ events when pairing $W$ bosons and $b$ jets,
resulting in possible charges of $|Q|$= 2$e$/3 or 4$e$/3 for the top quark.
An exotic quark with charge $-4$$e$/3 being the discovered particle at
the Tevatron instead of the standard model top quark would be
compatible with precision electroweak measurements if the right-handed
$b$ quark were to mix with the $-1$$e$/3 charged exotic doublet partner of
such an exotic top quark. The standard model top quark with charge
2$e$/3 would yet have to be discovered in this scenario, due to its
large mass of
$271^{+33}_{-38}$~GeV/c$^2$~\cite{Chang:1998pt,Chang:1999zc,Choudhury:2001hs}.

\begin{figure}[!t]
  \begin{center} 
    \includegraphics[width = 0.49\textwidth]{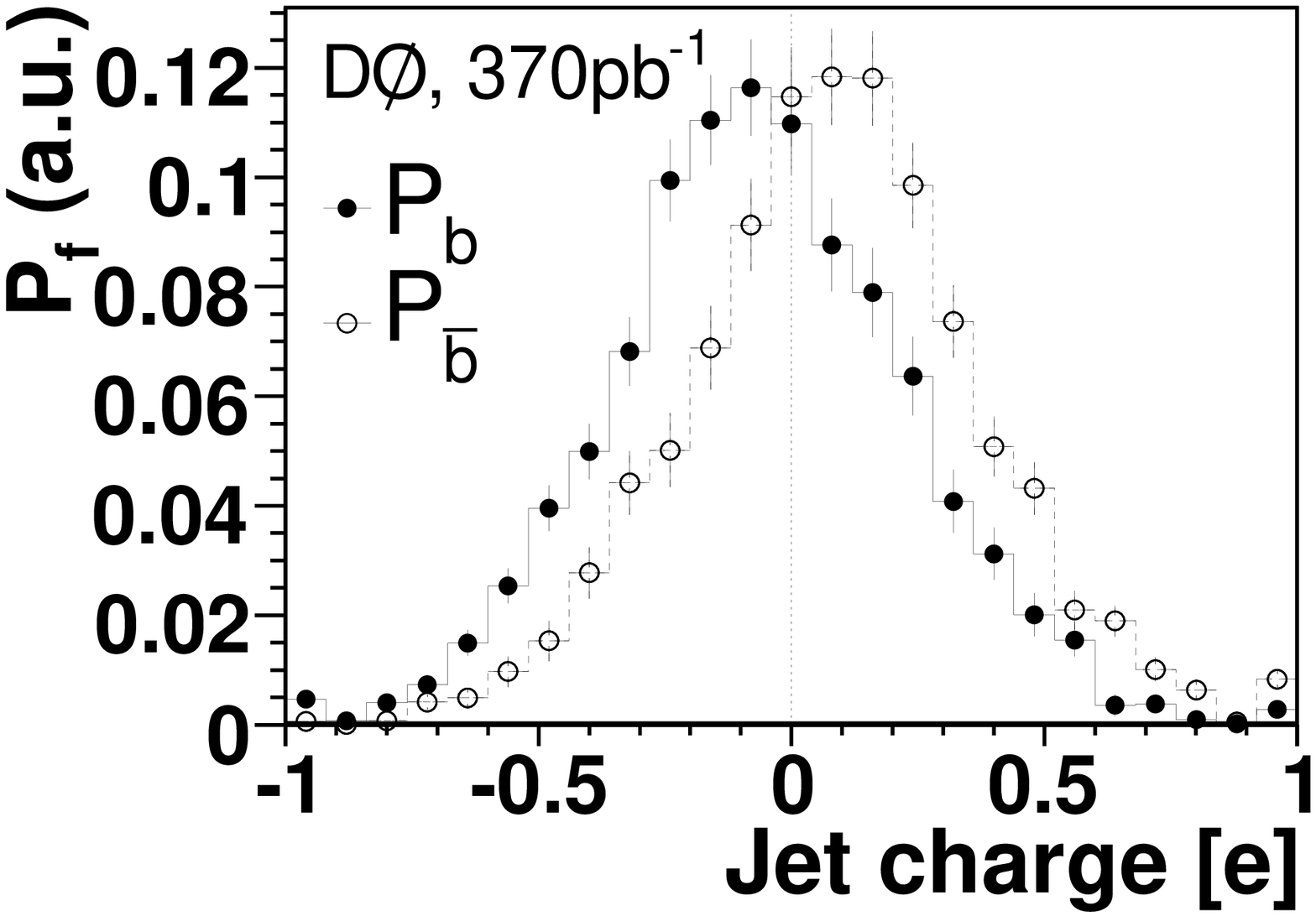}
    \includegraphics[width = 0.49\textwidth]{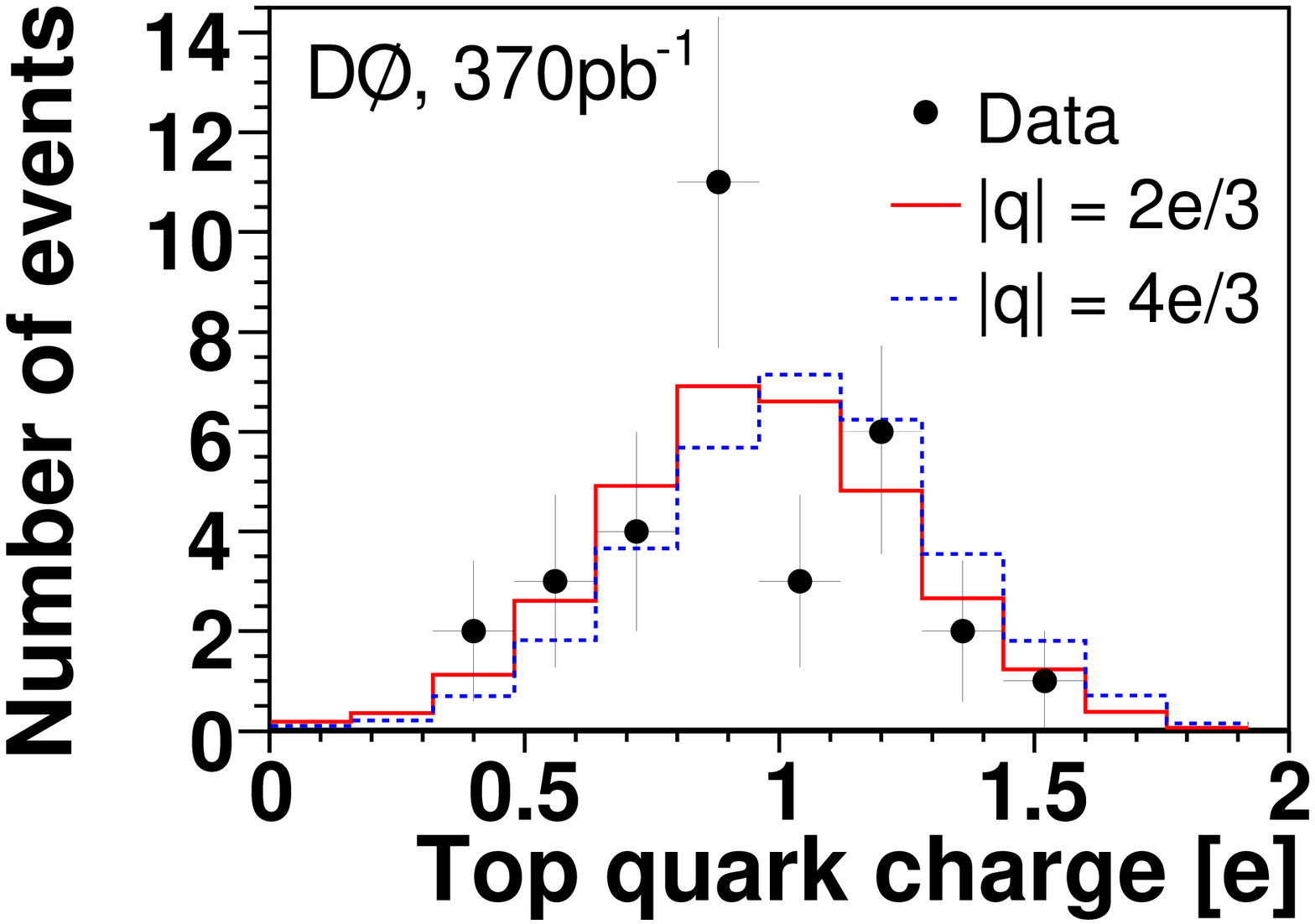}
    \caption{Left: $b/\bar{b}$ jet charge distributions obtained in
    dijet data. Right: Distribution of the top quark charge obtained
    in data, overlaid with the expectations from standard model and
    exotic model~\cite{Abazov:2006vd}.}
      \label{fig:topcharge}
  \end{center}
\end{figure}

D0 has published a first measurement discriminating between the 2$e$/3
and 4$e$/3 top quark charge scenarios in a 0.4 fb$^{-1}$ lepton +
jets dataset with $\geq 2~b$ tagged jets~\cite{Abazov:2006vd}.
The sample exhibits high signal purity, with the two
dominant background processes of $Wb\bar{b}$ and single top production
contributing only 5\% and 1\% to the selected events, respectively. Each \ttbar
event provides two measurements of the absolute value of the top quark
charge, one from the leptonic and one from the hadronic decay of the
$W$ boson in $t\to Wb$.
The charge of the $W$ boson is determined from the
(inverse) lepton charge for the leptonic (hadronic) $W$ boson decay.
The $b$ jet charge, discriminating between $b$ and $\bar{b}$ jets, is
determined using a jet charge algorithm, based on the $p_T$ weighted
average of the charges of the tracks associated with the $b$ tagged
jet. The corresponding distributions are derived from dijet collider
data~\cite{Abazov:2006vd} (see Fig.\ \ref{fig:topcharge}). The top quark charge observable
is then defined as the absolute value of the sum of the $W$ boson- and associated
$b$ jet charge, where the right pairing is determined through a
constrained kinematic fit. By comparing the distribution in
data with the expected shape from the standard model and
exotic model (see Fig.\ \ref{fig:topcharge}), respectively, D0 excludes the
hypothesis of only exotic quarks of charge $|Q|$= 4$e$/3 being produced
at up to 92\% C.L.\ and limits an exotic quark admixture in the sample to
at most 80\% at the 90\% C.L.

Using a similar analysis approach, CDF obtains a preliminary result on
the top quark charge using double $b$ tagged lepton + jets and $b$
tagged dilepton events in a 1.5~fb$^{-1}$ dataset ~\cite{CDF8967}. The
observed $2\ln$(Bayes Factor) is 12, meaning that the data favor very
strongly the SM top quark hypothesis over the exotic
model.

While the results from CDF and D0 are not directly comparable because of
different statistical approaches in the interpretation of their
results, both agree to favor the SM top quark charge
hypothesis. This is supported by the searches for new heavy top-like
quark pair production (see Section~\ref{sec:tprime}), which are starting to
exclude additional quark production in the mass range predicted by the
alternative model. A top quark charge measurement determining the $b$ jet
charge in soft-lepton tagged events from the soft-lepton charge rather
than the current track-based approach has not been performed
yet.

The measurement of the top quark charge in $t\bar{t}\gamma$ events will
be possible at the LHC because of the increased production rate and
the reduction of irreducible background from photons radiated
off the incoming quarks in \ttbar production via $q\bar{q}$
annihilation. Consequently, top quark charge measurements via photon
radiation in \ttbar events are predicted to achieve a precision of
10\%~\cite{Baur:2001si}. Using the top quark decay products to provide
an additional charge measurement will help to disentangle the
measurements of top quark electromagnetic coupling strength from the
top quark charge in the $t\bar{t}\gamma$ events. This will help to
rule out possible anomalous admixtures in the electromagnetic
interaction of the top quark.

\subsection{Top quark lifetime}
\label{sec:toplifetime}
The lifetime of the top quark $\tau_t =\hbar\, \Gamma_t^{-1} \approx
\hbar\,(1.3{\rm~GeV})^{-1}$ is approximately \mbox{$5\cdot
  10^{-25}$~s} in the framework of the standard model, as discussed in
Section~\ref{sec:tdecaywidth}. Consequently, the production and decay
vertices of the top quark are separated by ${\cal O}(10^{-16})$m,
which is orders
of magnitude below the spatial resolution of any detector. Also, the
top quark width is narrower than the experimental resolution at both
the Tevatron and the LHC. Consequently, a direct measurement of the
top quark lifetime or its width will be limited by
detector resolution. A measurement of the top quark lifetime is still
useful to confirm the standard model nature of the top quark, and
exclude new top quark production channels through long-lived particles. A
measurable lifetime of the top quark itself would imply a
correspondingly small $V_{tb}$ matrix element and render single top
quark production at the Tevatron undetectable, in contradiction with
the observed evidence described in Section~\ref{sec:ST}.

CDF has set limits on the top quark lifetime and width using
two approaches. One analysis uses a $b$ tagged lepton +
jets dataset of 0.3 fb$^{-1}$~\cite{CDF8104} to measure the impact
parameter (smallest distance) between the top quark production vertex
and the lepton track from the leptonic $W$ boson decay in the plane
orthogonal to the beam direction. Fitting the obtained distribution
with signal Monte Carlo templates for $c\tau_t$ between 0 and 500
$\mu$m and a background template, the template for 0 $\mu$m describes the
data best, which translates into a 95\% C.L.\ limit on $c\tau_t$ of $< 52.5\
\mu$m.

The second analysis uses a $b$ tagged lepton + jets dataset of 1
fb$^{-1}$\cite{Aaltonen:2008ir} to reconstruct the top quark mass in each
event using a kinematic fit. The observed distribution is compared in
a fit to \ttbar signal Monte Carlo templates of different widths for a
top quark mass of 175 GeV/c$^2$ and background templates. From the fit
result, at 95\% C.L.\ the top quark width is found to be smaller than
13.1 GeV, corresponding to a lower limit on the top quark
lifetime of $5\cdot10^{-26}$ s.

\subsection{Top quark mass}
\label{sec:topmass}
The top quark is set apart from all other known fundamental particles
by its large mass. Being the only particle with its Yukawa coupling
close to unity also raises the question whether it plays a special role in
the process of mass generation. Since the lifetime of the top quark is
so short (see Sections~\ref{sec:tdecaywidth} and
\ref{sec:toplifetime}), unlike the other quarks, it does not hadronize 
and properties like its mass can be determined directly without
the complication of having a quark embedded in a hadron. Being a
sensitive probe for physics beyond the standard model, it is also important
to measure its mass in different decay modes that could be
affected differently by novel physics contributions.

In the framework of the standard model, the top quark mass is a free
parameter. As discussed in Section~\ref{sec:topmassewprecision}, its
precise determination together with a precise $W$ boson mass
measurement provides a test of the self-consistency of the
framework and constrains the mass of the yet
undiscovered Higgs boson (or other new
particles~\cite{Heinemeyer:2003ud,Heinemeyer:2007bw}) via electroweak
radiative corrections. 

Measurements of the top quark mass have been performed thus far only in
\ttbar events, and mainly in the dilepton, lepton + jets and
all-hadronic final states. A complete kinematic reconstruction of the
\ttbar pair from the observed objects in the event can be performed
in the all-hadronic final state where no neutrinos are present.
Assuming \MET arises solely from the escaping neutrino in the lepton + jets
channel, a kinematic fit can be performed here constraining the
invariant mass of the charged lepton and neutrino to that of the $W$
boson, yielding a twofold ambiguity for the neutrino's longitudinal
momentum solution. Because of the two neutrinos contributing
to \MET in the dilepton final state, a direct kinematic
reconstruction of the \ttbar event is not possible without adding more information
or making other assumptions about the kinematics of the
objects in the event. 

Since the assignment of partons to reconstructed objects in an
event is not definite, combinatorial ambiguities arise in all
channels. Depending on the analysis technique, either all
combinations are used to extract the top quark mass, or the best
combinations are selected based on, for example, the lowest $\chi^2$ in a kinematic fit
to the \ttbar event hypothesis.
Identification of $b$ jets can be used to reduce the number
of combinations to consider in the lepton + jets and the
all-hadronic channels.
Even if both $b$ jets from the \ttbar decay are identified, 
four combinations remain in the lepton + jets channel
(including the neutrino $p_z$ ambiguity) and six in the all-hadronic
channel. In these channels, usually at least one $b$ tagged jet is
required to increase sample purity, or the data can be split on the
basis of $b$ tag multiplicity, and therefore purity, to optimize
overall sensitivity.

The techniques used in top quark mass analyses can be divided into
three categories:
\begin{romanlist}[(ii)]
\item {\bf Template Method (TM):} Observables sensitive to
the mass of the top quark such as the reconstructed top quark mass
$m_{\rm reco}$ or $H_T$ are evaluated in the dataset under
consideration. The resulting distribution is then compared in a fit
with expected contributions from \ttbar signal
(with varying top quark masses) and background processes.
\item {\bf Matrix Element Method (ME):} Based on the leading-order 
matrix elements of contributing signal and background
processes, the four-vectors of the
reconstructed objects in each event define a probability density
as a function of the top quark mass.
The total likelihood for the event sample is given as the
product of the individual event likelihoods. This method is also
referred to as the Dynamical Likelihood Method (DLM).
\item {\bf Ideogram Method (ID):} An event-by-event likelihood
depending on the assumed top quark mass is formed based on a
constrained fit of the event kinematics, taking all object
permutations and possible background contributions into account. As
mentioned above, this kind of constrained fit is only possible in the
all-hadronic and lepton + jets channels.
\end{romanlist}
Naturally, the analyses most sensitive to the top quark mass are also
very sensitive to the jet energy scale (JES) calibration. The
systematic uncertainty due to the external jet energy calibration (see
Section~\ref{sec:qgjets-JES}) then is usually the dominant
systematic uncertainty in such analyses. This can be reduced in
decay channels where at least one $W$ boson decays ``hadronically'' by
using the well-measured $W$ boson mass to constrain {\it in-situ} the
jet energy calibration in top quark events~\cite{Abulencia:2005ak,Abulencia:2005aj}.
Determining such an overall scale factor for jet energies, absorbs a large
part of this uncertainty into an uncertainty scaling with \ttbar sample
statistics, while residual uncertainties, for example, due to $\eta$ and
\pt dependence of JES corrections or differences between light-quark
and $b$-quark JES still remain. By performing an analysis simultaneously in
the dilepton and all-hadronic and/or lepton + jets channels, the
{\it in-situ} JES calibration can also be transferred to the dilepton
channel~\cite{CDF9206,Aaltonen:2008gj}.

Another approach to reduce the dependence of top quark mass
measurements on JES is to utilize observables with minimal JES
dependence that are still correlated with the top quark mass, such as the
mean \pt of the charged lepton from the $W$ boson decay ($p_T^\ell$) or
the mean transverse decay length of $b$ jets $L_{xy}$ in \ttbar
events~\cite{CDF9414}. While such measurements are statistically
limited at the Tevatron, their uncertainties are basically
uncorrelated with those of other statistically more sensitive
analyses. This helps to reduce the overall uncertainty on $m_t$
when all measurements are combined. Also, the observed signal event yield can 
provide an additional constraint on $m_t$ via the mass dependence of 
the \ttbar production cross section~\cite{Aaltonen:2007jw}.

Performance, calibration and statistical uncertainty derivation of
each mass analysis are checked using sets of simulated pseudo-experiments
(ensemble tests), based on mean and rms of the extracted mass and pull
distributions.

Measurements of $m_t$ were pioneered in Run~I, based on
0.1~fb$^{-1}$ of data in the
dilepton~\cite{Abe:1997iz,Abe:1998bf,Abe:1998bferr,Abbott:1997fv,Abbott:1998dn},
lepton +
jets~\cite{Abe:1997vq,Affolder:2000vy,Abachi:1997jv,Abbott:1998dc,Abazov:2004cs}
and all-hadronic~\cite{Abe:1997rh,Abazov:2004ng} channels. A
combination of the Run~I results yields $m_t = 178.0 \pm 2.7
{\rm(stat.)} \pm 3.3 {\rm (syst.)}$~GeV/c$^2$~\cite{Azzi:2004rc}. Also
in Run~II, results come mainly from the dilepton, lepton +
jets and all-hadronic channels, with the most precise measurements being
from lepton + jets samples. One analysis uses an inclusive \MET
+ jets signature, vetoing energetic isolated leptons and thereby
enhancing the $\tau$ + jets signal contribution to
44\%~\cite{Aaltonen:2007xx}. This result is listed together with
measurements in the all-hadronic channel in
Section~\ref{sec:allhadmassmeasurements}, and is consistent with the
world-averaged top quark mass. A top quark mass measurement with
explicit hadronic $\tau$ reconstruction has not been performed thus far,
but given the recent progress in the corresponding cross section
analyses discussed in Section~\ref{sec:ttbarxsecmeashadtau}, this
could still be feasible at the Tevatron.

In the following sections, the latest (final) Run~I results are
given, along with the current preliminary and published Run~II
top quark mass measurements using various analysis techniques for each
of the three main decay channels. Some of the most precise analyses
entering the world-averaged top quark mass will be
highlighted. A more detailed review of top quark mass analysis
techniques pursued at the Tevatron can be found
in Ref.~\cite{Fiedler:2007}. The final section presents the current world
average and some of its implications.

\subsubsection{Dilepton final state}
\label{sec:dilepmassmeasurements}
Analyses in the dilepton final state are performed either based on the
matrix element or the template method. For the template approach,
additional assumptions on the kinematics of the involved objects are
made in order to solve the otherwise underconstrained
system kinematics. Assuming several top quark masses, the consistency
of the observed event kinematics can be used to obtain weights for
each event as a function of $m_t$, based on input parton distribution
functions and the observed charged lepton energies (``matrix
weighting'' ${\cal M}$), or using simulated neutrino pseudo-rapidity
or azimuthal angle (``neutrino weighting'': $\nu_\eta$,
$\nu_\phi$), or \ttbar longitudinal momentum ($p_z^{t\bar{t}}$)
distributions~\cite{Abbott:1997fv,Abulencia:2006js}.
Top quark mass estimators are derived from the obtained weight
distributions, such as the peak mass position, or mean and rms of the
distributions. These values are then used in a template fit to data
to obtain the most likely $m_t$ from the sample.

The most precise top quark mass result in the dilepton channel
entering the world average is from D0, and has a precision of
2.2\%~\cite{D05743}. It combines results from neutrino weighting
($\nu_\eta$) obtained on 1~fb$^{-1}$ in the dielectron, dimuon and
lepton + track channels~\cite{D05746} with a measurement in the
$e\mu$ channel using 2.8~fb$^{-1}$ of data and the matrix element
method~\cite{D05743}.

\begin{table}[t!]
  \caption{Top quark mass measurements performed thus far at the
    Tevatron in the dilepton channel with their integrated
    luminosities, data selections ($\ell\ell$ = dilepton,
    $\ell$+trk = lepton + track, NN = neural network) and
    analysis methods used. The two analyses using 0.1 fb$^{-1}$ are
    from Run~I, and the references marked with an asterisk correspond to
    preliminary results.}
  \renewcommand{\arraystretch}{1.2}
  \addtolength{\tabcolsep}{-1pt}
  \begin{center}
    \begin{tabular*}{\textwidth}{@{\extracolsep{\fill}}|c|l|l|l|l|}
      \hline 
      $\int{\cal L}dt$   & \multirow{2}{*}{Selection} & \multirow{2}{*}{Method} & $m_t\pm({\rm stat.})\pm({\rm syst.})$  & \multirow{2}{*}{Ref.}\\ 
      ~[fb$^{-1}$] &&&[GeV/c$^2$]&\tabularnewline
      \hline
      \hline 
0.1 & $\ell\ell$ & TM:$\nu_\eta$ & $167.4 \pm 10.3  \pm 4.8 $ & \cite{Abe:1998bf,Abe:1998bferr}\\%
0.1 & $\ell\ell$ & TM:${\cal M}, \nu_\eta$ & $168.4 \pm 12.3  \pm 3.7 $ & \cite{Abbott:1998dn}\\
0.3 & $\ell\ell$ & ME & $165.2 \pm 6.1  \pm  3.4 $ & \cite{Abulencia:2006mi,Abulencia:2005uq}\\ %
0.4 & $\ell\ell$,$\ell$+trk & TM:$\nu_\eta, \nu_\phi, p_z^{t\bar{t}}$ & $170.1  \pm 6.0  \pm 4.1 $ & \cite{Abulencia:2006js}\\%
0.4 & $\ell\ell$,$\ell$+trk & TM:${\cal M}, \nu_\eta$ & $178.1 \pm 6.7  \pm 4.8 $ & \cite{Abazov:2006bg}\\ %
1.0 & $\ell\ell$ & ME & $164.5 \pm 3.9  \pm 3.9 $ & \cite{Abulencia:2006ry}\\
1.0 & $\ell\ell$ & TM:${\cal M}$ & $175.2 \pm 6.1  \pm 3.4 $ & \cite{D05463}*\\%
1.0 & $\ell\ell$,$\ell$+trk & TM:$\nu_\eta$ & $176.0 \pm 5.3  \pm 2.0 $ & \cite{D05746}*\\
1.2 & $\ell\ell$ & TM:$p_z^{t\bar{t}}$ & $169.7^{+5.2}_{-4.9} \pm 3.1 $ & \cite{Aaltonen:2007jw}\\
1.2 & $\ell\ell$ & TM:$p_z^{t\bar{t}}\oplus\sigma_{t\bar{t}}$ & $170.7^{+4.2}_{-3.9} \pm 2.6 \pm 2.4({\rm th.})$ & \cite{Aaltonen:2007jw}\\
1.8 & $\ell\ell$ & TM:$p_T^\ell$ & $156^{+22}_{-19} \pm 4.6$ & \cite{CDF8959}*\\
1.8 & $\ell\ell$ & TM:$\nu_\eta$ & $172.0 ^{+5.0}_{-4.9} \pm 3.6$ & \cite{CDF8955}*\\%
1.8 & $\ell\ell$ & ME & $170.4 \pm 3.1  \pm 3.0  $ & \cite{CDF8951}*\\
2.0 & $\ell\ell$ $\oplus$ NN & ME & $171.2 \pm 2.7  \pm 2.9 $ & \cite{Aaltonen:2008bd}\\%
2.8 & $\ell$+trk & TM:$\nu_\phi$ & $165.1^{+3.3}_{-3.2} \pm 3.1 $ & \cite{CDF9456}*\\%
2.8 & $e\mu$ & ME & $172.9 \pm 3.6  \pm 2.3 $ & \cite{D05743}*\\%
2.8 & $e\mu$/$\ell\ell$,$\ell$+trk & ME/TM:$\nu_\eta$ & $174.4 \pm 3.2  \pm 2.1 $ & \cite{D05743}*\\%
      \hline 
    \end{tabular*}
    \label{tab:mtopdil}
  \end{center}
\end{table}
The matrix element method evaluates the probability density for each
event ($P_{\rm evt}$) with measured object four-vectors $x$ to originate
from \ttbar production, depending on the top quark mass, or from the
dominant background arising from $Z\to\tau\tau$ + jets production
in the following linear combination, based on the known expected
signal fraction in the sample $f$:
\begin{equation}
P_{\rm evt}(x;m_t) = f \cdot P_{\rm sig}(x;m_t) + (1-f) \cdot P_{\rm bkg}(x).
\end{equation}
$P_{\rm sig}$ and $P_{\rm bkg}$ are the signal and background
probability densities for \ttbar and $Z\to\tau\tau$ + jets
production, based on the leading order matrix element for $q\bar{q}\to
t\bar{t}$ and the \vecbos~\cite{Berends:1990ax}
parametrization of the $Z$-production matrix element, respectively. The probability densities are
calculated by integrating over all unknown quantities, such as the
unmeasured neutrino energies and all parton states that can lead
to the $x$ objects observed in the detector:
\begin{equation}
P_{\rm sig}(x;m_t) = 1/\sigma_{\rm obs}(m_t) \int_{q_1,q_2,y} \sum_{\rm flavors} dq_1dq_2f(q_1)f(q_2)
\frac{(2\pi)^4|{\cal M}|^2}{q_1q_2 s}d\Phi_6W(x,y),
\end{equation}
where $q_1$ and $q_2$ are the momentum fractions of the colliding
partons from the proton and antiproton, $f(q_i)$ the corresponding PDFs,
${\cal M}$ is the matrix element for the signal process yielding the
partonic final state $y$, $s$ is the squared center-of-mass energy and
$d\Phi_6$ a six-body phase space element. The transfer function
$W(x,y)$, which incorporates the detector resolution, describes the
probability for a final state $x$ in the detector to be reconstructed
from the partonic state $y$. The two possible permutations from the
unknown jet-parton assignment are summed over, and the probability is
normalized to the expected observable production rate $\sigma_{\rm
obs}(m_t)$. The calculation of $P_{\rm bkg}(x)$ proceeds in an
analogous way, except that there is no dependence on $m_t$.

The top quark mass of an event sample can be obtained by
maximizing the total likelihood function, which is the product of the
individual event probabilities, with respect to $m_t$. In this way,
each event contributes according to its quality and inherent
resolution. While the ME technique exploits the full kinematic
information available, and usually yields the statistically most
sensitive measurements, it is also computationally intensive because of
the involved multidimensional integrations. The result of the 2.8~fb$^{-1}$
$e\mu$ analysis is given together with other mass
measurements performed in the dilepton channel at the Tevatron in
Table~\ref{tab:mtopdil}. The dominant systematic uncertainty in this
analysis arises, as expected, from systematic uncertainties on the
JES calibration, and response differences between light quarks and $b$
quarks.

\subsubsection{Lepton + jets final state}
\label{sec:ljmassmeasurements}
The precision of the world-averaged top quark mass is driven by
measurements in the lepton + jets channel that provides the best
compromise between sample purity and signal statistics. In this
channel, all three analysis methods (TM, ME, ID) have been deployed,
with the most precise results consistently coming from matrix element
analyses, starting with its first application at D0 in
Run~I~\cite{Abazov:2004cs}.

For the current world-averaged mass, CDF and D0 contribute measurements
in the lepton + jets channel based on the matrix element method, as
discussed in the previous section, simultaneously fitting the top quark mass
and an overall {\it in-situ} JES scale factor to the data. Events with one
energetic isolated lepton, large \MET and exactly four jets are
selected so as to best comply with the leading order matrix element used in the
calculations. Also, at least one of the jets is required to be $b$ tagged. Using
datasets of 2.7 and 2.2~fb$^{-1}$, for CDF~\cite{CDF9427} and
D0~\cite{D05750}, respectively, both measure the top quark mass with a precision
of 1.0\%, incidentally even yielding the same mass value of
172.2~GeV/c$^2$.
\begin{figure}[!t]
  \begin{center} 
    \includegraphics[width=\textwidth]{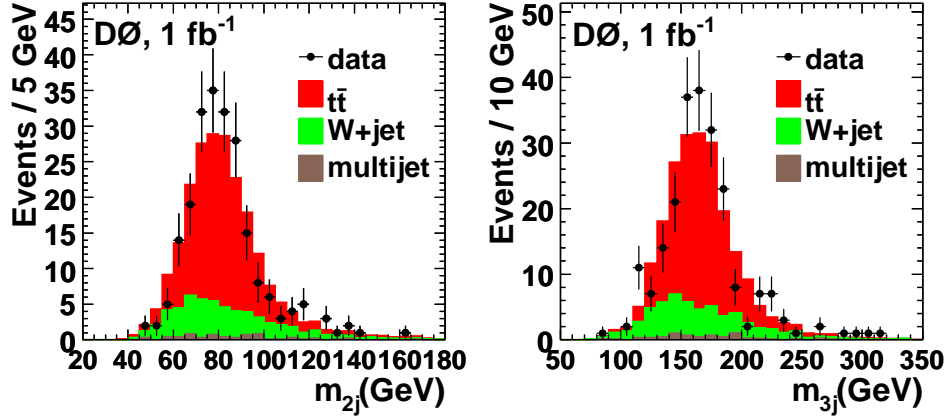}
    \caption{Dijet (left) and three-jet (right) invariant mass
      distributions in 1~fb$^{-1}$ of D0 data compared with simulated \ttbar signal (for $m_t = 170$~GeV/c$^2$) and
      background contributions~\cite{Abazov:2008ds}. 
      The results are for jet permutations of largest weight using the ME method.}
      \label{fig:mtopljets}
  \end{center}
\end{figure}
\begin{table}[h!]
  \caption{Top quark mass measurements performed thus far at the
    Tevatron in the lepton + jets channel with their integrated
    luminosities, data selections ($\ell$+jets = lepton
    + jets, $\ell\ell$ = dilepton) and analysis methods used. The
    two analyses using 0.1 fb$^{-1}$ are from Run~I, and the references
    marked with an asterisk correspond to preliminary results. Measurements
    marked with a cross contain the uncertainty from the
    {\it in-situ} JES calibration within the quoted statistical uncertainty.}
  \renewcommand{\arraystretch}{1.2}
  \addtolength{\tabcolsep}{-1pt}
  \begin{center}
    \begin{tabular*}{\textwidth}{@{\extracolsep{\fill}}|c|l|l|l|l|}
      \hline 
      $\int{\cal L}dt$   & \multirow{2}{*}{Selection} & \multirow{2}{*}{Method} & $m_t\pm({\rm stat.})\pm({\rm syst.})$  & \multirow{2}{*}{Ref.}\\ 
      ~[fb$^{-1}$] &&&[GeV/c$^2$]&\tabularnewline
      \hline
      \hline 
0.1 & $\ell$+jets & TM:$m_{\rm reco}$ & $176.1 \pm 5.1 \pm 5.3$ & \cite{Affolder:2000vy}\\
0.1 & $\ell$+jets & ME  & $180.1 \pm 3.6  \pm 3.9 $ & \cite{Abazov:2004cs}\\
0.3 & $\ell$+jets & DLM & $173.2^{+2.6}_{-2.4} \pm 3.2$ & \cite{Abulencia:2005pe,Abulencia:2005ak}\\%
0.3 & $\ell$+jets & TM:$m_{\rm reco}$ $\oplus$ JES & $173.5^{+3.7}_{-3.6} \pm 1.3^\dagger$ & \cite{Abulencia:2005ak,Abulencia:2005aj}\\%
0.4 & $\ell$+jets & ME $\oplus$ JES & $170.3^{+4.1}_{-4.5}\,^{+1.2}_{-1.8}\, ^\dagger$ & \cite{Abazov:2006bd}\\%
0.4 & $\ell$+jets & ME $\oplus$ JES & $169.2^{+5.0}_{-7.4}\,^{+1.5}_{-1.4}\, ^\dagger$ & \cite{Abazov:2006bd}\\%
0.4 & $\ell$+jets & ID $\oplus$ JES & $173.7 \pm 4.4^{+2.1}_{-2.0}\, ^\dagger$ & \cite{Abazov:2007rk}\\
0.7 & $\ell$+jets & TM:$L_{xy}$ & $180.7^{+15.5}_{-13.4} \pm 8.6$ & \cite{Abulencia:2006rz}\\
1.0 & $\ell$+jets & ME $\oplus$ JES & $171.5 \pm 1.8  \pm 1.1 ^\dagger$ & \cite{Abazov:2008ds}\\%
1.0 & $\ell$+jets & ME $\oplus$ JES  & $170.8 \pm 2.2  \pm 1.4 ^\dagger$ & \cite{Abulencia:2007br}\\
1.0 & $\ell$+jets & TM:$m_{\rm reco}$ & $168.9 \pm 2.2 \pm 4.2$ & \cite{CDF8669}*\\%
1.2 & $\ell$+jets & ME $\oplus$ JES & $173.0 \pm 1.9  \pm 1.0 ^\dagger$ & \cite{D05750}*\\%
1.7 & $\ell$+jets & DLM $\oplus$ JES & $171.6 \pm 2.0  \pm 1.3 ^\dagger$ & \cite{CDF9135}*\\
1.9 & $\ell$+jets & TM:$L_{xy},p_T^\ell$ & $175.3 \pm 6.2\pm 3.0$ & \cite{CDF9414}*\\
\multirow{2}{*}{1.9} & $\ell$+jets& TM:$m_{\rm reco}$ $\oplus$ JES & \multirow{2}{*}{$171.9 \pm 1.7 \pm 1.1^\dagger$} &\multirow{2}{*}{\cite{CDF9206,Aaltonen:2008gj}}\\%
&$\ell\ell$ &TM:$\nu_\eta,H_T$&&\\
1.9 & $\ell$+jets & ME $\oplus$ JES  & $172.7 \pm 1.8  \pm 1.2 ^\dagger$ & \cite{Aaltonen:2008mx}\\
2.2 & $\ell$+jets & ME $\oplus$ JES & $172.2 \pm 1.0  \pm 1.4 $ & \cite{D05750}*\\%
2.7 & $\ell$+jets & ME $\oplus$ JES  & $172.2 \pm 1.3  \pm 1.0 ^\dagger$ & \cite{CDF9427}*\\
      \hline 
    \end{tabular*}
    \label{tab:mtopljet}
  \end{center}
\end{table}
A comparison of the invariant dijet and three-jet mass distributions
based on the permutation of largest weight in data and
simulation is shown in Fig.\ \ref{fig:mtopljets} for a 1~fb$^{-1}$
subset~\cite{Abazov:2008ds} of D0's 2.2~fb$^{-1}$ ME analysis.

The largest systematic uncertainties on the measurement by CDF arise
from the \ttbar MC generator used to calibrate the result (\pythia\
versus \herwig) and the residual JES uncertainty. For D0, the dominant
uncertainty comes from the $b$-jet over light-jet calorimeter-response ratio and
the signal modeling uncertainty, taking the impact of
extra jets into account based on the observed four to at least five
jet event ratio in data. Both experiments are in the process of
streamlining their methods used to assess systematic uncertainties as
well as exploring new sources of uncertainties that start to become
important at the current level of precision~\cite{wwwtopsyst}.
Examples for the latter are differences arising from using NLO rather
than LO MC generators, or non-perturbative QCD effects such as color
reconnection~\cite{Skands:2007zg,Wicke:2008iz}.

Table~\ref{tab:mtopljet} summarizes the latest (final) Run~I
results and the current preliminary or published Run~II top
quark mass measurements in the lepton + jets channel.

\subsubsection{All-hadronic final state}
\label{sec:allhadmassmeasurements}
Analyses in the all-hadronic channel have been performed so far using
template and ideogram methods, and exhibit comparable sensitivity. The
precision in this channel by now is similar to that in the dilepton
final state, mainly as a result of using {\it in-situ} JES calibration, thereby
reducing the otherwise overwhelming systematic uncertainty from the
external JES.

CDF's best measurement in the all-hadronic channel entering the world
average has been performed using a template method on 2.1~fb$^{-1}$ of
data, yielding a precision of 2.4\%~\cite{CDF9165}. Similar to the
analysis~\cite{Aaltonen:2007qf} described in
Section~\ref{sec:ttbarxsecmeasallh}, events are required to have
between six and eight energetic central jets, no isolated energetic
lepton or significant \METns, and have to pass a selection based on the
output of a neural network discriminant. The dataset is split into
subsamples with exactly one and exactly two $b$ tagged jets.

\begin{figure}[!t]
  \begin{center} 
    \includegraphics[width=.42\textwidth]{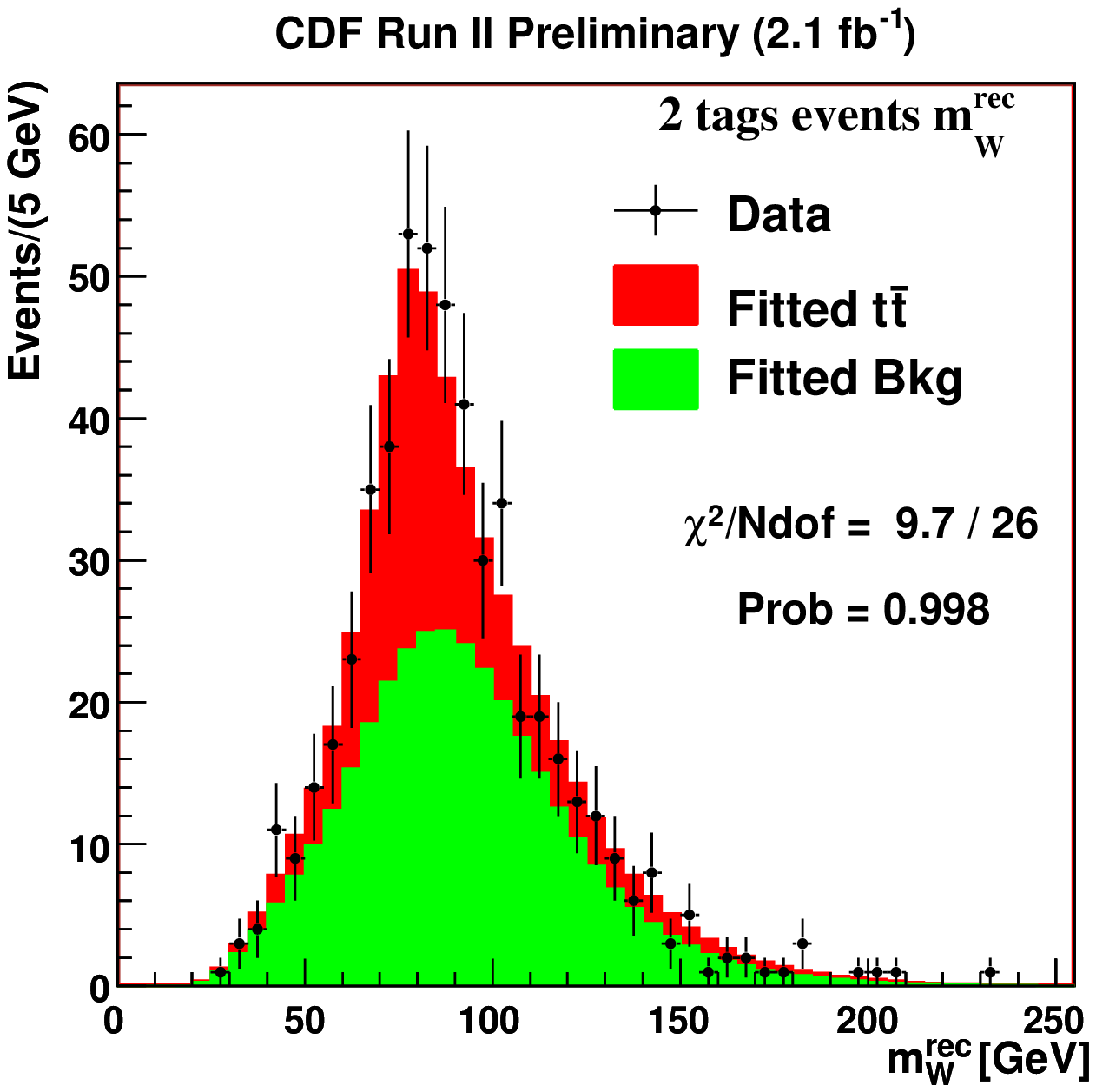}\hspace*{5mm} 
    \includegraphics[width=.42\textwidth]{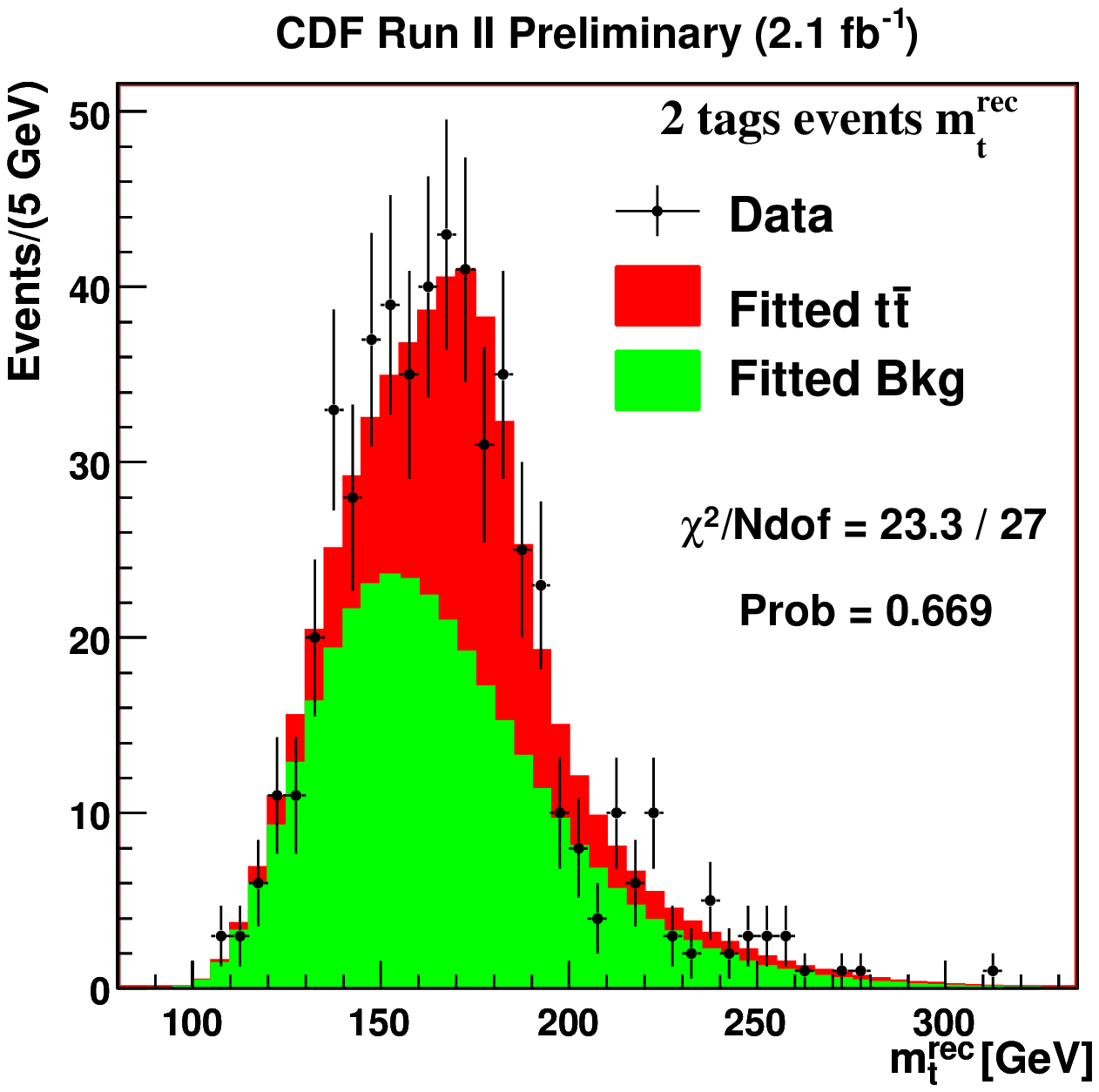}
    \caption{Reconstructed mass of the $W$ boson (left) and top quark (right)
      obtained in 2.1~fb$^{-1}$ of double $b$ tagged CDF
      data compared to the expected contributions from multijet
      background and 177 GeV/c$^2$ \ttbar signal, using the JES
      corresponding to the best fit~\cite{CDF9165}.}
      \label{fig:mtopalljet}
  \end{center}
\end{figure}

The leading six jets in each event define the \ttbar signal and
multijet background templates for the reconstructed mass of the top
quark and of the $W$ boson, with assignments based on a kinematic fit, where the permutation
with lowest $\chi^2$ is selected for further study. While the signal
templates depend on both $m_t$ and the JES scale factor, the
background templates are assumed to not depend on the top quark mass, and no JES
dependence is considered either. The measurement is then performed in
a two-dimensional fit to $m_t$ and {\it in-situ} JES scale factor,
using these templates and the observed distribution in data in both
subsamples, and applying a Gaussian constraint from the external JES
calibration. The reconstructed mass distributions obtained in data
with two $b$ tags, overlaid with templates for expected background and \ttbar signal
(for $m_t = 177$~GeV/c$^2$ and unchanged JES
with respect to the external calibration, as obtained from the fit) are
shown in Fig.\ \ref{fig:mtopalljet}.

The result of this analysis is given in Table~\ref{tab:mtopalljet}, together with other top quark mass
measurements performed in the all-hadronic channel at the Tevatron.
The dominant systematic uncertainties in
this analysis arise from uncertainties on shape and normalization of
the background templates and residual JES uncertainties. The
compatibility of this measurement with that using the ideogram
method~\cite{CDF9265} on an overlapping dataset of almost the same
size, which yields a central value $\approx$12 GeV/c$^2$ lower, is currently
being investigated.

\begin{table}
  \caption{Top quark mass measurements performed thus far at the
    Tevatron in the all-hadronic channel with their integrated
    luminosities, data selections and analysis methods
    used. The two analyses using 0.1 fb$^{-1}$ are from Run~I, and the
    references marked with an asterisk correspond to preliminary measurements.
    Results marked with a cross contain the uncertainty from
    the {\it in-situ} JES calibration within the quoted statistical uncertainty.}
  \begin{center}
    \begin{tabular*}{\textwidth}{@{\extracolsep{\fill}}|c|l|l|l|l|}
      \hline 
      $\int{\cal L}dt$   & \multirow{2}{*}{Selection} & \multirow{2}{*}{Method} & $m_t\pm({\rm stat.})\pm({\rm syst.})$  & \multirow{2}{*}{Ref.}\\ 
      ~[fb$^{-1}$] &&&[GeV/c$^2$]&\tabularnewline
      \hline
      \hline 
0.1 & jets only & TM:$m_{\rm reco}$ & $178.5 \pm 13.7  \pm  7.7 $ & \cite{Abazov:2004ng}\\%
0.1 & jets only & TM:$m_{\rm reco}$ & $186 \pm 10  \pm  5.7 $ & \cite{Abe:1997rh,Affolder:2000vy}\\%
0.3 & \METns+jets & TM:$H_T$ & $172.3^{+10.8}_{-9.6}  \pm 10.8 $ & \cite{Aaltonen:2007xx}\\%
0.3 & jets only & ID & $177.1 \pm 4.9  \pm  4.7 $ & \cite{Aaltonen:2006xc}\\%
0.9 & jets only & TM:ME $\oplus$ JES & $171.1 \pm 3.7  \pm 2.1^\dagger $ & \cite{CDF8709,Aaltonen:2008bg}\\%
1.0 & jets only & TM:$m_{\rm reco}$ & $174.0 \pm 2.2  \pm 4.8 $ & \cite{Aaltonen:2007qf}\\
1.9 & jets only & ID $\oplus$ JES & $165.2 \pm 4.4  \pm 1.9^\dagger $ &  \cite{CDF9265}*\\%
2.1 & jets only & TM:$m_{\rm reco}$ $\oplus$ JES& $176.9 \pm 3.8  \pm 1.7^\dagger$ & \cite{CDF9165}*\\%
      \hline 
    \end{tabular*}
    \label{tab:mtopalljet}
  \end{center}
\end{table}

\subsubsection{World-averaged top quark mass}
\label{sec:topmassaverage}
\label{sec:massfromxsec}
The top quark mass has been measured at the Tevatron in the three main
decay channels using various methods, as described in the past sections.
In the lepton + jets channel, precisions of 1\% are achieved in
single measurements, while in the dilepton and all-hadronic channels
the precision is $\approx$2\%. The different methods assume standard
model \ttbar production and decay, but still exhibit differences in
the strength of their model dependence. While ME methods provide the
best sensitivity, they also are strongly model dependent through their
implemented matrix elements. Template methods purely relying on the
measured event kinematics are more robust with respect to possible
deviations from the standard model, but in general exhibit lower
sensitivity.

Pursuing mass measurements in all \ttbar decay channels with different
methods is a valuable test of the self-consistency of the standard
model assumptions, and can also be used to probe for new phenomena 
\cite{Kane:1996ny}. While no top quark mass measurement
has been performed thus far in \ttbar decay modes involving hadronic $\tau$ decays,
the progress in the corresponding cross section analyses discussed in
Section~\ref{sec:ttbarxsecmeashadtau} indicates this could still be
possible at the Tevatron, completing the \ttbar decay channels for
measuring $m_{t}$.

CDF and D0 have combined their recent preliminary Run~II results with
their measurements obtained in Run~I, respectively. Based on the results of highest sensitivity in the
dilepton, lepton + jets and all-hadronic channels (CDF), and
in the dilepton and lepton + jets channels (D0), both
experiments attain a total precision of 0.9\% on their combined
measurements, respectively. Based on analyses using up to 2.7~fb$^{-1}$, CDF gets
$172.4 \pm 1.0 {\rm (stat.)} \pm 1.3 {\rm (syst.)}$ GeV/c$^2$
\cite{CDF9450}, while D0 obtains $172.8 \pm 0.9 {\rm (stat.)} \pm 1.3
{\rm (syst.)}$ GeV/c$^2$ \cite{D05747} using analyses on up to
2.8~fb$^{-1}$ of data.

\begin{figure}[t]
  \centering
  \includegraphics[width=0.5\textwidth,clip=]{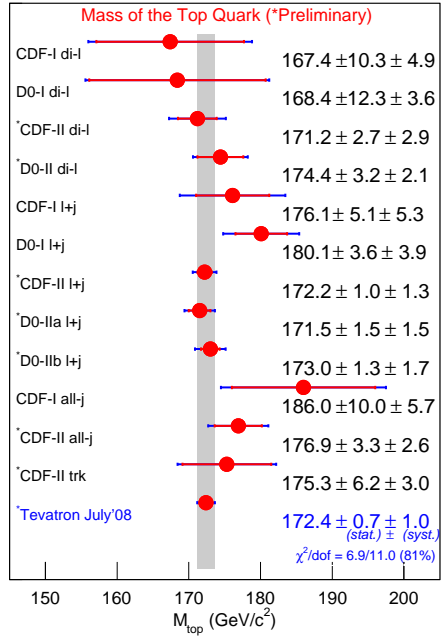}
  \caption{Measurements of the mass of the top quark used as input to the current
    preliminary world average~\cite{TEWWG:2008}.}
  \label{fig:topmass}
\end{figure}

The results of both experiments are in very good agreement, and their overall
combination yields $172.4 \pm 0.7 {\rm (stat.)} \pm 1.0 {\rm (syst.)}$
GeV/c$^2$ \cite{TEWWG:2008}, corresponding to an overall precision of
0.7\%, as illustrated in Fig.\ \ref{fig:topmass}. Combining the separate
results from the all-hadronic, the lepton + jets and the dilepton channels,
yields 177.5 $\pm$ 4.0~GeV/c$^2$, 172.2 $\pm$ 1.2~GeV/c$^2$
and 171.5 $\pm$ 2.6~GeV/c$^2$, respectively. These results are
consistent with each other, and have $\chi^2$ probabilities of at
least 17\% between any two of the channels.

All these combinations are calculated using the BLUE
method~\cite{Lyons:1988rp,Valassi:2003mu} and assume Gaussian
systematic uncertainties, with their correlations properly accounted
for. The different sources of systematic uncertainties are broken down
into twelve orthogonal categories. Six of them deal with uncertainties
related to the JES, while others address signal and background
modeling, fitting procedures, specifics of MC generation and lepton
energy scale. The main contributions to the 1.0~GeV/c$^2$ systematic
error on the world average of $m_{t}$ are (in units of GeV/c$^2$):
total JES ($\pm 0.8$), signal, background and MC model ($\pm 0.3$
each) and lepton scale and fitting procedure ($\pm 0.1$ each).

Having reached a precision of 0.7\%, the world-averaged top quark mass
is now limited by the systematic uncertainties that in turn are
dominated by JES-related uncertainties. Further improvements on the
JES can be expected since the increasing integrated luminosity will help
constrain the corresponding uncertainties better, especially the
significant contribution from {\it in-situ} JES calibration. While a final
absolute top quark mass uncertainty of $\Delta m_{t}\lesssim 1$~GeV/c$^2$
should be achievable by the end of Run~II, it will
still require a significant effort to determine the contributing
systematic uncertainties consistently among the experiments, and
evaluate any new contributions that should be considered at this 
level of precision.

This measurement marks the most precise determination of a quark mass
and will certainly provide a legacy well into the LHC era, where it will
serve as an important calibration signal until large datasets can produce
more refined measurements. However, improving the
precision by another order of magnitude can only be expected
from a threshold scan of \ttbar production at a future linear $e^+e^-$
collider~\cite{Brau:2007zza,Brau:2007sg,Behnke:2007gj}.

Before the impact of the current top quark mass measurement is
discussed, it should be noted that the value of this SM
parameter depends on the defining convention. For instance, the
$\overline{\sc MS}$ calculation gives a value lower than the pole mass of the top quark propagator
by $\approx$10~GeV/c$^2$ at ${\cal O} (\alpha_s^3)$. The pole mass itself exhibits an intrinsic
ambiguity of ${\cal O} (\Lambda_{\rm QCD}) \approx 0.2$~GeV (see for
example Ref.~\cite{Bernreuther:2008ju} and references therein).

The top quark mass measurements described in this review are usually interpreted
as representing the pole mass. However, they are calibrated using LO
MC simulations with higher orders approximated by parton showers, where
the top mass parameter does not follow a theoretically well-defined
convention. Hence, calculations and predictions using the measured
mass as the pole mass should be taken with a grain of
salt.

D0 has conducted consistency checks of the compatibility of
the direct top quark mass measurements at the
Tevatron with the pole mass extracted from the \ttbar production
rate~\cite{Abazov:2008gc,D05742}. Comparing the measured \ttbar
production cross section with SM predictions
derived at NLO, including soft-gluon resummations that are performed in
a well-defined renormalization scheme using the top quark pole mass,
constrains the mass of the top quark. The cross section
measurements depend less on MC modeling of signal kinematics
than direct mass measurements do. For cross sections, the MC is mainly needed for determining the signal
acceptance, which is expected to be rather insensitive to
higher order corrections: A comparison of NLO and LO predictions
shows that higher order corrections affect more the normalization than
the shape of the relevant kinematic distributions~\cite{Frixione:1995fj}.
\begin{figure}[!t]
  \begin{center} 
    \includegraphics[width=.48\textwidth]{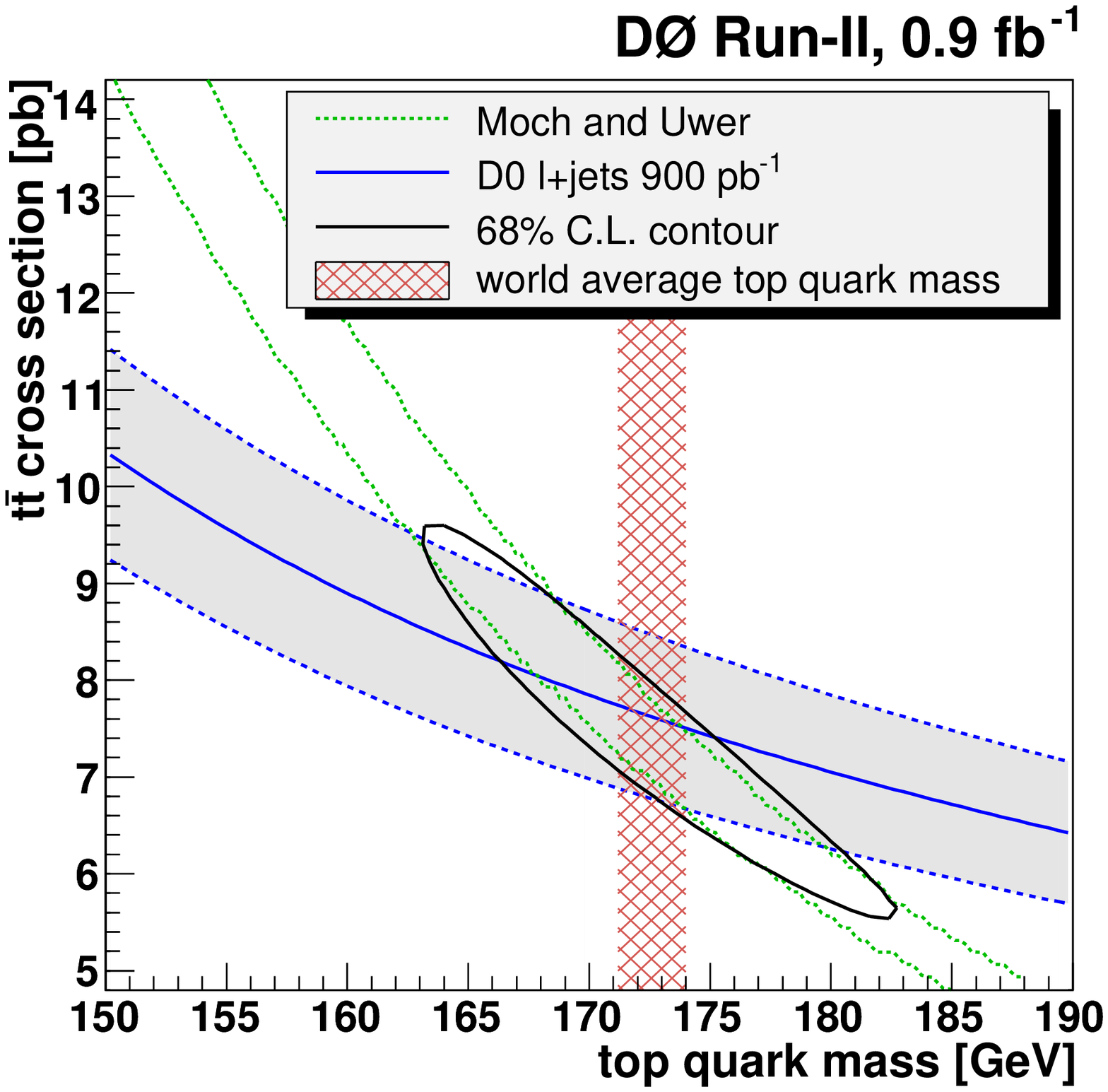}\hspace*{1mm}
    \includegraphics[width=.48\textwidth]{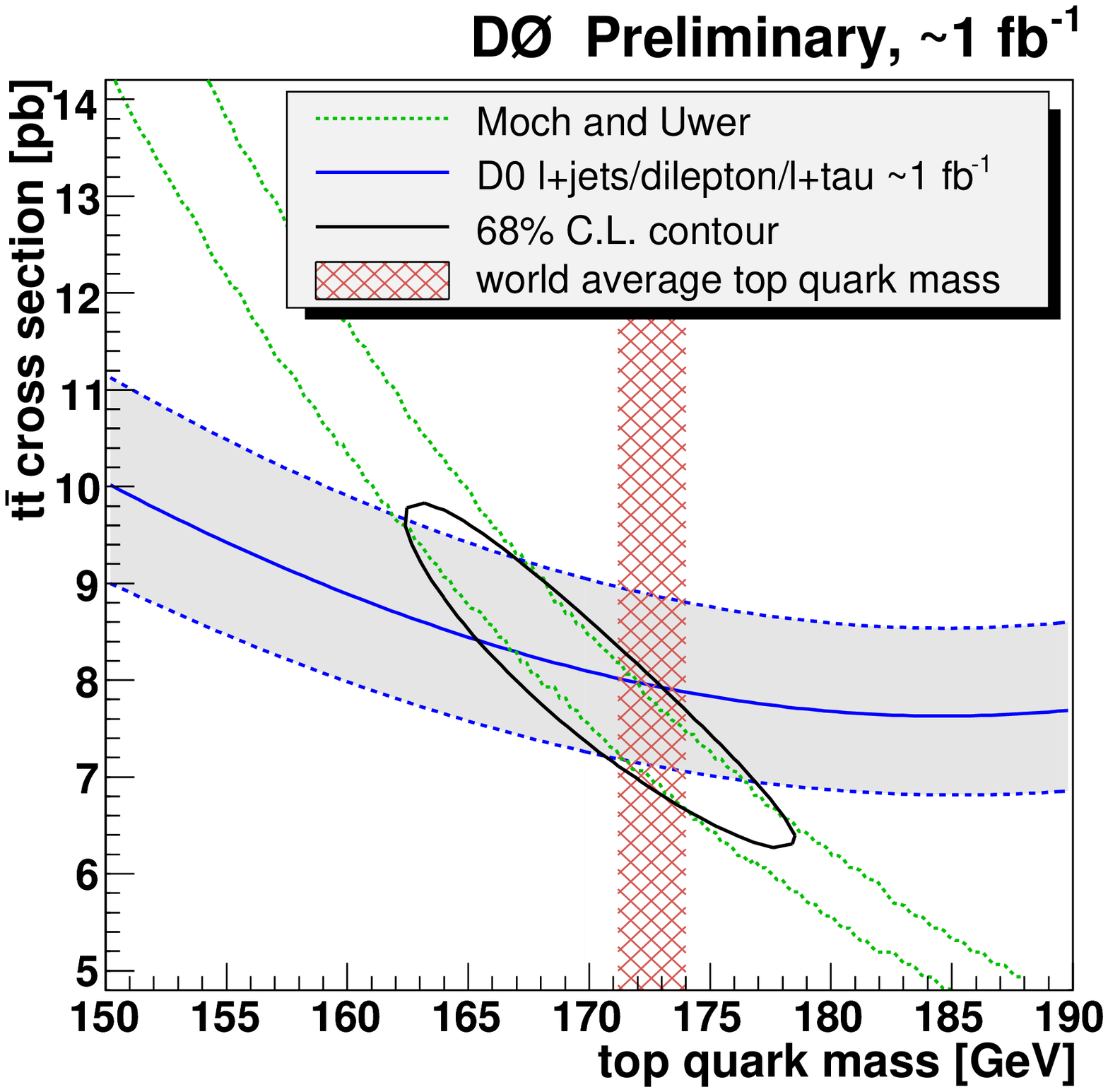} 
    \caption{Mass dependences of the \ttbar production cross sections
      measured by D0 in the lepton + jets channel (left,
      \cite{Abazov:2008gc}), and in a combination of lepton
      + jets, dilepton and $\tau$ + lepton channels (right,
      \cite{D05715}), compared with the theoretical prediction by Moch and
      Uwer~\cite{Moch:2008qy}, based on CTEQ6.6M~\cite{Nadolsky:2008zw}
      PDFs. The previous world-averaged $m_{t} = 172.6 \pm
      1.4$~GeV/c$^2$ \cite{TEWWG:2008spring}, as well as the
      68\% C.L.\ contours of the joint likelihoods resulting from the
      convolutions of measurement and prediction~\cite{D05742}, are also shown.}
      \label{fig:mtopfromxsec}
  \end{center}
\end{figure}

In a recent analysis, D0 uses two \ttbar cross section measurements to
extract constraints on the top quark mass by comparing measurement with
theoretical predictions~\cite{D05742}. One result is obtained in the
lepton + jets channel based on the combination of a counting
experiment using $b$ tagging and an analysis utilizing a topological
multivariate discriminant for 0.9~fb$^{-1}$ of
data~\cite{Abazov:2008gc}, as discussed in
Section~\ref{sec:ttbarxsecmeaslj}. The second result combines
measurements in the lepton + jets, dilepton and $\tau$ + lepton
channels obtained from approximately 1.0~fb$^{-1}$ of
data~\cite{D05715} (see Section~\ref{sec:ttbarxsecmeassummary}). Being
based on different analysis techniques and different final states, both results exhibit a
different dependence on the top quark mass, as illustrated in
Fig.\ \ref{fig:mtopfromxsec}.

One theoretical prediction used for the comparison is that of Moch and
Uwer~\cite{Moch:2008qy}, and is based on CTEQ6.6M~\cite{Nadolsky:2008zw} PDFs (see Section~\ref{sec:ttbarprod}).
A joint likelihood depending on $m_{t}$ and $\sigma_{t\bar{t}}$
is obtained as the product of the likelihood functions of the
measurement, including its total experimental uncertainty, and the
theoretical prediction, including scale and PDF uncertainties. The
contour of the joint likelihood's smallest region containing 68\% of
its integral is also shown in Fig.\ \ref{fig:mtopfromxsec} for
both measurements. By integrating over the \ttbar production rate, the
top quark mass can be extracted. For the lepton + jets channel
measurement a top quark mass of $171.2 ^{+6.5}_{-6.2}$~GeV/c$^2$ is
obtained. The combined lepton + jets, dilepton and $\tau$ +
lepton measurement yields $169.6 ^{+5.4}_{-5.5}$~GeV/c$^2$, which includes
an additional systematic uncertainty of 1~GeV/c$^2$ due to a smaller
mass dependence range available from this measurement. Both
results are in good agreement with the world-averaged $m_{t}$
obtained from the complementary direct measurements.

\begin{figure}[t!]
  \centering
  \includegraphics[width=0.495\textwidth]{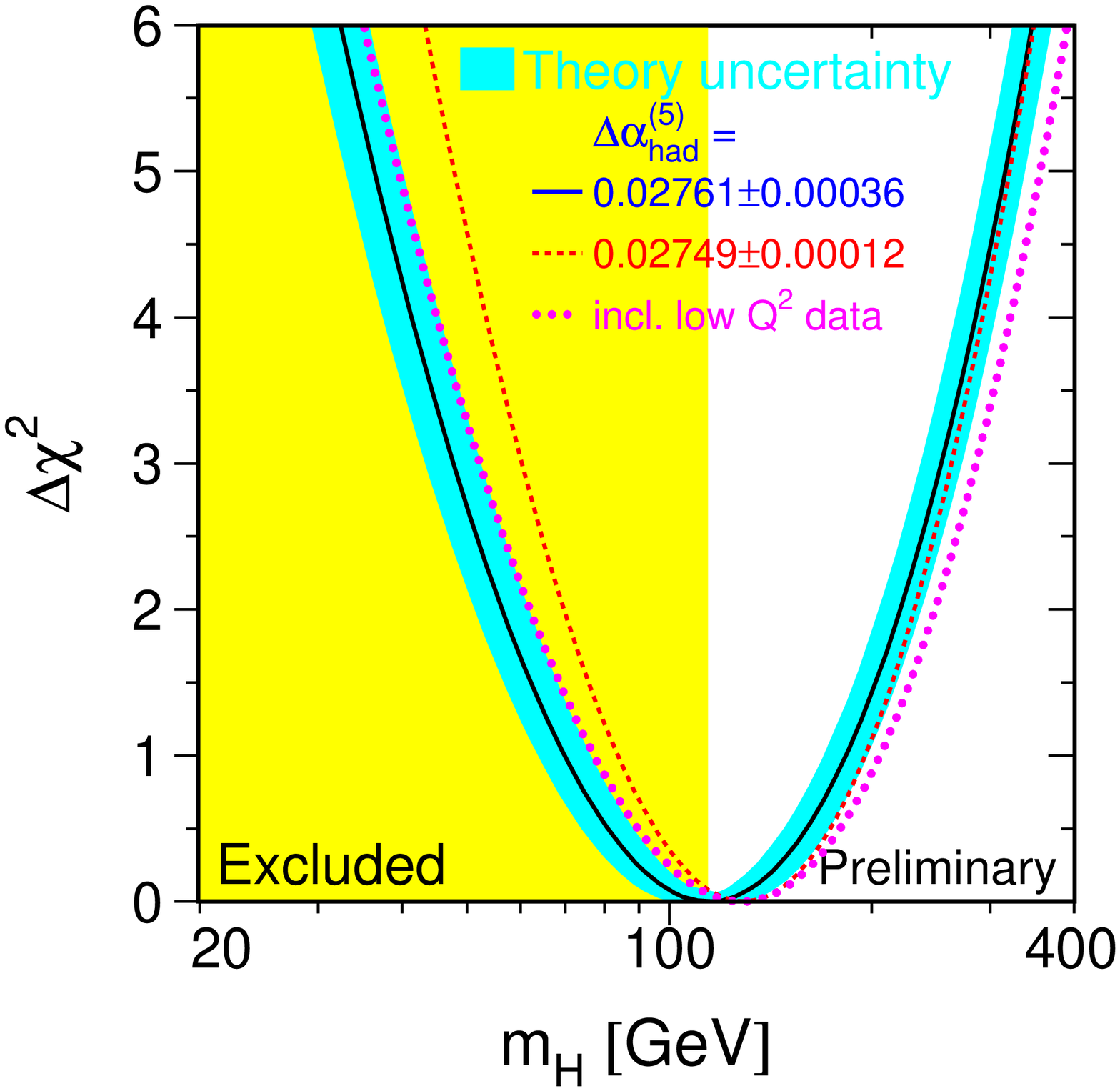}
  \includegraphics[width=0.495\textwidth]{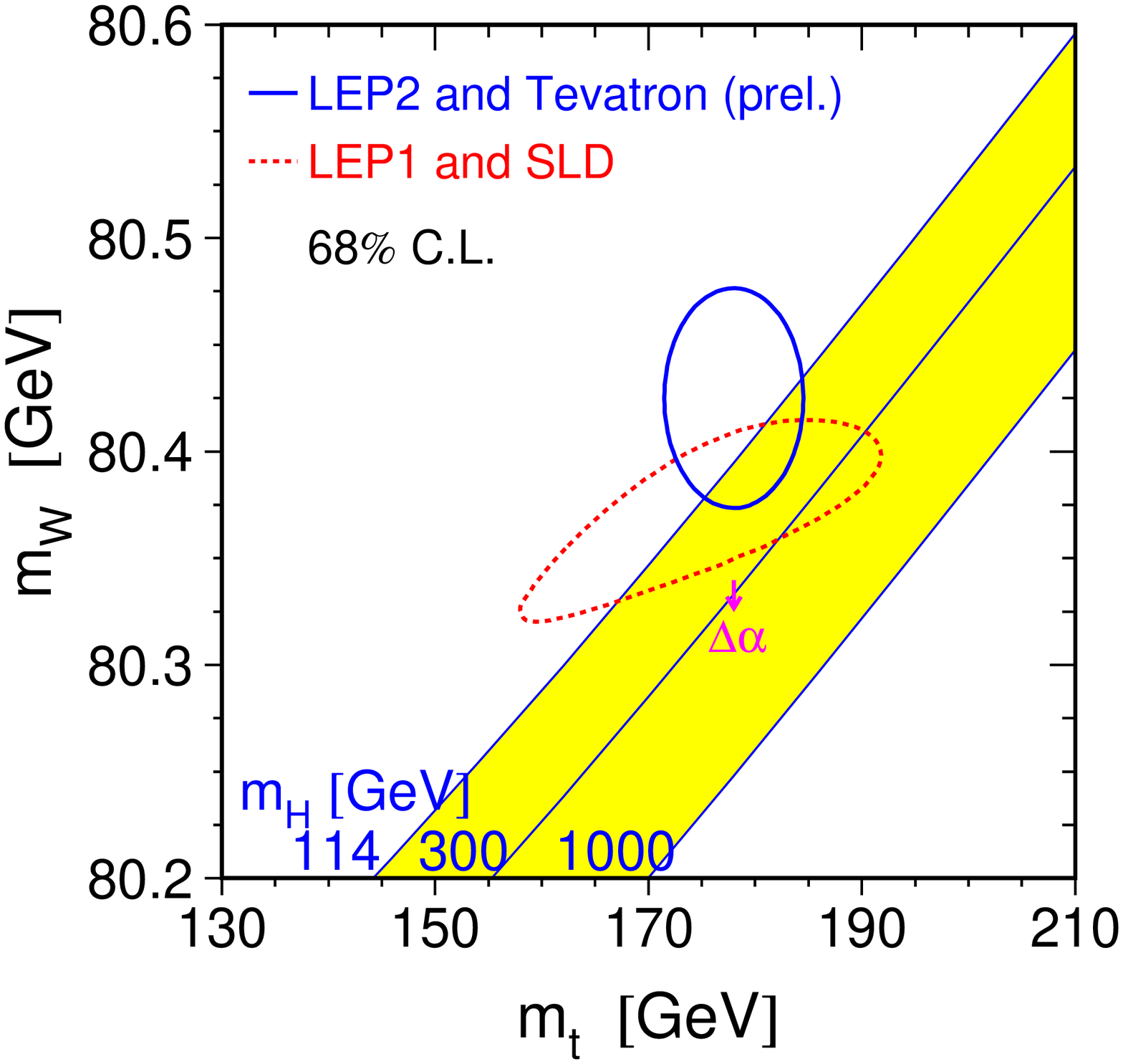}\\
  \includegraphics[width=0.495\textwidth]{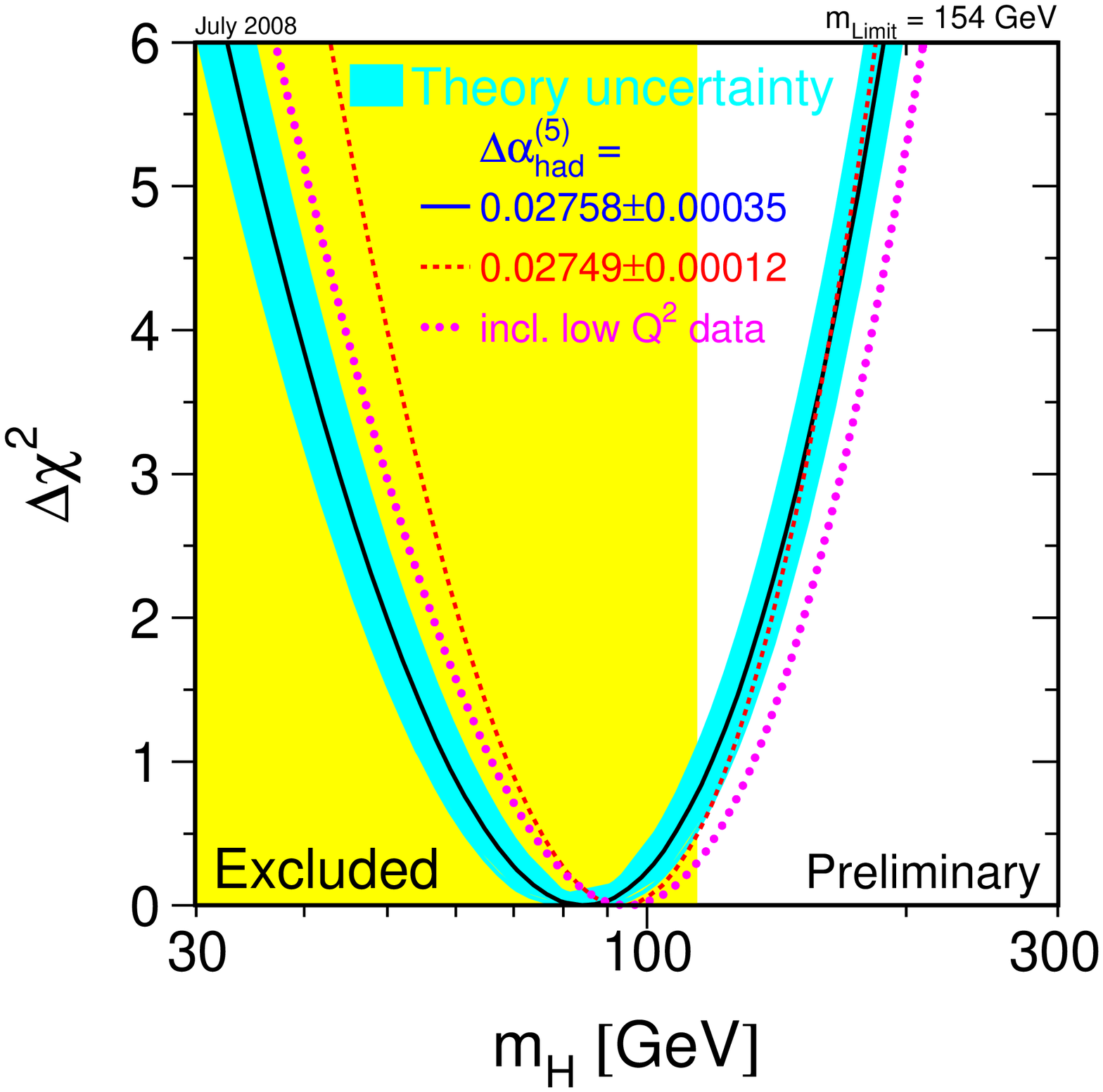}
  \includegraphics[width=0.495\textwidth]{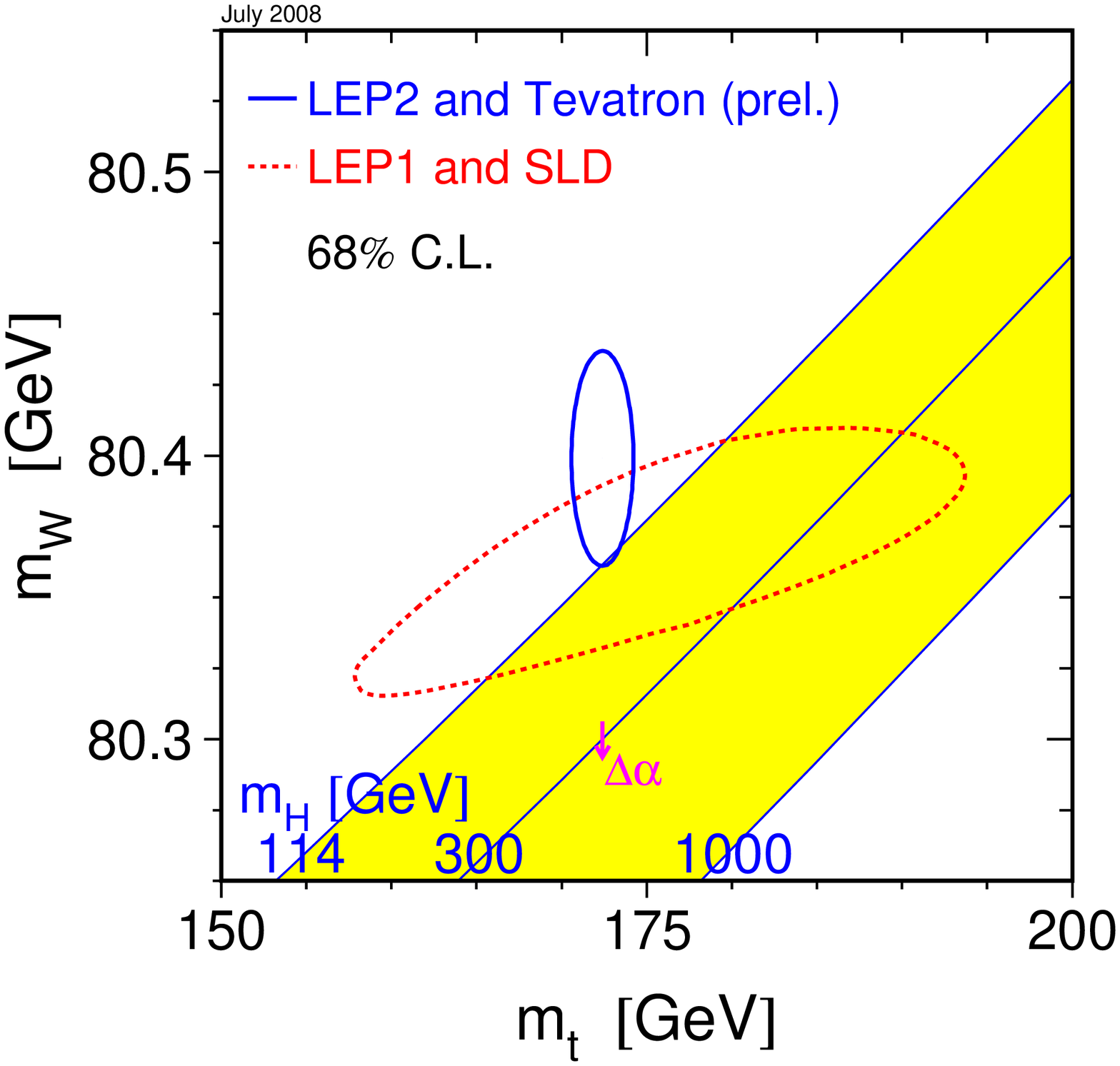}
  \caption{Mass constraints on the Higgs boson in 2004 (top, \cite{EWWG:2004}),
    using the Tevatron Run~I mass combination of $m_t = 178.0 \pm
    4.3$~GeV/c$^2$~\cite{Azzi:2004rc}, and now (bottom,
    \cite{EWWG:2008}), using the current preliminary world-averaged
    result $m_t = $\mtopwa.}
  \label{fig:mtvsmh}
\end{figure}

The current world-averaged $m_{t}$ of \mtopwa\ is also in good
agreement with its predicted value of 179$^{+12}_{-9}$ GeV/c$^2$~\cite{EWWG:2008} in the framework of the standard
model, based on precision electroweak data, as discussed in
Section~\ref{sec:topmassewprecision}. The top quark mass
measurement together with that of the $W$ boson mass
($80.398 \pm 0.025$~GeV/c$^2$ \cite{PDG2008}) can be used to obtain limits on the
Higgs boson mass via the radiative corrections on the $W$ boson mass
in a global electroweak fit. This yields $m_H = 84^{+34}_{-26} $
GeV/c$^2$~\cite{EWWG:2008}, as illustrated in Fig.\ \ref{fig:mtvsmh},
where the uncertainties are only from experiment. To demonstrate the
impact of the improvements in precision for the measurements of
both $m_{t}$ and $m_{W}$ since the beginning of Run~II,
the corresponding fit results in spring 2004~\cite{EWWG:2004} are also shown in Fig.\ \ref{fig:mtvsmh}.
The current resulting 95\% C.L.\ upper limit on
the Higgs boson mass is 154~GeV/c$^2$, which includes both experimental and
theoretical uncertainties.

The direct searches for the standard model Higgs boson
at LEP provide a 95\% C.L.\ lower bound of
114.4~GeV/c$^2$~\cite{Barate:2003sz}, as also illustrated in
Fig.\ \ref{fig:mtvsmh}. CDF and D0 have recently excluded a
SM Higgs boson of 170~GeV/c$^2$ mass at 95\% C.L., and a
mass range of about 165 to 175~GeV/c$^2$ at 90\% C.L.~\cite{D05754}
based on 3~fb$^{-1}$ of data. This is not reflected in
Fig.\ \ref{fig:mtvsmh}.

\section{Summary}
More than thirteen years after its discovery, the properties of the
top quark are being studied at the Tevatron with unprecedented
precision by the CDF and D0 collaborations. The Tevatron is operating
very smoothly, and has already delivered more than 4~fb$^{-1}$ of
integrated luminosity to each experiment. Before the end of Run~II,
it is expected to deliver 2~fb$^{-1}$ per year. CDF
and D0 have exploited these increasing luminosities together with novel
advanced analysis techniques to improve upon previous
measurements, but also to explore top quark properties that were not
accessible before such as its electroweak production, which provides first
direct measurements of the $|V_{tb}|$ CKM matrix element. Thus far all
results are consistent with standard model expectations and
between the experiments, which constrains specific extensions
of the standard model impacting the properties under consideration.

Top quark pair production is well established in the lepton + jets,
dilepton and all-hadronic final states, and these channels are also
used to study further properties of the top quark, such as its mass. The
top quark signal is also being established in final states involving
hadronically decaying $\tau$ leptons.
The observed production rates in all
final states are consistent with each other and with the standard
model expectation. No novel contributions to \ttbar production and
signal samples have yet been observed, and corresponding constraints are derived
on \ttbar production via resonances or massive gluons, and possible
contributions of a fourth fermion generation or scalar top quarks to
the selected signal samples. The kinematics of the observed events
also agree with the SM prediction. Since
contributions from physics beyond the standard model could affect the
observed \ttbar final states differently also via the top quark decay, for
example, as a result of top quark decays into charged Higgs bosons,
corresponding limits are derived as well.

The \ttbar production cross section has been measured to be $\approx$7.3~pb
for $m_{t} = 175$~GeV/c$^{2}$ with a precision of 10\%,
matching the uncertainties of the theoretical predictions. This provides
stringent tests of the corresponding perturbative QCD calculations.
Ultimately, the precision of the cross section at the
Tevatron might reach the 6\% level, dominated by the uncertainty on
the integrated luminosity. First measurements have been performed to
determine contributions of $q\bar{q}$ annihilation and gluon-gluon
fusion to \ttbar production, and are found to be consistent with
QCD predictions. Also, higher order effects such as the top quark charge
asymmetry, measured for the first time, agree with
SM expectation within statistical precision.

First evidence for electroweak single top quark production has been
found by both experiments, and observation at the five standard
deviation level appears imminent. The observed production rates are
consistent with the standard model expectation of $\approx$3~pb and provide
first direct measurements of the CKM matrix element $|V_{tb}|$. The
most stringent 95\% C.L.\ lower limit is found to be
\mbox{$|V_{tb}|>0.71$}. Searches for contributions to single top production via
mechanisms beyond the SM, e.g., mediated by
$W'$ or charged Higgs bosons, or flavor changing neutral interactions
with gluons, are also performed. The lack of any significant deviations
from the standard model has been used to set corresponding stringent limits
on such processes. Sensitivity of single top production to the
$Wtb$ vertex structure has provided constraints on more general $Wtb$
interactions, including first limits on left-
and right-handed tensor couplings.

The decay properties of top quarks have been studied using \ttbar samples
providing both sufficient statistics and sample purity. The $W$ boson
helicity in \ttbar decays can now be measured in a model independent
way by extracting the fractions of left-handed and longitudinally
polarized $W$ bosons simultaneously. This provides additional
information about the $Wtb$ vertex structure that can be used together
with constraints from single top quark production to fully specify
$Wtb$ couplings. The $W$ boson helicity measurements are found to be consistent
with SM expectation, and further studies in the lepton
+ jets and dilepton final states will scrutinize the observed $>2$~sd
discrepancies between the two channels.
The ratio of decays of $t$ to $b$ quarks versus any quarks, i.e., $R =
{\cal B}(t \rightarrow Wb) / {\cal B}(t \rightarrow Wq) = 0.97^{+0.09}_{-0.08}$, has
reached a precision of 9\%,
confirming the expectation of dominant $t \rightarrow Wb$ decay
assumed in most analyses. Top quark decays beyond the
standard model, mediated through neutral currents, invisible
decay modes, or decays to charged Higgs bosons have been sought,
but have yielded only upper limits on such processes.

Measurements of fundamental top quark properties such as its charge, lifetime and
mass based on \ttbar final states
thus far confirm the standard model nature of the top quark. First
measurements of charge and lifetime are consistent with
expectation, and help to constrain ideas beyond the SM. The top quark mass
has been measured in lepton + jets, dilepton and
all-hadronic final states, yielding consistent results among the
channels and between the CDF and D0 experiments. Combining
the results yields $m_{t} = 172.4 \pm 1.2$~GeV/c$^{2}$, which marks the
most precise measurement of the mass of any quark with a precision of 0.7\%. By
the end of Run~II, a measurement with an absolute precision of
$\lesssim 1$~GeV/c$^2$ should be achievable. This result
will provide an important calibration 
at the LHC, until there is sufficient
luminosity for further refined measurements.

While these mass measurements are usually interpreted as
representing the pole mass of the top quark, it should be recognized that their calibration through current
Monte Carlo simulations raises certain ambiguities of interpretation.
Nevertheless, indirect mass
measurements utilizing the mass dependence of \ttbar production
based on the pole mass are consistent with the direct measurements.

Utilizing radiative corrections to the $W$ boson mass in a global
electroweak fit to data that includes the world-averaged top quark and $W$
boson masses as inputs, the mass of the yet to be observed standard model
Higgs boson can be constrained. This provides a 95\% C.L.\ upper
limit on the Higgs boson mass of 154~GeV/c$^2$.

The large mass of the top quark does not only render it an ideal
window to new physics, but this most massive known fundamental object
also has a lifetime that is short compared to hadronization
times. Consequently, observables sensitive to the top quark spin
can be accessed undisturbed by hadronization processes. Spin-related
measurements have not as yet been performed in Run~II, but
it was shown in Run~I that a measurement of \ttbar spin
correlations is feasible, and this measurement will greatly benefit
from the increased statistics.
The observation of single top quark production in addition should 
enable first studies of the polarization of top quarks when produced
via the electroweak interaction.

For precision measurements such as those of top quark mass and pair
production cross section, the study of systematic uncertainties
(consistently across experiments) and evaluation of any possible new or smaller
contributions not considered in the past become a high priority. Other
measurements, particularly involving single top quark production,
will remain statistically limited throughout Run~II. The LHC will
be a ``top factory'', producing millions of top quarks per year thanks
to the two orders of magnitude increased production cross sections 
and enhanced luminosity relative to the Tevatron. A broad top quark
physics program is in preparation at the LHC \cite{Beneke:2000hk,
Bernreuther:2008ju} that will complement and further expand that of
the Tevatron.

A great many analyses are being pursued at the Tevatron, characterizing
top quark data samples both as signal and as background contributions for other possible processes
of similar signature that still remain to be studied. The top quark
serves as a probe into new physics, both in production and decay,
that could appear in the form of new particles or as modified
couplings relative to the standard model. While thus far
all measurements are in agreement with the standard model, there is
still much room for new physics to be explored both at the
Tevatron and the LHC.

\section*{Acknowledgements}
The author would like to thank all his colleagues working on the CDF
and D0 experiments and at the Fermilab accelerator complex who made this
review possible through their dedicated work and excellent results.
Support from the Alexander von Humboldt Foundation,
the University of Rochester and the University of Bonn is also gratefully
acknowledged.

The author is indebted to Tom Ferbel and Regina Demina at the
University of Rochester for introducing him to the world of top quark
physics and for their continued encouragement. Norbert Wermes and
Eckhard von T\"orne at the University of Bonn are gratefully
acknowledged for their support of the author's research and writing
process. Special thanks go to all the group members at the
Universities of Rochester and Bonn for their contribution to making
this work such a fruitful and inspiring experience for the author.

The author is grateful in particular to Florencia Canelli, Fr\'ed\'eric
D\'eliot, Tom Ferbel, Amnon Harel, Ann Heinson, Ulrich Heintz, Ulrich Husemann,
Michelangelo Mangano, Heather Pleier, Iris
Rottl\"ander, Christian Schwanenberger, Lisa Shabalina, Kirsten
Tollefson, Eckhard von T\"orne and Norbert Wermes for helpful
discussions and comments on the manuscript. {\it SDG}.
\bibliographystyle{h-physrev3.bst}
\addcontentsline{toc}{section}{References}
\bibliography{literature}

\end{document}